UNIVERSITÀ
DEGLI STUDI
FIRENZE

Scuola di Studi Umanistici e della Formazione
Corso di Laurea in Logica, Filosofia e Storia della Scienza

Tesi di Laurea

# THE FERTILE STEPPE: COMPUTABILITY LOGIC AND THE DECIDABILITY OF ONE OF ITS FRAGMENTS

Stella Spadoni

Supervisor: *Pierluigi Minari*
Co-supervisor: *Andrea Cantini*

# ACKNOWLEDGEMENTS

To whom it is about to concern.

Usually, I am not one for sappy pages or unfelt, courteous thank-you formulas[1].

Nevertheless, I believe that giving thanks, if really felt, makes any endeavour of this sort even *richer*, as much as the humus we have been planted into.

That is why it is important to contextualise, lower the altitude, give a proper time and space to what we write: not by providing numbers and coordinates, but *names* and *relationships* which have played different roles during the basting stage (one without end, really).

As a result, I feel the need to thank the people which I have brought (consciously or not) into the steppe with me.
Here is who made the special-expedition team.

First of all, I deeply thank my Supervisor, professor Pierluigi Minari. Without your continuous support and advice (anytime anywhere), this work would have never come out the way it did.

Secondly, I extend my sincere gratitude to Giorgi Japaridze. Thank you for your great availability while venturing into this ever-fertile play of threads.

As for the rest[2], I deeply thank my parents: you have always been there for me, even when it may have been more challenging.

To friends and relatives who have walked this steppe alongside the margins of these pages - you already know who you are.

And finally, to my ever-present co-pilot.

1 That is to say, if you are mentioned on this page it means I *really do* thank you!
2 As somebody would say. . . - on that note, you made the team too.

CONTENTS

# INTRODUCTION

*Arsen'ev*: We're exploring this region,
the big hills, the valleys, the rivers.
Do you want to be our guide?
*Dersu Uzala*: Me, sleep on it.
*Arsen'ev*: All right, you think it over.
It's time that we turned in too.

The next day, without saying a word, Dersu took his place at the head of the group.

*from* Dersu Uzala (1975),
*directed by Akira Kurosawa*

Finding out about Computability Logic, "CoL" for short, has been an unexpected *return to the land*, the ground, the soil whence life stems green and brown, the tangible experience which many inquiries have ever set foot from.

Indeed, after a long time of wrapping my head around some apparent abstractness, I got to touch grass once again with the advent of Giorgi Japaridze's relatively new research project. Georgian logician, Japaridze has produced an impressive amount of new and pulsating work around the system he has developed over the last twenty years.

However, what struck me as odd was that it had yet to receive the recognition it deserves, at least for what concerns the Italian logic literature. After some more digging, I found it even more thought-provoking to see how little it is known on a more global scale.

Therefore, ever since having first crossed paths with it, my goal has been to give CoL the appropriate academic presentation it calls for - and, eventually, contributing to the long list of open problems by providing a possible solution to one of them.

The core idea behind CoL is that:

> [. . . ] the ultimate purpose of logic is providing an intellectual tool for navigating real life.[3]

On many occasions, indeed, contemporary logic seems to be a bit far away from our day-to-day experience. Upon a closer look, however, one can easily assess that classical logic, for example, started out as real-life driven, with truth values assigned to formulas in line with the "real"[4] state of things. Indeed, logic is (or, at least, should be) far closer to us than we (or, at least, I) tend to remember.

What I mean by *return to the land*, then, is a reminder that when we are dealing with any relevant logical development of our times, despite the apparently distant nature, we are dealing with an earliest and simplest desire to guide our action, to return to what sparked interest in logic in the first place: reality.
Not a groundbreaking thought, I am aware, nor it aims to be[5].

The point is that CoL caught me off guard by prompting me to recall something I had long forgotten: the importance of the tangible (causes and) effects of logic.[6]

As a result, I felt like being taken down to the firm ground which I once so carelessly left. From this unexpected yet simple realisation, stems the work you are reading right now; indeed, for the last several months, I have been exploring this whole new eastern land, wrapped up in its hills, valleys and rivers in the attempt to learn the most out of it.

---

3 From [53], page 566.

4 Let us set aside, at least for the time being, the metaphysical question surrounding reality. As Kit Fine ([19], 2001) once put it, we should appeal to the primitive concept of reality, meaning the one which cannot be understood in fundamentally different terms. Henceforth, we shall take for granted an elementary and intuitive notion of what reality actually seems to be.

5 I know that things are not as simple as how I described them to be. However, I feel that it is important to give the reader a brief account of how and why I became so invested in Computability Logic.

6 As this may seem a bit off topic, this strikingly reminded me of the way the people of the Old Testament tend to regularly forget about God, which is something peculiar to the very least. Typically, the process begins in a careless manner, usually getting distracted by the things of the world; then it slowly (and deliberately) turns into a full neglect of the Lord: "But my people have forgotten me; they make offerings to false gods" (Jer. 18, 15a). Every time humanity enters this state of forgetfulness, which is not caused by unbelief nor unfaithfulness (curious), something big happens and shakes people up. Likewise, I guess, CoL woke me up from a deep and naive sleep of which I was not aware.

What is now needed is a sort of map of this ever-expanding steppe which unfolds from its Georgian spring. I will try to guide you through these fruitful surroundings starting from *das fruchtbare Bathos der Erfahrung*, just as Kant did in his *Prolegomena*[7].

As a sort of tiny (relocated) Dersu Uzala, hopefully with a much more extended vocabulary, my goal will be to provide the reader with a safe and clear journey through this new fertile environment, exposing the core concepts and moving from an informal to a structured and well-defined description of it.

Furthermore, following a thorough analysis of the whole system, we will have a look at a possible proof for the decidability of the CL15 fragment of CoL, an open problem which I have recently tried to tackle by means of an effective procedure.

We will indeed start with a more philosophical approach from to the fertile bathos of experience upwards, prodding first-hand the main ideas behind the rise of Computability Logic (and how the classical role of truth is here replaced with the less ambiguous - and perhaps more pragmatic - notion of computability).

Following this crescent movement, in the first chapter, we will travel the semantics-syntax route outlined by Japaridze, first by weighing up the germinal concepts and then by crystallising them in more formal, logical fashion.

After learning the official language of this new land, we will proceed with the second chapter, which introduces the reader to Cirquent Calculus: a

7 In [75], 1783, Kant used the term *bathos* as in *lowliness*, a word I really enjoy - some would say it has a *woody* ring to it. Indeed, he wanted to distance himself from one of his critics who accused him of speaking of higher idealism, thus he wrote: "My place is the fertile bathos of experience, and the word transcendental [...] does not mean something that goes beyond all experience, but rather something that precedes it (a priori), but is nevertheless intended for nothing more than simply making empirical knowledge possible". We will be doing the same, proceeding from the lowliness of the ground and making our way through this empirically-rooted system.

new, resource-conscious proof-theoretic approach sewn on CoL itself to axiomatise and, thus, obtain various fragments.
Upon acquiring these different pieces, we ought to review each fragment and, in full mosaic fashion, inlay them into a great birds-eye overview of CoL.

In the third chapter we will fully zoom in on fragment CL15, in order to better frame the question of its decidability and examine the solution articulated in this work.

Finally, before any conclusion is drawn, we will have a brief look into all the possible applications of Computability Logic, ranging from Artificial Intelligence and computer science to mathematics and, closing up full-circling back to, philosophy.

Before embarking upon such topographic-surveying expedition[8], a quick methodological reminder.

Throughout this entire inquiry, many aspects, reflections and considerations of various nature will come into play at different times. The land we need to peruse is, indeed, vast, and it may be disorienting every once in a while. This is why many key concepts will be repeated throughout this work, facilitating the reader's comprehension of what is being introduced and where to locate it in the grand scheme of things - our ever-growing map of these new grounds.

Furthermore, we will proceed through what will be referred to as a *wok-pan seasoning* method[9]: indeed, we will often briefly mention some new notions to acquaint us with, without giving exact definitions yet, in order to better understand the concept for which they were introduced in the first place and gain a fuller grasp of the whole system.

As a matter of fact, in this new and remote land, just as in nature, everything is connected to everything else; as such, there will be times in which mentioning something (that has yet to be fully introduced) for the purpose of understanding something else will be crucial to one's

8 Which, as such, will start with the aerial assessment trip, as explained in 1.1.1, in order to gain an overview of this fertile Georgian steppe.
9 There is no need to explain such name, at least for now, since it will become clear once the reader has gone through the first few sections.

comprehension.

Thus, the reader is advised (and will be helped) to keep these many different bits and pieces all together in a persisting state: they are never to be forgotten or left behind. We need to have everything laid out in front of our eyes, kept *afloat*, fresh and present: we certainly do not wish to overlook any detail while mapping this new, uncharted grassland.

Now that we are ready and rightly equipped for the journey ahead, we may as well depart from the very puerile question of *why*.

# 1

# COMPUTABILITY LOGIC: AN OVERVIEW

## 1.1 WHY AND HOW (OR RATHER, WHAT AND HOW)

CoL stands to *computability* just as classical logic stands to *truth*. Indeed, Computability Logic is an isomorphic extension of the latter, obtained by replacing the notion of truth with the less ambiguous one of computability. This means that a true formula in classical logic is an algorithmically solvable problem in CoL.

Before expanding on such basic informal concepts, though, we ought to ask why Japaridze focused on computability in the first place.

Historically, ever since the 1930s, logicians have been relentlessly working on the notion of computability. With many provably equivalent definitions provided, such as the ones of Turing, Church, Gödel and Kleene, to name a few, it has rapidly become one of the most fundamental concepts in mathematics and computer science.

Indeed, this interest quickly sparked the revolutionising development of computers, which were virtually conceived as theoretical devices known as Turing Machines[1].
As a matter of fact, much of what we are able to do today, both in the good and in the bad, we owe to the study of the notion of computability.

That is why I agree with Japaridze in considering it "one of the most interesting and fundamental concepts"[2] - I would also add "most useful" in terms of today's demands, resonating with the whole pragmatic take of the research project.

Consequently, just a few lines down, "[...] it is only natural to ask what logic it [the notion of computability] induces": CoL is the direct consequence of such question which, eventually, has led Japaridze to

1 As we will see, CoL refers to and expands the definition of such machines to more suitable versions.
2 [67], page 1.

expand the classical definition of computability to a more real-life useful account.

The purpose of Japaridze's new research project, then, lies on pragmatic and constructive grounds.
Indeed, as already mentioned, the core idea is that logic should be considered a real-life navigational tool[3], a powerful aid for our day-to-day activities (which, usually, entail some degree of interaction with something or someone else - more on that later). From the most banal, as asking for the time, to the more important ones, such as working out the best investment strategy, logic must be key to a sound and fruitful outcome of any daily task.

This means that logic must be actually useful; and on what more useful notion should we base our logic than computability?

In order to develop a real problem-solving tool that can help us in the most practical sense, a logic of computability should be able to provide two different answers.
When we find ourselves face to face with a certain problem and we seek help, we need to be told what the solution to our issue is and how we can implement such solution. This means that just knowing the answer to *what* is not going to cut it[4]: we also need to know the *how*. This is the only way to be actually helpful and useful in times of need.

In such constructivist fashion, in line with the intuitionistic attire, the purpose of CoL is to "provide a systematic, universal-utility tool for telling what can be computed and how"[5]; something Japaridze seems to have already touched upon back in 1997, while attempting to define a new game semantical approach for linear logic.

Indeed in [37], page 88, the notion of *effective truth* seems to already lay the groundwork for the firm bedrock of CoL:

---

3 As explained in [67], page 3: "Logic is meant to be the most basic, general-purpose formal tool potentially usable by intelligent agents in successfully navigating the real life".

4 Indeed, most of the times we already know what the solution to our problem is; however, we are at a loss as to how to act it out.

5 From [60], page 174.

> The gist of the semantics of effective truth is that sentences are considered as certain tasks, problems which are to be solved by a machine, that is, by an agent who has an effective strategy for doing this; **effective truth means existence of such a strategy.**[6]

As in real life, when trying to help somebody tackle a certain issue, a correct assessment of the problematic circumstances is key for giving great advice and finding a solution.

Thus, intuitively, in order to refine such adequate grasp of the situation at hand, our useful logic has to be resource-conscious (borrowing from Girard one of the most ground-breaking aspects of contemporary logic - more on this later in the chapter).

In order to redevelop logic away from the old vestiges of truth and into the newly-extended notion of computability, as declared in [43], we should start from an intuitive semantical approach and then formally make our way into the syntax of our system.

This is the route Japaridze believes every logic should take when being developed from scratch. Classical logic has, indeed, grown so, as opposed to intuitionistic and linear ones. Let us see why this path is the best one and what are the consequences of not following it.

### 1.1.1 *Logic's FDP*

Just as airplanes must follow the airport's security guidelines for take off and landing, known as ATP[7], Japaridze has historically outlined a path for us to safely tread when laying new logical foundations.

However, since our aim is to do an aerial recognition of this whole new and uncharted land, we need more than just an ATP in order to fly safely. Just as propositional logic is generalised to predicate logic, we need to similarly extend our flight directory from the ground level to higher types

6 Bold effect added here to highlight the concept.

7 Airfield Traffic Pattern, it is the path aircrafts need to follow in order to safely take off and land in either military or small airports - which is our case. Usually, they are indicated through arrows painted on the ground pointing to the direction the aircraft has to follow.

of routes - we certainly do not want to glide in the wrong direction[8].

Therefore, we will generalise the notion of ATP to a new one which includes FPs, meaning Flight Plans[9], which we need while air cruising. Consequently, we will extend Japaridze's directions from the wheeling ground to the higher airspace, in order to safely fly over these new hills and map a first impression of what is waiting down for us.
Thus, we will take these precious, extended instructions and name them our FDP, Flight Directive Plan.

As indicated by Japaridze, classical logic has been developed through the semantics/syntax pattern we have just mentioned beforehand, culminating with Gödel's completeness theorem. According to such guidelines, then, there should not be any syntax before semantics.

Indeed, when developing a new logic, we ought to start from the latter and end with the former[10]. As explained in [67], syntax owes to semantics in that it establishes the ultimate real-life meaning of logic. When a logical system lacks soundness, the whole syntax blows up while the semantics remains unscathed, indicating a somewhat prior and quasi-transcendental state.

Consequently, when adopting this approach, there is no fixed syntactic construction to semantically serve: indeed, we are free to develop our syntax right from the formalised, once-raw semantical intuitions that have laid the fertile soil for our grass to grow.

Hence, the safety of this suggested path lies in the fact that it prevents difficult situations and contradictions from rising while travelling. We

8 Not only because we do not want to get lost, but mostly because the way we approach something for the first time, also in geometrical and geographical terms, has the potential to completely change our perception of such object. Indeed, we want to perceive this piece of land for the first time just as Japaridze suggests us to - by following his footsteps and travelling the secure path he has already traced for us.

9 Documents filed by a pilot which include basic information such as departure ad arrival points, estimated time *en route*, alternative airports in case of bad weather and so on.

10 Syntax as in "the study of axiomatisations or other string-manipulation systems", from [67] page 3. One could also argue that this a-priori semantics/syntax distinction we usually take for granted has no solid theoretical ground and we could, indeed, forbear from using it. However, let us keep going through this approach, which seems to be quite effective in presenting CoL in a clear and polished manner.

will later have a taste of such risks while analysing the wrong direction intuitionistic and linear logics have taken.

Visually, this is the FDP we ought to follow from our small airport, with a crescent arc inbetween the first two steps (take off and highest cruising point) and a gibbous arc right afterwards, which ends with the third step (landing):

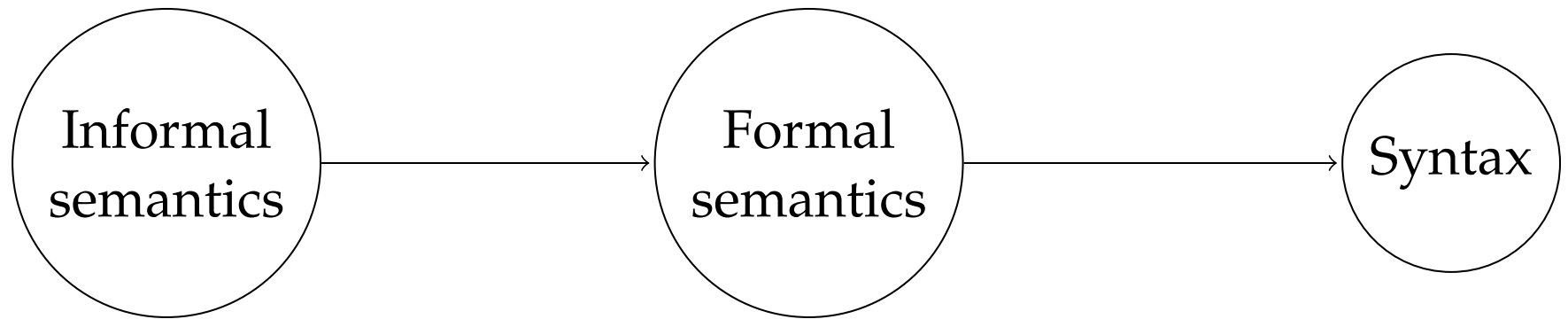

Indeed, classical logic, as well as Hintikka's IF (independence-friendly) logic, according to Japaridze [69], have followed this line of development. CoL has joined the team, which is why its syntax, being at the last step, is still flourishing and developing to this day.

On the other hand, this is the direction the other two above-mentioned logics have wrongly taken:

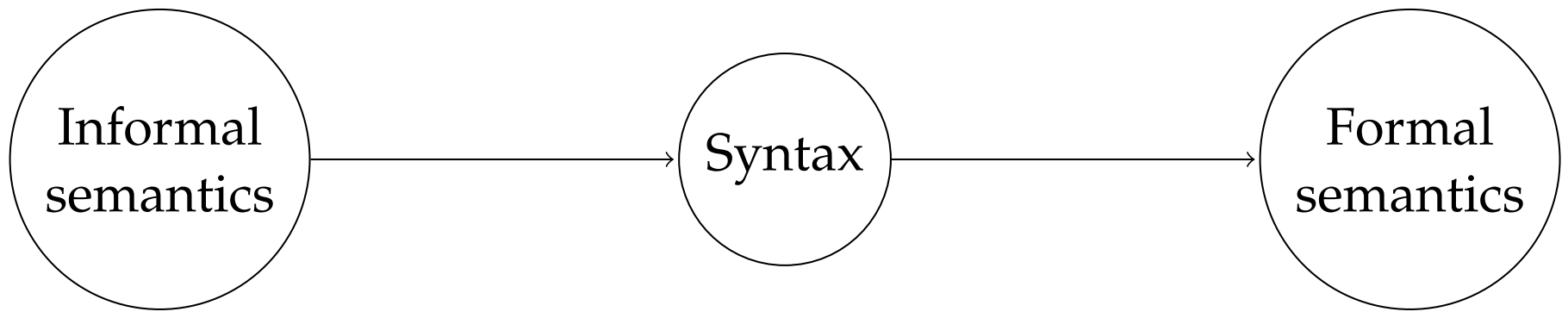

As you can see, intuitionistic and linear logics have "skipped" the middle formal semantics stage, leaving it as a problem to later solve according to a specific syntax that has already been determined - not an easy task for sure!

Historically, as briefly recalled by Japaridze in [69], Heyting's intuitionistic logic started out by examining proof systems for classical logic and removing postulates that did not fit the informal intuitionistic point of view (e.g. the principles of excluded middle and double negation[11]).

On the other hand, Girard's linear logic was obtained by Gentzen's sequent calculus for classical logic by deleting certain rules incompatible with the resource philosophy linear logic went for (i.e. contraction and weakening[12]).

Clearly, there is a great downfall in cruising the syntax-semantics route: we can never be sure of having an adequate proof system in line with our informal semantics, meaning our starting philosophical intuitions.

Indeed, we cannot even properly mention adequacy here, since the comparison would be between a formal object (syntax) and an informal one (semantics) - quite an unlogical move, as Japaridze points out. For this reason, formal semantics is key to prevent inadequate syntaxes (or sterile semantics) from developing.

Since this comparison cannot be soundly made, intuitionistic and linear logics have been facing a common problem: finding an adequate, natural and intuitive formal semantics to the already fixed syntax.

Many have tried to provide some solutions, for instance Blass ([6], 1991), Abramsky and Jagadeesan ([1], 1994) for linear logic and Lorenzen ([83], 1959), Kleene ([76], 1957) and Gödel ([26], 1933) for intuitionistic logic. However, none of these seemed to be a good match for the two logics, which remain incomplete[13].

---

11 In intuitionistic logic reigns a constructive and proof-centric point of view, according to which for something to be classically true means for it to have a logical proof. This means that the formulas $A \wedge \neg A$ and $A \longleftrightarrow \neg\neg A$ are not valid in Heyting's logic.
The principle of excluded middle, dubbed by Brouwer as the *principle of omniscience*, constructively means that, either way, we hold an actual proof for $A$ or for not $A$ (with negation interpreted as reduction to absurd): however, this rarely happens, as, indeed, we are not omniscient - clearly, intuitionistic logic is an epistemic logic.
As for the principle of double negation, only the weaker direction is logically valid, meaning $A \rightarrow \neg\neg A$, and not the other way round. Indeed, pardon this brutal simplification for space reasons, $\neg\neg A$ can be interpreted as '$A$ is not wrong', while $A$ as '$A$ is true'. Hence, stating that $A$ is not wrong if $A$ is true is a valid sentence; still, $A$ is true if $A$ is not wrong, is not - again, we are dealing with an epistemic logic!

12 As explained in [37], page 91, contraction tells us that we can always double our resources, which collides with linear logic's approach of one single use per formula: indeed, we cannot use more than we possess. Conversely, weakening tells us that we can always reduce our resources; however, according to linear logic, we have to use all the resources we possess. The result of allowing the weakening rule in MALL (multiplicative-additive linear logic) is called *affine logic* (also known as BCK logic for the three combinators it adopts).

Summarising much of what has been said in this section, I would like to directly quote Japaridze from [51], page 253, who brilliantly encapsulated the undoubtedly superiority of the semantics-syntax route in just a few lines:

> The reason for the failure of $P \sqcup \neg P$ [principle of excluded middle] in CoL is not that this principle... is not included in its axioms. Rather, the failure of this principle is exactly the reason why this principle, or anything else entailing it, would not be amongst the axioms of a sound system for CoL.

Having shown our trajectory, and warmed up the engines, let us have a more in-depth look at the relationships between CoL and its neighbouring logics in order to have a clear picture of the surroundings borders before taking off.

### 1.1.2 *CoL-thy-neighbour*

In a 2002 paper, just before the official emergence of Computability Logic, Japaridze was already addressing what he called a "logic of tasks", a sort of early version of CoL. Here, in [38], this logic of tasks is defined in contrast with classical logic, which is described as a "logic of facts" (page 262).

Indeed, while classical logic focuses on facts which predetermine the truth values (true/false) that we assign to formulas, a logic of tasks focuses on imperative sentences (as "Make the bed!") and its values do not concern truth, but, rather, the accomplishment or failure of said tasks. Moreover, these values are not predetermined in any way, unlike classical logic through facts and reality.

This is the semantic difference between classical logic and CoL, both fruitful products of the same course of development. Indeed, CoL is a conservative extension of the former; as explained in [39], page 95:

13 Indeed, they are fragments of CoL and cannot prove every valid formula therein - examples provided below. However, Computability Logic seems to provide an intuitive formal semantics to both intuitionistic and linear logics.

> As it turns out, classical truth is nothing but a special case of computability - in particular, computability restricted to elementary problems (predicates), so that computability logic is a natural conservative extension, generalization and refinement of classical logic and, for this reason, it could come to replace the latter in virtually every aspect of its applications. Whatever we can say or do by the means that classical logic offers, we can automatically also say (without changing the meaning or even the notation) and do with our universal logic. At the same time, the universal logic enriches classical logic by adding very strong constructive elements to its expressive power.

We will properly address and define the aforementioned notions in the chapters to come. However, this excerpt gives us a preparatory taste in our mouths, just like a wok pan needs to be seasoned before being used to cook.

One thing worth mentioning is that in the above-quoted work, Japaridze uses *universal logic* to refer to the set of all valid formulas of CoL, meaning the ones that are always computable.
As explained in the first page: "The name 'universal' is related to the potential of this logic to integrate, on the basis of one semantics, classical, intuitionistic and linear logics, with their seemingly unrelated or even antagonistic philosophies".

This is why CoL becomes essential in setting the record straight between these three different logics. As Japaridze himself titled his 2019 paper [69], Computability Logic seems to succeed in *Giving Caesar what belongs to Caesar* and solving the apparent logical contradictions rising between these fragments. A sort of peace-maker of the entire neighbourhood.

Indeed, even though linear and intuitionistic logics have taken the wrong path, they are still fragments of CoL, just as classical logic is. Visually, as shown in the first Japaridze's university lectures[14]:

14 The different colours stand for the different developmental approaches: CL and CoL from semantics to syntax, IL and LL from syntax to semantics.

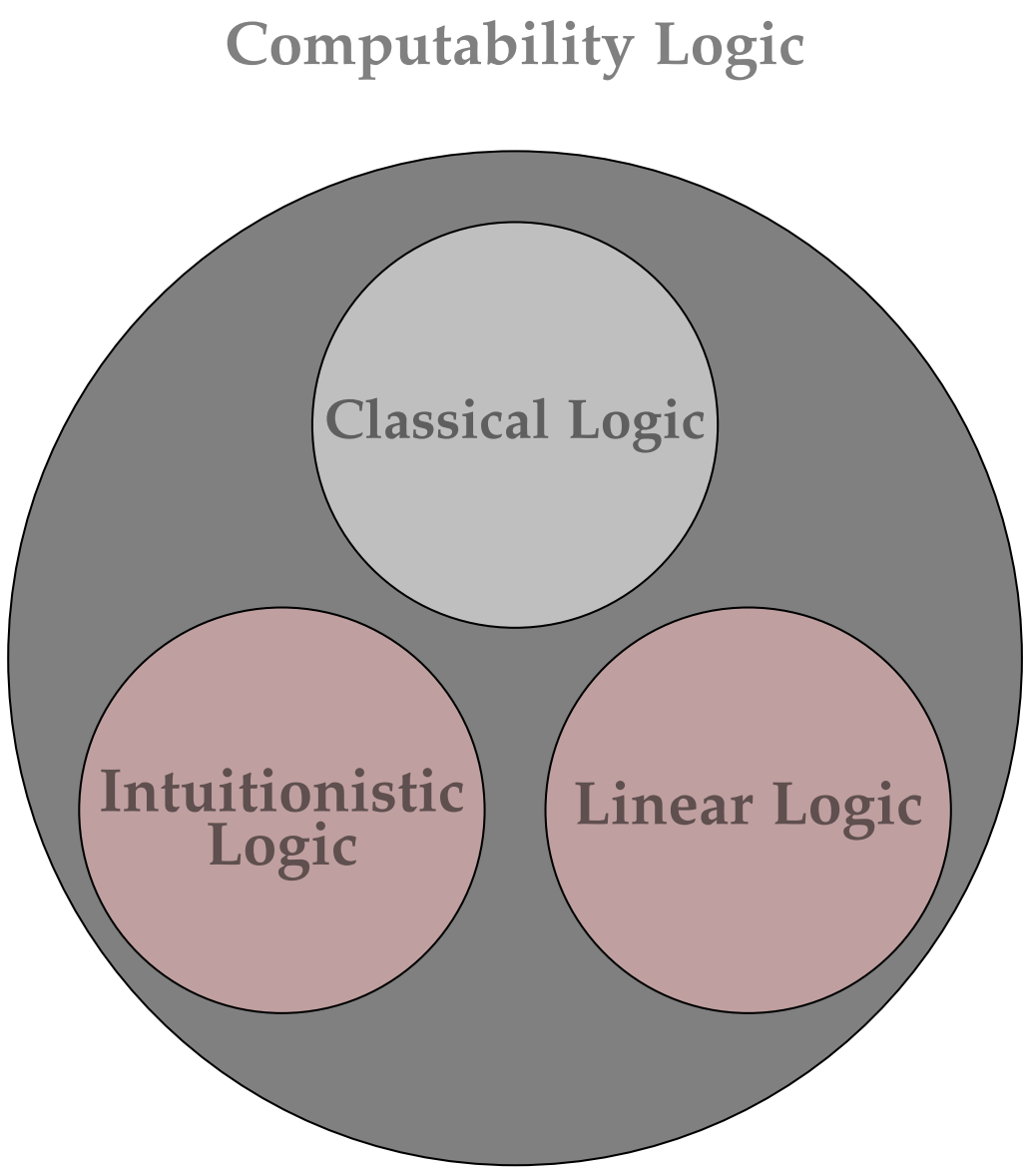

Before venturing ourselves into the partial reCoLciliation of these three logics, let us first briefly touch upon the relationships between the two reddish fragments and Computability Logic. Indeed, as Japaridze has illustrated in many of his works, CoL seems to provide a suitable formal semantics for both logics.

Concerning the intuitionistic fragment, a constructive notion of truth can be derived from CoL's semantics, which, in turn, offers a formal solution to Kolmogorov's thesis in [77] - according to which Heyting's intuitionistic calculus, INT, is (or should be) a logic of problems.

In [46], Japaridze proves the soundness of INT rephrased using a signature of CoL's language; thus, it becomes a problem-solving tool, meaning that a solution to a given problem can be found if and only if the corresponding proof can be found in the theory.

As shown in [46], this is possible because the intuitionistic implication, meaning the one that allows the agent to reuse the resource (antecedent) any number of times, is entirely captured by $\circ\!\!-$ (called "brimplication", from *branching recurrence implication*), an interactive algorithm reduction operator which acts as a generalisation of the Turing-reducibility from the traditional input/output problems to computational interactive tasks[15].

This operator, which we will later introduce and observe in full motion, is key to the translation of the fragment INT into CoL's syntax and its soundness proof. Thus, intuitionistic logic becomes a sound, but still incomplete, fragment of Computability Logic[16].

Regarding linear logic, on the other hand, it shares a resource-conscious approach with CoL, even though intuitive and naive in the former and structured and formal in the latter. Indeed, linear logic's syntax is much simpler than that of CoL, but it lacks a formal attire for the semantical take on formulas meant as resources.

When no exponentials[17] are involved, the two logics agree on many simple formulas[18], mainly thanks to $\sqcap$ ("chand") and $\sqcup$ ("chor") operators for additive constructive conjunction and disjunction; but still disagree on many more evolved ones, such as Blass' Principle[19], as Japaridze named it.

Furthermore, CoL has been equipped with a new, axiomatising calculus which is even more resource conscious than that of linear logic and from which, according to Japaridze, the latter could benefit - as we will later see.

In an attempt to encompass all these logics we have touched upon, we could sketch a summarising diagram which emphasises the core and focal notions they focus on:

15 More on the role of interaction in the next subsection - a fundamental concept of the whole CoL research project.

16 Through the signature $\{ \circ\!\!-, \sqcap, \sqcup, \bigsqcap, \bigsqcup \}$, which will be much clearer to the reader once manoeuvred the syntax chapter.

17 ⫰, *brecurrence*, and ⫯, *cobrecurrence*, are respectively CoL's counterparts for linear logic's exponentials ! and ?.

18 For example, both CoL and linear logic reject $(P \to P \wedge P)$ and accept $(P \to P \sqcap P)$, with the classical-shape propositional connectives standing for the corresponding multiplicative operators of linear logic, and square-shape operators as additives ($\sqcap$ = 'with', $\sqcup$ = 'plus').

19 As explained in [69], page 103: "I call the latter Blass's principle as Andreas Blass [Blass, 1992] was the first to study it as an example of a game-semantically valid principle underivable in linear logic". Formally, with connectives understood in the multiplicative sense, it is written as such: $(A \wedge B) \vee (C \wedge D) \to (A \vee C) \wedge (B \vee D)$.

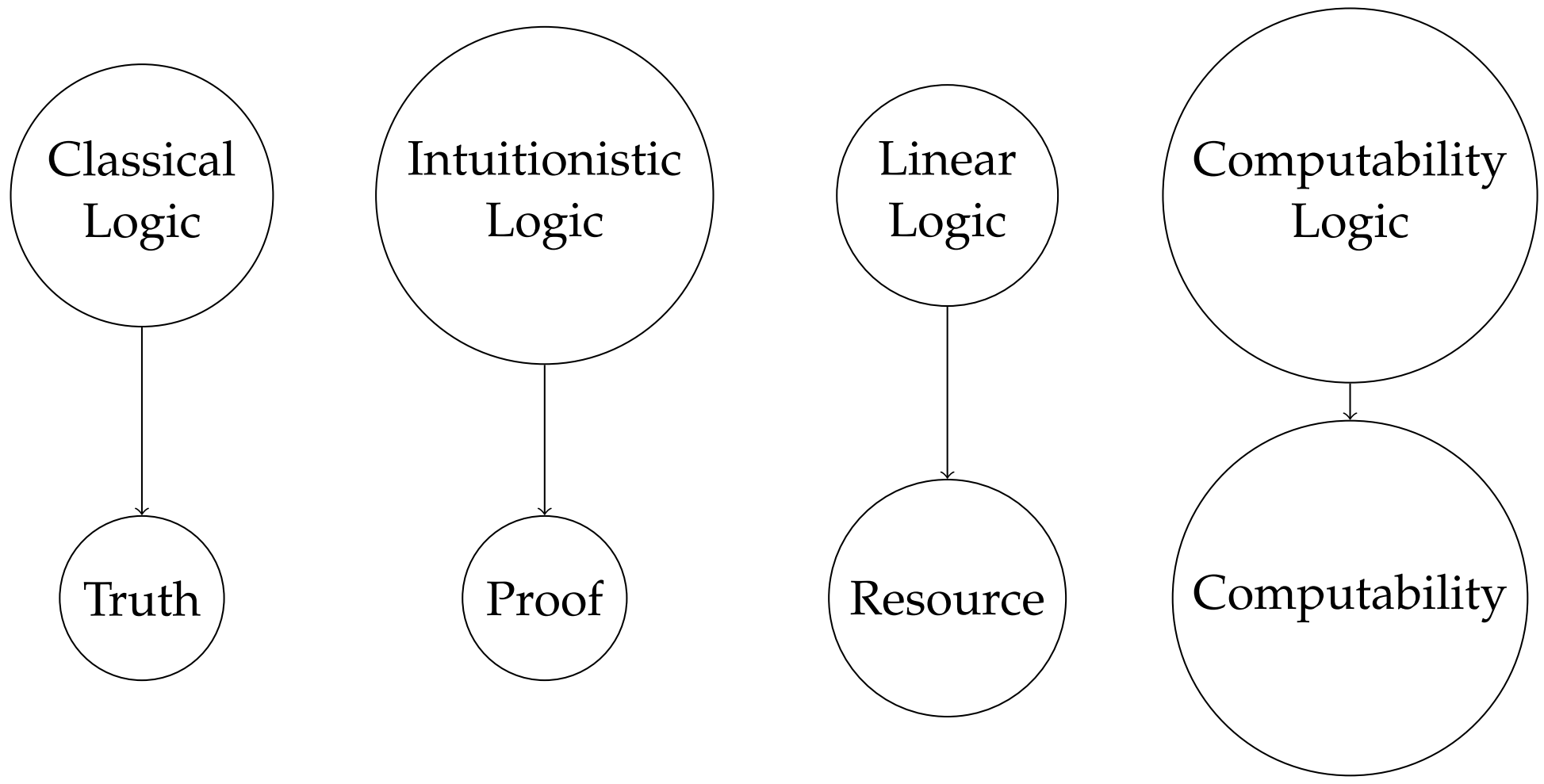

Let us witness firsthand how CoL manages to make amends between classical logic and the other two.

---

*Classical and intuitionistic logics*

Clearly, the main issue betwixt classical and intuitionistic logics regards the principle of excluded middle, as already mentioned. Indeed, classical logic affirms that $(A \vee \neg A)$ is valid, while the constructivistically-minded beg to differ for intuitive, semantical reasons (no omniscience, as already explained in the last section).

However, we soon realise that these two logics have been associating two different meanings to $\vee$ all along, all the while using the same symbol. Here enters CoL, which semantically neutralises the issue: classical logic is right in affirming that $(A \vee \neg A)$ is valid, just as intuitionistic logic is right in stating that $(A \sqcup \neg A)$ is not valid.

To each their own: there is no reason for arguing anymore. The crux of the matter, then, seems to have been a semantical misunderstanding of logical operators.

---

*Classical and linear logics*

Similarly, CoL intervenes between classical and linear logics. Indeed, while classical logic accepts the formula[20] $(\neg A \wedge \neg A) \vee A$ as a tautology, linear logic deems it invalid (with $\vee$ and $\wedge$ seen as multiplicative).

The problem, here, lies on a misunderstanding of logical variables. Indeed, the two occurrences of $\neg A$, in the eyes of classical logic, are just two copies of the same object, as in $(\neg A \wedge \neg A) = \neg A$. On the contrary, linear logic sees these two occurrences as two, quantitatively different objects, two resources that can be each used only once: so $(\neg A \wedge \neg A) \neq \neg A$.

Luckily, CoL's language is equipped with two sorts of **problem letters**[21]: elementary and general ones.
**Elementary**, *non-logical* atoms are meant to be interpreted as elementary games, meaning moveless games of 0-depth (thus, only $\top$ and $\bot$)[22]; **general**, *non-logical* atoms represent any games, including elementary ones, with possibly infinite legal runs[23]. We use the lowercase $p, q, r \ldots$ for elementary atoms and the uppercase $P, Q, R \ldots$ for general atoms.

Indeed, this distinction allows us to to account for the different meanings both quarreling neighbours assign to variables: elementary games end up being represented by $\top$ and $\bot$, for which classical logic is right in considering $(\neg p \wedge \neg p) \vee p$ valid - the letter $p$ refers to one and the same outcome, which cannot be any different from another such occurrence.

On the other hand, general games are a generalisation of elementary ones, thus linear logic fairly deems $(\neg P \wedge \neg P) \vee P$ invalid - as we will see, games are defined by the set of legal runs (the structure) and the winner function (the content).

20 Formula which, for the record, fails in CoL unless $A$ is elementary - more on this in the next lines.
21 As explained in [39], page 79, they correspond to the predicate letters of classical logic - it will be clearer in the next section.
22 In [72], page 2, we read: "This naturally makes the classical concept of truth a special case of computability - computability by doing nothing. Further, all operators of classical logic are conservatively generalized from moveless games to all games, which eventually makes classical logic a conservative fragment of the otherwise much more expressive CoL". Indeed, the game semantics of CoL is a generalising, refining and conservative extension of the semantics of classical logic.
23 Indeed, CoL's formal semantics is already peeping out the corner, but we do not want to spoil it all yet: thus, we will keep things to the bare, minimum necessary.

Indeed, different *P*s can have different outcomes, even though they refer to the very same gameframe and may also share the same winner (think of the game *Chess*, for instance: if played on two separate boards, according to the same rules, the same two players may go for different moves and end up winning in one and losing in the other).

Bottom line is: different occurrences of *P* mean totally distinct copies of *P* and not two naming forms of the same game. Yet again, Computability Logic managed to satisfy everyone and appease the neighbourhood.

Now that we have provided a rough sketch of these relevant dynamics, let us address one last, main aspect of CoL before reaching the Informal Semantics runway.

### 1.1.3 *Bursting the bubble: interaction*

Surely, we are now aware that Computability Logic aims to have some real-life relevance. Indeed, logic should fit the day-to-day physique and, for such purpose, some informal, primal theoretical adjustments have to be made beforehand.

As seen in the above section, CoL's fundamental and stemming notion is, indeed, computability. However, Japaridze, just as Turing had already observed in [94], feels that its classical notion, rising from the well-known Church-Turing thesis, is too narrow and restrictive for our purposes. This is why CoL adopts a renewed, extended understanding of it, which Japaridze calls *interactive computability*.

Before jumping to some definitions, though, let us recall why interaction is so important for this research project.

Our daily routine involves some degree of interaction, which is not restricted to humans as it is also object-oriented - just turning the stove on is considered to be an interaction of sorts here.
In view of the semantics we are due to examine, we may understand the situation as a game: by pressing and turning the corresponding knob, we ask a specific stove to turn on. If such stove turns on correctly, then it wins; otherwise, it loses - unless I have been pressing and turning another knob the entire time.

However, there are many more elaborate tasks which entail a higher degree of interaction and a larger amount of inputs and outputs: take medical diagnosing, for instance.
There could be multiple possible explanations to a given set of symptoms a patient is showing. In trying to determine the cause, the doctor has to rule the wrong diagnoses out by asking questions and running tests. Many inputs are given in this long and rich interaction; it is worth highlighting that not every input requires a correspondent output from the doctor.

Problem is that the classical notion of computability cannot account for all of these natural and organic aspects of interactive computational problems. Hence, we outline Japaridze's proposal, which aims at overcoming three different limits the Church-Turing thesis has imposed over such paramount notion.

Indeed, such well-known thesis takes computable problems as algorithmically-solvable problems, in particular through Turing Machines. The task performed by a Turing Machine, however, is nothing but receiving an input $x$ and generating the correct output $f(x)$ for some function $f$.

This approach is undeniably unfit for real-life situations: as a matter of fact, most tasks that computers and computer networks perform are indeed interactive, with strategies not always allowing themselves to be adequately understood as functions. On such note, Japaridze further explains in [67], page 4:

> To understand this, it would be sufficient to just reflect on the behaviour of one's personal computer. The job of your computer is to play one long game against you. Now, have you noticed your faithful servant getting slower every time you use it? Probably not. That is because the computer is smart enough to follow a non-functional strategy in this game. If its strategy was a function from positions (interaction histories) to moves, the response time would inevitably keep worsening due to the need to read the entire, continuously lengthening interaction history every time before responding. Defining strategies as functions of only the latest moves is also not a way out. The actions of your computer certainly depend on more than your last keystroke.

Therefore, the first obstacle we need to face is **functionality** itself. In light of CoL's "relaxed semantics, in striving to get rid of 'bureaucratic pollutants' and only deal with the remaining true essence of games" (*ibid.*, page 44), we ought to bring more generality and flexibility to a natural and organic view of computational problems, or rather, games. Hence, there should be no regulations on which player can or should move in a given situation. Furthermore, there are many problems which allow more than one correct answer (for example finding the root of a quadratic equation).

This brings us to the second restriction that needs to be tackled: the **input/output scheme**. As seen in the medical example, many natural situations have to do with some inputs which do not obligate the Machine to generate an output (e.g. inputs on which the supposed strategy-function $f$ is undefined).

The third and final restraint we need to loosen up is **depth**-related. Usually, computational problems are thought of games of depth 2: there is an input (1) to which an adequate output (2) has to be provided.
As Japaridze points out in his lecture notes, why should we restrict ourselves to very short "dialogues" (referencing Lorenzen's 1959 game semantics for intuitionistic logic, in [83]), as in games, when we can capture problems with high degrees of interactivity? This has to change.

In our game semantics, computability means **winnability**, as in the existence of a machine that wins the game against any possible (behaviour of the) environment.
As we will see, interactivity is given between two players[24], one of which is called **Machine** (also addressed as $\top$) and the other one **Environment** ($\bot$). The former has to act according to an effective (algorithmic) and fully determined behaviour, while the latter is free to act arbitrarily.

However, since we want to encompass a bigger notion of computability in terms of interaction, player $\top$ cannot be described as a typical, traditional Turing Machine (which lies on the restricted notion of computational problems we have just discussed).
This intuition is captured by the model of interactive computation where $\top$ is formalised as what we call **HPM**, *Hard Play Machine*, which we will

24 Japaridze first introduced these two ideal players in [37], back in 1997, with the names of *proponent* and *opponent*.

later address together with **EPM** (*Easy Play Machine*).

Altogether, the two main concepts of CoL's semantics are those of *static games*, which we will later define, and their *winnability*.
Correspondingly, CoL's philosophy relies on two beliefs that, put together, present what can be considered an **interactive version of the Church-Turing thesis**[25], which we will accept henceforth:

**Thesis 1.1.3.1**

1. The concept of static games is an adequate formal counterpart of our intuition of ("pure", speed-independent) interactive computational problems;

2. The concept of winnability is an adequate formal counterpart of our intuition of algorithmic solvability of such problems.

We have now reached the end of our preliminary circuit, which means we are finally ready to begin the aerial recognition-journey we have been preparing for.
The plane will fly the parabolic semantics-syntax route in order to conduct a first survey of this new land from a bird's eye view; this, in turn, will help us deal with the asperity of the terrain later on, in a thorough and physical topographic endeavour.

## 1.2 THE CRESCENT ARC: HARVESTING SEMANTICS

### 1.2.1 *The informal womb*

> In the beginning was Semantics, and Semantics was Game Semantics, and Game Semantics was Logic. Through it all concepts were conceived; for it all axioms are written, and to it all deductive systems should serve... This is not an evangelical story, but the story and philosophy of computability logic (CL), the recently introduced mini-religion within logic.[26]

25 Provided by Japaridze in [67], page 4.
26 From [51], page 250. Note that in the earlier papers, Japaridze used CL and not CoL as an abbreviation of Computability Logic. Today CoL is the official shortened name.

Taking off strong with the well-known opening of John's gospel[27], Japaridze sums up pretty effectively the philosophical basis of CoL.
Let us follow his footsteps and deliver a clear and synthetic recap of the philosophical approach at hand, in order to deepen some aspects of the informal game semantics still intentionally left a bit blurry.

Indeed, as we have tried to unfold in the last sections, Computability Logic focuses on an interactive notion of computability through the safe FDP development: firstly from a semantical stance, then, after some needed formalisation, from a syntactic point of view.

The core idea is that logic should be as natural and real-life resembling as possible: thus, the classical notion of computational problems as input/output functions must be extended to include interaction.

This way, we obtain a generalisation of the Church-Turing thesis, through which computability is defined as winnability: a problem is computable if and only if there is a machine that always wins the game (meaning, it always finds the adequate solution to the problem)[28].

Interactive problems are meant as games (or dialogues, in Lorenzen's terms) between Machine ($\top$) and Environment ($\bot$). Even though it is often a human user who acts as $\bot$, since it has no restriction on its behaviour, our sympathies are with $\top$.

27 It may be interesting to mention Martin Luther's sermon of this biblical passage, as presented in [13], in which he emphasises how God is one with the Word, but not as in *logos*. The Word, meaning circularity that ends the $\alpha$ and $\omega$, is the first ever pronounced word and it is uttered by a mute speaker: God gives himself the form of the Word, he becomes one with the Word while also being distinct from it. It is a creative form and not a semantical one: creature and creator originally coincide. God is not naked but coated with and alive in the verb: *Deus sigillatus non vagus*. Similarly, in Japaridze's version, we may view Logic as all coated in Semantics, the word it has soundlessly uttered, the form which is not yet semantical *per se* but, rather, creative. Indeed, Semantics becomes semantical (meaningful) only when used to create ("through it all concepts were conceived"), which, in our case, is in Game Semantics.

28 As a result, computational tasks in the traditional, functional sense now become special cases of games with only two moves: the first move ("input") is by $\bot$ and the second move ("output") by $\top$. Elementary games, which have zero depth, are automatically won by Machine, which is why it bears the symbol $\top$. It is interesting to note how without interaction we have immediate understanding of what is true (or false); otherwise, interaction entails complexity - not complication! - and makes understanding where the truth went even harder. I enjoy this thought.

Indeed, as Japaridze explains in [51], page 255, we do not want to win against Machine. After all, "[. . . ] the machine is a tool, and what makes it valuable as such is exactly its winning the game, i.e. its not being malfunctional (it is precisely losing by a Machine the game that it was supposed to win what in everyday language is called malfunctioning)".

How does CoL account for such interaction? As we will later formally see, the observable actions of an interaction are called **moves**, while interaction histories (meant as the list of observable actions recorded between the two players) are named **runs**: the possible ones are said **legal runs**[29]. Typically, a computational problem is comprised of the set of all possible runs and the subset of all the successful ones.

Drawing this quick summary of CoL's philosophy up to a close, we have seen how a useful logic should be resource conscious and problem-solving.

Concerning the former, CoL's language (as seen in the contrast between linear and classical logics) and its proof system (*Cirquent Calculus*, which we will introduce later on) allow us to crisply account for the actual resources an agent disposes of in a given situation - and it does so even better than linear logic, as explained in the next section[30].

Regarding the latter, Japaridze builds Computability Logic up from a constructivist basis[31]. Indeed, there is a big difference between truth and being able to actually find what is true, which is key to constructing (proving) a certain solution to a certain problem. As the title of the first section reads, we need to know the *what* and the *how* of a computational problem.

It is time, now, to introduce some new details around the informal concepts we have carefully laid altogether on the ground from which we took off.

29 A player who makes an illegal move is said to be an **offender**.

30 Mainly due to the fact that CoL provides two types of resources: elementary and general. The former ones are reusable, while the latter ones may not be reusable. Indeed, formulas of linear logic cannot account for the nature of resources, their sharing or the absence of thereof. See [55] for an example that highlights the differences between CoL and ARS - *Abstract Resource Semantics.*

31 The clear advantage of starting from semantics lies on this very stone: one can develop a new system by choosing firsthand which characteristics it must showcase.

As we have said earlier, CoL's philosophy is based upon two concepts: winnability and static games. Let us have a look at the latter, while introducing some new notions which will contribute to forge a well-defined and robust skeleton in the sections to come.

**Static games**, as appealed in [39], are *procedural-rule-free* sort of games, since they transcend the traditional, functional approach we have tried to overcome.
To this day, such restricting game approach has been exalted in many game semantics proposals, which seem to have not yet considered any other alternative approaches. We call these conventional function-strategy-type games **strict games**. The strictness derives from the fact that, at most, only one of the players is allowed to have legal moves in a given position.

However, Japaridze has extended his semantics to **free games**, which do not suffer from this restriction[32], thus being untied from rules and procedures: any player can make any move at any time. As such, free games are the "most powerful and universal [. . . ] modeling tool for various sorts of situations of interactive nature" ([39], 13).

As of now, this concept of games seems to be a complete formal counterpart to the intuitive notion of interactive problems.
However, it would be harder to argue that it is also sound. Indeed, as it may appear general enough to model any interactive computational problem we deem as such, it is a bit too general. There are some free games that may not represent any meaningful or well-defined computational problems (according to someone's, or most people's, perception).

To obviate such trouble, Japaridze considers a subset of these free games, the so-called "pure" problems (i.e. static games): these are interactive tasks that remain meaningful and well defined without any particular assumption on speed, time or any other detail - tasks where, for being successful, only matters *what* to do rather than *how* quickly it is done[33].

Indeed, recalling Thesis 1.1.1, Japaridze feels that these static games are

32 In [51], page 256, Japaridze defines positions in which both players have legal moves as **heterogeneous**.
33 Quoting from [72], page 24: "Intuitively, static games are games where speed is irrelevant because, using Blass's words, 'it never hurts a player to postpone making moves'".

an adequate formal counterpart to our intuitive "pure" computational problems.

Let us now get into the groove of formal semantics, leaving behind (*Aufheben*[34]-style) the first step of our route in order to approach the vibrant and tentacular world of universes, interpretations, valuations and formal definitions.

### 1.2.2 *Bones to this flesh!*

With the aim of providing a formal structure to the informal semantics we have introduced thus far, let us start from the very notion of games. However, in order to accomplish this, we first need to agree on some technical terms and conventions.

---

*Basic notions*

We assume that the universe of discourse[35] is always the set **NAT** = {0, 1, 2, 3, . . . } of natural numbers[36], both for classical logic and CoL. The concept of a universe in CoL is somewhat more evolved than in classical logic[37] and nothing is affected by limiting universes to the standard one; nevertheless, a lot is gained by the resulting simplicity.

---

34 In Hegel's *Wissenschaft der Logik* ([31], first published in 1812), the term refers to the process of overcoming without losing. Indeed, what is surpassed is actually kept alive in something new, which is its realisation, its completion; what is apparently left behind is actually maintained and *lifted up*, as one of the meanings listed in today's German vocabulary.

35 Briefly, a **universe of discourse** is a pair $U = (Dm, Dn)$, where the first element is the **domain** of $U$, a nonempty set, while the second is the **denotator** of $U$, a total function of the type $Constants \rightarrow Dm$. Elements of the domain are said **individuals**. As presented in [72], page 5, the intuitive meaning of $d = Dn(c)$ is that the individual $d$ is the **denotat** of the constant $c$, thus $c$ is a **name** or *code* for $d$.

36 With some abuse of terminology, as explained by Japaridze in his lecture notes, natural numbers are identified with their decimal representation. The denotator of NAT is the bijective function that sends each constant to the number it represents in standard decimal notation. Bijection is not a necessary requisite: many individuals may have many names, just think of Frege's Morning Star. The universes which possess a bijective denotator are said **ideal**.

37 Indeed, as explained in [67], page 5: "Classical logic exclusively deals with individuals of a universe without a need to also consider names for them, as it is not concerned with decidability or computability. CoL, on the other hand, with its computational semantics, inherently calls for being more careful about differentiating between individuals and

We then fix two sets of expressions:

- **Constants**, whose elements are *constants*; these are all the possible element names of the universe of discourse. We assume that Constants = $\{0, 1, 2, \ldots\}$. As a matter of fact, constants have no analogue in classical logic: indeed, a constant in CoL is not a 0-ary function letter, but a decimal string;
- **Variables**, whose elements (or strings) are *variables*. These will range over constants. We assume that Variables = $\{v_0, v_1, v_2, \ldots\}$.

The letters $x$, $y$, $z$, ... will be used as *metavariables* for variables, and $a$, $b$, $c$, ... for constants. Furthermore, the set of **Terms** is defined as the union of Variables and Constants.

**Valuations** is the set of all *valuations*, meaning functions, symbolised with the letter $e$, of the type Variables $\rightarrow$ Constants.
For the sake of clarity, let us also define CoL's intensional approach[38] of functions:

**Definition 1.2.2.1**
Let $n$ be a natural number. An $n$-ary **function** is a pair $\boldsymbol{f} = (Vr, f)$ where:

1. $\boldsymbol{Vr}$ is a set of $n$ distinct variables, meaning the variables of $f$ from which it depends from;
2. $f$ (or the *extension* of $f$) is a mapping that assigns a constant $f(e)$ to each $Vr$-valuation of $e$.

A **move**, then, meaning the interaction between Machine and Environment, is any finite string over the keyboard alphabet: we will use $\alpha, \beta, \gamma$ as metavariables for moves.

A move prefixed with $\bot$ or $\top$ will be called **labelled move**, where the prefix (label) $\bot$ or $\top$ indicates which player has made the move[39].

their names, and hence for explicitly considering universes in the form $(Dm, Dn)$ rather than just $Dm$ as classical logic does".

38 Similar to the approach on predicates we have mentioned before; we will look deeper into it once we have all the notions needed.

39 We could have gone with the colourful approach of the lecture notes, where green stands for $\top$ and red for $\bot$; however, it seemed best to prefer the symbolic approach in order to avoid any sort of visual annoying cacophony. Thus, any colour-labelled example provided by Japaridze will be here displayed through $\top$ and $\bot$ labels.

Sequences of labelled moves are **runs**, for which we will be using $\Gamma, \Delta$ as metavariables. A **position** is a finite run, symbolised through $\Phi, \Psi, \Xi, \Omega$ metavariables.

Runs and positions will often be delimited by "$\langle$" and "$\rangle$", with $\langle\rangle$ standing for the **empty position**.
For instance, the expression $\langle \Phi, \Psi, \bot\beta, \Gamma \rangle$ is the run consisting of the labelled moves of position $\Phi$, followed by the moves of position $\Psi$, succeeded by the labelled move $\bot\beta$ and then by the moves of the (possibly infinite) run $\Gamma$.

Furthermore, a set $S$ of runs is said to be **prefix-closed** iff, whenever a run is in $S$, so are all of its initial segments. The **limit-closure** of a set $S$ of runs is the result of adding to $S$ every infinite run $\Gamma$ such that all finite initial segments of $\Gamma$ are in $S$.

We call **gamestructure** the nonempty, prefix- and limit-closed set **Lr** of **legal runs**[40]. This means that a (finite or infinite) run is an element of Lr iff all of its finite initial segments are in Lr. The empty position $\langle\rangle$ is thus always legal.

Lastly, we need to define the notion of **content** in order to provide a formal account of CoL's fundamental notion of game.

**Definition 1.2.2.2**
Let Lr be a gamestructure. A **content** on Lr is defined as the function **Wn**: Lr $\rightarrow \{\top, \bot\}$.

When Wn $\langle\Gamma\rangle$ = $\top$, we say that $\Gamma$ is **won** by $\top$ (and lost by $\bot$); and when Wn $\langle\Gamma\rangle$ = $\bot$, we say that $\Gamma$ is won by $\bot$ (and **lost** by $\top$).
We extend the domain of Wn to all runs by stipulating that an illegal run is always lost by the offender.

40 The runs that are not in Lr are said **illegal**. As Japaridze explains in [67], page 10: "An **illegal move** in a given position $\langle\Phi\rangle$ is a move $\lambda$ such that $\langle\Phi, \lambda\rangle$ is illegal. The player who made the first illegal move in a given run is said to be the **offender**. Intuitively, illegal moves can be thought of as moves that cannot or should not be made. Alternatively, they can be seen as actions that cause the system crash (e.g., requesting to divide a number by 0)".

We can now finally tackle the most basal definition in CoL's formal semantics.

**Definition 1.2.2.3**
A **game** is a pair $\boldsymbol{A} = (\mathrm{Lr}^A, \mathrm{Wn}^A)$ where:

1. $\mathbf{Lr}^A$ is a gamestructure;

2. $\mathbf{Wn}^A$ is a content on $\mathrm{Lr}^A$.

Hence, we call the Lr component of a game its *structure* and the Wn component its *contents*.

Finite games can be visualised as upside-down trees[41] stemming from a common **root**, where **nodes** represent positions and **edges** stand for labelled legal moves. Each **branch** is a legal run, namely, the sequence of the edges' labels. An example is provided in [72], page 3:

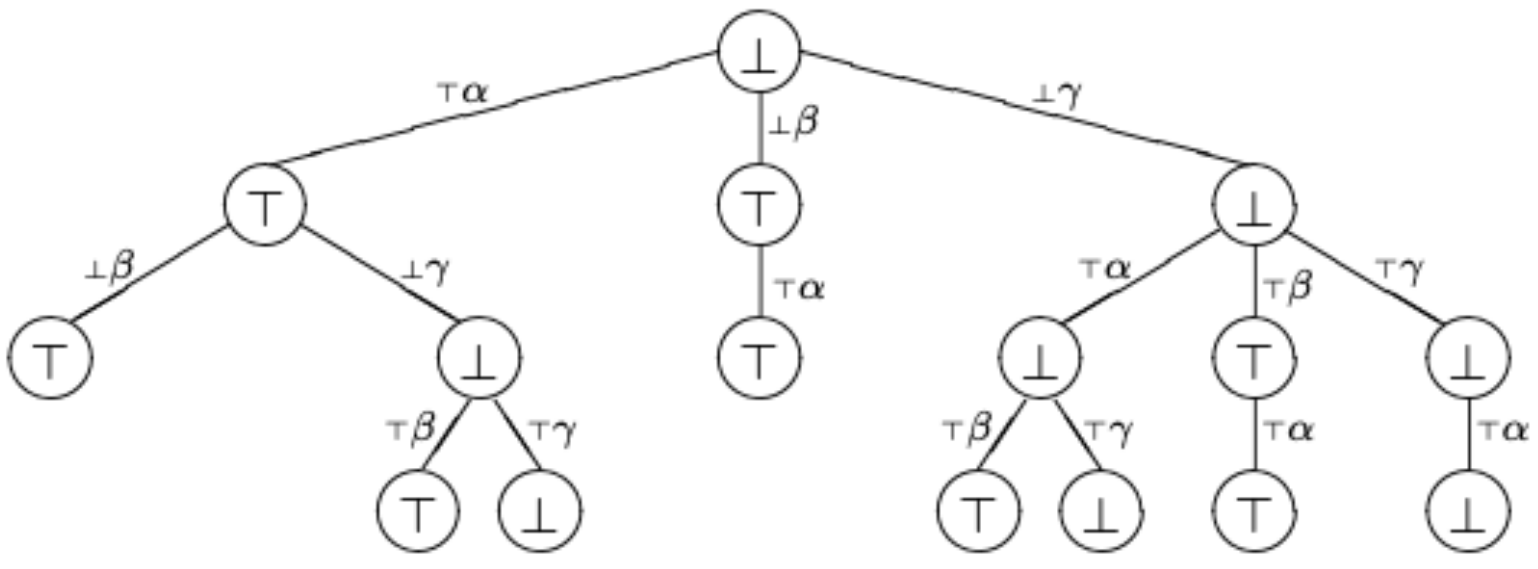

Figure 1: Illustration of a game in tree form.

**Definition 1.2.2.4**
Let $A$ be a game and $\Phi$ be a legal position of $A$ (i.e. $\langle\Phi\rangle \in \mathrm{Lr}^A$). The game $\langle\boldsymbol{\Phi}\rangle\boldsymbol{A}$, called the **$\boldsymbol{\Phi}$-prefixation** of $A$, is defined by:

1. $\mathrm{Lr}^A = \{\langle\Gamma\rangle | \langle\Phi,\Gamma\rangle \in \mathrm{Lr}^A\}$;

2. $\mathrm{Wn}^{\langle\Phi\rangle A}\langle\Gamma\rangle = \mathrm{Wn}^A\langle\Phi,\Gamma\rangle$.

41 Intuitively, the set of edges and "empty" nodes - meaning the little circles without any symbol inside, constitutes the structural part of the game. The $\top$ and $\bot$ symbols drawn inside of each position is, indeed, the content of the game - meaning the winner of each finite run.

Intuitively, $\langle\Phi\rangle A$ is the game playing which means playing $A$ starting (continuing) from position $\Phi$. Indeed, $\langle\Phi\rangle A$ is the game to which $A$ evolves to (will be "brought down to") after the moves of $\Phi$ have been made.
Visually, this means that the game starts from the node corresponding to $\Phi$, having already proceeded through the given run (branch) thus far.

---

*Game properties*

We may now have a look at some *structural game properties*, meaning properties which depend only on the structure of a game and not on its contents.

Let $G$ be a game.

Let $e$ be a $(Vr, Dm)$-valuation[42], meaning a total function of the type $Vr \rightarrow Dm$, where $Vr$ is the set of variables on which the game depends and $Dm$ is its domain[43].

As previously mentioned, $G$ is **elementary** iff, for every such valuation $e$, $\text{Lr}_e^A = \{\langle\rangle\}$.

Furthermore, $G$ is said to be **finite-depth** iff there is a (smallest) integer $d$, called the **depth** of $G$, such that, for every valuation $e$, the length of every legal run of $G$, meaning each element of $\text{Lr}_e^A$, is $\leq d$[44].

In addition, we say that $G$ is **perifinite-depth** iff, for every $e$, every legal run (even if there are arbitrarily long ones) is finite.

A legal run $\Gamma$ of $G$ is **maximal** iff $\Gamma$ is not a proper initial segment of any other legal run of $G$. We say that $G$ is **finite-breadth** iff the total number of maximal legal runs of $G$, called the **breadth** of $G$, is finite[45].

42 Which we will simply call *valuation* from now on, as to simplify our syntax.
43 As explained in [72], when $Vr$ is finite, $e$ can simply be written as an $n$-tuple $(a_1, \ldots, a_n)$ of individuals, meaning that $e(x_1) = a_1, \ldots, e(x_n) = a_n$ (with $x_1, \ldots, x_n$ being the list of variables lexicographically ordered).
44 Visually, it means that each and every branch of a finite-depth-game tree has a maximum length of $d$.
45 Generally, the breadth of a game may not only be infinite, but also uncountable.

Moreover, $G$ is said to be **finite** iff it only has a finite number of legal runs. Clearly, $G$ is finite if and only if it is finite-breadth.
We may summarise as such:

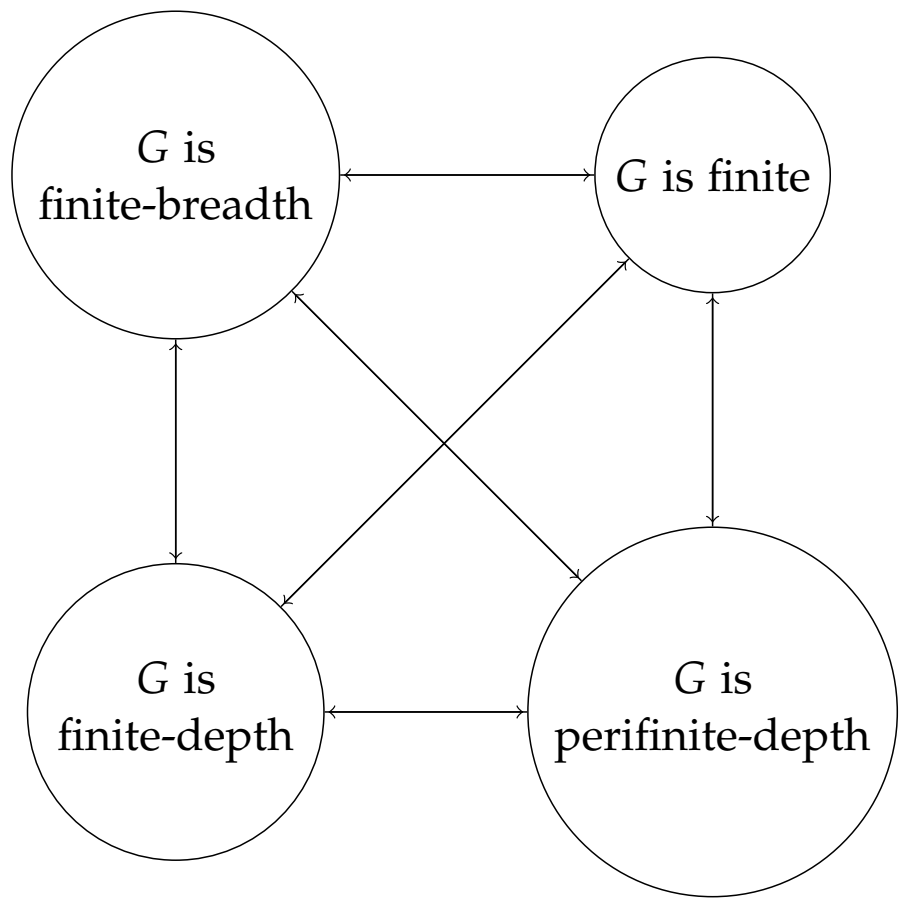

The games on which we are going to focus are said **static games**, as we already know by now. In these particular games, it is a contest of intellect rather than speed: the strategy has to be good no matter the speed of the adversary in making moves.

Formally, the static property is defined in terms of *delays*. We say that a run $\Delta$ is a $\top$ (respectively $\bot$) **delay** of run $\Gamma$ iff the following two conditions are satisfied:

1. The moves of either player are arranged in $\Gamma$ in the same order as in $\Delta$;
2. For any $n, k \geq 1$, if the $k$-th $\bot$ (respectively $\top$) move is made earlier than the $n$-th $\top$ (respectively $\bot$) move in $\Gamma$, then so is it in $\Delta$.

In other words, $\Delta$ is the result of shifting to the right (*delaying*) some $\top$ (respectively $\bot$) moves in $\Gamma$ without violating their order.

For example[46]:

$$\langle \bot 0, \bot 2, \top \mathbf{1}, \top 3, \bot 4, \top 5, \bot 7, \bot 8, \top \mathbf{6}, \bot 10, \top \mathbf{9} \rangle$$

is a $\top$ delay of:

46 As seen in [67], page 18.

$$\langle \bot 0, \top \mathbf{1}, \bot 2, \top 3, \bot 4, \top 5, \top \mathbf{6}, \bot 7, \bot 8, \top \mathbf{9}, \bot 10 \rangle.$$

The former is obtained from the latter by shifting to the right some $\top$ moves. When doing so, a $\top$ move can jump over a $\bot$ move, but it cannot jump over another $\top$ move.

**Definition 1.2.2.5**
A game $G$ is **static** iff, for either player $\wp$ and any runs $\Gamma$ and $\Delta$ such that $\Delta$ is a $\wp$ delay of $\Gamma$, in the context of $G$, the following two conditions are satisfied:

1. If $\Gamma$ is won by $\wp$, then so is $\Delta$;
2. If $\wp$ does not offend in $\Gamma$, then neither does it in $\Delta$.

The static property extends to all games, as we will later see, by stipulating that any game is static iff all of its instances are[47]. Furthermore, all game operations studied in CoL are provably closed under the static property of games[48].

Here are some more technical facts about static games, as shown in [39], page 18.

**Lemma 1.2.2.6**
Suppose $\Delta$ is a $\wp$-delay of $\Gamma$ and $\neg\wp\alpha$ is the first labelled move of $\Gamma$. Then $\neg\wp\alpha$ is the first labelled move of $\Delta$ as well.

**Lemma 1.2.2.7**
If $\Delta$ is a $\wp$-delay of $\Gamma$, then $\Gamma$ is a $\neg\wp$-delay of $\Delta$.

**Lemma 1.2.2.8**
Assume $A$ is a static game, $e$ is any valuation and $\Delta$ is a $\wp$-delay of $\Gamma$. Then we have:

1. If $\Delta$ is a $\wp$-illegal run of $A$ with respect to $e$, then so is $\Gamma$;
2. If $\Gamma$ is a $\neg\wp$-illegal run of $A$ with respect to $e$, then so is $\Delta$.

47 As we are going to later explain in detail, games can be extended to their generalisation, which we call gameframes. Indeed, games then become *instances* of gameframes thanks to a valuation that replaces each variable of the extension of the gameframe with a constant. The formal definition will be provided in page 38 - the *wok-pan* strikes again.
48 Proof provided in [39] for each CoL operator.

It goes without saying that elementary games are static: their only legal run $\langle\rangle$ does not have any proper $\wp$-delays; moreover, if we consider a non-empty run $\Gamma$ won by $\wp$, the adversary of $\wp$ must have made the first (illegal) move in $\Gamma$.

Consequently, by Lemma 1.2.2.6, any $\wp$-delay of $\Gamma$ will have the same first labelled move and will be won by $\wp$ for the same reason as $\Gamma$ was won.

Clearly, we have:

**Proposition 1.2.2.9**
Every strict game is static.

However, not all static games are strict.

**Definition 1.2.2.10**
A game $G$ is **strict** iff, for every $(Vr, Dm)$-valuation $e$ and $\Phi$, with **Lm** standing for the set of a player's legal moves, either $\mathrm{Lm}\bot_e^A\langle\Phi\rangle = \varnothing$ or $\mathrm{Lm}\top_e^A\langle\Phi\rangle = \varnothing$.

Thus, in a strict game, at most one of the players is allowed to have legal moves in a given position.

Disregarding any time or speed-related details (for the static property), here is a strategy[49], meaning a series of moves, which ensures $\top$ a win in the strict game of Figure 1:

> Regardless of what the adversary is doing or has done, go ahead and make move $\alpha$; make $\beta$ as your second move if and when you see that the adversary has made move $\gamma$, no matter whether this happened before or after your first move[50].

---

49 Again, we are giving up the traditional functional approach and expanding it to longer interactions.
50 Quoting from [72], page 4.

*Interactive Turing Machines*

Formally, $\top$'s algorithmic strategies can be understood as what CoL calls **HPMs**, *Hard Play Machines*.

An HPM, which we have already mentioned, is a Turing Machine with the capability of making moves. While an ordinary Turing Machine halts after generating an output, an HPM generally continues after making a move, allowing longer interactions with many subsequent moves.

These HPMs are equipped with an additional[51], read-only tape called the **run tape**, initially empty. Every time an HPM makes a move $\alpha$, the string $\top\alpha$ is automatically appended to the content of this tape.

At any time, any nondeterministic $\bot$-labelled move $\bot\beta$ may also be appended to the content of the run tape: this means that Environment has just made move $\beta$. This way, at any step of the process, the run tape spells the current position of the play, keeping a sort of transcript of the whole run.

For a run $\Gamma$ and a computation branch $B$ of an HPM or EPM (which we are about to introduce), we say that $B$ **cospells** $\Gamma$ iff $B$ spells $\neg\Gamma$ ($\Gamma$ with all labels reversed).

Intuitively, when a Machine $\mathcal{M}$ plays as $\bot$ rather than $\top$, the run that is generated by a given computation branch $B$ of $\mathcal{M}$ is the run cospelled (rather than spelled) by $B$, for the moves that $\mathcal{M}$ makes have the label $\bot$, and the moves that its adversary makes have the label $\top$.

Our extended version of the Church-Turing thesis naturally expands from ordinary Turing Machines to HPMs, which adequately correspond to what we intuitively perceive as algorithmic strategies[52].

---

51 We already assume a typical Turing Machine tapes asset, meaning at least one ordinary read/write **work tape**.

52 This means that rather than attempting to formally describe an HPM playing a given game, we can simply describe its work in a relaxed and informal way, as we did for the above-seen strategy for the game of Figure 1. There is also no need to define $\bot$'s strategies. All possible behaviours of $\bot$ are accounted for by the above-mentioned nondeterministic updates of the run tape, including $\bot$'s relative speed (since there are no restrictions on when or with what frequency the updates can take place).

Indeed, just like an ordinary Turing Machine, an HPM has a finite set of states, one of which is the **start state**. There are no accept, reject, or halt states, but there are specifically designated states called **move states**. Each tape of the Machine has beginning but no end, being divided into infinitely many cells arranged from left to right.

At any time, each cell contains one **symbol** out of a fixed finite set of tape symbols. The blank symbol - is amongst the tape symbols.
Each tape has its own **scanning head**, at any given time located on one of the tape cells[53].
We also assume a **move buffer**, with unlimited size, allowing the machine to construct a move piece-by-piece before officially making such move.

A transition from one **computation step**, or **clock cycle**, to another is regulated by the fixed *transition function* of the machine. This is represented through the typical *graph* or *automaton* assigned to the transition functions of Turing Machines.

For instance, dipping my cursor in the old waters of my Bachelor's thesis:

| Starting state | Symbol on tape | Written symbol | Direction | Following state |
|---|---|---|---|---|
| $q_0$ | 1 | 1 | $R$ | $q_0$ |
| $q_0$ | $B$ | 1 | $L$ | $q_1$ |
| $q_1$ | 1 | 1 | $L$ | $q_1$ |
| $q_1$ | $B$ | $B$ | $R$ | $q_2$ |

Here is the same Turing Machine represented as an automaton:

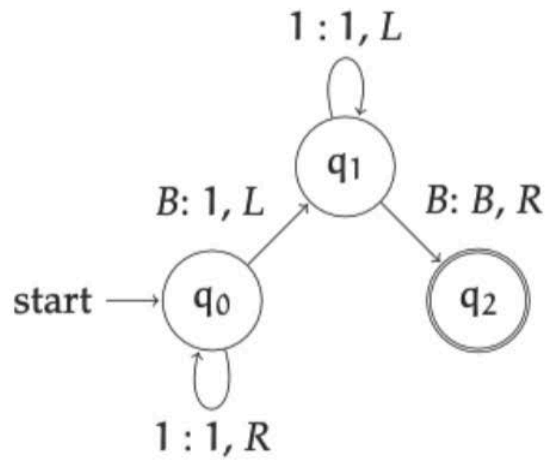

Figure 2: Example of a Turing Machine whose set of quintuples, meaning the set of transitions, is first represented through a *table of behaviours* and then through a *graph* (or *automaton*).

53 As we will see in a few lines, the scanning head moves according to the instructions the machine gives out. Indeed, the head reads the cell on its right if it receives the R instruction; reads the one on the left if the L instruction is given; remains on the same cell if - is instructed.

Below is an example of an HPM provided by Japaridze in his lecture notes number 11 (even though the move buffer is not represented):

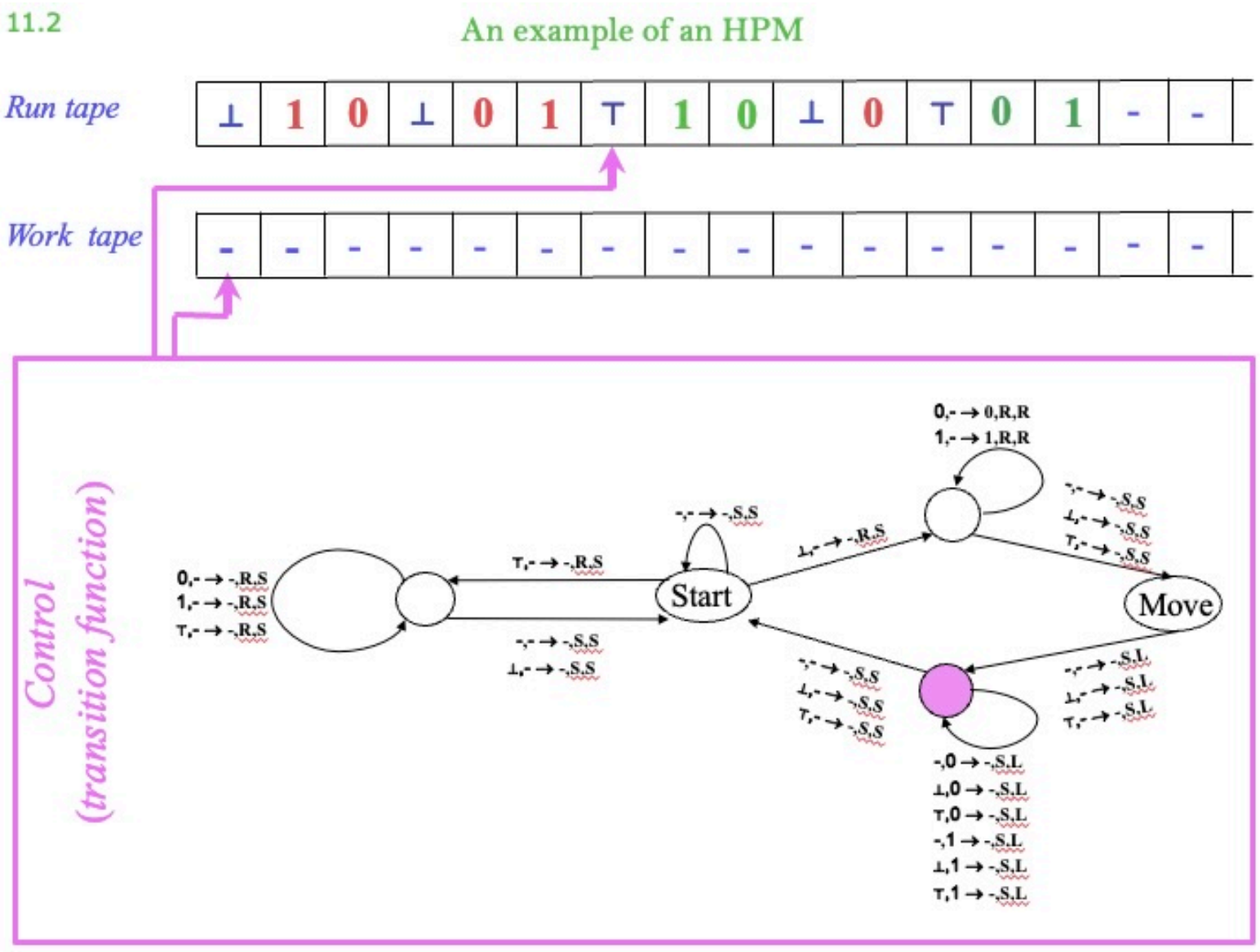

Figure 3: The HPM reads the run tape (i.e. what is happening in the game) and, in line with the behaviour it has been given (i.e. the transition function), writes the appropriate symbols on the work tape and moves the scanning head in the prescribed direction. This is how Machine plays the game against Environment.

Furthermore, a **configuration** is a full description of the situation at some given computational step. It consists of records of the ("current") contents of the work and run tapes, the content of the buffer, the location of each scanning head, and the state of the machine.

Depending on which $\bot$-moves (and when they) occur in the run tape of an HPM $\mathcal{M}$, different runs will be eventually spelled on $\mathcal{M}$'s run tape (within the limit). We call any such run a **run generated by** $\mathcal{M}$.

**Definition 1.2.2.11**
We say that an HPM $\mathcal{M}$ **computes** a game $G$ iff every run generated by $\mathcal{M}$ is a $\top$-won run of $G$. Such an $\mathcal{M}$ is said to be a **solution** or an *algorithmic winning strategy* for $G$.

We may rewrite the expanded, *interactive* Church-Turing thesis as follows:

**Thesis 1.2.2.12**
For every (interactive) computational problem $G$, we have:

1. If some HPM computes $G$, then $G$ has an algorithmic solution (or effective solution) according to anyone's reasonable intuition;
2. If $G$ has an algorithmic solution according to anyone's reasonable intuition, then there is an HPM that computer $G$.

An **EPM**, or *Easy Play Machine*, is slightly different from an HPM in that:

- There is an additional special state called **permission state**. Entering this state means *granting permission*;
- In the EPM model, Environment can, but may not, make a single move only when the Machine is in a permission state;
- Machine wins a game $G$ iff, for every computation branch $B$ of the Machine, where $\Gamma$ is the run spelled by $B$, we have:
  1. $\mathrm{Wn}^G\langle\Gamma\rangle = \top$;
  2. As long as Environment does not offend in $\Gamma$, Machine grants permission infinitely many times in $B$.

In the EPM model, Machine has full control over the speed of its adversary, which has to patiently wait for explicit permissions from Machine to move.
The only fairness condition that Machine is expected to satisfy is that, as long as the adversary plays legal, Machine has to grant permission every once in a while (in which case, the EPM is said to be **fair**). How long that "while" lasts, however, is totally up to Machine.

What happens if we limit the relative speed of Environment? Nothing, as long as static games[54] are concerned.

54 Meaning the very entities that we agree to call computational problems.

The EPM model takes the idea of limiting the speed of Environment to the extreme, yet, according to the following theorem, it yields the same class of computable problems as the HPM model.

As Japaridze explains in the same lecture:

> Even though the EPM model is equivalent to the HPM model, one of the reasons why we still want to have it is that describing winning strategies in terms of EPMs is much easier than in terms of HPMs.

**Theorem 1.2.2.13**
For every static game $G$, the following conditions are equivalent:

1. There is an HPM that computes $G$;
2. There is an EPM that computes $G$.

Moreover, every HPM $H$ can be effectively converted into an EPM $E$ such that $E$ wins every static game that $H$ wins. And vice versa: every EPM $E$ can be effectively converted into an HPM $H$ such that $H$ wins every static game that $E$ wins.

Therefore, we may agree on the fact that:

**Definition 1.2.2.14**
Let $G$ be a computational problem. An **algorithmic solution**, or an *algorithmic winning strategy* for $G$, is an HPM or EPM that computes $G$. Furthermore, $G$ is said to be **computable** (winnable) iff it has an algorithmic solution.

We write $\vDash G$ to say that $G$ is computable.

---

*From games to gameframes*

Indeed, games are CoL's analogue of classical propositions. Since the classical propositional fragment is notoriously unable to express more complex formulas, first order classical logic extends its consideration from propositions to predicates.

We need to extend our consideration too, then, from games to their generalisation, which we will call **gameframes**. These generalise predicates in the same sense as game generalise propositions.

For such reason, CoL needs to be upgraded from the propositional level to the first order.

We first define a ***Vr*-valuation**, for $Vr \subseteq$ Variables[55], as a function $e$ of type $Vr \rightarrow$ Constants. Where $x_1, x_2, \ldots, x_n$ are all the variables of $Vr$ listed in a lexicographic order, any such valuation $e$ can be written as an $n$-tuple $(c_1, c_2, \ldots, c_n)$ meaning that $e(x_1) = c_1$, $e(x_2) = c_2$, ..., $e(x_n) = c_n$.

**Definition 1.2.2.15**
Let $n$ be a natural number. An $n$-ary **gameframe** is a pair $(Vr, G)$ where:

- ***Vr*** is a set of $n$ distinct variables - meaning the *variables* the gameframe depends on;
- ***G***, e.g. the *extension* of the gameframe, is a mapping that assigns a game $G(e)$ to each $Vr$-valuation $e$. We say that $G(e)$ is an **instance** of $G$.

As previously mentioned, predicates in Computability Logic are approached in a different way than classical logic.

Indeed, under an intensional[56] point of view, CoL's $n$-ary predicates are defined through an analogous definition of $n$-ary gameframes. The only difference is that the extension function returns propositions in the case of predicates, rather than games as in the definition of gameframes.

Furthermore, just like propositions are nothing but 0-ary predicates, games will be seen as 0-ary gameframes. Therefore, gameframes generalise games in the same way predicates do with propositions in classical logic.
Thus, the key takeaway is this: games are going to be seen as generalised propositions, while gameframes will be generalised predicates.

55 As seen in Definition 1.2.2.1.
56 Meaning a variable-sensitive, concept-centered understanding, the same one adopted for the notion of function we have introduced before.

Consequently, relations become predicates and operations turn into functions. Here are two examples, provided in the lecture notes number 4:

- The relation "$var_3 < var_7$" on NAT shall be formally understood as the predicate:
$$p = (\{var_3, var_7\},\ p)$$
where: $p$ (5, 32) = $\top$ means "$p$ is true at valuation (5, 32)"; $p$ (5, 1) = $\bot$ means "$p$ is false at (5, 1)" and so on;
- The operation "$var_1 + var_2$" on NAT can be formally understood as the function:
$$f = (\{var_1, var_2\}, f)$$
where: $f$ (5, 32) = 37; $f$ (5, 1) = 6; and so on.

A gameframe is **elementary** iff all of its instances are; thus, gameframes generalise elementary games in the same sense that games generalise $\top$ and $\bot$.

Since the Lr component of all instances of elementary games is trivial (being $\langle\rangle$ the only legal run), and since depending on runs is the only thing that differentiates gameframes from propositions, we are going to use "predicate" and "elementary gameframe" as synonyms.

Specifically, we understand each predicate *(Vr, p)* as the elementary gameframe *(Vr, G)* such that, for any *Vr*-valuation *e*, $\text{Wn}_e^G\langle\rangle = \top$ iff $p$ is true at *e*.
Vice versa: we understand every elementary gameframe *(Vr, G)* as the predicate *(Vr, p)* such that, for any *Vr*-valuation *e*, p is true at *e* iff $\text{Wn}_e^G\langle\rangle = \top$.

---

*Further definitions*

We are now going to sketch some more geographical details onto our map in order to be rightly equipped for the syntax exploration of this land.

Consider a gameframe *G*. A legal run of all instances of *G* is said to be a **unilegal run** of *G*.

In addition, $G$ is said to be **unistructural** iff every legal run of every instance of $G$ is a unilegal run of $G$. In other words, $G$ is unistructural iff all of its instances have the same structure, meaning the same Lr component[57].
Furthermore, a game $A$ is ***x*-unistructural**, or *unistructural in x*, iff, for any two valuations $e_1$ and $e_2$ that agree on all variables except $x$, $\mathrm{Lr}^A_{e_1} = \mathrm{Lr}^A_{e_2}$.

Thus, in a unistructural gameframe, the legality of runs does not depend on the values assigned by $e$ to the variables of the gameframe.
All elementary gameframes and all games (which are non other than 0-ary gameframes) are unistructural. The wide class of unistructural gameframes is known[58] to be closed under all gameframe operations studied in CoL.

Unistructural gameframes, even when non 0-ary, can be visualised as trees in the earlier seen style. The only difference is that the nodes will now be labelled with predicates (and not propositions) which only depend on the variables of the gameframe.
For any given valuation $e$, each such label $p$ is telling us the player that makes the move: $\top$ if *p(e)* is true, $\bot$ if false.

To make things clearer, here are two visual examples taken from [67]. This is the tree for the "successor" game:

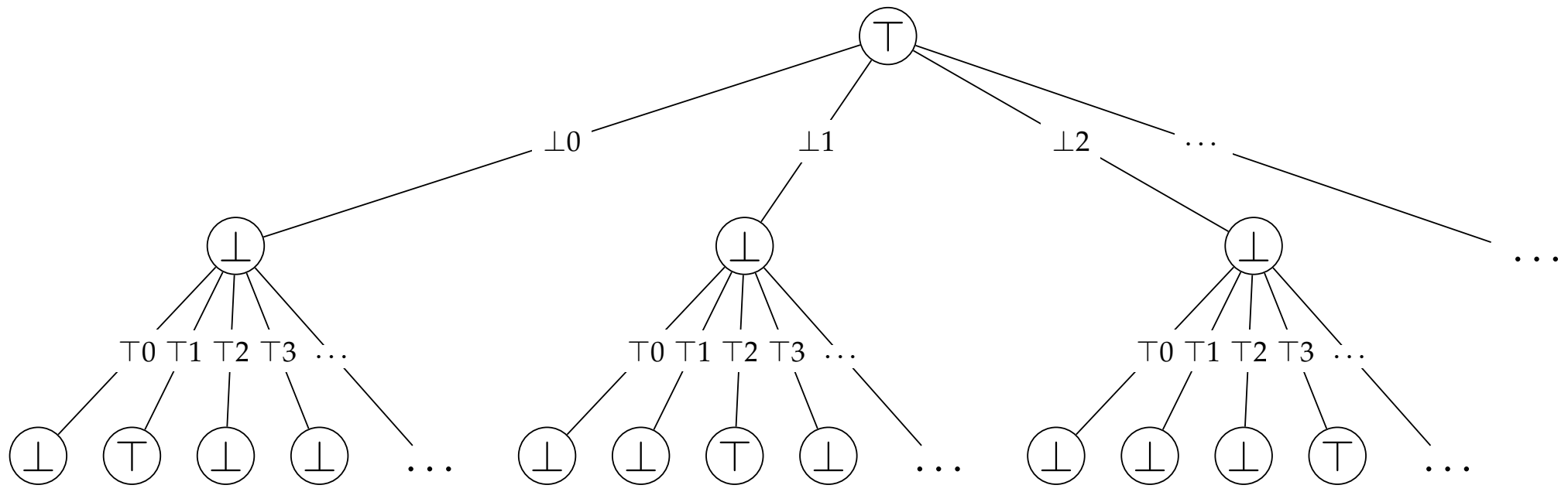

This is the tree for the "generalised successor" game (with only two branches shown for space reasons):

57 In other words, taking from [39], page 12, "We say that a game A is [...] *unistructural* iff, for every $e$, $\mathrm{Lr}^A_e = \mathrm{LR}^A$", with $\mathbf{LR}^A$ being the set of all unilegal runs of $A$.
58 See [39].

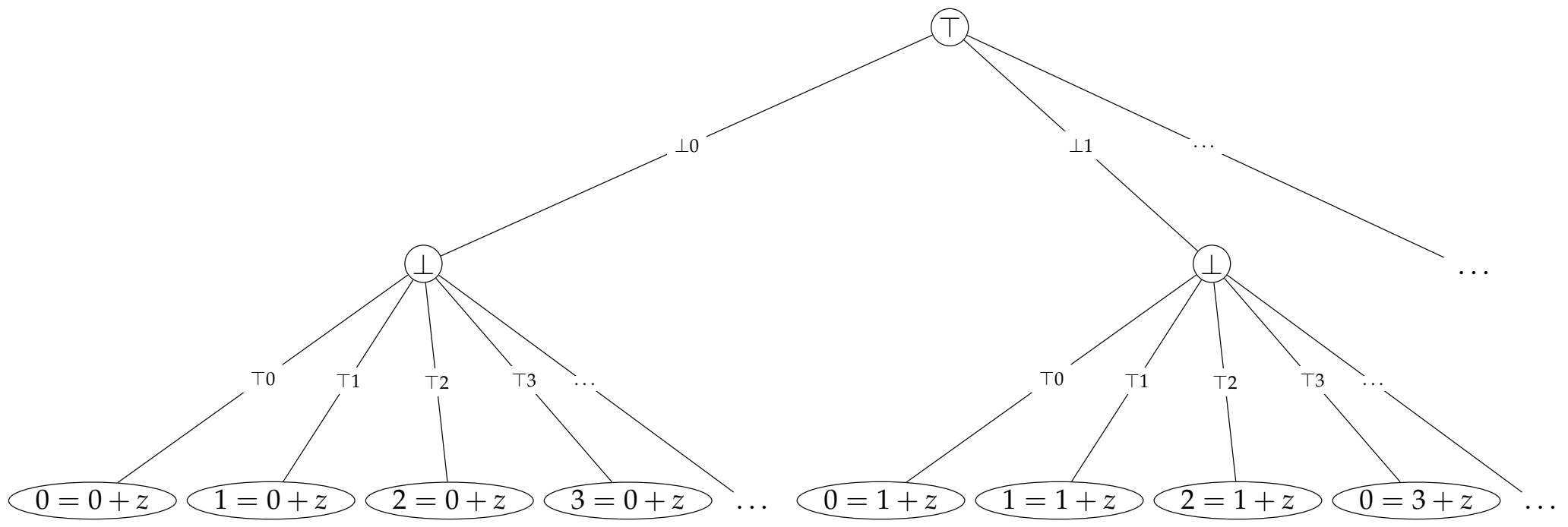

The standard operation of substitution of variables through terms in predicates naturally extends to gameframes as generalised predicates.

**Convention 1.2.2.16**
Consider a gameframe or function $(Vr, K)$, a set $Vr'$ of variables with $Vr \subset Vr'$, and a $Vr'$-valuation $e'$. Even though $e'$ is not a $Vr$-valuation, we may still write ***K(e′)*** to mean ***K(e)***, where $e$ is the restriction of $e'$ to $Vr$. In other words, $e$ is the $Vr$-valuation which agrees with $e'$ on all variables of $Vr$.

**Definition 1.2.2.17**
Consider a gameframe or function $(Vr, K)$. Following the standard practice established in the literature for predicates and functions, we may fix a tuple $(x_1,\ldots,x_n)$ of pairwise distinct variables when first mentioning the gameframe or function $\boldsymbol{K}$, and write $K$ as $K(x_1,\ldots,x_n)$. When doing so, we do not necessarily mean that $\{x_1,\ldots,x_n\} = Vr$.

Writing $K$ as $K(x_1,\ldots,x_n)$ sets a context in which later, for whatever functions $f_1 = (Vr_1, f_1)$, $\ldots$, $f_n = (Vr_n, f_n)$, we can write $K(f_1, \ldots, f_n)$ to mean the gameframe or function $(Vr', K')$ where:

- $Vr' = (Vr \setminus \{x_1, \ldots, x_n\}) \cup Vr_1 \cup \cdots \cup Vr_n$;
- For any $Vr'$-valuation $e'$, $K'(e') = K(e)$, where $e$ is the $Vr$-valuation such that $e(x_1) = f_1(e')$, $\ldots$, $e(x_n) = f_n(e')$ and $e$ agrees with $e'$ on all other variables from $Vr$.

Here is a substitution example taken from the same lecture.

Let $K(var_3)$ be the elementary gameframe *(*$\{var_3, var_4\}$*, K)* such that, for any constants $a$ and $b, K(a,b) = \top$ iff $a < b$. In other words, $K(var_3)$ is the predicate: $var_3 < var_4$.

Let $f$ be the function *(*$\{var_1, var_2\}$*, f)* such that, for any constants $a$ and $b, f(a,b) = a + b$. In other words, $f$ is the function: $var_1 + var_2$.
Then, $K(f)$ is the elementary gameframe $(\{var_1, var_2, var_3, var_4\}$***, K')*** such that, for any constants $a$, $b$ and $c, K'(a,b,c) = \top$ iff $a + b < c$. In other words, *K(f)* is the predicate: $var_1 + var_2 < var_4$.

We also may define constants and variables as functions.

**Definition 1.2.2.18**
Where $c \in$ Constants and $x \in$ Variables, $c^F$ and $x^F$ are functions defined as follows:

- $c^F = (\emptyset, f)$, where $f() = c$;
- $x^F = (\{x\}, f)$, where, for any constant $b$, $f(b) = b$.

We will usually omit the superscript $^F$ for readability. For example, where $G(x,y)$ is a gameframe and $h(x,y)$ a function, the expression $G(c, h(z,x))$ is our lazy way to write $G(c^F,\ h(z^F, x^F))$.

Bottom line is: substitution of variables[59] in CoL's gameframes (respectively functions) works in the same way as for predicates (respectively functions) in classical logic. Correspondingly, CoL uses the same notational conventions for gameframes (respectively functions) as classical logic does for predicates (respectively functions).

### 1.2.3 *"Speak, thou vast and venerable head"*

Having provided some bone structure to our informal creature, which is nothing but our increasing comprehension of Computability Logic, we now need to put some words into its mouth, build its proper language[60]

59 As previously mentioned, in [39], page 22, Japaridze proves that the operation of substitution of variables is static.

60 Even though this section should probably belong to the Syntax step, we preferred to introduce these definitions a little earlier, just as Japaridze does, in order to better

- indeed, learn to speak the local dialect.

CoL's language is still *in fieri*, with an open-ended formalism. It successfully extends the language of classical predicate calculus, while abandoning both identity and functional symbols.
Indeed, this new language is far more expressive than the latter, not only because it showcases many non-classical operators[61], but also due to having two sorts of atoms: **elementary** and **general**.

Let us take one step back and start from the very basal bricks, the firm founding definitions upon which this Tower of *Behirut*[62] longs to be built.

Other than variables and constants, which we have already defined, we have infinitely many **extralogical function letters**, **elementary gameframe letters** and **general gameframe letters**, each letter with a fixed **arity**.

**Terms** are inductively defined as follows:

- All variables are terms;
- All constants are terms;
- If $t_1, \ldots, t_n$, with $n \geq 0$, are terms and $f$ is an $n$-ary function letter, then $f(t_1, \ldots, t_n)$ is a term.

**Atoms** are inductively defined as follows:

- $\top$ and $\bot$ are atoms;
- If $t_1$ and $t_2$ are terms, then $t_1 = t_2$ is an atom;
- If $t_1, \ldots, t_n$, with $n \geq 0$, are terms and $L$ is an extralogical $n$-ary gameframe letter, then $L(t_1, \ldots, t_n)$ is an atom.

understand the following interpretation and validity conditions which belong to the Formal Syntax step.

61 For instance $\sqcap$, $\bigsqcap$, ⫰ and so on. These new, non-classical operators are the direct consequence of our choice to deal with more natural, unrestrained interactions (as explained in 1.1.3).

62 *Behirut* as in Hebrew for *clarity*, בהירות. Our aim is to steer clear from the biblical linguistic jumble of the Tower of *Babel* (from בבל, which, counter to *behirut*, means *confusion*). Indeed, we will try to achieve the exact contrary (not opposite), not only to avoid any linguistic confusion, which goes without saying, but primarily to establish a common language that can be used by other logics as well (as implied in the *CoL-thy-neighbour* section concerning classical, linear and intuitionistic logics). One common tower as in one common language to encompass them all.

Elementary atoms represent elementary gameframes, i.e. predicates, while general atoms represent any computational problems, i.e. any (not-necessarily-elementary) static gameframe.

As briefly mentioned in [51], page 285, elementary problems are, indeed, interesting and meaningful, since they can validate principles that may not be valid in the general case.
Consequently, we want to be able to analyse games at a reasonably fine level of granularity, which is amongst the main reasons for having two sorts of atoms in our language.

This means that for each integer $n \geq 0$, our language has infinitely many $n$-ary **elementary letters** and $n$-ary **general letters**. The former ones correspond to the *predicate letters* of classical logic. As already mentioned in 1.1.2, we will use the lowercase $p, q, r, \ldots$ as metavariables for elementary letters and uppercase $P, Q, R, \ldots$ as metavariables for general letters.

A **non-logical atom** is $L(t_1, \ldots, t_n)$ where $L$ is an $n$-ary elementary or general letter, and each $t_i$ is a term (i.e. a variable or a constant). Such an atom $L(t_1, \ldots, t_n)$ is said to be ***L*-based**.
When $L$ is 0-ary, the only $L$-based atom will be usually written as $L$ rather than $L()$. An $L$-based atom is said to be elementary, general, $n$-ary and so on, if $L$ is so.

We have two **logical atoms**: $\top$ and $\bot$. Still, there are three elementary gameframe letters: $\top$, $\bot$ and =.

**Formulas** are standardly obtained from atoms and variables by applying to them:

- Unary connectives: ¬, ∘¬, ≻¬, ▷¬, ⤚¬, ⩚, ⩛, ⫰, ⫯, ⋏, ⋎, ⍋, ⍒;
- Binary connectives: →, ⩾ , ∘— , ▷ , ⊐, ≻— , ▷— , ⤚;
- Variable - ($\geq$ 2) arity connectives: ∧, ∨, ⊓, ⊔, △ , ▽ , ⩓, ⩔;
- Quantifiers: ∀, ∃, ⨅, ⨆, ⋀, ⋁, ⩓, ⩔, △, ▽.

Consequently, we inductively define[63] them as follows:

63 Henceforth, many of these definitions are going to mention the non-classical operators which we are just about to introduce next. Again, we proceed through the *wok-seasoning* method.

- All atoms are formulas;
- If E is a formula, then so are ¬(E), ○¬(E), ≻¬(E), ▷¬(E), ⤚¬(E), ⩚(E), ⩛(E), ⫰(E), ⫯(E), ⋏(E), ⋎(E), ⍋(E), ⍒(E);
- If E and F are formulas, then so are (E) ∧ (F), (E) ∨ (F), (E) ⊓ (F), (E) ⊔ (F), (E) △ (F), (E) ▽ (F), (E) ⩓ (F), (E) ⩔ (F), (E) → (F), (E) ○— (F), (E) ⩾ (F), (E) ▷ (F), (E) ≻— (F), (E) ⊐ (F), (E) ≻— (F), (E) ▷— (F);
- If E is a formula and $x$ is a variable, then $\forall x$(E), $\exists x$(E), $\bigsqcap x$(E), $\bigsqcup x$(E), $\bigwedge x$(E), $\bigvee x$(E), ⩓$x$(E), ⩔$x$(E), $\triangle x$(E), $\bigtriangledown x$(E) are formulas.

Throughout the rest of this work, unless otherwise specified, "formula" will always mean a formula of this language, and letters $E, F, G, H, \ldots$ will be used as metavariables for formulas. We also continue using $x$, $y$, $z$ as metavariables for variables, $c$ for constants and $t$ for terms.

The definitions of **bound occurrence** and **free occurrence** of a variable are standard. They extend from variables to terms by stipulating that the occurrence of a constant is always free.

When an occurrence of a variable $x$ is within the scope of $Qx$ for several quantifiers $Q$, then $x$ is considered to be bound by the "closer", next-in-line quantifier.

For instance, the occurrence of $x$ within $Q(x, y)$ in:

$$\forall x(P(x) \vee \bigsqcap x \bigwedge y Q(x, y))$$

is bound by $\bigsqcap x$ rather than $\forall x$, for the latter is overridden by the former.

An occurrence of a variable that is bound by $\forall x$ or $\exists x$ is said to be **blindly bound**. No formula contains both free and bound occurrences of the same variable. **Sentences**, or *closed formulas*, are formulas with no free occurrences of variables.

We may introduce other specifications for occurrences.
With $F \rightarrow G$ meaning $(\neg F) \vee G$, a **positive** (respectively **negative**) **occurrence** of a subexpression is an occurrence that is in the scope of an even (respectively odd) number of occurrences of $\neg$.
A **surface occurrence**[64] of a subexpression means an occurrence that is not in the scope of choice operators.

In line with the similar notational practice already established for games, sometimes we will represent a formula $F$ as $F(x_1, \ldots, x_n)$, where the $x_i$ are pairwise distinct variables.
Thus, $F(t_1, \ldots, t_n)$ is the result of simultaneously replacing in $F$ all free occurrences of each variable $x_i$ ($1 \leq i \leq n$) by term $t_i$. In case each $t_i$ is a variable $y_i$, it may be not clear whether $F(x_1, \ldots, x_n)$ or $F(y_1, \ldots, y_n)$ was originally meant to represent $F$ in a given context.

Our disambiguating convention is that the context is set by the expression that was used earlier. That is, if we first mention $F(x_1, \ldots, x_n)$ and then use the expression $F(y_1, \ldots, y_n)$, the latter should be understood as the result of replacing variables in the former rather than vice versa.
Moreover, when representing $F$ as $F(x_1, \ldots, x_n)$, we do not necessarily mean that $x_1, \ldots, x_n$ are exactly the variables that have free occurrences in $F$.

It is worth to point out that, on many different occasions, Japaridze introduces the following terminology, notations and conventions:

- We often need to differentiate between **subformulas** as such, and particular occurrences of subformulas. We will be using the term **osubformula** ("o" for "occurrence") to mean a subformula together with a particular occurrence. The prefix "o" will be used with a similar meaning in words such as **oatom**, **oliteral**[65] and so on[66];
- An occurrence of an osubformula is **positive** iff it is not in the scope of $\neg$. Otherwise it is **negative**. According to our conventions regarding the usage of $\neg$, only oatoms may be negative;

64 As we will further see, the **elementarisation** of a formula F is the result of replacing in F every surface occurrence of the form G $\sqcup$ H by $\bot$. As a result, a formula is said to be **stable** iff its elementarisation is a valid formula (tautology) of classical logic - otherwise, it is said **instable**.

65 Where a **literal** is an atom with or without negation $\neg$. As defined in [67], page 64, "A literal is $L(t_1, \ldots, t_n)$ or $\neg L(t_1, \ldots, t_n)$, where $L(t_1, \ldots, t_n)$ is an atom".

66 So, for instance, the formula $P \wedge Q \rightarrow P$ has two atoms but three oatoms. Yet, we may still say "the oatom $P$", assuming that it is clear from the context which of the possibly many occurrences of $P$ is meant. Similarly for osubformulas and oliterals.

- A **politeral** is a positive oliteral;
- A ⩚**-(sub)formula** is a (sub)formula of the form $F_1 ⩚ \ldots ⩚ F_n$. Similarly for ⩛, $\wedge, \vee$, $\triangle$, $\triangledown$, $\sqcap, \sqcup$, some of which are shown below;
- A **sequential (sub)formula** is one of the form $F_1 \triangle \ldots \triangle F_n$ or $F_1 \triangledown \ldots \triangledown F_n$. We say that $F_1$ is the **head** of such a (sub)formula, and $F_2, \ldots, F_n$ form its **tail**;
- Similarly, a **parallel (sub)formula** is one of the form $F_1 \wedge \ldots \wedge F_n$ or $F_1 \vee \ldots \vee F_n$, a **toggling (sub)formula** is one of the form $F_1 ⩚ \ldots ⩚ F_n$ or $F_1 ⩛ \ldots ⩛ F_n$, and a **choice (sub)formula** is one of the form $F_1 \sqcap \ldots \sqcap F_n$ or $F_1 \sqcup \ldots \sqcup F_n$;
- A formula is said to be **quasielementary** iff it contains no general atoms and no operators other than $\neg, \top, \bot$, ⩚, ⩛;
- A formula is said to be **elementary** iff it is a formula of classical propositional logic, i.e. contains no general atoms and no operators other than$\neg, \top, \bot, \wedge, \vee$;
- A **semisurface osubformula** (or *occurrence*) is an osubformula (or occurrence) that is not in the scope of any choice connectives (i.e. $\sqcap$ and $\sqcup$);
- A **surface osubformula** (or *occurrence*) is an osubformula (or occurrence) that is not in the scope of any connectives other than $\neg, \wedge, \vee$;
- The **quasielementarisation** of a formula $F$ , denoted by $|F|$, is the result of replacing in $F$ every sequential osubformula by its head, every $\sqcap$-osubformula by $\top$, every $\sqcup$-osubformula by $\bot$, and every general politeral by $\bot$ (the order of these replacements does not matter).

  For instance:

$$|((P ⩛ q) \vee ((p \wedge \neg P) \triangle (Q \wedge R))) ⩚ (q \sqcap (r \sqcup s))| =$$
$$((\bot ⩛ q) \vee (p \wedge \bot)) ⩚ \top;$$

- The **elementarisation** of a quasielementary formula $F$, denoted by $F$, is the result of replacing in $F$ every $\wedge$ - osubformula by $\top$ and every $\vee$-osubformula by $\bot$ (again, the order of these replacements does not matter).

For example:

$$||(s \wedge (p \mathbin{\wedge\!\!\!\wedge} (q \mathbin{\vee\!\!\!\vee} r))) \vee (\neg s \vee (p \mathbin{\vee\!\!\!\vee} r))|| = (s \wedge \top) \vee (\neg s \vee \bot);$$

- A quasielementary formula $F$ is said to be **stable** iff its elementarisation $F$ is a tautology of classical logic. Otherwise $F$ is **instable**.

These definitions, which may come as a repetition of previously introduced notions, are mostly employed while sketching CoL's many fragments, as shown in the second chapter.

---

*Interpretation and validity conditions*

An **interpretation** is a mapping * such that:

- * sends every $n$-ary function letter $f$ to an $n$-ary function $f^*(var_1, \ldots, var_n)$ whose variables are the first $n$ variables of Variables;
- * sends every extralogical $n$-ary gameframe letter $L$ to an $n$-ary static gameframe $L^*(var_1, \ldots, var_n)$[67] whose variables are the first $n$ variables of Variables. Besides, is the letter $L$ is elementary, then so is the gameframe $L^*$.

Interpretations are meant to turn formulas into games. Not every interpretation is equally good for every formula though, and some precaution is necessary to avoid confusing variables, as well as to guarantee that $\forall x$, $\exists x$ are only applied to games for which they are defined, i.e. games unistructural in $x$.
For this reason, we restrict interpretations to "admissible" ones.

An interpretation * is said to be **admissible** for a formula E iff, whenever $E$ has an occurrence of an atom $L(t_1, \ldots, t_n)$ in the scope of $\forall x$ or $\exists x$ and one of the terms $t_i$ (with $1 \leq i \leq n$) contains the variable $x$,

67 We denote $(var_1, \ldots, var_n)$ as the **canonical tuple** of $L^*$. When we do not care about the canonical tuple, simply $L^*$ can be written instead of $L^*(var_1, \ldots, var_n)$. As explained in [51], page 286, according to our earlier conventions, $x_1, \ldots, x_n$ have to be neither all nor the only variables on which the game $L^* = L^*(x_1, \ldots, x_n)$ depends. In fact, $L^*$ does not even have to be finitary here. The canonical tuple is only used for setting a context, in which $L^*(t_1, \ldots, t_n)$ can be conveniently written later for $L^*(x_1/t_1, \ldots, x_n/t_n)$. This eliminates the need to have a special syntactic construct in the language for the operation of substitution of variables.

$L^*(var_1, \ldots, var_n)$ is unistructural in $var_i$.

We extend * to a mapping which sends each term $t$ to a function $t^*$ and each formula $E$ for which it is admissible to a game $E^*$ as follows:

- Where $c$ is a constant, $c^*$ is $c^F$, simply written as $c$;
- Where $x$ is a variable, $x^*$ is $x^F$, simply written as $x$;
- Where $f$ is an $n$-ary function letter and $t_1, \ldots, t_n$ are terms, $f(t_1, \ldots, t_n)^*$ is $f^*(t_1{}^*, \ldots, t_n{}^*)$;
- $\top^*$ is $\top$, $\bot^*$ is $\bot$, and, where $t_1$ and $t_2$ are terms, $(t_1 = t_2)^*$ is $t_1{}^* = t_2{}^*$;
- Where $L$ is an $n$-ary gameframe letter and $t_1, \ldots, t_n$ are terms, $(L(t_1, \ldots, t_n))^*$ is $L^*(t_1{}^*, \ldots, t_n{}^*)$;
- * commutes with all logical operators, seeing them as the corresponding gameframe operations[68]: $(\neg E)^*$ is $\neg(E)^*$, $(E \sqcap F)^*$ is $(E)^* \sqcap (F)^*$, $(\forall x E)^*$ is $\forall x (E)^*$ etc.

When $F^* = A$, we say that * interprets $F$ as $A$, and that $F^*$ is an interpretation of $F$.

Concerning validity, we have:

**Definition 1.2.3.1**
We say that a sentence $F$ is:

- **logically** (or **uniformly**) **valid** iff there is an HPM $\mathcal{M}$ such that, for every interpretation * admissible for $F$, $\mathcal{M}$ solves $F^*$. Such an $\mathcal{M}$ is said to be a **logical** (or **uniform**) **solution** of $F$;
- **extralogically** (or **multiformly**) **valid** iff for every interpretation * admissible for $F$, there is an HPM $\mathcal{M}$ such that $\mathcal{M}$ solves $F^*$.

The above concepts extend from sentences to all formulas by identifying each formula $F$ with its $\sqcap$-closure $\sqcap F$, i.e. with $\sqcap x_1, \ldots, \sqcap x_n F$, where $x_1, \ldots, x_n$ are all free variables of $F$ listed lexicographically.

68 Which we are going to later thoroughly define.

An interpretation * gives meanings to formulas: a formula $E$ is just a string, while $E^*$ is a computational problem.

Intuitively, a logical solution $\mathcal{M}$ for a sentence $F$ is an interpretation-independent winning strategy: the "intended" or "actual" interpretation * is not visible nor known to Machine, so $\mathcal{M}$ has to play in some standard, uniform way that would be successful for any possible interpretation of $F$, i.e. any possible meaning of its atoms.

Indeed, which of our two versions of validity is more important depends on the motivational standpoint.
Extralogical validity tells us what can be computed in principle. So, a computability-theoretician would focus on extralogical validity.

On the other hand, it is logical rather than extralogical validity that is of interest in more applied areas[69].
In these applications we want a logic on which a universal problem-solving machine can be based. Such a machine would (or should) be able to solve problems represented by logical formulas without any specific knowledge of the meanings of their atoms, i.e. without knowledge of the actual interpretation.

However, for most fragments of the language of CoL, it is known or conjectured that the two sorts of validity coincide extensionally: a sentence is logically valid iff it is extralogically valid.
Notable exceptions are sentences containing elementary letters. For instance, where $p$ is a 0-ary elementary gameframe letter, $p \sqcup \neg p$ is valid extralogically but not logically.

Indeed, when we focus on extralogical validity, CoL is a logic of computability; otherwise, when we look at logical validity, CoL is a logic of "knowability".
Since our interest in CoL's application in computer science, we are going to focus only on the logical sort of validity.

We may also present some closure theorems for our notions of interpretation and validity. In order to do this, we fix these notational conventions:

69 Such as knowledgebase systems, systems for planning and action, constructive applied theories or declarative programming languages, as we will see later on.

- $\models F^*$ means "$F^*$ is computable";
- $\Vdash F$ means "$F$ is extralogically valid";
- $\Vvdash F$ means "$F$ is logically valid".

**Theorem 1.2.3.2**
Computability, logical validity and extralogical validity are closed under all three versions of Modus Ponens. Namely, for any formulas $E$, $F$ and any interpretation $*$, where $\supset \in \{\rightarrow, \geqslant, \circ\!\!-\!\!-\}$, we have:

1. If $\models E^*$ and $\models E^* \supset F^*$, then $\models F^*$;
2. If $\Vdash E$ and $\Vdash E \supset F$, then $\Vdash F$;
3. If $\Vvdash E$ and $\Vvdash E \supset F$, then $\Vvdash F$.

Furthermore, we have:

**Theorem 1.2.3.3**
For any formula $F$, variable $x$ and interpretation $*$, we have:

1. If $\models F^*$, then:
   - $\models$ ⅄$F^*$;
   - $\models$ ⫰$F^*$;
   - $\models \sqcap x F^*$;
   - $\models \bigwedge x F^*$.
2. If $\Vdash F$, then:
   - $\Vdash$ ⅄$F$;
   - $\Vdash$ ⫰$F$;
   - $\Vdash \sqcap x F$;
   - $\Vdash \bigwedge x F$.
3. If $\Vvdash F$, then:
   - $\Vvdash$ ⅄$F$;
   - $\Vvdash$ ⫰$F$;
   - $\Vvdash \sqcap x F$;
   - $\Vvdash \bigwedge x F$.

On the other hand, unlike classical validity, we do not have closure under the rule "From $E$ conclude $\forall xE$". For example:

$$\vDash Even(x) \sqcup Odd(x) \text{ and } \nvDash \forall x(Even(x) \sqcup Odd(x))$$

Providing two other notational conventions, we present a stronger form for clauses 1. and 3. of the past two theorems.

- $\mathcal{M} \vDash F^*$ means "$\mathcal{M}$ computes $F^*$";
- $\mathcal{M} \Vdash F$ means "$\mathcal{M}$ is a logical solution for $F$".

We are equipped with a new, stronger version of closure: **uniform-constructive closure**.

**Theorem 1.2.3.4**
For each $\supset \in \{\rightarrow, \succ, \circ\!\!-\}$, there is an effective function $f$:{HPMs} $\times$ {HPMs} $\rightarrow$ {HPMs} such that, for any formulas $E$, $F$, interpretation * and HPMs $\mathcal{M}$, $\mathcal{N}$, we have:

1. If $\mathcal{M} \vDash E^*$ and $\mathcal{N} \vDash E^* \supset F^*$, then $f(\mathcal{M}) \vDash F^*$;
2. If $\mathcal{M} \Vdash E$ and $\mathcal{N} \Vdash E \supset F$, then $f(\mathcal{M}, \mathcal{N}) \Vdash F$.

**Theorem 1.2.3.5**
There are effective functions $f_1, f_2, f_3, f_4$: {HPMs} $\rightarrow$ {HPMs} such that, for any formula $F$, interpretation * and HPM $\mathcal{M}$, we have:

1. If $\mathcal{M} \vDash F^*$, then:
   - $f_1(\mathcal{M}) \vDash \curlywedge F^*$;
   - $f_2(\mathcal{M}) \vDash \circ\!\!\mid F^*$;
   - $f_3(\mathcal{M}) \vDash \sqcap F^*$;
   - $f_4(\mathcal{M}) \vDash \bigwedge F^*$.
2. If $\mathcal{M} \Vdash F$, then:
   - $f_1(\mathcal{M}) \Vdash \curlywedge F$;
   - $f_2(\mathcal{M}) \Vdash \circ\!\!\mid F$;
   - $f_3(\mathcal{M}) \Vdash \sqcap F$;
   - $f_4(\mathcal{M}) \Vdash \bigwedge F$.

Having touched upon the key formal semantical concepts, we have reached half of our aerial journey. It is now time to enter the final stage of our descent: Syntax.

## 1.3 THE GIBBOUS ARC: LANDING SYNTAX

### 1.3.1 *Operators descent*

As we already know, logical operators in CoL stand for operations on games. There is an open-ended pool of operations of potential interest, ready to be studied according to all kinds of needs and tastes.

Here is a first preview in two tables of the *CoL zoo of game operations* that has been introduced thus far:

| | - | Blind | Choice | Parallel |
|---|---|---|---|---|
| Negation | $\neg$ ("not") | | | |
| Conjunction | | | $\sqcap$ ("chand") | $\wedge$ ("pand") |
| Disjunction | | | $\sqcup$ ("chor") | $\vee$ ("por") |
| Universal quantifier | | $\forall$ ("blall") | $\bigsqcap$ ("chall") | $\bigwedge$ ("pall") |
| Existential quantifier | | $\exists$ ("blexists") | $\bigsqcup$ ("chexists") | $\bigvee$ ("pexists") |
| Implication | | | $\sqsupset$ ("chimplication") | $\rightarrow$ ("pimplication") |
| Recurrence | | | | ⅄ ("precurrence") |
| Corecurrence | | | | Y ("coprecurrence") |
| Rimplication | | | | ⤚ ("primplication") |
| Refutation | | | | ⤚¬ ("prefutation") |

| | Sequential | Toggling | Branching |
|---|---|---|---|
| Negation | $\neg$ ("not") | | |
| Conjunction | $\triangle$ ("sand") | ⩚ ("tand") | |
| Disjunction | $\triangledown$ ("sor") | ⩛ ("tor") | |
| Universal quantifier | $\bigtriangleup$ ("sall") | ⩚ ("tall") | |
| Existential quantifier | $\bigtriangledown$ ("sexists") | ⩛ ("texists") | |
| Implication | $\triangleright$ ("simplication") | $\succ$ ("timplication") | |
| Recurrence | ⅄ ("srecurrence") | ⅄ ("trecurrence") | ⫯ ("brecurrence") |
| Corecurrence | Y ("cosrecurrence") | Y ("cotrecurrence") | ⫰ ("cobrecurrence") |
| Rimplication | ▷— ("srimplication") | ⤚ ("trimplication") | ○— ("brimplication") |
| Refutation | ▷¬ ("srefutation") | ⤚¬ ("trefutation") | ○¬ ("brefutation") |

The basic operations are the "propositional" connectives: $\neg$, $\wedge$, $\vee$, $\sqcap$, $\sqcup$, ⅄, Y, ⫰, ⫯ along with implication-style connectives $\rightarrow$, $\succ\!\!-$, $\circ\!\!-$ .

Inside this often-referred-as "zoo" by Japaridze (e.g. in [72]), we see the operators of classical logic; our choice of the classical notation for them is no accident. Indeed, classical logic is nothing but the elementary, zero-interactivity fragment of Computability Logic.

After having analysed the relevant definitions, each of the classically-shaped operations, when restricted to elementary games, can be easily seen to be equivalent to the corresponding operator of classical logic[70].

However, in general (not-necessarily-elementary) cases, $\neg$, $\wedge$, $\vee$, $\rightarrow$ become more reminiscent of the corresponding multiplicative operators of linear logic.
Of course, here we are comparing apples with oranges: we need to remind ourselves that linear logic is a syntax, while Computability Logic is a semantics; it may be not clear in what precise sense one can talk about similarities or differences.

In similar fashion, $\sqcap$, $\sqcup$, $\bigsqcap$, $\bigsqcup$ can be perceived as relatives of the additive connectives and quantifiers of linear logic; $\bigwedge$, $\bigvee$ as "multiplicative quantifiers"; ⫰, ⫯, ⅄, Y as "exponentials" - even though it is hard to guess which of the two groups (⫰, ⫯ or ⅄, Y) would be closer to an orthodox linear logician's heart. Still, the blind, sequential and toggling groups of operators have no counterparts in linear logic.

Throughout this section, we are going to see intuitive explanations as well as formal definitions of all of the above-listed operators.

We agree that in the definitions to come, $\Omega$ ranges over any runs, $\Gamma$ over legal runs of the gameframe $G = (Vr, G)$ that is being defined and $e$ ranges over $Vr$-valuations[71].
Each definition has two clauses: one stating when a position is a legal position of the compound game, the other one declaring who wins any given legal run.

70 For instance, if $A$ and $B$ are elementary games, then so is $A \wedge B$, and the latter is exactly the classical conjunction of $A$ and $B$ understood as an elementary game

71 Thus, in these definitions "such that" should be understood as "such that, for all $e$, $\Gamma \ldots$".

This section also provides many examples of particular games.
Let us agree that, unless otherwise suggested by the context, in all those cases we have the ideal universe in mind.

Frequently, we let non-numerals such as people, Turing Machines and so on, act in the roles of "constants". These should be understood as abbreviations of the corresponding decimal numerals that encode these objects in some fixed reasonable encoding. It should also be remembered that algorithmicity is a minimum requirement on $\top$'s strategies.

Some of our examples implicitly assume stronger requirements, such as efficiency or ability to act with imperfect knowledge.
For instance, as Japaridze exemplifies, the problem of telling whether there is or has been life on Mars is, of course, decidable, for this is a finite problem. Yet our knowledge does not allow us to actually solve the problem.

Similarly, chess is a finite game and (after ruling out the possibility of draws) one of the players does have a winning strategy in it. However, we do not specifically know what (and which player's) strategy is a winning one.

When omitting parentheses in compound expressions, we assume that all unary operators (negation, refutations, recurrences, corecurrences and quantifiers) take precedence over all binary operators (conjunctions, disjunctions, implications, rimplications), amongst which implications and rimplications have the lowest precedence[72].

Lastly, as proven in [39], and also partially in [50] and [54], let us not forget that all operators listed in this subsection preserve the static property of games (i.e. when applied to static games, the resulting game is also static).

72 This means that $A \to \neg B \vee C$ should be understood as $A \to ((\neg B) \vee C)$ rather than, say, as $(A \to \neg B) \vee C$ or as $A \to (\neg(B \vee C))$.

---

*Negation operation* $(\neg)$

For a run $\Gamma$, by $\neg\Gamma$ we mean the negative image of $\Gamma$ (with $\top/\bot$ interchanged). For instance, $\neg\langle\bot\alpha, \top\beta, \bot\gamma\rangle = \neg\langle\top\alpha, \bot\beta, \top\gamma\rangle$.

**Definition 1.3.1.1** Let $A = (Vr, A)$ be a gameframe. Then $\neg A$ (read "**not** $A$") is the gameframe $G = (Vr, G)$ such that:

1. $\text{Lr}_e^G = \{\Omega | \neg\Omega \in \text{Lr}_e^A\}$;
2. $\text{Wn}_e^G\langle\Gamma\rangle = \top$ iff $\text{Wn}_e^A\langle\neg\Gamma\rangle = \bot$.

Intuitively, negation is pretty easy to visualise. Take chess, for example. On board $A$ ("Chess"), player one moves white while player two moves black; applying the negation operator $\neg$ to $A$ means taking the board in our hands and rotate it by a full 180° ("¬Chess"). The result is the same game played through inverted roles: player one now moves black, while player two moves white.

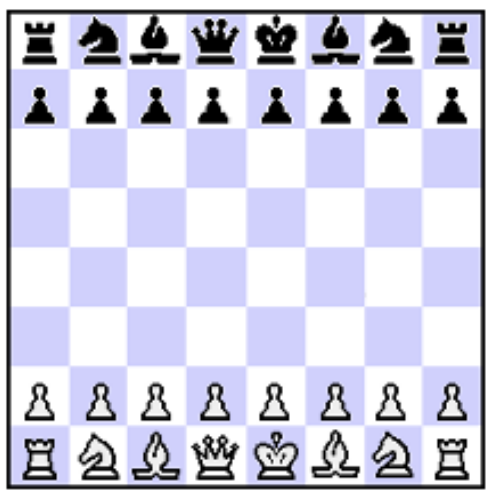

Figure 4: Chess

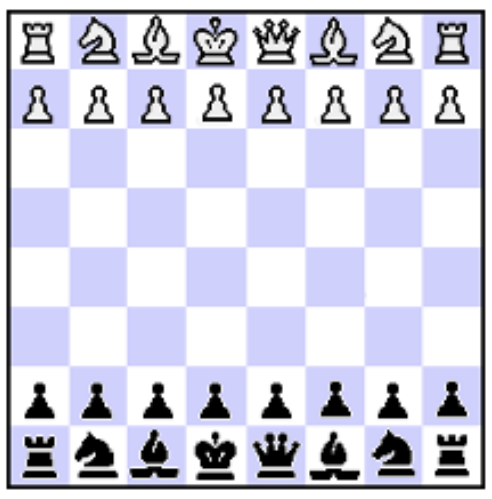

Figure 5: ¬Chess = ssǝɥƆ

As one may have already assessed, the principle of double negation still holds (unlike intuitionistic logic): $\neg\neg A = A$. Indeed, interchanging the players' roles twice restores the original roles of the players.

---

*Choice operations ($\sqcap$, $\sqcup$, $\sqsupset$, $\bigsqcap$, $\bigsqcup$)*

**Choice conjunction** ("chand") $A \sqcap B$ and **choice disjunction** ("chor") $A \sqcup B$ combine games in the following way:

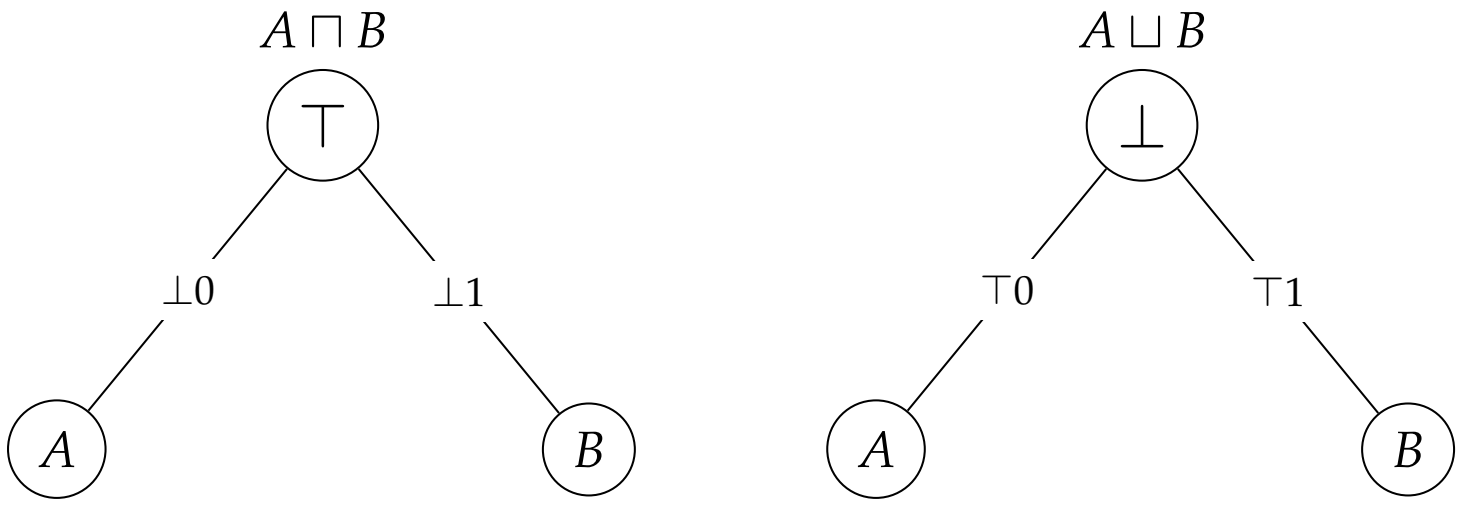

$A \sqcap B$ is a game where, in the initial (empty) position, only Environment has legal moves. Such a move should be either "0" or "1", which simply indicate the number of the choice one makes. If Environment moves 0, the game continues as A (meaning that $\langle 0 \rangle (A \sqcap B) = A$); if it fails to make either move (or "choice"), then it loses.

$A \sqcup B$ is similar, with the only difference that here it is Machine who has initial moves and who loses if no such move is made.

In these two definitions, and in the next to come, since we preferred to avoid a colour-based approach as in [67] and in the lecture notes, we took the liberty to set $i_\perp$ and $i_\top$ for $i$ and $i$, respectively.

**Definition 5.9.a**
Let $A_0 = (Vr_0, A_0)$ and $A_1 = (Vr_1, A_1)$ be gameframes. Then $A_0 \sqcap A_1$ (read "$A_0$ **chand** $A_1$") is the gameframe $G = (Vr_0 \cup Vr_1, G)$ such that:

1. $\mathrm{Lr}^G = \{\langle\rangle\} \cup \{\langle i_\perp, \Omega\rangle | i_\perp \in \{0,1\}, \Omega \in \mathrm{Lr}_e^{A_{i_\perp}}\}$;
2. $\mathrm{Wn}_e^G\langle\rangle = \perp;\ \mathrm{Wn}_e^G\langle i_\perp, \Gamma\rangle = \mathrm{Wn}_e^{A_{i_\perp}}\langle\Gamma\rangle$.

**Definition 1.3.1.2**
Let $A_0 = (Vr_0, A_0)$ and $A_1 = (Vr_1, A_1)$ be gameframes. Then $A_0 \sqcup A_1$ (read "$A_0$ **chor** $A_1$") is the gameframe $G = (Vr_0 \cup Vr_1, G)$ such that:

1. $\mathrm{Lr}^G = \{\langle\rangle\} \cup \{\langle i_\top, \Omega\rangle | i_\top \in \{0,1\}, \Omega \in \mathrm{Lr}_e^{A_{i_\top}}\}$;

2. $\mathrm{Wn}_e^G\langle\rangle = \bot$; $\mathrm{Wn}_e^G\langle i_\top, \Gamma\rangle = \mathrm{Wn}_e^{A_{i_\top}}\langle\Gamma\rangle$.

As one can clearly see, De Morgan classical laws are still valid[73]. Indeed we always have:

$$\neg(A \sqcap B) = \text{∀} \sqcup \text{ꓭ} = \neg A \sqcup \neg B$$
$$\neg(A \sqcup B) = \neg A \sqcap \neg B.$$

Together with the double negation principle, we have that:

$$A \sqcup B = \neg(\neg A \sqcap \neg B)$$
$$A \sqcap B = \neg(\neg A \sqcup \neg B).$$

Similarly, this is also valid for choice quantifiers, which we are going to introduce right away.

Given a game $A(x)$, its **choice universal quantification** ("chall") $\bigsqcap x A(x)$ is nothing but the *infinite choice conjunction* $A(0) \sqcap A(1) \sqcap A(2) \sqcap \ldots$, while its **choice existential quantification** ("chexists") $\bigsqcup x A(x)$ is the *infinite choice disjunction* $A(0) \sqcup A(1) \sqcup A(2) \sqcup \ldots$ .

Visually we have:

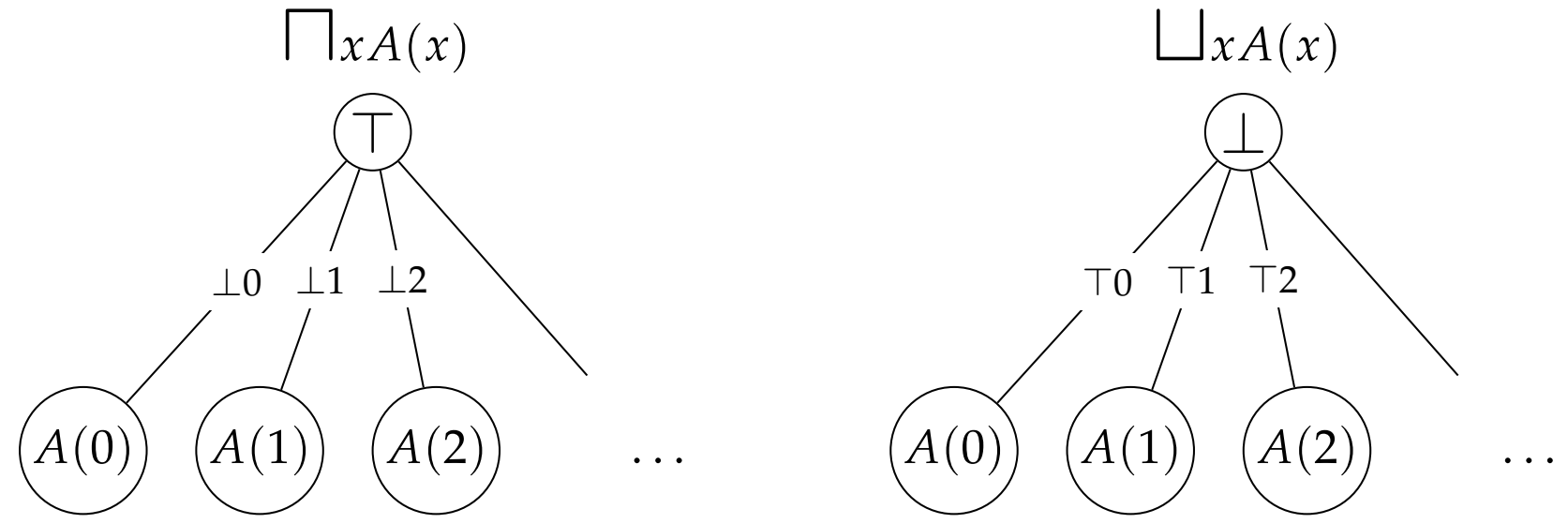

73 Notice the perfect symmetry/duality between $\wedge$ and $\vee$ or $\sqcap$ and $\sqcup$: the definition of each of these operations can be obtained from the definition of its dual by interchanging $\top$ with $\bot$.

Specifically, $\sqcap x A(x)$ is a game where, in the initial position, only Environment has legal moves, and such a move should be one of the constants. If Environment moves $c$, meaning as a certain number move, then the game continues as $A(c)$, and if Environment fails to make an initial move/choice, then it loses.

$\sqcup x A(x)$ is similar, with the difference that Machine has initial moves and loses if no such move is made. So, we always have $\langle c_\perp \rangle \sqcap x A(x) = A(c)$ and $\langle c_\top \rangle \sqcup x A(x) = A(c)$.
Below is a formal definition of all choice operations.

**Definition 1.3.1.3**
Let $A(x) = (Vr, A)$ be a gameframe. Then $\sqcap x A(x)$ (read "**chall** $x$ $A(x)$") is the gameframe $G = (Vr \setminus \{x\}, G)$ such that:

1. $\mathrm{Lr}^G = \{\langle\rangle\} \cup \{\langle c_\perp, \Omega\rangle | c_\perp \in \text{Constants}, \Omega \in \mathrm{Lr}_e^{A_{c_\perp}}\}$;
2. $\mathrm{Wn}_e^G\langle\rangle = \top;\ \mathrm{Wn}_e^G\langle c_\perp, \Gamma\rangle = \mathrm{Wn}_e^{A_{c_\perp}}\langle\Gamma\rangle$.

**Definition 1.3.1.4**
Let $A(x) = (Vr, A)$ be a gameframe. Then $\sqcup x A(x)$ (read "**chexists** $x$ $A(x)$") is the gameframe $G = (Vr \setminus \{x\}, G)$ such that:

1. $\mathrm{Lr}^G = \{\langle\rangle\} \cup \{\langle c_\top, \Omega\rangle | c_\top \in \text{Constants}, \Omega \in \mathrm{Lr}_e^{A_{c_\top}}\}$;
2. $\mathrm{Wn}_e^G\langle\rangle = \perp;\ \mathrm{Wn}_e^G\langle c_\top, \Gamma\rangle = \mathrm{Wn}_e^{A_{c_\top}}\langle\Gamma\rangle$.

Also in this case, De Morgan's laws still hold:

$$\neg \sqcap x A = \sqcup x \underline{V} = \sqcup x \neg A$$
$$\neg \sqcup x A = \sqcap x \neg A$$

Furthermore, with the principle of double negation we obtain:

$$\sqcap x A = \neg \sqcup x \neg A$$
$$\sqcup x A = \neg \sqcap x \neg A$$

Now we are already able to express traditional computational problems using formulas.

Traditional problems come in two forms: the problem of computing a function $f(x)$, or the problem of deciding a predicate $p(x)$. The former can be written as $\sqcap x \sqcup y A(y = f(x))$, and the latter as $\sqcap x(\neg p(x) \sqcup p(x))$.

So, for instance, the "successor" game will be written as $\sqcap x \sqcup y A(y = x + 1)$, and the unary "generalised successor" game as $\sqcap x \sqcup y A(y = x + z)$.

The following game, which is about deciding the "evenness" predicate, could be written as[74]

$$\sqcap x(\neg \exists y(x = 2y) \sqcup \exists y(x = 2y))^{75}.$$

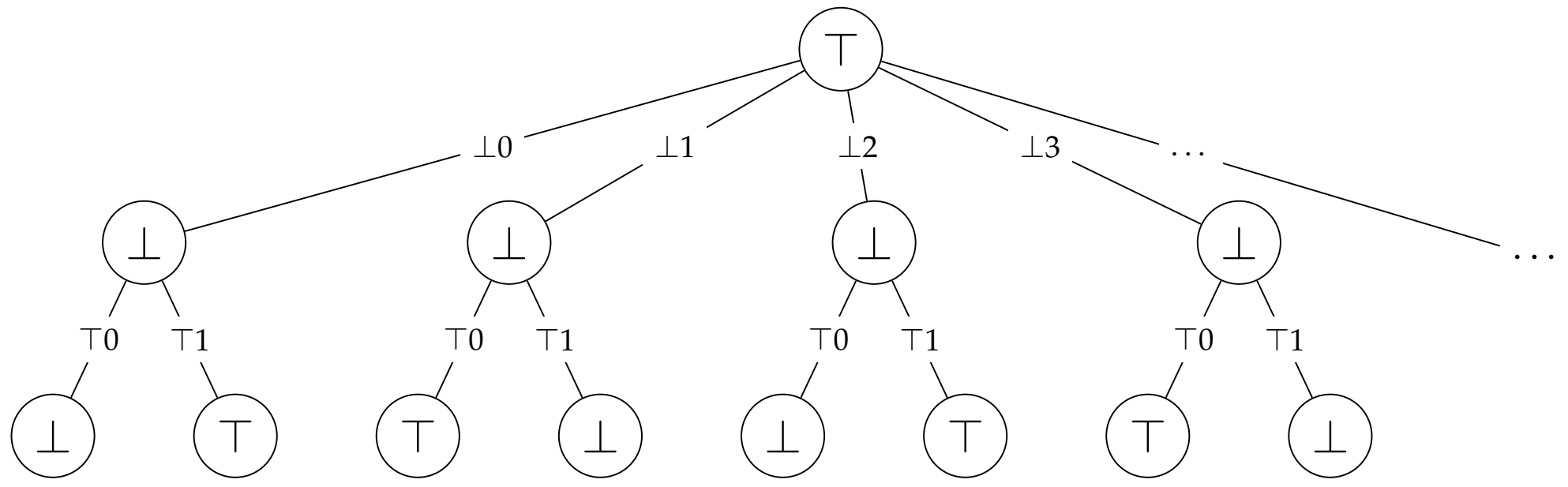

Classical logic has been repeatedly criticised for its operators not being constructive.

Consider, for example, $\forall x \exists y(y = f(x))$. It is true in the classical sense as long as $f$ is a total function. Yet its truth has little (if any) practical import, as $\exists y$ merely signifies *existence of* $y$, without implying that such a $y$ can actually be found. And, indeed, if $f$ is an incomputable function, there is no method for finding $y$.

On the other hand, the choice operations of CoL are constructive. Computability ("truth") of $\sqcap x \sqcup y A(y = f(x))$ means more than just existence of $y$; it means the possibility of actually finding (computing, constructing) a corresponding $y$ for every $x$, as we have already explained at length.

Similarly, let $Halts(x, y)$ be the predicate "Turing Machine $x$ halts on input $y$". Consider:

74 Or, alternatively, simply $\sqcap x(\neg Even(x) \sqcup Even(x))$.

75 $\exists$ will be officially defined later, but, as promised, its meaning is going to be exactly classical when applied to an elementary game like $x = 2y$.

$$\forall x \forall y(\neg Halts(x,y) \vee Halts(x,y)).$$

This is true in classical logic, yet not in a constructive sense.
Its truth means that, for all $x$ and $y$, either $\neg Halts(x,y)$ or $Halts(x,y)$ is true, but it does not imply existence of an actual way to tell which of these two is true after all. And such a way does not really exist, as the Halting problem is undecidable.

This means that

$$\sqcap x \sqcap y(\neg Halts(x,y) \sqcup Halts(x,y))$$

is not computable.
Generally, as pointed out earlier, the principle of the excluded middle "$\neg A$ *OR* $A$", validated by classical logic and causing the indignation of the constructivistically-minded, is not valid in computability logic with *OR* understood as choice disjunction.

The following is an example of a game of the form $\neg A \sqcup A$ with no algorithmic solution:

$$\neg \sqcap x \sqcap y(\neg Halts(x,y) \sqcup Halts(x,y)) \sqcup$$
$$\sqcap x \sqcap y(\neg Halts(x,y) \sqcup Halts(x,y))$$

(Chess $\sqcup$ $\neg$Chess)[76], on the other hand, is an example of a computable-in-principle yet practically incomputable problem, with no real computer anywhere close to being able to handle it[77] - impossible to have full guarantee of success.

There is no need to give a direct definition of the remaining choice operation of **choice implication** ("chimplication") $A \sqsupset B$, for it can be defined in terms of $\neg$ and $\sqcup$ in the standard way[78]:

---

76 Which means choosing between playing white or black and then winning the chosen game - much harder than parallel disjunction, which, taken from classical logic, has two boards onto which try and win in at least one of them.

77 No human and no modern computer can tackle this task with a full guarantee of success. Indeed, probably no future machine will be able to succeed either, even though Chess is a finite game and, theoretically, there should be an algorithmic winning strategy here (excluding the possibility of draw outcomes, of course; they can be considered wins for the black player, for example).

78 Each of the other sorts of disjunctions (parallel, sequential and toggling) generates the corresponding implication the same way.

**Definition 1.3.1.5**
$A \sqsupset B =_{def} \neg A \sqcup B$.

---

*Parallel operations* ($\wedge, \vee, \bigwedge, \bigvee, \rightarrow$, ⅄, ⋎, $\succ\!-$, $\succ\!\neg$)

**Parallel conjunction** ("pand") $A \wedge B$ and **parallel disjunction** ("por") $A \vee B$ are games playing which means playing the two games simultaneously. In order to win in $A \vee B$ (respectively $A \wedge B$), $\top$ needs to win in both (respectively at least one) of the components $A, B$.

For instance, $\neg$Chess $\vee$ Chess is a two-board game, where $\top$ plays black on the left board and white on the right board, and where it needs to win in at least one of the two parallel sessions of chess.
A win can be easily achieved here just by mimicking in Chess the moves that the adversary makes in $\neg$Chess, and vice versa.

This *copycat strategy* guarantees that the positions on the two boards always remain symmetric, as illustrated below, and thus $\top$ eventually loses on one board but wins on the other.

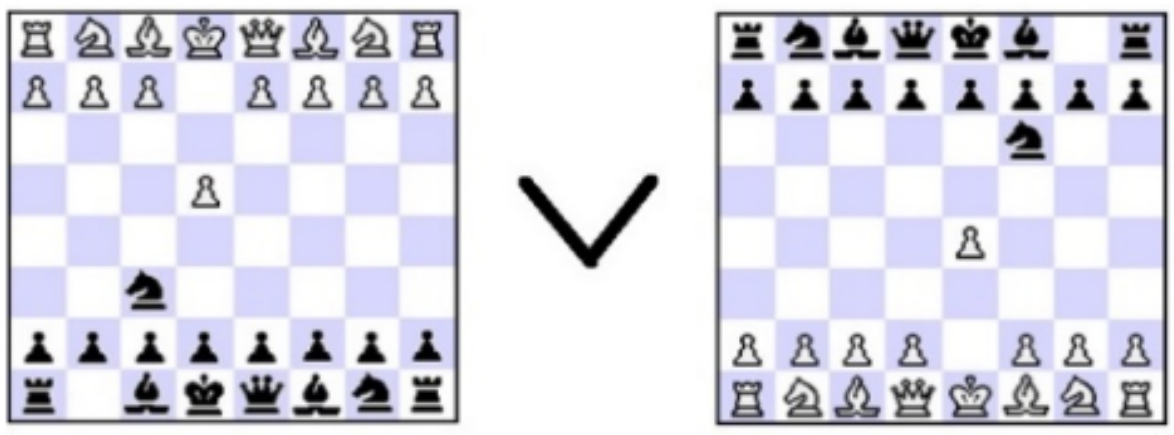

Figure 6: The *copycat strategy* guarantees a win for $\top$ in $\neg$Chess $\vee$ Chess.

This is very different from $\neg$Chess $\sqcup$ Chess. Here $\top$ needs to choose between the two components and then win the chosen one-board game, which makes $\neg$Chess $\sqcup$ Chess essentially as hard to win as either $\neg$Chess or Chess.

Indeed, a game of the form $A \vee B$ is generally easier (at least, not harder) to win than $A \sqcup B$, the latter is easier to win than $A \sqcap B$, and the latter in turn is easier to win than $A \wedge B$.

Technically, a move $\alpha$ in the left (respectively right) $\wedge$-conjunct or $\vee$-disjunct is made by prefixing $\alpha$ with "0.".
For instance, in (the initial position of) $(A \sqcup B) \vee (C \sqcap D)$, the move "1.0" is legal for $\bot$, meaning choosing the left $\sqcap$-conjunct in the second $\vee$-disjunct of the game. If such a move is made, the game continues as $(A \sqcup B) \vee C$. The player $\top$, too, has initial legal moves in $(A \sqcup B) \vee (C \sqcap D)$, which are "0.0" and "0.1".[79]

In the formal definitions of this section and throughout the rest of our topographical-recognition, we use the important notational convention according to which:

**Notation 1.3.1.6**
For a run $\Gamma$ and a string $\alpha$, $\Gamma^{\alpha}$ means the result of removing from $\Gamma$ all moves except those that start with string $\alpha$, and then further deleting the prefix $\alpha$ in the remaining moves[80].

Let us now define the parallel conjunction and disjunction operators.

**Definition 1.3.1.7**
Let $A_0 = (Vr_0, A_0)$ and $A_1 = (Vr_1, A_1)$ be gameframes. Then $A_0 \wedge A_1$ (read "$A_0$ **pand** $A_1$") is the gameframe $G = (Vr_0 \cup Vr_1, G)$ such that:

1. $\Omega \in \mathrm{Lr}_e^G$ iff every move of $\Omega$ starts with 0. or 1. and, for both $i \in \{0,1\}$, $\Omega^{i.} \in \mathrm{Lr}_e^{A_{i}.}$;

2. $\mathrm{Wn}_e^G\langle\Gamma\rangle = \top$ iff $\mathrm{Wn}_e^{A_0}\langle\Gamma^{0.}\rangle = \mathrm{Wn}_e^{A_1}\langle\Gamma^{1.}\rangle = \top$.

**Definition 1.3.1.8**
Let $A_0 = (Vr_0, A_0)$ and $A_1 = (Vr_1, A_1)$ be gameframes. Then $A_0 \vee A_1$ (read "$A_0$ **por** $A_1$") is the gameframe $G = (Vr_0 \cup Vr_1, G)$ such that:

1. $\Omega \in \mathrm{Lr}_e^G$ iff every move of $\Omega$ starts with 0. or 1. and, for both $i \in \{0,1\}$, $\Omega^{i.} \in \mathrm{Lr}_e^{A_{i}.}$;

2. $\mathrm{Wn}_e^G\langle\Gamma\rangle = \bot$ iff $\mathrm{Wn}_e^{A_0}\langle\Gamma^{0.}\rangle = \mathrm{Wn}_e^{A_1}\langle\Gamma^{1.}\rangle = \bot$.

---

79 Thus, a player makes move $\alpha$ in the $i$-th component of a parallel combination of games by prefixing $\alpha$ with "$i$.". Any other moves are considered illegal.
80 For instance $\langle\bot 1.0, \bot 2.1, \top 1.1.2\rangle^{1.} = \langle\bot 0, \top 1.2\rangle$.

Notice now the perfect symmetry between $\wedge$ and $\vee$, leading us to conclude the validity of De Morgan's laws:

$$\neg(A \wedge B) = \neg A \vee \neg B \qquad A \wedge B = \neg(\neg A \vee \neg B)$$
$$\neg(A \vee B) = \neg A \wedge \neg B \qquad A \vee B = \neg(\neg A \wedge \neg B)$$

When $A$ and $B$ are finite games, the depth of $A \wedge B$ or $A \vee B$ is the sum of the depths of $A$ and $B$, as seen below.

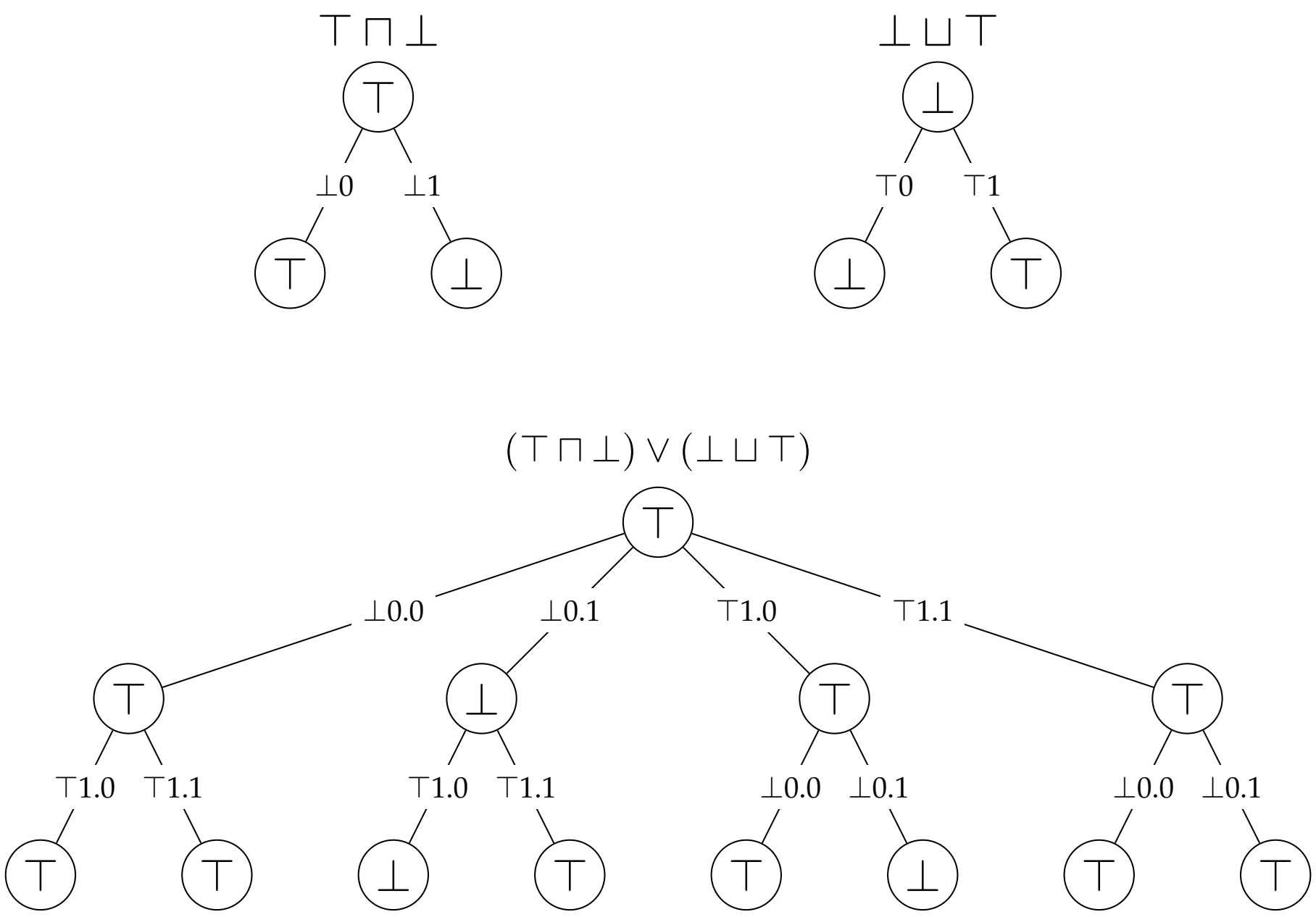

There is an exponential growth of the breadth, meaning that, once we have reached the level of parallel operations, continuing drawing trees in the earlier style becomes tricky[81]. After all, making it possible to express large or infinite-size game trees in a compact way is what our game operators are all about.

Whether trees are either helpful or not in visualising parallel combinations

81 It may be helpful to draw what Japaridze calls **evolution trees**. An evolution tree for a game $G$ is obtained from the game tree of $G$ by replacing every node (position) $\Phi$ with the game $\langle\Phi\rangle G$ to which our game $G$ has evolved in position $\Phi$. Each legal run induces an **evolution sequence** - the sequence of the games from the corresponding branch of the evolution tree. To have a taste of such visualisation, see lecture number 6.

of unistructural games, prefixation is still very much so if we think of each unilegal position $\Phi$ of $A$ as the game $\langle\Phi\rangle A$. Consequently, every unilegal run $\Gamma$ of $A$ becomes a sequence of games, as illustrated in the following example.

Here is an example of how the moves of $\Gamma$ affect and modify the game that is being played. Let us take the game:

$$A = \sqcup x \sqcap y(y \neq x^2) \vee \sqcap x \sqcup y(y = x^2)$$

and the unilegal run $\Gamma = \langle \bot 1.7, \top 0.7, \bot 0.49, \top 1.49 \rangle$.

This is the induced sequence:

1. $\sqcup x \sqcap y(y \neq x^2) \vee \sqcap x \sqcup y(y = x^2)$ i.e. $A$;
2. $\sqcup x \sqcap y(y \neq x^2) \vee \sqcup y(y = 7^2)$ i.e. $\langle \bot 1.7 \rangle A$;
3. $\sqcap y(y \neq 7^2) \vee \sqcup y(y = 7^2)$ i.e. $\langle \bot 1.7, \top 0.7 \rangle A$;
4. $49 \neq 7^2 \vee \sqcup y(y = 7^2)$ i.e. $\langle \bot 1.7, \top 0.7 \bot 0.49 \rangle A$;
5. $49 \neq 7^2 \vee 49 = 7^2$ i.e. $\langle \bot 1.7, \top 0.7 \bot 0.49, \top 1.49 \rangle A$;

The run hits the true proposition $49 \neq 7^2 \vee 49 = 7^2$, hence Machine wins.

As previously mentioned, classical logic is resource-blind: it sees no difference between, say, $A$ and $A \wedge A$. Therefore, the formula $\neg A \vee (A \wedge A)$ is a tautology as $\neg A \vee A$ is.
CoL, on the other hand, is resource-conscious: $A$ is by no means the same as $A \wedge A$ or $A \vee A$. Thus, the principle $\neg A \vee (A \wedge A)$, unlike $\neg A \vee A$, is not valid.[82]

However, there are formulas which are valid in CoL but not provable in linear or affine logic. For example:

82 If we visualise it as three Chess boards, where $\top$ moves black in the first and white in the other "conjuncted" two, we can easily conclude that the copycat strategy between the two disjuncts does not work anymore. Indeed, it is impossible to synchronise the first board with both the other two; even though originally the last two boards are the same game Chess, they may evolve in different ways and thus generate different runs - one won and one lost. The best that $\top$ can do in this three-board game is to synchronise Chess with one of the two conjuncts of Chess $\wedge$ Chess and hope that both Chess and the unmatched Chess are not lost, otherwise the whole game will be lost.

$$((\neg P \vee \neg P) \wedge (\neg P \vee \neg P)) \vee ((P \vee P) \wedge (P \vee P))$$

As a matter of fact, if we imagine each $P$ as a Chess board, as long as the right pairs of disjuncted boards are chosen for mutual synchronisation (matching), the game can be easily won through the copycat strategy.

As we may guess, the **parallel universal quantification** ("pall") $\bigwedge xA(x)$ of $A(x)$ is nothing but $A(0) \wedge A(1) \wedge A(2) \wedge \ldots$, and the **parallel existential quantification** ("pexists") $\bigvee xA(x)$ of $A(x)$ is nothing but $A(0) \vee A(1) \vee A(2) \vee \ldots$ .

**Definition 1.3.1.9**
Let $A(x) = (Vr, A)$ be a gameframe. Then $\bigwedge xA(x)$ (read "**pall** $x$ $A(x)$") is the gameframe $G = (Vr \setminus \{x\}, G)$ such that:

1. $\Omega \in \mathrm{Lr}_e^G$ iff every move of $\Omega$ starts with $c.$ for some $c \in$ Constants and, for all such $c$, $\Omega^{c.} \in \mathrm{Lr}_e^{A_{c.}}$;
2. $\mathrm{Wn}_e^G\langle\Gamma\rangle = \top$ iff, for all $c \in$ Constants, $\mathrm{Wn}_e^{A_c}\langle\Gamma^{c.}\rangle = \top$.

**Definition 1.3.1.10**
Let $A(x) = (Vr, A)$ be a gameframe. Then $\bigvee xA(x)$ (read "**pexists** $x$ $A(x)$") is the gameframe $G = (Vr \setminus \{x\}, G)$ such that:

1. $\Omega \in \mathrm{Lr}_e^G$ iff every move of $\Omega$ starts with $c.$ for some $c \in$ Constants and, for all such $c$, $\Omega^{c.} \in \mathrm{Lr}_e^{A_{c.}}$.
2. $\mathrm{Wn}_e^G\langle\Gamma\rangle = \bot$ iff, for all $c \in$ Constants, $\mathrm{Wn}_e^{A_c}\langle\Gamma^{c.}\rangle = \bot$.

Notice the perfect symmetry between $\wedge$ and $\vee$, $\bigwedge$ and $\bigvee$, $\top$ and $\bot$. Therefore, just as for choice operators, De Morgan's laws still hold:

$$\neg\bigwedge xA = \bigvee x\neg A \qquad \bigwedge xA = \neg\bigvee x\neg A$$
$$\neg\bigvee xA = \bigwedge x\neg A \qquad \bigvee xA = \neg\bigwedge x\neg A$$

The next group of parallel operators are **parallel recurrence** ("precurrence") $⅄A$ and **parallel corecurrence** ("coprecurrence") $\curlyvee A$.

Before skipping to the explanation and definition of these new operators, let us quickly introduce the family of game operations called **recurrence** **operations**, which, as we will see, will also have their own sequential, toggling and branching versions[83] (as seen in the "zoo" introduction).

Generally speaking, when recurrence operators are applied to a game $A$, they turn it into a game playing which means repeatedly playing A.
In terms of resources, recurrence operations generate multiple copies of $A$, thus making $A$ a reusable and recyclable resource.

Recurrence-style operations in classical logic would be meaningless (redundant), since, as we know, it is resource-blind and detects no difference between one and multiple copies of $A$.

However, in Computability Logic, recurrence operations are not only meaningful, but also necessary to achieve a satisfactory level of expressiveness and realise its full potential.

Indeed, hardly any computer program is used only once; rather, it is run over and over again. Loops within such programs also assume multiple repetitions of the same subroutine.
As a result, real-life tasks performed by computers, robots or humans are typically recurring ones or, at least, they entail recurring subtasks to some degree.

Let us formally take into consideration the two recurrence operators we have just introduced.
$\curlywedge A$ is the infinite parallel conjunction $A \wedge A \wedge A \wedge \dots$, while $\curlyvee A$ is the infinite parallel disjunction $A \vee A \vee A \vee \dots$ . Equivalently, $\curlywedge A$ and $\curlyvee A$ can be understood as $\bigwedge x A$ and $\bigvee x A$, where $x$ is a dummy variable on which $A$ does not depend.

Intuitively, playing $\curlywedge A$ means simultaneously playing infinitely many copies of $A$; $\top$ is the winner iff it wins $A$ in all copies.
$\curlyvee A$ is similar, with the only difference that here winning in just one copy is sufficient.

Visually, we may put the $\curlywedge$ operator in terms of Chess boards. Indeed, playing $\curlywedge$Chess means playing, right back from the beginning, possibly

83 Indeed, there are various naturally-emerging recurrence operations. The differences between them are related to how repetition or reusage is exactly understood.

infinitely many Chess sessions in parallel boards, without destroying the old, completed ones[84].

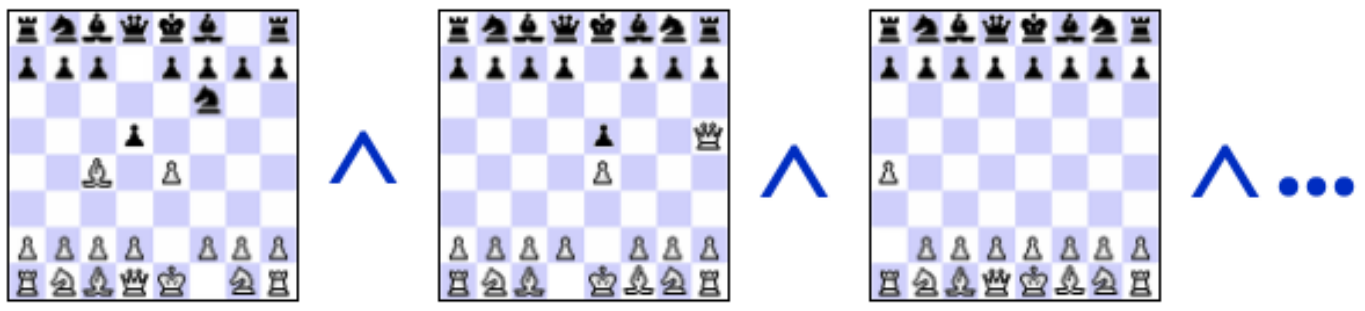

Figure 7: The game $\curlywedge$Chess. If we put $\vee$ instead of every $\wedge$, we have $\curlyvee$Chess.

**Definition 1.3.1.11**
Let $A = (Vr, A)$ be a game. Then $\curlywedge A$ (read "**precurrence** $A$") is the game $G = (Vr, G)$ such that:

1. $\Omega \in \mathrm{Lr}_e^G$ iff every move of $\Omega$ starts with $c.$ for some $c \in$ Constants and, for each such $c$, $\Omega^{c.} \in \mathrm{Lr}_e^{A_c}$.

2. $\mathrm{Wn}_e^G\langle\Gamma\rangle = \top$ iff, for all $c \in$ Constants, $\mathrm{Wn}_e^{A_c}\langle\Gamma^{c.}\rangle = \top$.

**Definition 1.3.1.12**
Let $A = (Vr, A)$ be a game. Then $\curlywedge A$ (read "**precurrence** $A$") is the game $G = (Vr, G)$ such that:

1. $\Omega \in \mathrm{Lr}_e^G$ iff every move of $\Omega$ starts with $c.$ for some $c \in$ Constants and, for each such $c$, $\Omega^{c.} \in \mathrm{Lr}_e^{A_c}$;

2. $\mathrm{Wn}_e^G\langle\Gamma\rangle = \bot$ iff, for all $c \in$ Constants, $\mathrm{Wn}_e^{A_c}\langle\Gamma^{c.}\rangle = \bot$.

As for choice operations, the definition of each parallel operation seen so far can be obtained from the definition of its dual by just interchanging $\top$ with $\bot$.
Thus, we witness a perfect symmetry which allows De Morgan's laws to be valid once again:

$$\neg\curlywedge A = \curlyvee\neg A \qquad \curlywedge A = \neg\curlyvee\neg A$$
$$\neg\curlyvee A = \curlywedge\neg A \qquad \curlyvee A = \neg\curlywedge\neg A$$

84 Which, on the contrary, usually happens on our computers when we play Chess. Selecting a new game means destroying the current board and start playing afresh on a new one.

Indeed, each parallel operation is definable in the standard way in terms of its dual operation and negation.

Three other parallel operations that share this same property are **parallel implication** ("pimplication") $A \rightarrow B$, **parallel rimplication** ("primplication") $A \succ\!\!- B$ and **parallel refutation** ("prefutation") $\succ\!\neg A$. Here, the prefix *p-* stands for parallel, while the *r* of primplication stands for recurrence.

**Definition 1.3.1.13**

- $A \rightarrow B =_{def} \neg A \vee B$;
- $A \succ\!\!- B =_{def} \curlywedge A \vee B$;
- $\succ\!\neg A =_{def} A \succ\!\!- \bot$.

Just like negation, and unlike choice operations, parallel operations preserve the elementary property of games. When restricted to elementary games, the meanings of $\wedge$, $\vee$ and $\rightarrow$ coincide with those of classical conjunction, disjunction and implication.

Furthermore, as long as all individuals of the universe have naming constants, the meanings of $\bigwedge$ and $\bigvee$ coincide with those of classical universal quantifier and existential quantifier.

The same preservation of classical meaning (still without any conditions on the universe) is also the case with the blind quantifiers $\forall$, $\exists$, which will be later defined.
Thus, at the elementary level, when all individuals of the universe have naming constants, $\bigwedge$ and $\bigvee$ are respectively indistinguishable from $\forall$ and $\exists$.
As for the parallel recurrence and corecurrence, for an elementary $A$ we simply have $\curlywedge A = \curlyvee A = A$.

While all classical tautologies automatically remain valid when parallel operators are applied to elementary games, in the general case the class of valid, always computable principles shrinks.
For instance, as already mentioned, $P \rightarrow P \wedge P$, i.e. $\neg P \vee (P \wedge P)$, is not valid.

The principle $P \rightarrow P \wedge P$ is valid in classical logic because the latter sees no difference between $P$ and $P \wedge P$.
On the other hand, in virtue of its resource-conscious semantics, CoL does not validate $P \rightarrow P \wedge P$, unlike $P \succ\!\!- P \wedge P$, which is a valid principle. Here, in the antecedent, we have infinitely many copies of P[85]. Pick any two copies and, via copycat, synchronise them with the two conjuncts of the consequent. A win is guaranteed.

At this point, it seems only reasonable to briefly introduce reductions and their relations with rimplications in order to better understand these operations.

---

*Reductions*

Intuitively, $A \rightarrow B$ is the problem of **reducing** $B$ to $A$: solving $A \rightarrow B$ means solving $B$ while having $A$ as a computational resource.

Specifically, $\top$ may observe how $A$ is being solved (by Environment), and utilise this information in its own solving $B$. As already pointed out, **resources** are symmetric to **problems**: a problem to solve for one player is a resource that the other player can use, and vice versa.

Since $A$ is negated in $\neg A \vee B$ and negation means switching the roles, $A$ is a resource for $\top$ in $A \rightarrow B$, rather than a problem.
Our copycat strategy for $\neg$Chess $\vee$ Chess was an example of reducing Chess to $\neg$Chess. The same strategy was underlying the example of $A = \bigsqcup x \bigsqcap y (y \neq x^2) \vee \bigsqcap x \bigsqcup y (y = x^2)$, page 63, where $\bigsqcap x \bigsqcup y (y = x^2)$ was reduced to itself.

Let us look at a more meaningful example in which the Acceptance problem is reduced to the Halting one, both notably undecidable problems. The former, as a decision problem, will be written as:

$$\bigsqcap x \bigsqcup y (\neg A(x,y) \sqcup A(x,y))$$

with $A(x,y) =$ "Turing Machine $x$ accepts input $y$".

85 Indeed, rimplications are weak sorts of reductions - which we will now tackle, but it felt best to already spoil something in order to grasp it better after. The difference between $\succ\!\!-$, $\circ\!\!-$ (which we will tackle later) and the ordinary reduction $\rightarrow$ is that, in an rimplicative reduction of problem $B$ to problem $A$, the resource $A$ can be (re)used as many times as we like.

Similarly, the Halting problem is written as:

$$\sqcap x \sqcup y(\neg H(x,y) \sqcup H(x,y))$$

with $H(x,y) =$ "Turing Machine $x$ halts on input $y$".

Indeed, neither problem has an algorithmic solution, yet the following pimplication does:

$$\sqcap x \sqcup y(\neg H(x,y) \sqcup H(x,y)) \to \sqcap x \sqcup y(\neg A(x,y) \sqcup A(x,y))$$

Here is $\top$'s winning strategy for the above game. Machine waits until Environment makes moves $\bot 1.m$ and $\bot 1.n$ for some $m$ and $n$. Making these moves essentially means asking the question "Does Machine $m$ accept input $n$?". If such moves are never made, Machine ($\top$) wins. Otherwise, the moves bring the game down to:

$$\sqcap x \sqcup y(\neg H(x,y) \sqcup H(x,y)) \to \neg A(m,n) \sqcup A(m,n)$$

Machine moves $\top 0.m$ and $\top 0.n$, thus asking the counterquestion "Does Machine $m$ halt on input $n$?". $\top$'s moves bring the game down to:

$$\neg H(m,n) \sqcup H(m,n) \to \neg A(m,n) \sqcup A(m,n)$$

Environment will have to answer this question, or else it loses. If it answers by moving $\bot 0.0$ (meaning "No, $m$ does not halt on $n$"), Machine makes the move $\top 1.0$ ("No, $m$ does not accept $n$").
Consequently, the game will be brought down to:

$$\neg H(m,n) \to \neg A(m,n)$$

Indeed, Machine wins: if $m$ does not halt on $n$, then it does not accept $n$, either.

Otherwise, if Environment answers with move $\bot 0.1$ ("Yes, $m$ halts on $n$"), Machine starts simulating $m$ on $n$ until $m$ halts. If $m$ accepted $n$, Machine then makes the move $\top 1.1$ ("Yes, $m$ accepts $n$"); otherwise makes the move $\top 1.0$ ("No, $m$ does not accept $n$").

Of course, it is possible that the simulation goes on forever. However, it would mean that Environment has lied when saying "Yes, $m$ halts on $n$"; in other words, the antecedent is false, and $\top$ still wins.

As a matter of fact, Machine manages to reduce the Acceptance problem to the Halting problem: it solved the former using an external (Environment-provided) solution to the latter.

There are many natural concepts of reduction; we may strongly agree on the thesis that the sort of reduction captured by $\rightarrow$ is the most basic one. For this reason, we agree that if we simply say "reduction" it always means the sort of reduction captured by $\rightarrow$.

A great variety of other reasonable concepts of reduction is expressible in terms of $\rightarrow$. Out of those is **Turing reduction**.
A predicate $q(x)$ is said to be Turing reducible to a predicate $p(x)$ iff $q(x)$ can be decided by a Turing Machine equipped with an **oracle**[86], a resource for $p(x)$. For a positive integer $n$, $n$-bounded Turing reducibility is similarly defined, with the only difference that the oracle is allowed to be used only $n$ times.

Indeed, parallel rimplication and branching implication, which we will later see, are conservative generalisations of such Turing reduction.
Namely, when $p(x)$ and $q(x)$ are elementary games (i.e. predicates), $q(x)$ is Turing reducible to $p(x)$ iff the problem

$$\sqcap x(\neg p(x) \sqcup p(x)) \succ\!\!- \sqcap x(\neg q(x) \sqcup q(x))$$

has an algorithmic solution. If we change $\succ\!\!-$ back to $\rightarrow$ , we get the same result for 1-bounded Turing reducibility[87].

More generally, as one might guess, $n$-bounded Turing reduction will be captured by:

$$\sqcap x(\neg p(x_1) \sqcup p(x_1)) \wedge \ldots \wedge \sqcap x(\neg p(x_n) \sqcup p(x_n)) \rightarrow \sqcap x(\neg q(x) \sqcup q(x))$$

86 In theory of computation, an oracle is an ideal machine, usually modelled as a black box, able to answer specific *Entscheidungsprobleme*. In the case of Turing Machines, we can visualise it as an extra tape on which the query is typed and the answer is instantaneously read. The machine can, thus, pause its computation to query the oracle and receive an answer, which is used to continue its computation. Indeed, the oracle is assumed to be able to solve certain problems that might be difficult or impossible for a standard Turing machine.

87 In other words, when restricted to the traditional sorts of problems (such as deciding a predicate or computing a function), the Turing reducibility of $B$ to $A$ coincides with the computability of $A \succ\!\!- B$ as well as of $A \circ\!\!- B$ (which we will later see). The differences between $\succ\!\!-$ and $\circ\!\!-$ become more relevant when these operations are applied to non-traditional, properly interactive problems.

If, instead, we write:

$$\sqcap x_1 \ldots \sqcap x_n((\neg p(x_1) \sqcup p(x_1)) \wedge \ldots \wedge (\neg p(x_n) \sqcup p(x_n))) \to \sqcap x(\neg q(x) \sqcup q(x))$$

we get a conservative generalisation of $n$-bounded **weak truth-table reduction**. The latter differs from $n$-bounded Turing reduction in that all $n$ oracle-queries should be made at once, before seeing responses to any of those queries.

What is called **mapping** (or **many-one**) **reducibility** of $q(x)$ to $p(x)$ is nothing but computability of $\sqcap x \bigsqcup y(q(x) \longleftrightarrow p(y))$, where $A \longleftrightarrow B$ abbreviates $(A \to B) \wedge (B \to A)$.

The list goes on and on. However, many other natural concepts of reduction expressible in the language of CoL may have no established names in the literature.

For instance, we can show that a certain reducibility-style relation holds between the previously mentioned predicates $A(x,y)$ and $H(x,y)$ in an even stronger sense than the algorithmic winnability of:

$$\sqcap x \bigsqcup y(\neg H(x,y) \sqcup H(x,y)) \to \sqcap x \bigsqcup y(\neg A(x,y) \sqcup A(x,y)).$$

As a matter of fact, this problem has an algorithmic solution; however, also the generally harder-to-solve problem does:

$$\sqcap x \bigsqcup y(\neg H(x,y) \sqcup H(x,y)) \to \neg A(x,y) \sqcup A(x,y).$$

Indeed, Computability Logic provides the suitable formalism and deductive machinery to systematically express and study computation-theoretic relations such as reducibility, decidability, enumerability and so on, taking into account all kinds of variations of such concepts.

Let us now continue where we left our reducibility account off.
While the standard approaches only allow us to talk about reducibility as relation between problems, reduction becomes an operation on problems in CoL, with reducibility as relation simply meaning computability of the corresponding combination (such as $A \to B$) of games.

Same goes for other relations and properties already mentioned, such as the property of decidability. Indeed, this property becomes the operation of deciding whether we define the problem of deciding a predicate (or any computational problem) $p(x)$ as the game $\sqcap x(\neg p(x) \sqcup p(x))$ or not.

Consequently, we are now able to ask meaningful questions such as: "Is the reduction of the problem of deciding $q(x)$ to the problem of deciding $p(x)$ always reducible to the mapping reduction of $q(x)$ to $p(x)$?".
This question is equivalent to asking whether the following formula is valid or not in CoL:

$$\sqcap x \sqcup y(q(x) \longleftrightarrow p(y)) \to (\sqcap x(\neg p(x) \sqcup p(x)) \to \sqcap x(\neg q(x) \sqcup q(x))).$$

The answer turns out to be "Yes", meaning that mapping reduction is at least as strong as reduction.

Here is a strategy which wins this game no matter what particular predicates $p(x)$ and $q(x)$ are:

1. Wait until, for some $m$, Environment brings the game down to:
   $\sqcap x \sqcup y(q(x) \longleftrightarrow p(y)) \to (\sqcap x(\neg p(x) \sqcup p(x)) \to \neg q(m) \sqcup q(m));$
2. Bring the game down to:
   $\sqcup y(q(m) \longleftrightarrow p(y)) \to (\sqcap x(\neg p(x) \sqcup p(x)) \to \neg q(x) \sqcup q(x));$
3. Wait until, for some $n$, Environment brings the game down to:
   $(q(m) \longleftrightarrow p(n)) \to (\sqcap x(\neg p(x) \sqcup p(x)) \to \neg q(x) \sqcup q(x));$
4. Bring the game down to:
   $(q(m) \longleftrightarrow p(n)) \to (\neg p(n) \sqcup p(n) \to \neg q(x) \sqcup q(x));$
5. Wait until Environment brings the game down to one of the following:
   - $(q(m) \longleftrightarrow p(n)) \to (\neg p(n) \to \neg q(x) \sqcup q(x));$

   In this case, further bring it down to:

   $$(q(m) \longleftrightarrow p(n)) \to (\neg p(n) \to \neg q(x))$$

   and you have won.
   - $(q(m) \longleftrightarrow p(n)) \to (p(n) \to \neg q(x) \sqcup q(x));$

   In this case, further bring it down to:

   $$(q(m) \longleftrightarrow p(n)) \to (p(n) \to q(x))$$

   and you have won, again.

We may also ask: "Is the mapping reduction of $q(x)$ to $p(x)$ always reducible to the reduction of the problem of deciding $q(x)$ to the problem of deciding $p(x)$?".
This question is equivalent to asking whether the following formula is valid or not in CoL:

$$(\sqcap x(\neg p(x) \sqcup p(x)) \to \sqcap x(\neg q(x) \sqcup q(x))) \to \sqcap x \sqcup y(q(x) \longleftrightarrow p(y)).$$

The answer here turns out to be "No", meaning that mapping reduction is properly stronger than reduction.
This can be easily proven by showing that the above formula is not provable in the deductive system CL4[88], which we are introducing later.

To summarise, CoL offers not only a convenient language for specifying computational problems and relations or operations on them, but also a systematic tool for answering questions in the above style and beyond.

Let us now return to our syntax overview and move onto Blind operators.

---

*Blind operations* $(\forall, \exists)$

Our definition of **blind universal quantifier** ("blall") $\forall x$ and **blind existential quantifier** ("blexists") $\exists x$ assumes that the game $A(x)$ to which they are applied satisfies the weaker condition of unistructurality in $x$ - as previously defined.

Intuitively, unistructurality in $x$ means that the structure of the game does not depend on (how the valuation evaluates) the variable $x$. Every unistructural game is also unistructural in $x$, but not vice versa.

Informally, playing $\forall x A(x)$ or $\exists x A(x)$ means just playing $A(x)$ "blindly", without knowing the actual value of $x$.
In $\forall x A(x)$, Machine must win the game for every possible value of $x$ from the domain, while in $\exists x A(x)$ is sufficient that Machine succeeds

88 CL4 is sound and complete with respect to validity. Its completeness implies that any formula which is not provable in it (such as the above formula) is not valid. And the soundness of CL4 implies that every provable formula is valid. Indeed, had our *ad hoc* attempt to come up with a strategy for $\sqcap x \sqcup y(q(x) \longleftrightarrow p(y)) \to (\sqcap x(\neg p(x) \sqcup p(x)) \to \sqcap x(\neg q(x) \sqcup q(x)))$ failed, its validity could have been easily established by finding a CL4-proof of it.

for just one value.

When applied to elementary games, blind quantifiers act exactly like classical logic quantifiers, even though not all individuals of the universe may have naming constants.

**Definition 1.3.1.14**
Let $A(x) = (Vr, A)$ be a $x$-unistructural game. Then $\forall x A(x)$ (read "**blall** $x$ $A(x)$") is the game $G = (Vr \smallsetminus \{x\}, G)$ such that:

1. $\Omega \in \mathrm{Lr}_e^G$ iff every (some) constant $c$, $\Omega \in \mathrm{Lr}_e^{A_c}$;

2. $\mathrm{Wn}_e^G\langle\Gamma\rangle = \top$ iff, for all constants $c$, $\mathrm{Wn}_e^{A_c}\langle\Gamma\rangle = \top$.

**Definition 1.3.1.15**
Let $A(x) = (Vr, A)$ be a $x$-unistructural game. Then $\exists x A(x)$ (read "**blexists** $x$ $A(x)$") is the game $G = (Vr \smallsetminus \{x\}, G)$ such that:

1. $\Omega \in \mathrm{Lr}_e^G$ iff every (some) constant $c$, $\Omega \in \mathrm{Lr}_e^{A_c}$;

2. $\mathrm{Wn}_e^G\langle\Gamma\rangle = \bot$ iff, for all constants $c$, $\mathrm{Wn}_e^{A_c}\langle\Gamma\rangle = \bot$.

Per usual, there lies a perfect symmetry between $\forall$ and $\exists$. Therefore, as with the other quantifiers seen so far, the standard De Morgan laws and interdefinabilities hold:

$$\begin{array}{ll} \neg\forall x A = \exists x \neg A & \forall x A = \neg\exists\neg A \\ \neg\exists x A = \forall x \neg A & \exists x A = \neg\forall x \neg A \end{array}$$

Unlike $\bigwedge x A(x)$, which is a game on infinitely many boards, both $\sqcap x A(x)$ and $\forall x A(x)$ are one-board games. Yet, they are very different from each other.

To see this difference, let us compare the problems $\sqcap x(Even(x) \sqcup Odd(x))$ and $\forall x(Even(x) \sqcup Odd(x))$.

The former is an easily winnable game of depth 2: Environment selects a number and Machine tells whether that number is even or odd. Evenness is, indeed, a decidable problem.

The latter, on the other hand, is a game which is impossible to win. This is a game of depth 1, where the value of $x$ is not specified by either player, and only Machine moves by choosing whether (the unknown) $x$ is even or odd. Whatever Machine says, it loses, because there is always a value for $x$ that makes the answer wrong.

This suggests that nontrivial $\forall$-games can never be won.

For instance, the problem:

$$\forall x(Even(x) \sqcup Odd(x) \to \sqcap y(Even(x+y) \sqcup Odd(x+y)))$$

has an easy solution.
The idea for a winning strategy is that, for any given $y$, in order to tell whether $x+y$ is even or odd, it is not really necessary to know the value of $x$. Indeed, simply knowing whether $x$ is even or odd is sufficient; moreover, we can assess this just from the antecedent.

In other words, for any known $y$ and unknown $x$, the problem of telling whether $x+y$ is even or odd is reducible to the problem of telling whether $x$ is even or odd. Specifically, if both $x$ and $y$ are even or both are odd, then $x+y$ is even; otherwise $x+y$ is odd.

Below is the evolution sequence induced by the run $\langle \bot 1.5, \bot 0.0, \top 1.1 \rangle$, where Machine uses such strategy:

1. $\forall x(Even(x) \sqcup Odd(x) \to \sqcap y(Even(x+y) \sqcup Odd(x+y)))$;
2. $\forall x(Even(x) \sqcup Odd(x) \to Even(x+5) \sqcup Odd(x+5)))$;
3. $\forall x(Even(x) \to Even(x+5) \sqcup Odd(x+5))$;
4. $\forall x(Even(x) \to Odd(x+5))$.

Machine wins because the play hits the true $\forall x(Even(x) \to Odd(x+5))$.

Notice how $\forall x$ persisted throughout the sequence. Generally, the $(\forall, \exists)$-structure of a game will remain unchanged in such sequences. Same goes for parallel operations such as $\to$ in this case.

We may put in relation our quantifiers as such[89]:

89 Parallel and blind quantifiers are, indeed, incomparable since the different number of game boards they each assume.

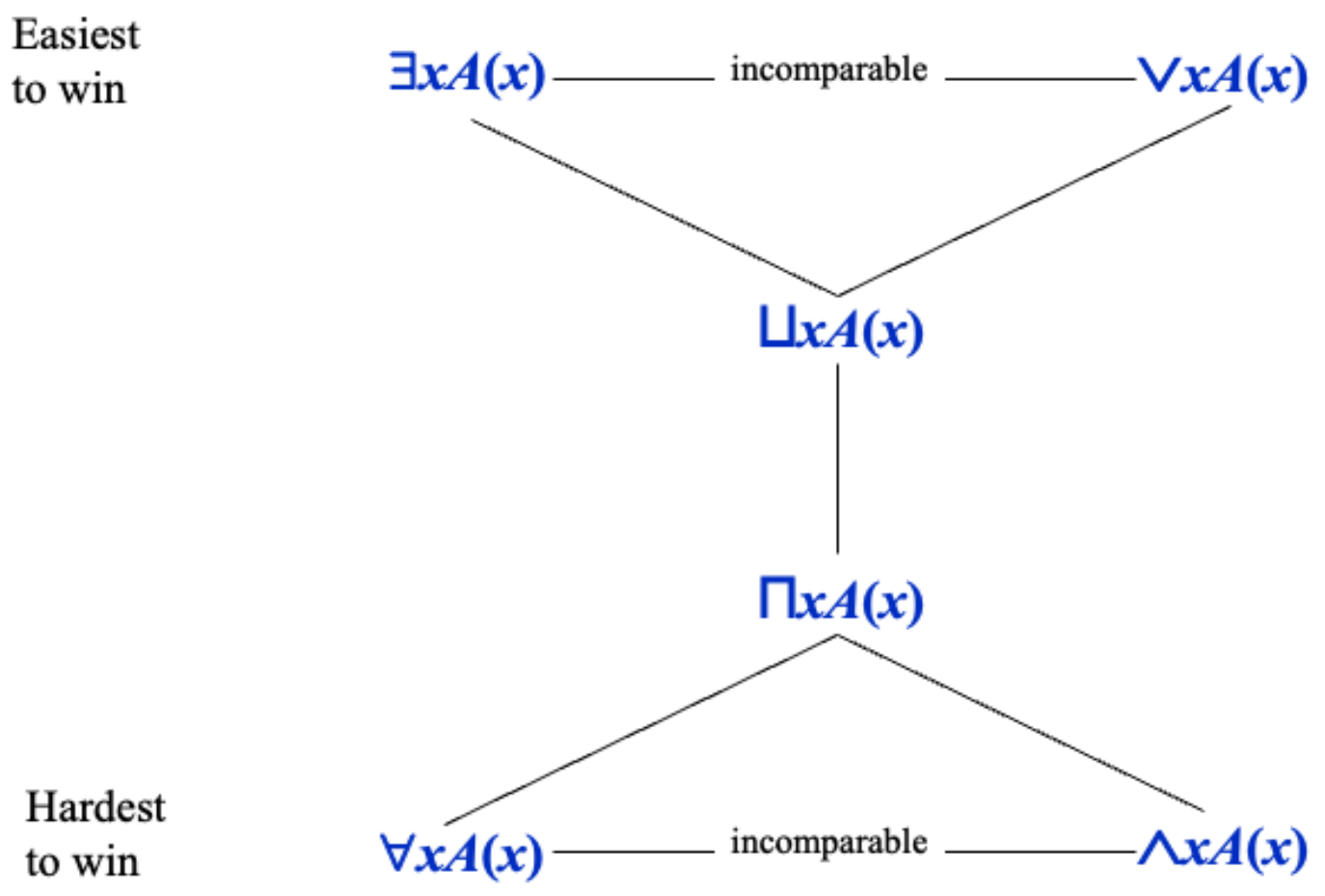

Figure 8: Hierarchy of quantifiers, from lecture number 8.

Furthermore, turning a game tree for a unistructural gameframe $A$ into a gameframe tree for $\forall xA$ just means prefixing each node with "$\forall x$". This also applies to $\exists xA$.

Here is an example:

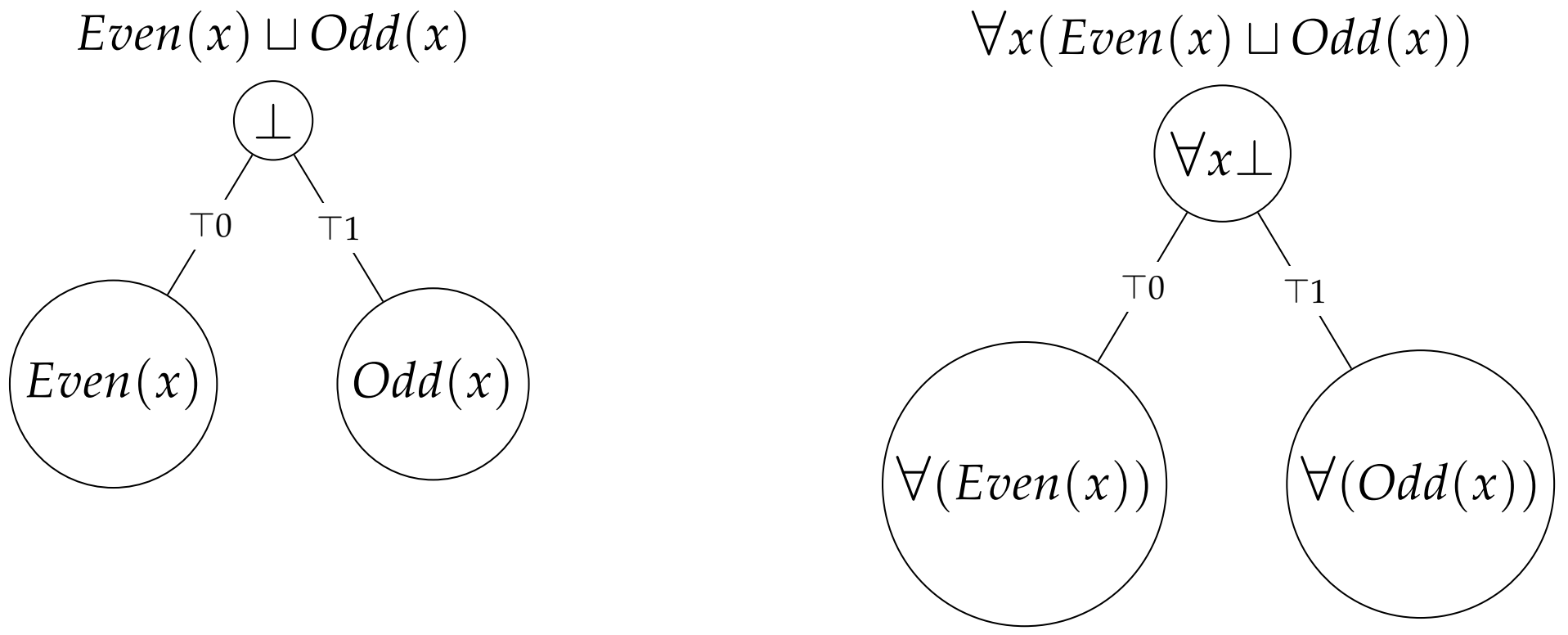

---

*Branching operations* (⫰, ⫯, ○—, ○⌐)

Up to this point, we have generally introduced recurrence operations by examining the parallel one.

We now introduce **branching recurrence** ("brecurrence") ⫰$A$ as the operator that, just as ⅄, allows the user to start new sessions of the same game $A$ without destroying old ones.
Furthermore, ⫰ makes it possible to branch and replicate any particular stage of any particular session, i.e. to create any number of copies of any already reached positions in the multiple, parallel plays of $A$. This allows the user to test different evolutions beginning from the same position.

Intuitively, the difference between ⫰ and ⅄ is that in ⫰$A$, unlike ⅄$A$, $\perp$ does not have to restart $A$ from the very beginning each time it wants to reuse it as a resource.
Indeed, Environment is allowed to backtrack to any of the previous, not-necessarily-starting positions and try a new continuation from there, all the while depriving Machine of the possibility to reconsider the moves it has already made in that position.

As a matter of fact, this is the type of reusage every purely software resource allows - or, at least, would allow - to an advanced operating system with unlimited memory.
One can start running process $A$; then fork it at any stage and create two threads that have a common past but possibly diverging futures (with the possibility of treating one of the threads as a backup copy and preserve it for backtracking purposes); then further fork any of the branches at any point, and so on.

The less flexible type of reusage of $A$ assumed in ⅄$A$, on the other hand, as Japaridze explains, is closer to what infinitely many autonomous physical resources would naturally offer, such as an unlimited number of limited-memory computers each one performing task $A$ and only capable of running a single thread of it.
Indeed, the effect of forking a certain advanced stage of $A$ cannot be achieved here, unless, by slim chance, there are two identical copies of such stage, meaning that the corresponding two computers have acted precisely the same way so far.

The formal definitions of ⫰$A$ and its dual **branching corecurrence** ("cobrecurrence") ⫯$A(= \neg$⫰$\neg A)$ in early papers on CoL, as [39] and [48], were direct formalisations of the above intuitions, with an explicit presence of **replicative moves** used by players to fork a given thread of $A$ and create two threads out of one.
Later, in [58], another definition of brecurrence was found. Furthermore, it was proven to be equivalent to the old one in the sense of mutual reducibility of the old and the new versions of ⫰$A$ (and, consequently, ⫯$A$).

These new definitions less directly correspond to the above intuitions, but are technically simpler, thus we choose them as our canonical definitions of branching operations.

First, we agree on the following:

**Notation 1.3.1.16**
Where $\langle\Gamma\rangle$ is a run and $w$ is a **bitstring**, meaning a finite or infinite sequence of 0s and 1s, $\Gamma^{\leq w}$ means the result of deleting from $\Gamma$ all moves except those that look like $u.\beta$ for some initial segment $u$ of $w$, and then further deleting the prefix "$u.$" from such moves.

**Definition 1.3.1.17**
Let $A = (Vr, A)$ be a gameframe. Then ⫰$A$ (read "**brecurrence** $A$")is the game $G = (Vr, G)$ such that:

1. $\Omega \in \mathrm{Lr}_e^G$ iff every move of $\Omega$ has the prefix '$u.$' for some finite bitstring $u$, and, for every infinite bitstring $w$, $\Omega^{\leq w} \in \mathrm{Lr}_e^A$;
2. $\mathrm{Wn}_e^G\langle\Gamma\rangle = \top$ iff, for every infinite bitstring $w$, $\mathrm{Wn}_e^A\langle\Gamma^{\leq w}\rangle = \top$.

**Definition 1.3.1.18**
Let $A = (Vr, A)$ be a gameframe. Then ⫯$A$ (read "**cobrecurrence** $A$")is the game $G = (Vr, G)$ such that:

1. $\Omega \in \mathrm{Lr}_e^G$ iff every move of $\Omega$ has the prefix '$u.$' for some finite bitstring $u$, and, for every infinite bitstring $w$, $\Omega^{\leq w} \in \mathrm{Lr}_e^A$;
2. $\mathrm{Wn}_e^G\langle\Gamma\rangle = \bot$ iff, for every infinite bitstring $w$, $\mathrm{Wn}_e^A\langle\Gamma^{\leq w}\rangle = \bot$.

The direct intuitions underlying the above definitions are as follows.

To play ⫰$A$ or ⫯$A$ means to simultaneously play in multiple, parallel copies (threads) of $A$.
Each such thread is denoted by an infinite bitstring $w$ (consequently, there are uncountably many threads).

Every legal move by either player looks like $u.\beta$[90] for some finite bitstring $u$; this means that the move $\beta$ is simultaneously made in all threads $w$ such that $u$ is an initial segment of $w$.
As a result, $\Gamma^{\leq w}$ is the run of a given thread $w$ of $A$, where $\Gamma$ is the overall run of ⫰$A$ or ⫯$A$.

In order to win ⫰$A$, Machine needs to win $A$ in all threads, while for winning ⫯$A$ it is sufficient to win in just one thread.

Each game tree starts off with only one root, whose bitstring (address) is the **empty string** $\varepsilon$. Creating new branches (threads) in ⫰$A$ is exclusively Environment's privilege (while in a ⫯$A$ it is Machine's).

A position represented by a bitstring $w$[91] is branched by the replicative move "$\perp w$ :". Consequently, the leaf $w$ is split into two leaves $w0$ and $w1$, thus creating two branches out of one. The positions in these two new leaves will be the same as the position in the old leaf.

However, we also need to understand how this extends to infinite runs as well. Indeed, we need a more general characterisation of branching game trees.

Let us call the binary trees seen so far **bitstring trees**. Each branch of such a tree can be understood as the corresponding bit string.
Branches may be finite or infinite. Of course, infinite branches can only be generated by infinite runs where replicative moves have been made infinitely many times.
The branches that are not initial segments of some other, longer branches are said to be **complete**.

Here is an example of a ⫰-game tree taken from lecture number 9. We will attempt to show the dynamic evolution step by step just like Japaridze has done slide by slide. Machine and Environment are playing ⫰Chess:

90 Indeed, every time a player makes a (legal) move $\alpha$ of $A$, it should indicate for which of the many positions of $A$ the move is meant. This is done by prefixing $\alpha$ with "$w$.".
91 Note: $w$, in a replicative move of the form $w$ :, cannot be an internal node of the game tree: it has to be a leaf. On the other hand, in a **non-replicative (ordinary) move** $w.\alpha$, it does not have to be a leaf. If $w$ is a leaf, $w.\alpha$ means making the move $\alpha$ in the position of $A$ that is found at that leaf. If, otherwise, $w$ is an internal node, $w.\alpha$ means making the move $\alpha$ in all leaves that descend from $w$.

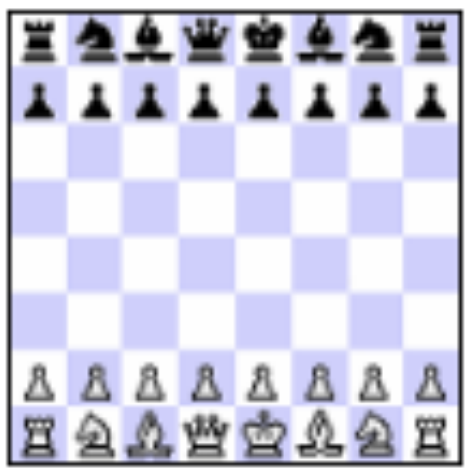

Figure 9: The two players start off with a clear board.

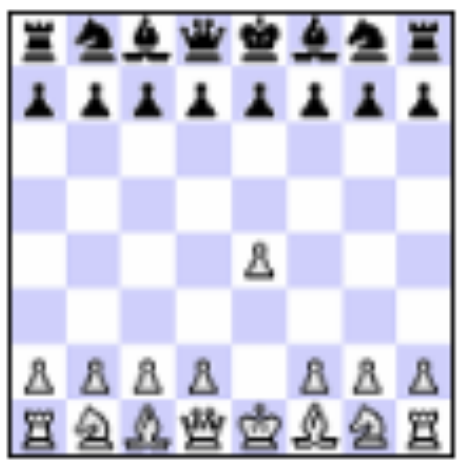

Figure 10: Machine opens with E2E4, thus the bitstring will read: $\top\varepsilon$.E2E4.

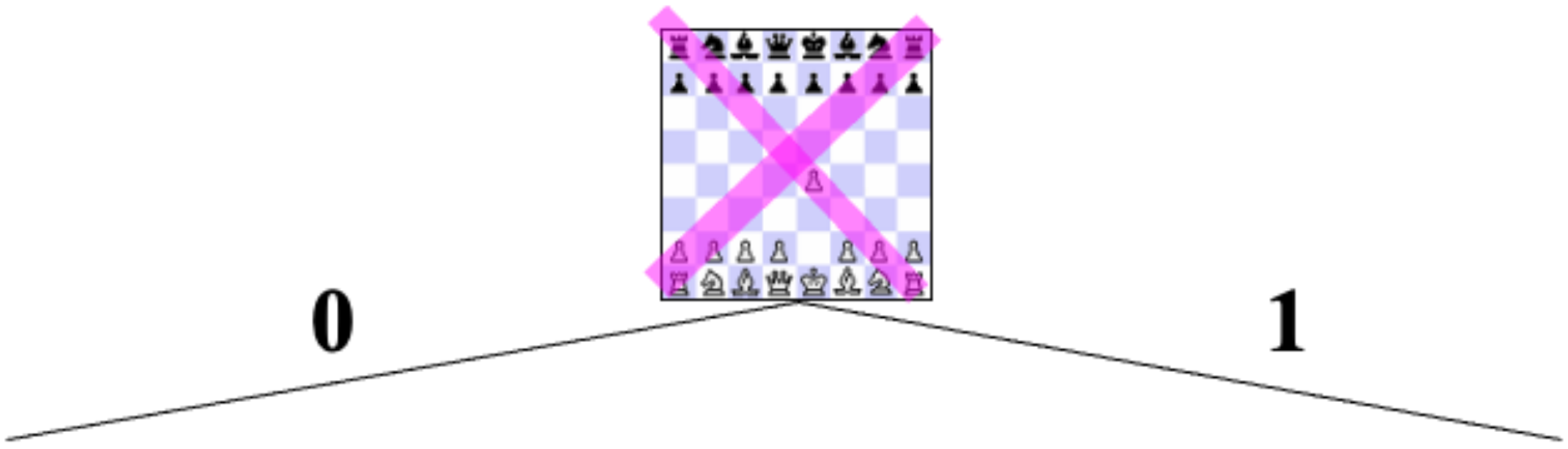

Figure 11: Environment uses a replicative move to create two branches from the same root. The bitstring now is: $\top\varepsilon$.E2E4, $\bot\varepsilon$ : .

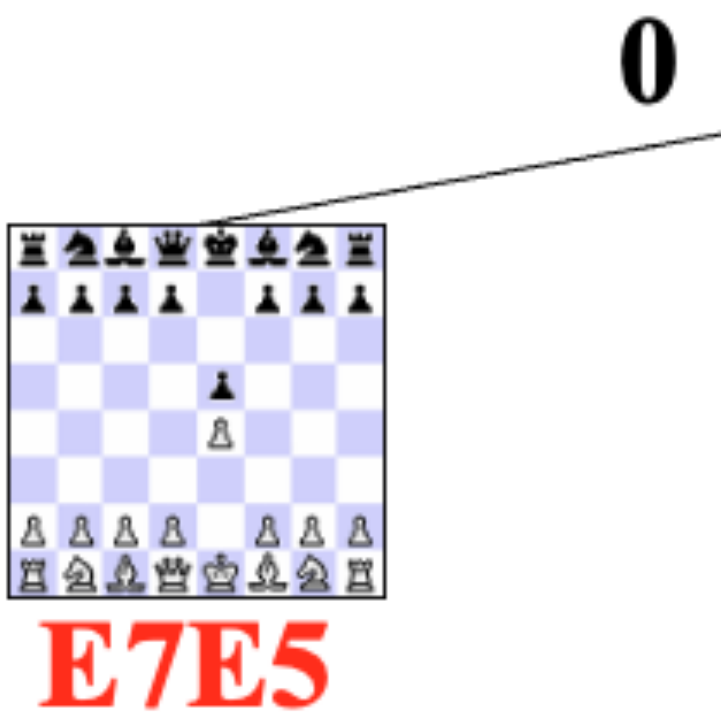

Figure 12: Environment moves E7E5 on the board in branch 0. The bitstring now is: ⊤ε.E2E4, ⊥ε :, ⊥0.E7E5.

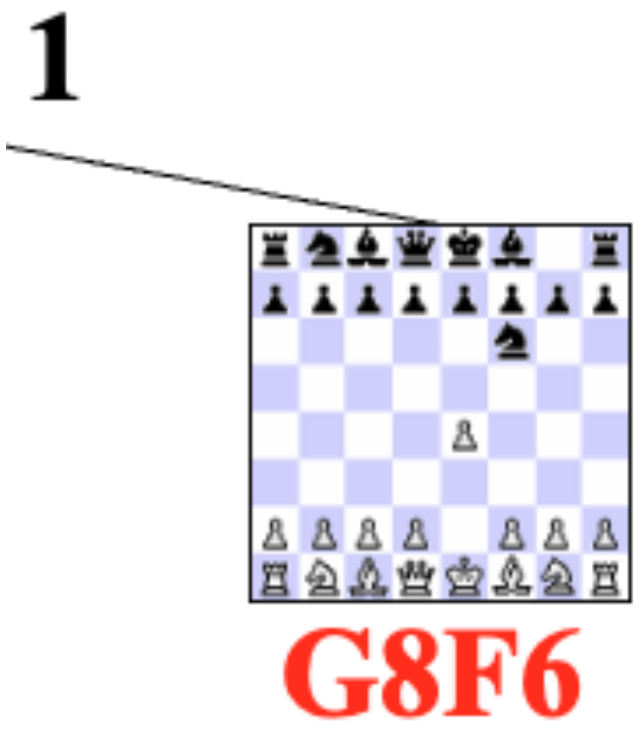

Figure 13: Environment moves G8F6 on the board of branch 1. The bitstring now reads: ⊤ε.E2E4, ⊥ε :, ⊥0.E7E5, ⊥1.G8F6.

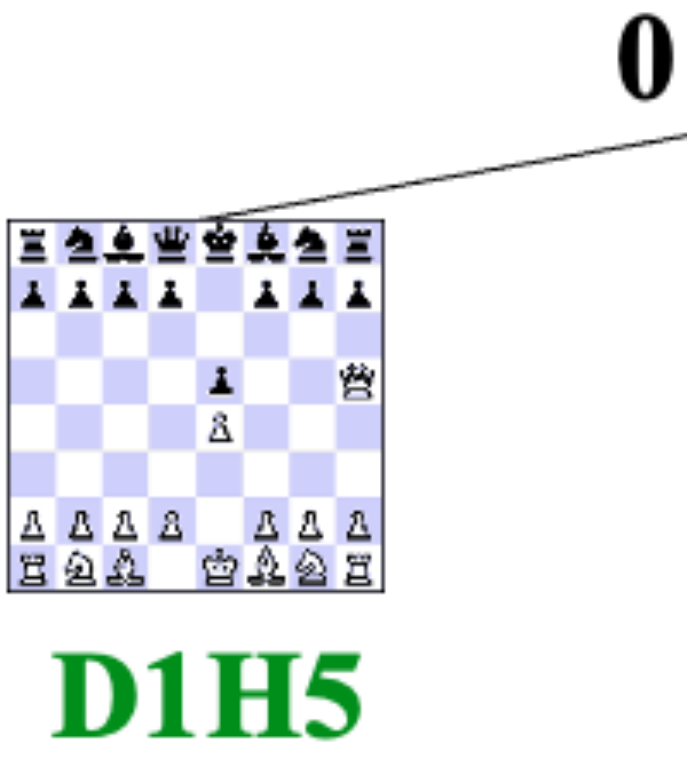

Figure 14: Machine moves D1H5 on the first board. The bitstring is: ⊤ε.E2E4, ⊥ε :, ⊥0.E7E5, ⊥1.G8F6, ⊤0.D1H5.

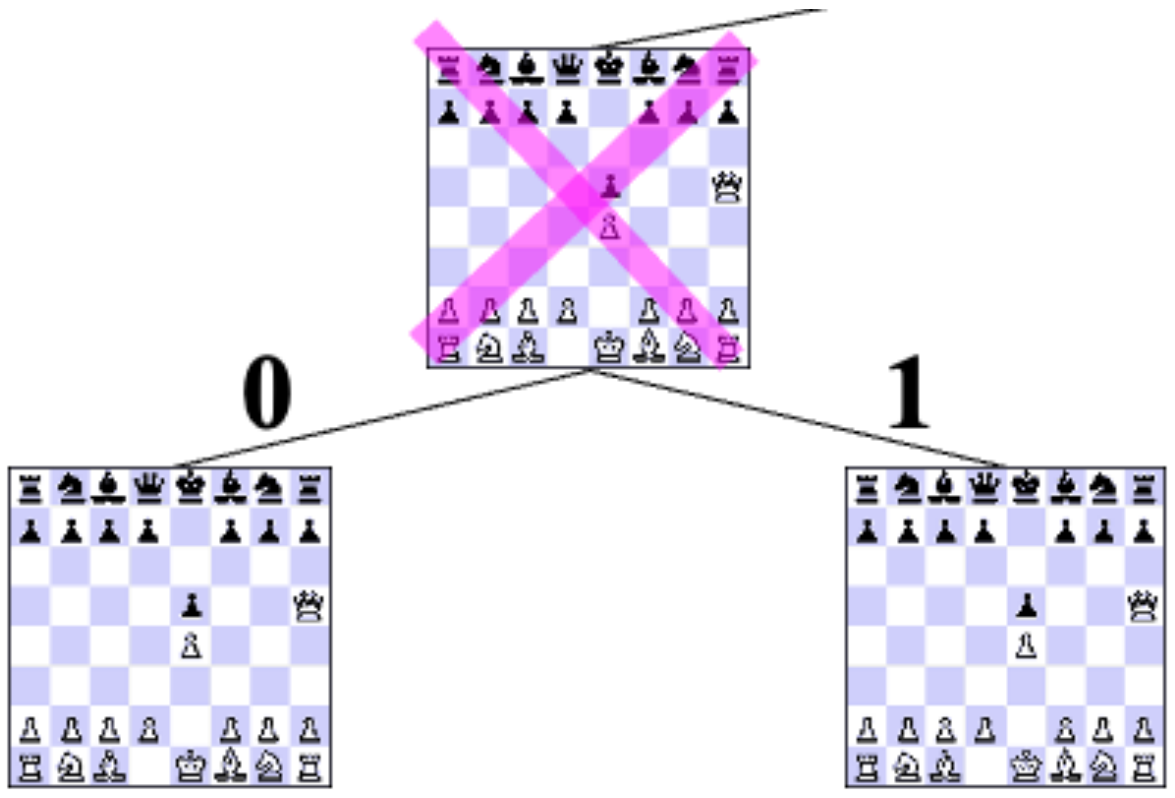

Figure 15: Environment makes a replicative move from the board in branch 0. The bitstring becomes: ⊤ε.E2E4, ⊥ε :, ⊥0.E7E5, ⊥1.G8F6, ⊤0.D1H5, ⊥0: .

And so on and so forth, until we reach:

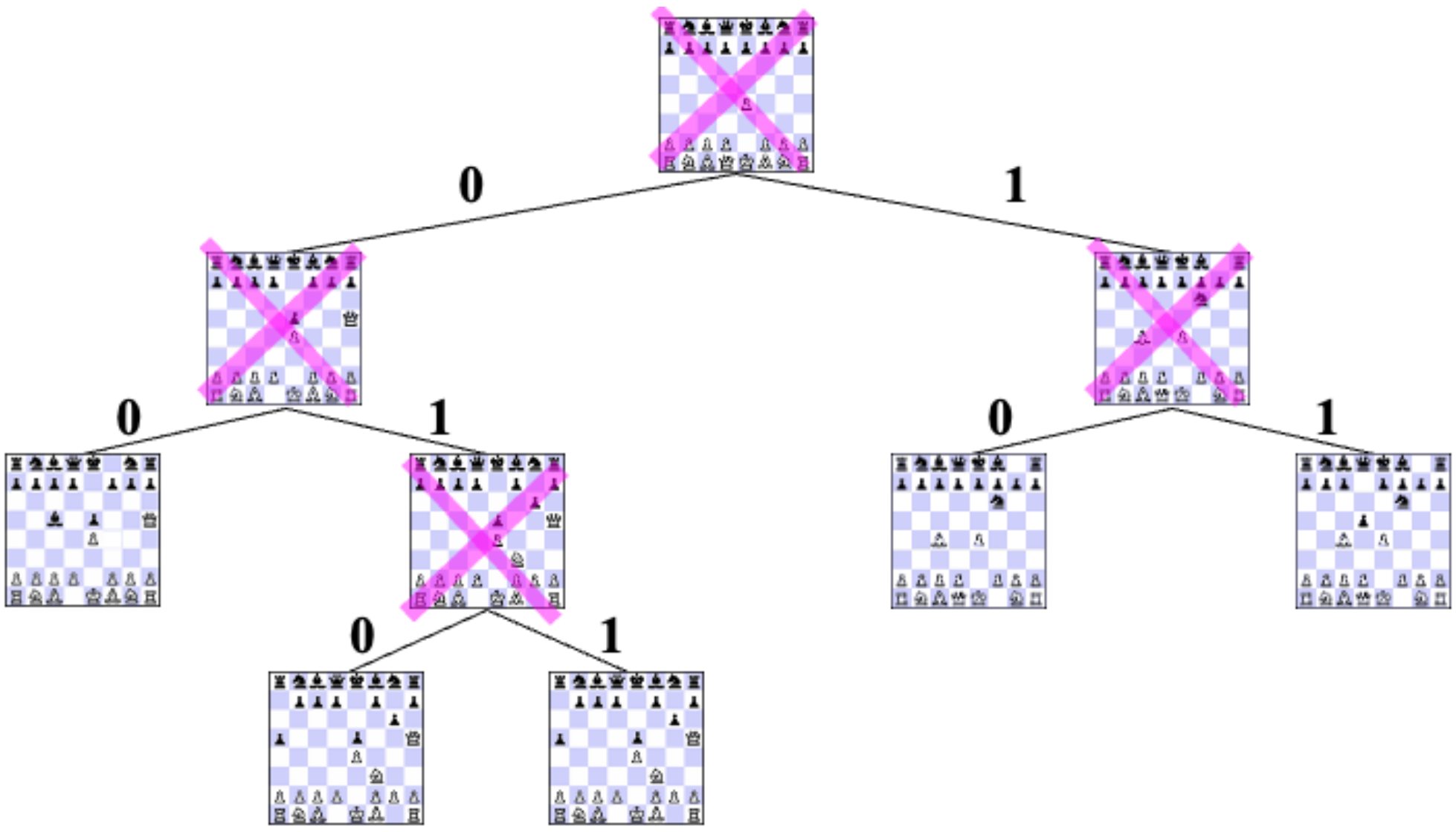

Figure 16: The bitstring will be: ⊤ε.E2E4, ⊥ε :, ⊥0.E7E5, ⊥1.G8F6, ⊤0.D1H5, ⊥0:, ⊤1.F1C4, ⊥01.G7G6, ⊤01.G1F3, ⊥00.H8C5, ⊥01:, ⊥1:, ⊥11.D7D5, ⊥01.A7A5.

Which can be summarised as:

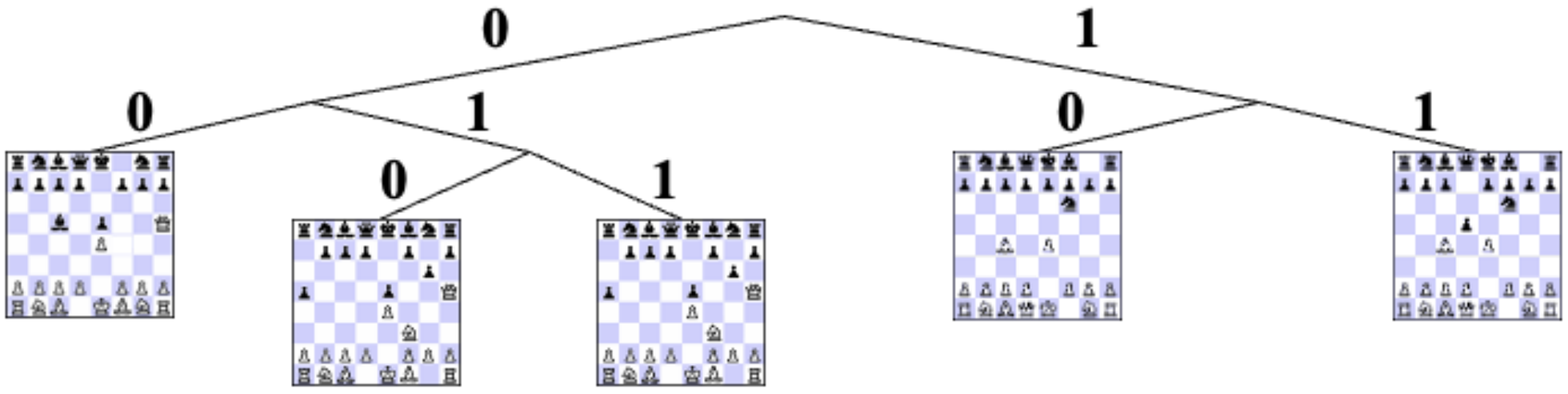

Figure 17: Tree representation of ⫰Chess.

At the end of this tree we see four different positions corresponding to the four final leaves. Each position $\Omega$ of Chess can be thought of as a game: specifically, the game $\langle\Omega\rangle$Chess, playing which, as we already know, means playing Chess starting from position $\Omega$.

Let us call the four games $A$, $B$, $C$ and $D$ for a more compact representation:

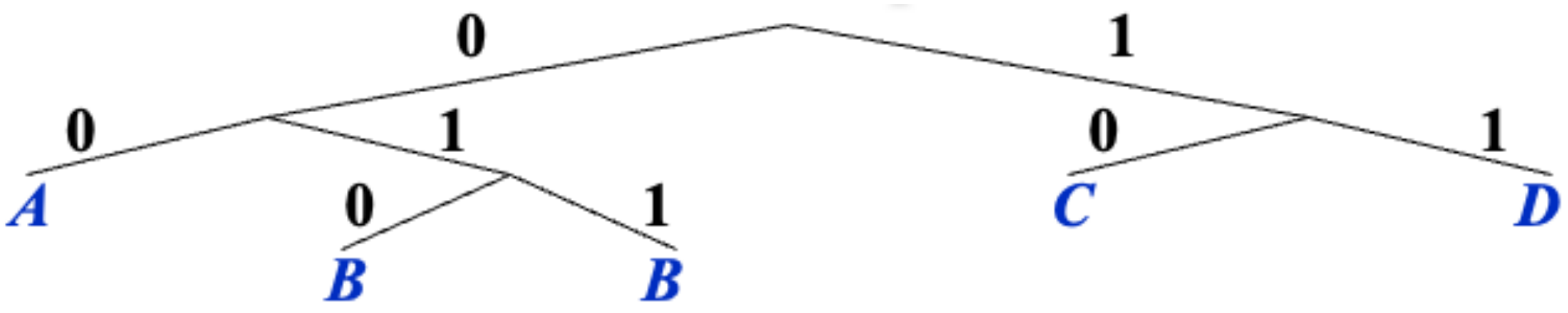

Figure 18: A more compact way of representing ⫰Chess.

For the very same purpose, we may use the **circle notation** and just write ⫰$((A \circ (B \circ B)) \circ (C \circ D))$ or ⫯$((A \circ (B \circ B) \circ (C \circ D))$, depending on whether this is a play of ⫰Chess or ⫯Chess.

It should be clear how to represent any other finite tree in the above style. The circle notation will be very handy in visualising evolution sequences of ⫰- and ⫯-games.

Here is an example of circle notation for the evolution sequence of a branching recurrence game. Let us represent the game ⫰$\sqcap x \sqcup y(y = x^2)$:

| Move | Game |
|---|---|
| $\varepsilon$: | $⫰\sqcap x \sqcup y(y = x^2)$ |
| $\bot 0.7$ | $⫰\sqcap x \sqcup y(y = x^2) \circ \sqcap x \sqcup y(y = x^2)$ |
| $\top 0.49$ | $⫰\sqcup y(y = 7^2) \circ \sqcap x \sqcup y(y = x^2)$ |
| $\bot 1$: | $⫰49 = 7^2 \circ \sqcap x \sqcup y(y = x^2)$ |
| $\bot 10.3$ | $⫰49 = 7^2 \circ (\sqcap x \sqcup y(y = x^2) \circ \sqcap x \sqcup y(y = x^2))$ |
| $\bot 11.5$ | $⫰49 = 7^2 \circ (\sqcup y(y = 3^2) \circ \sqcap x \sqcup y(y = x^2))$ |
| $\top 10.9$ | $⫰49 = 7^2 \circ (\sqcup y(y = 3^2) \circ \sqcup y(y = 5^2))$ |
| $\top 11.25$ | $⫰49 = 7^2 \circ (9 = 3^2 \circ \sqcup y(y = 5^2))$ |
| | $⫰49 = 7^2 \circ (9 = 3^2 \circ 25 = 5^2)$ |

At the end, we obtain 3 leaves with $\top$ in each one, so Machine wins.

Clearly, given the duality between $⫰$ and $⫯$, De Morgan's laws are still valid:

$$\neg⫰A = ⫯\neg A \qquad ⫰A = \neg⫯\neg A$$
$$\neg⫯A = ⫰\neg A \qquad ⫯A = \neg⫰\neg A$$

Branching recurrence $⫰$ can be shown to be stronger than its parallel counterpart $⅄$, in the sense that the principle $⫰A \to ⅄A$ is valid while $⅄A \to ⫰A$ is not.

Indeed, in isolation from each other, these operators validate different principles. For instance:

- $A \wedge ⅄(A \to A \wedge A) \to ⅄A$ is valid, while $A \wedge ⫰(A \to A \wedge A) \to ⫰A$ is not;
- $⫰(A \sqcup B) \to ⫰A \sqcup ⫰B$ is valid, while $⅄(A \sqcup B) \to ⅄A \sqcup ⅄B$ is not;
- $⫯\sqcap x(\neg A(x) \sqcup A(x))$ is valid while $Y\sqcap x(\neg A(x) \sqcup A(x))$ is not.

In addition, branching operations have a number of natural sharpenings, amongst which are the countable and uncountable versions.

Indeed, $⫰$ has a series of weaker versions obtained by imposing various restrictions on the quantity and form of reusages.

Out of the interesting and natural weakenings of ⫰ is **countable branching recurrence** $⫰^{\aleph_0}$ in the style of Blass's repetition operation $R$[92]. The definition remains the same, only difference being the $\aleph_0$ symbol at the top, which acts as a limitation to the number of reusages.

Indeed, the divergence between countable ($⫰^{\aleph_0}$) and uncountable (⫰) branching recurrences appears to be much more subtle than the one between parallel and branching recurrences.

As a matter of fact, the differences between ⫰ and ⅄ are the same between $⫰^{\aleph_0}$ and ⅄; however, $⫰^{\aleph_0}$ and ⫰ turn out to induce dramatically different logics, even though such logics coincide when $⟜^{\aleph_0}$ or ⟜ (or ≻—) is the only connective in the logical vocabulary[93].

**Branching rimplication** ("brimplication") $A$ ⟜ $B$ and **branching refutation** ("brefutation) ⫰¬$A$ are defined in terms of ⫰, $\rightarrow$ and $\bot$ the same way as parallel rimplication ≻— and refutation ≻¬ are defined in terms of ⅄, $\rightarrow$ and $\bot$:

**Definition 1.3.1.19**

- $A$ ⟜ $B =_{def}$ ⫰$A \rightarrow B$;
- ○¬$A =_{def} A$ ⟜ $\bot$.

We may also define **countable brimplication** as:

$$A ⟜^{\aleph_0} B =_{def} ⫰^{\aleph_0} A \rightarrow B.$$

Just like ≻—, ⟜ is a conservative generalisation of Turing reduction. Specifically, for any predicates $p(x)$ and $q(x)$, the problem

$$\sqcap x(\neg p(x) \sqcup p(x)) ⟜ \sqcap x(\neg q(x) \sqcup q(x))$$

92 See [5], page 162. In Blass' words: "[. . . ] we play a more complicated game which we call $R(A)$. This game resembles the tensor product of $\omega$ copies of $A$, in that player 0 may abandon plays of $A$ to start new ones and may later resume previously abandoned plays. [. . . ] However, $R(A)$ differs from this tensor product in that 1 (the oracle) is required to answer the same moves the same way in all plays".

93 For instance, as mentioned in [52], page 16, the principle: ⫰⫯$P \rightarrow$ ⫯⫰$P$, where ⫯ abbreviates ¬⫰¬ and ¬ is the role switch operation, is shown to be valid with ⫰ but not with $⫰^{\aleph_0}$ or ⅄.

is computable iff $q(x)$ is Turing reducible to $p(x)$.

This means that, when restricted to traditional sorts of problems, such as decision problems, the behaviours of $\circ\!\!-$ and $\succ\!\!-$ are indistinguishable - as previously mentioned in the reductions paragraph. This stops being the case when these operators are applied to problems with higher degrees of interactivity though.

For instance, the following problem is computable; however, it becomes incomputable with $\succ\!\!-$ instead of $\circ\!\!-$ :

$$\bigsqcup y \bigsqcap x(\neg H(x,y) \sqcup H(x,y)) \circ\!\!- \bigsqcup y(\bigsqcap x(\neg H(x,y) \sqcup H(x,y)) \wedge \bigsqcap x(\neg H(x,y) \sqcup H(x,y))).$$

Overall, $(A \succ\!\!- B) \to (A \circ\!\!- B)$ is valid but $(A \circ\!\!- B) \to (A \succ\!\!- B)$ is not.

While both $\succ\!\!-$ and $\circ\!\!-$ are weaker than $\to$ and, thus, more general, $\circ\!\!-$ remains a more interesting operation of reduction than $\succ\!\!-$. This one, in turn, assumes that $\circ\!\!\!\mid$ (and by no means $\curlywedge$) is the operation allowing to reuse its argument in the strongest algorithmic sense possible.
What makes $\circ\!\!-$ special is the following belief.

Let $A$ and $B$ be computational problems (games). We say that $B$ is **brimplicationally** (respectively primplicationally, pimplicationally, etc.) **reducible** to $A$ iff $A \circ\!\!- B$ (respectively $A \succ\!\!- B, A \to B$, etc.) has an algorithmic solution (winning strategy).

**Thesis 1.3.1.20**
Brimplicational reducibility, i.e. algorithmic solvability (computability) of $A \circ\!\!- B$, is an adequate mathematical counterpart of our intuition of reducibility in the weakest (and thus most general) algorithmic sense possible. Specifically:

1. Whenever a problem $B$ is brimplicationally reducible to a problem $A$, $B$ is also algorithmically reducible to $A$ according to anyone's reasonable intuition;

2. Whenever a problem $B$ is algorithmically reducible to a problem $A$ according to anyone's reasonable intuition, $B$ is also brimplicationally reducible to $A$.

This is pretty much in the same style as the Church-Turing thesis, which holds that a function $f$ is computable by a Turing Machine iff $f$ has an algorithmic solution according to everyone's reasonable intuition.

One final remark about $\circ\!\!-$ has yet to be made. As explained in [46], [47], [48] and [51], branching rimplication may be seen as the **intuitionistic implication**.

Indeed, CoL's intuitionistic fragment $INT^{\circ\!-}$, whose signature is $\{\circ\!\!-, \sqcap, \sqcup, \bigsqcap, \bigsqcup, \$\}$[94], has been shown to be sound and complete[95] with respect to Kripke's semantics[96].

As shown in [47], both $\succ\!\!-$ and $\circ\!\!-$ behave intuitionistically. However, while the former is exactly intuitionistic only in isolation and not when combined with other operators, the latter behaves intuitionistically under every circumstance.

Indeed, $A \succ\!\!- B$ is less flexible than $A \circ\!\!- B$, as we have already assessed. As a matter of fact, branching rimplication allows the player not to start all over again everytime: indeed, he can backtrack to any previous position. This makes $\circ\!\!-$ the weakest, most natural form of reduction which allows unlimited reusage.

Japaridze has managed to find a constructive proof[97] for the soundness

94 With $ standing for "absurd", as described in [48], or even "stronger problem" and "logical letter" in [46].

95 This proof is also a proof of the soundness and completeness of $INT^{\circ\!-\,\aleph_0}$, as stated in [47].

96 Just to give a glimpse of what Japaridze refers to, a **Kripke model** is a triple $\mathcal{M} = (\mathcal{W}, \mathcal{R}, \models)$ where, according to his definition in [47]:

- $\mathcal{W}$ is (here) a finite set of what are called worlds of $\mathcal{M}$;
- $\mathcal{R}$ is a transitive and reflexive relation between worlds. When $p\mathcal{R}q$, we say that world $q$ is accessible in $\mathcal{M}$ from world $p$;
- $\models$ is a relation between worlds and $INT^{\circ\!-}$ formulas, satisfying the following two conditions for all formulas $E, F$ and worlds $p, q$:
  1. If $p \models E$ and $p\mathcal{R}q$, then $q \models E$;
  2. $p \models E \circ\!\!- F$ iff, whenever $p\mathcal{R}q$ and $q \models E$, we have $q \models F$.

97 In [46], page 99, we read: "Consider an arbitrary sequent $S$ with $INT \vDash S$. By induction on the $INT$-derivation of $S$, we are going to show that $S$ has a uniform solution $E$.

of $INT^{\circ\!-}$ , therefore turning CoL into a possibly successful semantics for Heyting's intuitionistic calculus, as previously mentioned.

---

*Sequential operations (* $\triangle$ , $\triangledown$ , $\bigtriangleup$, $\bigtriangledown$, $\rhd$ , $\overset{|}{\triangle}$, $\rhd\!\!-$ , $\rhd\!\neg$*)*

The **sequential conjunction** ("sand") $A \triangle B$ of games $A$ and $B$ starts and proceeds as $A$. It will also end as $A$ unless, at some point, Environment decides to switch to the next component, in which case $A$ is abandoned and the game restarts, continues and ends as $B$.

The **sequential disjunction** ("sor") $A \triangledown B$ of $A$ and $B$ is similar, with the only difference that it is Machine who decides whether and when to switch from $A$ to $B$.

The original, formal definitions of $A \triangle B$ and $A \triangledown B$ found in [50] were a direct formalisation of the above description.
The definitions given below, however, while different and less direct, still faithfully formalise the above intuitions as long as only static games are considered. We opt for them since they are technically simpler.

Specifically, these new definitions allow either player to continue making moves in $A$ even after a switch takes place; such moves are meaningless and harmless. Similarly, it allows either player to make moves in $B$ without waiting for a switch to take place, even though a smart player would start making such moves only when a switch happens.

**Definition 1.3.1.21**
Assume $A_0 = (Vr^{A_0}, A_0)$ and $A_1 = (Vr^{A_1}, A_1)$ are gameframes. Then:

1. $A_0 \triangle A_1$ (read "$A_0$ **sand** $A_1$") is defined as the game G = $(Vr^{A_0} \cup Vr^{A_1}, G)$ such that:
   - $\Omega \in \mathrm{Lr}^G_e$ iff $\Omega = \langle \Xi, \Phi \rangle$ or $\Omega = \langle \Xi, \bot 1, \Phi \rangle$, where every move of $\langle \Xi, \Phi \rangle$ has the prefix "0." or "1." and, for both $i \in \{0, 1\}$, $\langle \Xi, \Phi \rangle^{i.} \in \mathrm{Lr}^{A_i}_e$;

---

This is sufficient to conclude that $INT$ is "uniformly sound". The theorem also claims "constructive soundness", i.e. that such an $E$ can be effectively built from a given $INT$-derivation of $S$". The idea is to consider 15 cases, corresponding to the 15 possible rules that might have been used at the last step of an $INT$-derivation of $S$, with $S$ being the conclusion of the rule.

- If $\Gamma$ does not contain the ("switch") move $\bot 1$, then $\mathrm{Wn}_e^G\langle\Gamma\rangle = \mathrm{Wn}_e^{A_0}\langle\Gamma^{0.}\rangle$; otherwise, $\mathrm{Wn}_e^G\langle\Gamma\rangle = \mathrm{Wn}_e^{A_1}\langle\Gamma^{1.}\rangle$.

2. $A_0 \bigtriangledown A_1$ (read "$A_0$ **sor** $A_1$") is defined as the game G = $(Vr^{A_0} \cup Vr^{A_1}, G)$ such that:
   - $\Omega \in \mathrm{Lr}_e^G$ iff $\Omega = \langle\Xi,\Phi\rangle$ or $\Omega = \langle\Xi,\top 1,\Phi\rangle$, where every move of $\langle\Xi,\Phi\rangle$ has the prefix "0." or "1." and, for both $i \in \{0,1\}$, $\langle\Xi,\Phi\rangle^{i.} \in \mathrm{Lr}_e^{A_i}$;
   - If $\Gamma$ does not contain the ("switch") move $\top 1$, then $\mathrm{Wn}_e^G\langle\Gamma\rangle = \mathrm{Wn}_e^{A_0}\langle\Gamma^{0.}\rangle$; otherwise, $\mathrm{Wn}_e^G\langle\Gamma\rangle = \mathrm{Wn}_e^{A_1}\langle\Gamma^{1.}\rangle$.

Given the duality between $\bigtriangleup$ and $\bigtriangledown$, De Morgan's laws still hold:

$$\neg(A \bigtriangleup B) = \neg A \bigtriangledown \neg B \qquad A \bigtriangleup B = \neg(\neg A \bigtriangledown \neg B)$$
$$\neg(A \bigtriangledown B) = \neg A \bigtriangleup \neg B \qquad A \bigtriangledown B = \neg(\neg A \bigtriangleup \neg B)$$

Recall that, for a predicate $p(x)$, $\sqcap x(\neg p(x) \sqcup p(x))$ is the problem of deciding $p(x)$. The similar looking $\sqcap x(\neg p(x) \bigtriangledown p(x))$ is, indeed, the problem of **semideciding** $p(x)$.
Machine has a winning strategy for this game if and only if $p(x)$ is semidecidable, i.e. recursively enumerable.

Namely, if $p(x)$ is recursively enumerable, a winning strategy for Machine would be to wait until Environment eventually brings the game down to $\neg p(n) \bigtriangledown p(n)$ for some particular $n$.
After that, Machine starts looking for evidence of $p(n)$'s truthfulness. If and when such evidence is found (meaning that $p(n)$ is indeed true), Machine makes a switch move turning $\neg p(n) \bigtriangledown p(n)$ into the true $p(n)$; however, if no evidence exists (meaning that $p(n)$ is false), Machine keeps endlessly looking for a non-existent object while never making any moves. Consequently, the game ends as $\neg p(n)$, which, again, is true.

Vice versa: any effective winning strategy for $\sqcap x(\neg p(x) \bigtriangledown p(x))$ can be seen as a semidecision procedure for $p(x)$ which accepts an input $n$ iff the strategy includes a switch move in the scenario where $\bot$ has initially chosen $n$ as a value for $x$.

The existence of an effective winning strategy for a given game has been shown to be closed under the following rules (see [67]):

- "from $A \to B$ and $A$ conclude $B$";
- "from $A$ and $B$ conclude $A \wedge B$";
- "from $A$ conclude $\sqcap x A$";
- "from $A$ conclude ⫰$A$".

Let us have a look at some relevant, valid formulas that we are now able to understand. Needless to say, these examples show once again how CoL can be employed as a systematic tool for defining new properties and relations between computational problems, while discovering an infinite variety of new facts.

The valid formula:

$$\sqcap x(\neg p(x) \triangledown p(x)) \wedge \sqcap x(\neg p(x) \triangledown p(x)) \to \sqcap x(\neg p(x) \sqcup p(x))$$

expresses the well known fact that, if both a predicate $p(x)$ and its complement $\neg p(x)$ are recursively enumerable, then $p(x)$ is decidable.

However, the validity of this formula actually means something more: the problem of deciding $p(x)$ is reducible to the ($\wedge$-conjunction of) the problems of semideciding $p(x)$ and $\neg p(x)$.
Indeed, an even stronger reducibility holds, expressed by the formula:

$$\sqcap x(\neg p(x) \triangledown p(x)) \wedge (p(x) \triangledown \neg p(x)) \to \neg p(x) \sqcup p(x)).$$

The formula:

$$\sqcap x \bigsqcup y(\neg q(x) \longleftrightarrow p(y)) \wedge \sqcap x(\neg p(x) \triangledown p(x)) \to \sqcap x(\neg q(x) \sqcup q(x))$$

is also valid and implies the known fact that, if a predicate $q(x)$ is mapping reducible to a predicate $p(x)$ and $p(x)$ is recursively enumerable, then so is $q(x)$.

Again, the validity of this formula means, in fact, something more: the problem of semideciding $q(x)$ is reducible to the problems of mapping reducing $q(x)$ to $p(x)$ and semideciding $p(x)$.

Certain other reducibilities hold exclusively in the sense of rimplications rather than implications. Here is an example.

Two Turing Machines are said to be equivalent iff they accept exactly the same inputs. Let $N(x,y)$ be the predicate "Turing Machines $x$ and $y$ are not equivalent". This predicate is neither semidecidable nor co-semidecidable.
However, its semidecision is problem primplicationally (and brimplicationally) reducible to the Halting problem.

Specifically, Machine has an effective winning strategy for the following game:

$$\sqcap z \sqcap t(\neg H(z,t) \sqcup H(z,t)) \succ\!\!- \sqcap x \sqcap y(\neg N(x,y) \bigtriangledown N(x,y))$$

A strategy here is to wait until Environment specifies some values $m$ and $n$ for $x$ and $y$. Then, create a variable $i$, initialise it to 1 and do the following.
Specify $z$ and $t$ as $m$ and $i$ in one yet unused copy of the antecedent, and as $n$ and $i$ in another yet unused copy. That is, ask Environment whether $m$ halts on input $i$ and whether $n$ halts on the same input.
Environment will have to provide the correct pair of answers, or else it loses:

1. If the answers are "No, No", increment $i$ to $i+1$ and repeat the step;
2. If the answers are "Yes, Yes", simulate both $m$ and $n$ on input $i$ until they halt. If both Machines accept or both reject, increment $i$ to $i+1$ and repeat the step. Otherwise, if one accepts and one rejects, make a switch move in the consequent and celebrate victory;
3. If the answers are "Yes, No", simulate $m$ on $i$ until it halts. If $m$ rejects $i$, increment $i$ to $i+1$ and repeat the step. Otherwise, if $m$ accepts $i$, make a switch move in the consequent and you win;
4. Finally, if the answers are "No, Yes", simulate $n$ on $i$ until it halts. If $n$ rejects $i$, increment $i$ to $i+1$ and repeat the step. Otherwise, if $n$ accepts $i$, make a switch move in the consequent and you win.

The **sequential universal quantification** ("sall") $\bigtriangleup x A(x)$ of $A(x)$ is essentially nothing but the infinite sequential conjunction $A(0) \bigtriangleup A(1) \bigtriangleup A(2) \bigtriangleup \ldots;$ the **sequential existential quantification** ("sexists") $\bigtriangledown x A(x)$ of $A(x)$ is $A(0) \bigtriangledown A(1) \bigtriangledown A(2) \bigtriangledown \ldots$ .

The **sequential recurrence** ("srecurrence") $\overset{|}{\vartriangle} A$ of $A$ is $A \vartriangle A \vartriangle A \vartriangle \ldots,$ while the **sequential corecurrence** ("cosrecurrence") $\underset{|}{\triangledown} A$ of $A$ is $A \triangledown A \triangledown A \triangledown \ldots$ .

Formally, we have:

**Definition 1.3.1.22**
Assume $x$ is a variable and $A(x) = (Vr, A)$ is a game. Then:

1. $\bigtriangleup xA(x)$ (read "**sall** $x$ $A(x)$") is defined as the game G = $(Vr \setminus \{x\}, G)$ such that:
   - $\Omega \in \mathrm{Lr}_e^G$ iff $\Omega = \langle \Xi_0, \bot 1, \Xi_1, \bot 2, \Xi_2, \ldots, \bot n, \Xi_n \rangle (n \geq 0)$, where every move of $\langle \Xi_0, \Xi_1, \Xi_2, \ldots, \Xi_n \rangle$ has the prefix "$c$." for some constant $c$ and, for every constant $c$, $\langle \Xi_0, \Xi_1, \Xi_2, \ldots, \Xi_n \rangle^{c.} \in \mathrm{Lr}_e^{A(c)}$;
   - Call the moves $\bot 1, \bot 2, \bot 3, \ldots$ switch moves. If $\Gamma$ does not contain any switch moves, then $\mathrm{Wn}_e^G \langle \Gamma \rangle = \mathrm{Wn}_e^{A_0} \langle \Gamma^{0.} \rangle$. If $\Gamma$ contains infinitely many switch moves, then $\mathrm{Wn}_e^G \langle \Gamma \rangle = \top$; otherwise, where $\bot n$ is the last switch move of $\Gamma$, $\mathrm{Wn}_e^G \langle \Gamma \rangle = \mathrm{Wn}_e^{A(n)} \langle \Gamma^{n.} \rangle$.

2. $\bigtriangledown xA(x)$ (read "**sexists** $x$ $A(x)$") is defined as the game G = $(Vr \setminus \{x\}, G)$ such that:
   - $\Omega \in \mathrm{Lr}_e^G$ iff $\Omega = \langle \Xi_0, \top 1, \Xi_1, \top 2, \Xi_2, \ldots, \top n, \Xi_n \rangle (n \geq 0)$, where every move of $\langle \Xi_0, \Xi_1, \Xi_2, \ldots, \Xi_n \rangle$ has the prefix "$c$." for some constant $c$ and, for every constant $c$, $\langle \Xi_0, \Xi_1, \Xi_2, \ldots, \Xi_n \rangle^{c.} \in \mathrm{Lr}_e^{A(c)}$;
   - Call the moves $\top 1, \top 2, \top 3, \ldots$ switch moves. If $\Gamma$ does not contain any switch moves, then $\mathrm{Wn}_e^G \langle \Gamma \rangle = \mathrm{Wn}_e^{A_0} \langle \Gamma^{0.} \rangle$. If $\Gamma$ contains infinitely many switch moves, then $\mathrm{Wn}_e^G \langle \Gamma \rangle = \bot$; otherwise, where $\top n$ is the last switch move of $\Gamma$, $\mathrm{Wn}_e^G \langle \Gamma \rangle = \mathrm{Wn}_e^{A(n)} \langle \Gamma^{n.} \rangle$.

Given the duality between the two sequential quantificators, we know that these laws are valid:

$$\neg \bigtriangleup xA = \bigtriangledown x \neg A \qquad \bigtriangleup xA = \neg \bigtriangledown \neg A$$
$$\neg \bigtriangledown xA = \bigtriangleup x \neg A \qquad \bigtriangledown xA = \neg \bigtriangleup x \neg A$$

**Definition 1.3.1.23**
Assume that $A(x) = (Vr, A)$ is a game. Then:

1. $\overset{|}{\triangle}A$ (read "**srecurrence** $A$") is defined as the game G = $(Vr, G)$ such that:
   - $\Omega \in \mathrm{Lr}_e^G$ iff $\Omega = \langle \Xi_0, \bot 1, \Xi_1, \bot 2, \Xi_2, \ldots, \bot n, \Xi_n \rangle (n \geq 0)$, where every move of $\langle \Xi_0, \Xi_1, \Xi_2, \ldots, \Xi_n \rangle$ has the prefix "$c$." for some constant $c$ and, for every constant $c$, $\langle \Xi_0, \Xi_1, \Xi_2, \ldots, \Xi_n \rangle^{c.} \in \mathrm{Lr}_e^A$;
   - Call the moves $\bot 1, \bot 2, \bot 3, \ldots$ switch moves. If $\Gamma$ does not contain any switch moves, then $\mathrm{Wn}_e^G \langle \Gamma \rangle = \mathrm{Wn}_e^A \langle \Gamma^{0.} \rangle$. If $\Gamma$ contains infinitely many switch moves, then $\mathrm{Wn}_e^G \langle \Gamma \rangle = \top$; otherwise, where $\bot n$ is the last switch move of $\Gamma$, $\mathrm{Wn}_e^G \langle \Gamma \rangle = \mathrm{Wn}_e^A \langle \Gamma^{n.} \rangle$.

2. $\underset{|}{\triangledown}A$ (read "**cosrecurrence** $A$") is defined as the game G = $(Vr, G)$ such that:
   - $\Omega \in \mathrm{Lr}_e^G$ iff $\Omega = \langle \Xi_0, \top 1, \Xi_1, \top 2, \Xi_2, \ldots, \top n, \Xi_n \rangle (n \geq 0)$, where every move of $\langle \Xi_0, \Xi_1, \Xi_2, \ldots, \Xi_n \rangle$ has the prefix "$c$." for some constant $c$ and, for every constant $c$, $\langle \Xi_0, \Xi_1, \Xi_2, \ldots, \Xi_n \rangle^{c.} \in \mathrm{Lr}_e^A$;
   - Call the moves $\top 1, \top 2, \top 3, \ldots$ switch moves. If $\Gamma$ does not contain any switch moves, then $\mathrm{Wn}_e^G \langle \Gamma \rangle = \mathrm{Wn}_e^A \langle \Gamma^{0.} \rangle$. If $\Gamma$ contains infinitely many switch moves, then $\mathrm{Wn}_e^G \langle \Gamma \rangle = \bot$; otherwise, where $\top n$ is the last switch move of $\Gamma$, $\mathrm{Wn}_e^G \langle \Gamma \rangle = \mathrm{Wn}_e^A \langle \Gamma^{n.} \rangle$.

Again, De Morgan's laws are valid:

$$\neg \overset{|}{\triangle}A = \underset{|}{\triangledown}\neg A \qquad \overset{|}{\triangle}A = \neg \underset{|}{\triangledown}\neg A$$
$$\neg \underset{|}{\triangledown}A = \overset{|}{\triangle}\neg A \qquad \underset{|}{\triangledown}A = \neg \overset{|}{\triangle}\neg A$$

Intuitively, $\overset{|}{\triangle}$ captures the simplest operating system that allows to start a session of game $A$; then, after finishing it or abandoning and destroying it, a new run starts and the process continues.

The game that such a system plays, i.e. the resource that it supports and provides, is $\overset{|}{\triangle}$A, which assumes an unbounded number of plays of $A$ in a sequential manner.

Let $k(x)$ mean "The Kolmogorov complexity of $x$". Indeed, $k(x)$ of a number $x$ is the size of the smallest Turing Machine which outputs $x$ on input 0.
More formally, let us assume that all Turing Machines are listed in the lexicographic order of their descriptions:

$$\mathcal{M}_0, \mathcal{M}_1, \mathcal{M}_2, \mathcal{M}_3, \ldots, \mathcal{M}_i, \ldots \quad .$$

Number $i$ can thus be considered the **code** of Machine $\mathcal{M}_i$.

**Definition 1.3.1.24**
Let $m$ be a natural number. The **Kolmogorov complexity** of $m$ is the size (logarithm) $|i|$ of the smallest number $i$ such that Machine $\mathcal{M}_i$ outputs $m$ on input 0.

Indeed, a Turing Machine outputting a number $m$ on a fixed input (such as 0) can be seen as a description of $m$. So, the Kolmogorov complexity $k$ of $m$ is the size of the smallest possible description of $m$: the greater the $k$, the harder, more complex it is to describe $m$.

Indeed, Kolmogorov complexity can be seen as the mathematical counterpart of the intuitive concepts of "randomness" or "amount of information". The greater the Kolmogorov complexity of a given object, the more random it is and the more information it contains.

The Kolmogorov complexity problem $\bigsqcap x \bigsqcup y(y = k(x))$ has no algorithmic solution, nor is it reducible to the Halting problem in the strong sense of $\to$, meaning that the problem:

$$\bigsqcap x(\neg H(x) \sqcup H(x)) \to \bigsqcap x \bigsqcup y(y = k(x))$$

has no algorithmic solution, either.
However, it is reducible to the Halting problem in the weaker sense of $\circ\!\!-$ , meaning that Machine has a winning strategy for

$$\bigsqcap x(\neg H(x) \sqcup H(x)) \circ\!\!- \bigsqcap x \bigsqcup y(y = k(x)).$$

Here is one such strategy provided by Japaridze in lecture number 2.

After receiving the question "What is the Kolmogorov complexity of $m$?", initialise variable $i$ to 0, and do the following:

1. Ask the oracle (resource/antecedent) if the Machine $M_i$ halts on input 0;
2. If the oracle says "No", increment $i$ to $i + 1$, and go back to Step 1;
3. If the oracle says "Yes", simulate $M_i$ on input 0 until it halts. If you see that the output of $M_i$ is $m$, return the size (logarithm) $|i|$ of $i$ as your output. Otherwise, increment $i$ to $i + 1$ and go back to Step 1.

Indeed, the oracle is queried more than once. The number of queries, however, can be shown to be bounded by a certain linear function of $m$.

**Theorem 1.3.1.25**
The Kolmogorov complexity problem is not computable.

Let us now return to our sequential recurrence and corecurrence operators. As we were saying, $k(x)$ is the function associated to the Kolmogorov complexity problem. The value of $k(x)$ is known to be bounded, not exceeding $x$ or a certain constant $const$, whichever is greater.

While $\sqcap x \sqcup y(y = k(x))$ is not computable, Machine does have a winning strategy for the problem $\sqcap x ⩛ \sqcup y(y = k(x))$. It goes like this[98]:

1. Wait until Environment specifies a value $m$ for $x$, thus asking "What is the Kolmogorov complexity of $m$?" and bringing the game down to $⩛ \sqcup y(y = k(m))$;
2. Answer that it is $m$, i.e. specify $y$ as $m$;
3. Start simulating, in parallel, all Machines $n$ smaller than $m$ on input 0. Whenever you find a Machine $n$ that returns $m$ on input 0 and is smaller than any of the previously found such Machines, make a switch move and, in the new copy of $⩛ \sqcup y(y = k(m))$, specify $y$ as the size $|n|$ of $n$ or as $const$, whichever is greater.

This obviously guarantees success: sooner or later the real Kolmogorov complexity $c$ of $m$ will be reached and named. Even though the strategy will never be sure that $k(m)$ is not something yet smaller than $c$, it will never really find a reason to further reconsider its latest claim that $c = k(m)$.

98 From [67], page 37.

As expected, **sequential implication** ("simplication") $A \rhd B$, **sequential rimplication** ("srimplication") $A \rhd\!\!- B$ and **sequential refutation** ("srefutation") $\rhd\!\!\neg A$ are defined as follows:

**Definition 1.3.1.26**

- $A \rhd B =_{def} \neg A \bigtriangledown B$;
- $A \rhd\!\!- B =_{def} \stackrel{|}{\vartriangle} A \rightarrow B$;
- $\rhd\!\!\neg A =_{def} A \rhd\!\!- \bot$.

---

*Toggling operations ( $\curlywedge$, $\curlyvee$, $\pmb{\curlywedge}$, $\pmb{\curlyvee}$, $\gtrdot$ , $\succ\!\!-$, $\succ\!\!\neg$)*

As all other sorts of conjunctions and disjunctions, **toggling conjunction** ("tand") $A \curlywedge B$ and **toggling disjunction** ("tor") $A \curlyvee B$ are dual to each other: the definition of one is obtained from the definition of the other by interchanging the roles of the two players.

Let us first focus on disjunction. One of the ways to characterise $A \curlyvee B$ is the following.

The game starts and proceeds as a play of $A$. It will end as an ordinary play of $A$ unless, at some point, Machine decides to switch to $B$, after which the game becomes and continues as $B$.
It will also end as $B$ unless, at some point, Machine decides to switch back to $A$. In that case, the game again becomes $A$, where $A$ resumes from the position in which it was abandoned (rather than from its start position, as would be the case, say, in $A \bigtriangledown B \bigtriangledown A$).
Later, Machine may again switch to (the abandoned position of) $B$, and so on. Machine wins the overall play iff it switches from one component to another (it may "change its mind"[99] or correct its mistake) at most a finite number of times and wins in its final choice, meaning the component which was chosen last to switch to.

---

99 Expression used by Japaridze to describe the dynamics of this sort of games. Even though we are going to use it in order to convey the same idea Japaridze wanted to, we are not at all convinced that it is appropriate to reference a sort of mind on Machine's part - the name speaks for itself. It would surely better fit Environment's description, being a "capricious" (almost *deranged*, as in behaving unrigorously) adversary who is able to follow a non-algorithmic strategy upon its own volition.

An alternative way to characterise $A \curlyvee B$ is to say that it is played exactly like $A \vee B$, with the only difference that Machine is allowed to make a "choose $A$" or "choose $B$" move some finite number of times.
If infinitely many choices are made, Machine loses. Otherwise, the winner will be the player who wins in the component that was chosen last (the **eventual choice**, as Japaridze calls it).

The case of Machine having made no choices at all is treated as if it had chosen $A$. Thus, as in sequential disjunction, the leftmost component is the default, or automatically made, initial choice.

The adversary (or perhaps even Machine itself) never knows whether a given choice of a component of $A \curlyvee B$ is the last choice or not. Otherwise, if Machine was required to declare that it has made its final choice, the resulting operation, for static games, would essentially be the same as $A \sqcup B$.

Indeed, recalling Blass' words, it never hurts a player to postpone making moves in static games: Environment could just inactively wait until the last choice is declared, and start playing the chosen component only after that, as in the case of $A \sqcup B$. Under these circumstances, making some temporary choices before the final one would not make any sense for Machine, either.

If we did not require Machine to change its mind only a finite number of times, there would be no eventual choice. Thus, the only natural winning condition would be to say that Machine wins iff it simply wins in one of the components.

However, the resulting operation would essentially be the same as $\vee$, as a smart Machine would always opt for keep switching between components forever. Indeed, allowing infinitely many choices would amount to not requiring any choices at all, as is the case of $A \vee B$.

What would happen if we allowed Machine to make an arbitrary initial choice between $A$ and $B$ and then reconsider its choice only (at most) once? Such an operation on games, albeit meaningful, would not be basic. That is because it can be expressed through our primitives as $(A \bigtriangledown B) \sqcup (B \bigtriangledown A)$.

All of this shows us that ⩛ is a weak sort of choice: the kind that, in real life, one would ordinarily call, as mentioned in [67] page 38, "choice after trial and error"[100].

Generally, a problem is considered to be solved after some trial and error if, after perhaps coming up with several wrong solutions, a true solution is eventually found. That is, mistakes are tolerated and forgotten as long as they are eventually corrected.

It is however necessary that new solutions stop coming at some point, so that there is a last solution whose correctness determines the success of the effort. Otherwise, if answers keep changing all the time, no answer is really given after all.[101]

As we recall, for a predicate $p(x)$, $\sqcap x(\neg p(x) \sqcup p(x))$ is the problem of deciding $p(x)$, and $\sqcap x(\neg p(x) \triangledown p(x))$ is the weaker (easier to solve) problem of semideciding $p(x)$.
Not surprisingly, $\sqcap x(\neg p(x) ⩛ p(x))$ is also a decision-style problem, but still weaker than the problem of semideciding $p(x)$.

This problem has been studied in the literature under several names, the most common of which is **recursively approximating** $p(x)$. It means telling whether $p(x)$ is true or not in the same style as semideciding acquires in negative cases: by correctly saying "Yes" or "No" at some point (after perhaps taking back previous answers several times) and never reconsidering the answer afterwards.

Semideciding $p(x)$ can be seen as always saying "No" at the beginning and then, if the answer is incorrect, changing it to "Yes" at some later time. Otherwise, when the answer is negative, the game begins with the answer "No" and ends without ever taking it back - however, without ever indicating that the answer is final and unchangeable.

100 In other words, [54], page 971: "The toggling operations can be characterized as lenient versions of choice operations where choices are retractable, being allowed to be reconsidered any finite number of times. This way, they model trial-and-error style decision steps in interactive computation."

101 As Japaridze colourfully exemplifies in the same page: "Or, imagine Bob has been married and divorced several times. Every time he said "I do", he probably honestly believed that this time, at last, his bride was "the one", with whom he would live happily ever after. Bob will be considered to have found his Ms. Right after all if and only if one of his marriages indeed turns out to be happy and final."

As a result, the difference between semideciding and recursively approximating is that, unlike a semidecision procedure, a recursive approximation[102] procedure can reconsider both negative and positive answers, and do so several times rather than only once.

A predicate $p(x)$ is recursively approximable (meaning the problem of its recursive approximation has an algorithmic solution) iff $p(x)$ has the arithmetical complexity $\Delta_2$, i.e. if both $p(x)$ and its complement $\neg p(x)$ can be written in the form $\exists z \forall y\, s(z,y,x)$, where $s(z,y,x)$ is a decidable predicate[103].

Indeed, let us show that algorithmic solvability of $\sqcap x(\neg p(x) \sqcup p(x))$ is equivalent to $p(x)$'s being of complexity $\Delta_2$:

$$\sqcap x(\neg p(x) \sqcup p(x)) \longleftrightarrow \exists z \forall y\, s(z,y,x)$$

**From right to left ($\leftarrow$ direction):**
First, assume $p(x)$ is of complexity $\Delta_2$, specifically, for some decidable predicates $q(z,y,x)$ and $r(z,y,x)$ we have $p(x) = \exists z \forall y\, q(z,y,x)$ and $\neg p(x) = \exists z \forall y\, r(z,y,x)$.

Then $\sqcap x(\neg p(x) \sqcup p(x))$ is solved by the following strategy:

1. Wait until Environment specifies a value $m$ for $x$, thus bringing the game down to $\neg p(m) \sqcup p(m)$;
2. Create the variables $i$ and $j$, initialise both to 1, choose the $p(m)$ component and do the following:
   - Step 1: Check whether $q(i,j,m)$ is true. If yes, increment $j$ to $j+1$ and repeat. If not, switch to the $\neg p(m)$ component, reset $j$ to 1, and go to Step 2;

102 According to Schönfield's Limit Lemma, a predicate $p(x)$ is recursively approximable (i.e. the problem of its recursive approximation has an algorithmic solution) iff $p(x)$ is of Turing degree $\leq \varnothing'$, that is, $p(x)$ is Turing reducible to the Halting problem.

103 This is because of Schönfield's Lemma. Indeed, $p(x)$ being Turing reducible to the Halting problem only means that $p(x)$ has an arithmetical complexity of $\Delta_2$, so that both $p(x)$ and its negation can be written in the form $\exists z \forall y\, s(z,y,x)$, where $s(z,y,x)$ is a decidable predicate. In the theory of computability-in-principle (as opposed to, say, complexity theory), by importance, the class of predicates of complexity $\Delta_2$ is only next to the classes of decidable, semidecidable and co-semidecidable predicates. This class also plays a crucial role in logic: a formula of classical predicate logic is valid iff it is true in every model where all atoms of the formula are interpreted as predicates of complexity $\Delta_2$.

- Step 2: Check whether $r(i,j,m)$ is true. If yes, increment $j$ to $j+1$ and repeat. If not, switch to the $p(m)$ component, reset $j$ to 1, increment $i$ to $i+1$, and go to Step 1.

One can easily see that the above algorithm indeed solves $\sqcap x(\neg p(x) \sqcup p(x))$.

**From left to right ($\rightarrow$ direction):**
Assume a given algorithm $Alg$ solves $\sqcap x(\neg p(x) \sqcup p(x))$. Let $q(z,y,x)$ be the predicate such that, in the scenario where Environment has initially specified $x$ as $m$ and the game has been brought down to $\neg p(m) \sqcup p(m)$, $q(i,j,m)$ is true iff:

1. At the $i$-th computation step, $Alg$ chose the $p(m)$ component;

2. At the $j$-th computation step, $Alg$ did not move.

Quite similarly, let $r(z,y,x)$ be the predicate such that, in the scenario where Environment has initially specified $x$ as $m$ and the game has been brought down to $\neg p(m) \sqcup p(m)$, $r(i,j,m)$ is true iff:

1. Either $i = 1$ or, at the $i$-th computation step, $Alg$ chose the $\neg p(m)$ component;

2. At the $j$-th computation step, $Alg$ did not move.

Of course, both $q(z,y,x)$ and $r(z,y,x)$ are decidable predicates, and so are $y > z \rightarrow q(z,y,x)$ and $y > z \rightarrow r(z,y,x)$. Clearly,

$$p(x) = \exists z \forall y \; (y > z \rightarrow q(z,y,x))$$

and

$$\neg p(x) = \exists z \forall y \; (y > z \rightarrow r(z,y,x))$$

hold.
Consequently, $p(x)$ is indeed of complexity $\Delta_2$. ■

As a real-life example of a recursively approximable predicate which is neither semidecidable nor co-semidecidable, consider the predicate $k(x) < k(y)$, which states that a certain number $x$ is simpler than a certain number $y$ in the sense of Kolmogorov complexity.

As previously mentioned, $k(z)$ (the Kolmogorov complexity of $z$) is bounded, never exceeding $max(z, const)$ for a certain constant $const$.

Here is an algorithm which recursively approximates the predicate $k(x) < k(y)$, i.e. solves the problem $\sqcap x \sqcap y(\neg k(x) < k(y) \curlyvee k(x) < k(y))$:

1. Wait until Environment brings the game down to $(\neg k(m) < k(n)) \curlyvee (k(m) < k(n))$ for some $m$ and $n$;

2. Start simulating, in parallel, all Turing Machines $t$ satisfying $t \leq max(m, n, const)$ on input 0;

3. Whenever you see that a Machine $t$ returns $m$ and the size of $t$ is smaller than that of any other previously found Machines that return $m$ or $n$ on input 0, choose $k(m) < k(n)$;

4. Whenever you see that a Machine $t$ returns $n$ and the size of $t$ is smaller than that of any other previously found Machine that returns $n$ on input 0, as well as smaller or equal to the size of any other previously found Machines that return $m$ on input 0, choose $\neg k(m) < k(n)$.

Clearly, the correct choice between $\neg k(m) < k(n)$ and $k(m) < k(n)$ will be made sooner or later and never reconsidered afterwards. This will happen when the procedure hits, in the role of $t$, a smallest-size Machine that returns either $m$ or $n$ on input 0.

Here is the formal account of the toggling conjunction and disjunction operators.

**Definition 1.3.1.27**
Assume $A_0 = (Vr^{A_0}, A_0)$ and $A_1 = (Vr^{A_1}, A_1)$ are gameframes. Then:

1. $A_0 \curlywedge A_1$ (read "$A_0$ **tand** $A_1$") is defined as the game G = $(Vr^{A_0} \cup Vr^{A_1}, G)$ such that:
   - $\Omega \in \mathrm{Lr}_e^G$ iff $\Omega = \langle \Xi_0, \bot i_1, \Xi_1, \bot i_2, \Xi_2, \ldots, \bot i_n, \Xi_n \rangle (n \geq 0)$, where $i_1, i_2, \ldots, i_n \in \{0, 1\}$, every move of $\langle \Xi_0, \Xi_1, \Xi_2, \ldots, \Xi_n \rangle$ has the prefix "0." or "1." and, for both $i \in \{0, 1\}$, $\langle \Xi_0, \Xi_1, \Xi_2, \ldots, \Xi_n \rangle^{i.} \in \mathrm{Lr}_e^{A_{i.}}$;
   - Call $\bot 0$ and $\bot 1$ switch moves. If $\Gamma$ does not contain any switch moves, then $\mathrm{Wn}_e^G \langle \Gamma \rangle = \mathrm{Wn}_e^{A_0} \langle \Gamma^{0.} \rangle$; if $\Gamma$ has infinitely many occurrences of switch moves, then $\mathrm{Wn}_e^G \Gamma \rangle = \top$; otherwise, where $\bot i$ is the last switch move, $\mathrm{Wn}_e^G \langle \Gamma \rangle = \mathrm{Wn}_e^{A_1} \langle \Gamma^{1.} \rangle$.

2. $A_0 \curlyvee A_1$ (read "$A_0$ **tor** $A_1$") is defined as the game G = $(Vr^{A_0} \cup Vr^{A_1}, G)$ such that:
   - $\Omega \in \mathrm{Lr}_e^G$ iff $\Omega = \langle \Xi_0, \top i_1, \Xi_1, \top i_2, \Xi_2, \ldots, \top i_n, \Xi_n \rangle (n \geq 0)$, where $i_1, i_2, \ldots, i_n \in \{0,1\}$, every move of $\langle \Xi_0, \Xi_1, \Xi_2, \ldots, \Xi_n \rangle$ has the prefix "0." or "1." and, for both $i \in \{0,1\}$, $\langle \Xi_0, \Xi_1, \Xi_2, \ldots, \Xi_n \rangle^{i.} \in \mathrm{Lr}_e^{A_{i}.}$;
   - Call $\top 0$ and $\top 1$ switch moves. If $\Gamma$ does not contain any switch moves, then $\mathrm{Wn}_e^G\langle\Gamma\rangle = \mathrm{Wn}_e^{A_0}\langle\Gamma^{0.}\rangle$; if $\Gamma$ has infinitely many occurrences of switch moves, then $\mathrm{Wn}_e^G\Gamma\rangle = \bot$; otherwise, where $\top i$ is the last switch move, $\mathrm{Wn}_e^G\langle\Gamma\rangle = \mathrm{Wn}_e^{A_1}\langle\Gamma^{1.}\rangle$.

As always, we know that duality entails De Morgan's laws:

$$\neg(A \curlywedge B) = \neg A \curlyvee \neg B \qquad A \curlywedge B = \neg(\neg A \curlyvee \neg B)$$
$$\neg(A \curlyvee B) = \neg A \curlywedge \neg B \qquad A \curlyvee B = \neg(\neg A \curlywedge \neg B)$$

The **toggling universal quantification** ("tall") $\curlywedge x A(x)$ of $A(x)$ is essentially nothing but $A(0) \curlywedge A(1) \curlywedge A(10) \curlywedge A(11) \curlywedge \ldots$, while the **toggling existential quantification** ("texists") $\curlyvee x A(x)$ of $A(x)$ is $A(0) \curlyvee A(1) \curlyvee A(10) \curlyvee A(11) \curlyvee \ldots$ .

Formally, we have:

**Definition 1.3.1.28**
Assume $x$ is a variable and $A(x) = (Vr, A)$ is a game. Then:

1. $\curlywedge x A(x)$ (read "**tall** $x$ $A(x)$") is defined as the game G = $(Vr \setminus \{x\}, G)$ such that:
   - $\Omega \in \mathrm{Lr}_e^G$ iff $\Omega = \langle \Xi_0, \bot c_1, \Xi_1, \bot c_2, \Xi_2, \ldots, \bot n, \Xi_n \rangle (n \geq 0)$, where every move of $\langle \Xi_0, \Xi_1, \Xi_2, \ldots, \Xi_n \rangle$ has the prefix "$c$." for some constant $c$ and, for every constant $c$, $\langle \Xi_0, \Xi_1, \Xi_2, \ldots, \Xi_n \rangle^{c.} \in \mathrm{Lr}_e^{A(c)}$;
   - Call the moves $\bot 1, \bot 2, \bot 3, \ldots$ switch moves. If $\Gamma$ does not contain any switch moves, then $\mathrm{Wn}_e^G\langle\Gamma\rangle = \mathrm{Wn}_e^{A_0}\langle\Gamma^{0.}\rangle$. If $\Gamma$ contains infinitely many switch moves, then $\mathrm{Wn}_e^G\langle\Gamma\rangle = \top$; otherwise, where $\bot n$ is the last switch move of $\Gamma$, $\mathrm{Wn}_e^G\langle\Gamma\rangle = \mathrm{Wn}_e^{A(n)}\langle\Gamma^{n.}\rangle$.

2. $\Vvee x A(x)$ (read "**texists** $x$ $A(x)$") is defined as the game G = $(Vr \setminus \{x\}, G)$ such that:
   - $\Omega \in \mathrm{Lr}_e^G$ iff $\Omega = \langle \Xi_0, \top 1, \Xi_1, \top 2, \Xi_2, \ldots, \top n, \Xi_n \rangle (n \geq 0)$, where every move of $\langle \Xi_0, \Xi_1, \Xi_2, \ldots, \Xi_n \rangle$ has the prefix "$c$." for some constant $c$ and, for every constant $c$, $\langle \Xi_0, \Xi_1, \Xi_2, \ldots, \Xi_n \rangle^{c.} \in \mathrm{Lr}_e^{A(c)}$;
   - Call the moves $\top c_1, \top c_2, \top c_3, \ldots$ switch moves. If $\Gamma$ does not contain any switch moves, then $\mathrm{Wn}_e^G \langle \Gamma \rangle = \mathrm{Wn}_e^{A_0} \langle \Gamma^{0.} \rangle$. If $\Gamma$ contains infinitely many switch moves, then $\mathrm{Wn}_e^G \langle \Gamma \rangle = \bot$; otherwise, where $\top n$ is the last switch move of $\Gamma$, $\mathrm{Wn}_e^G \langle \Gamma \rangle = \mathrm{Wn}_e^{A(n)} \langle \Gamma^{n.} \rangle$.

Given the duality, the following laws hold:

$$\neg \bigwedge\!\!\!\!\bigwedge x A = \Vvee x \neg A \qquad \bigwedge\!\!\!\!\bigwedge x A = \neg \Vvee \neg A$$
$$\neg \Vvee x A = \bigwedge\!\!\!\!\bigwedge x \neg A \qquad \Vvee x A = \neg \bigwedge\!\!\!\!\bigwedge x \neg A$$

As an example to show how toggling quantifiers work, remember that Kolmogorov complexity $k(x)$ is not a computable function: the problem $\sqcap x \sqcup y (y = k(x))$ has no algorithmic solution. However, locally replacing $\sqcup y$ with $\Vvee y$ yields an algorithmically solvable problem.

A solution for $\sqcap x \Vvee y (y = k(x))$ goes like this:

1. Wait until Environment chooses a number $m$ for $x$, thus bringing the game down to $\Vvee y (y = k(m))$, which is essentially nothing but $0 = k(m) \veebar 1 = k(m) \veebar 2 = k(m) \veebar \ldots$;
2. Initialise $i$ to a sufficiently large number, such as $i = max(|m| + const)$ where $const$ is the constant mentioned earlier;
3. Switch to the disjunct $i = k(m)$ of $0 = k(m) \veebar 1 = k(m) \veebar 2 = k(m) \veebar \ldots$ and then start simulating on input 0, in parallel, all Turing Machines whose sizes are smaller than $i$;
4. If and when you see that one of such Machines returns $m$, update $i$ to the size of that Machine and repeat the previous step.

We close this section with the formal definitions of **toggling recurrence** ("trecurrence") $\curlywedge A$, **toggling corecurrence** ("cotrecurrence") $\curlyvee A$, **toggling**

**implication**("timplication") $A \geqslant B$, **toggling rimplication** ("trimplication") $A \succ\!\!- B$ and **toggling refutation** ("trefutation") $\succ\!\!\neg A$.

**Definition 1.3.1.29**
Assume that $A(x) = (Vr, A)$ is a game. Then:

1. $\curlywedge A$ (read "**trecurrence** $A$") is defined as the game G = $(Vr, G)$ such that:
   - $\Omega \in \mathrm{Lr}_e^G$ iff $\Omega = \langle \Xi_0, \bot c_1, \Xi_1, \bot c_2, \Xi_2, \ldots, \bot c_n, \Xi_n \rangle (n \geq 0)$, where every move of $\langle \Xi_0, \Xi_1, \Xi_2, \ldots, \Xi_n \rangle$ has the prefix "$c$." for some constant $c$ and, for every constant $c$, $\langle \Xi_0, \Xi_1, \Xi_2, \ldots, \Xi_n \rangle^{c.} \in \mathrm{Lr}_e^A$;
   - Call the moves $\bot 1, \bot 2, \bot 3, \ldots$ switch moves. If $\Gamma$ does not contain any switch moves, then $\mathrm{Wn}_e^G\langle\Gamma\rangle = \mathrm{Wn}_e^A\langle\Gamma^{0.}\rangle$. If $\Gamma$ contains infinitely many switch moves, then $\mathrm{Wn}_e^G\langle\Gamma\rangle = \top$; otherwise, where $\bot n$ is the last switch move of $\Gamma$, $\mathrm{Wn}_e^G\langle\Gamma\rangle = \mathrm{Wn}_e^A\langle\Gamma^{n.}\rangle$.
2. $\curlyvee A$ (read "**cotrecurrence** $A$") is defined as the game G = $(Vr, G)$ such that:
   - $\Omega \in \mathrm{Lr}_e^G$ iff $\Omega = \langle \Xi_0, \top c_1, \Xi_1, \top c_2, \Xi_2, \ldots, \top c_n, \Xi_n \rangle (n \geq 0)$, where every move of $\langle \Xi_0, \Xi_1, \Xi_2, \ldots, \Xi_n \rangle$ has the prefix "$c$." for some constant $c$ and, for every constant $c$, $\langle \Xi_0, \Xi_1, \Xi_2, \ldots, \Xi_n \rangle^{c.} \in \mathrm{Lr}_e^A$;
   - Call the moves $\top 1, \top 2, \top 3, \ldots$ switch moves. If $\Gamma$ does not contain any switch moves, then $\mathrm{Wn}_e^G\langle\Gamma\rangle = \mathrm{Wn}_e^A\langle\Gamma^{0.}\rangle$. If $\Gamma$ contains infinitely many switch moves, then $\mathrm{Wn}_e^G\langle\Gamma\rangle = \bot$; otherwise, where $\top n$ is the last switch move of $\Gamma$, $\mathrm{Wn}_e^G\langle\Gamma\rangle = \mathrm{Wn}_e^A\langle\Gamma^{n.}\rangle$.

Again, De Morgan's laws are valid:

$$\neg\curlywedge A = \curlyvee\neg A \qquad \curlywedge A = \neg\curlyvee\neg A$$
$$\neg\curlyvee A = \curlywedge\neg A \qquad \curlyvee A = \neg\curlywedge\neg A$$

**Definition 1.3.1.30**

- $A \geqslant B =_{def} \neg A \veebar B$;
- $A \succ\!\!- B =_{def} \curlywedge A \rightarrow B$;

- $\succ\!\neg A =_{def} A \succ\!\!-\bot$.

At last, we have reached the end of our long semantics-syntax FDP, which allowed us to review the extent and characteristics of this uncharted portion of steppe. We now need to press down on our *cloche* and safely land on the (now less) unknown, in order to physically inspect the overviewed grounds and touch first-hand all their asperities.

Let us get to the heart of this rich framework right from its proof-system: Cirquent Calculus.

# 2

## NAMING THE LAND: AXIOMATISATION

Our ground level, topographical survey has just begun.

We have managed to land on a lush, vast amber green carpet with nothing more than just a sketch of what is expecting us.

Out of the things we caught glimpse of from the high teal ceiling, a peculiar-flowing mountain range that swirls in the midst of our rolling steppe seems to be the main land portion we ought to inspect first.

This apparent collective *inselberg* seems to be inhabited somehow, with visible tiny shelters cloistered away in its ever-concealing crevices.
Its peaks are not tremendously high - after all, we are in the *chartreuse*-grassed portion of the Georgian steppe. Nevertheless, we can clearly discern each coated crest with absolute clarity - a pointy pinch for each carpet lump.

Strangely enough, as seen during our aerial survey, there is a layer of low clouds, specifically a *stratus*, hovering in the distance, over and after some mountaintops.
However, as odd as it may be - since these grasslands are known to usually tumble under wide clear skies, we decide to start our expedition right away. Our goal is to learn all about the nooks and crannies this high chain is keeping quiet about.

We are going to thoroughly and physically explore this ridge step by step, little by little, thanks to the climbing gear Japaridze has provided us with: Cirquent Calculus.

This will allow us not only to reconnoitre and safely map these remote summits, but also get to the "bottom" of them by a deep-reaching approach: this allows us to apply our tools and exhaustively inspect the composition and structure of each mountain.

Furthermore, our journey further comprises a toponymical effort.

As we are used to do ever since we are born, pointing our finger in a deictic prelinguistic gesture, we are drawn to require what our attention is focused on and, at the same time, *call* it, give it a proper *name*, term it according to our attentive intentionality in a self-distancing act, a sort of preliminary aesthetic *sensus sui*[1]

Consequently, we will be assigning names to each mountain. For rigour's sake, we first have to agree upon a standard naming pattern as part of our geographical-recognition expedition.

Indeed, we will be starting from "CL1" - as in Computability Logic, and go all the way through "CL17" - which seems to be the last, visible peak, at least for the time being[2]. Indeed, just after this one, visibility seems to dramatically dim out.

Let us first understand how the powerful tool of CC works and carefully follow Japaridze's guiding tips[3], in order to inspect the tangible high creases of our fertile steppe.

---

1 Which is what separates man from animals, according to the Genesis. Indeed, Adam proves his humanity through the act of naming things around him. However, not all things were named by humans, from a biblical perspective. Indeed, God creates the universe piece by piece and names every *good* thing created (e.g. "day", "night", "sky" and so on). As Walter Benjamin writes in [4] (specifically in *On Language as Such and on the Language of Man*), "In God, name is creative because it is word, and God's word is cognizant because it is name. The absolute relation of name to knowledge exists only in God; only there is name, because it is inwardly identical with the creative word, the pure medium of knowledge. This means that God made things knowable in their names. Man, however, names them according to knowledge". Indeed, we will be naming each mountain according to the knowledge we have of them - for instance, the order in which we put them. For the sole purpose of weaving some red threads in-between these pages, it is interesting to point out the remarkable resemblance of the notion of name in Benjamin to the one of Word in Luther, which we have previously mentioned. Both constitute the core of what Benjamin calls *pure language*, meaning the one that does not communicate; indeed, this language only reveals itself *in* the Word, *in* the name.

2 This approach reminds me of Sesto's Sundial up in the Italian Dolomites, even though we are in a totally different environment. There, each mountain has its own numerical name, meaning the hour each peak shadows on the ground: each specific altitude allows us to acquire information on the current time. On our part, we will be calling each crest - also referred to as "fragment", with a numerical name indeed. However, the information acquisition is going to go the other way round, as we will gain knowledge on the mountains themselves, rather than something else, through external resources - meaning the appropriate equipment we have been provided with.

3 Which will already lightly describe certain fragments without, however, spoiling the entire thing before our expedition. Indeed, it is not a vain exploration, as it is meant

## 2.1 CIRQUENT CALCULUS, NEW ALTITUDES

First introduced in [42], 2005, Cirquent Calculus is a new proof system elaborated by Japaridze for resource-conscious logics; it was initially conceived as a deductive tool for Computability Logic, which "stubbornly resisted" all axiomatisation attempts within the framework of traditional syntactic approaches.

Indeed, as explained in [73], the word **cirquent** is a *portmanteau* of "circuit" and "sequent": as a matter of fact, Cirquent Calculus deals with circuit-style constructs rather than tree-like structures (such as sequents and formulas).
Consequently, we might say that Cirquent Calculus, "CC" from now on, is an elegant refinement of sequent calculus, with cirquents replacing formulas.

Unlike more traditional syntactic approaches that manipulate forest-like objects, such as formulas or sequents, CC deals with circuit-style constructs where children may be *shared* between different parent nodes.

Indeed, the main advantage of this calculus lies in the ability to account for the sharing of subformulas between different resources.

Gentzen's sequent calculus, on one hand, involves "disjoint and independent sequences of formulas" ([42], page 489), whose information cannot be grasped in its full scope through the rules we are provided with - mainly because it is a shallow-inference system.

On the other hand, CC can combine, within a single cirquent, different parallel nodes of a traditional proof tree - being intrinsically a deep-inference system, as we will later see.

Indeed, many situations rising from the world of computing witness prohibitively long formulas, which typically owe their sizes to reoccurring subformulas, and explosively large proof trees, which often emerge as a result of necessarily repeating the same steps over and over again.

As a result, sharing allows us to achieve higher efficiency, whether it be

---

to give a comprehensive account of all the fragment put together - which has yet to be done. Japaridze gives us the relevant directions to draw this map.

compactness of representation or number of steps in derivations and proofs.

Let us now introduce this new calculus from an informal stance and then proceed to a more formal account through definitions, rules and applications.

### 2.1.1 *No stone left unturned*

Just as in [49] and [73], we should start by asking ourselves the following question (first raised by Alessio Guglielmi in the preliminary discussion of [49]):

> What is the most natural representation of Boolean functions[4], formulas or circuits?

and by "natural", as Japaridze clears up right away, we mean direct, reasonable and efficient.

To see why we choose circuits over formulas, here is an example[5]:

$(A \wedge B) \vee (A \wedge C) \vee (B \wedge C)$

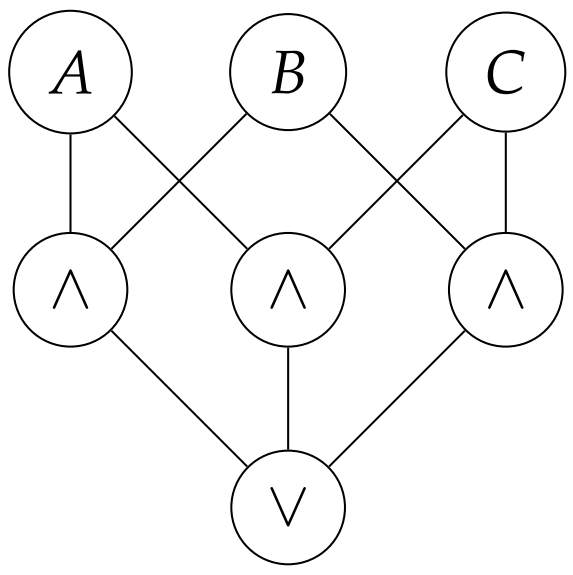

On the right we have the circuit for the Boolean function expressed by the formula on the left.
As you can see, we have a sharing situation in the circuit that is not explicitly shown in the formula: input $B$ is shared between the first and the third conjunctive gates.

4 Let us recall that a Boolean function is a function which, together with its arguments, always assumes two possible values, usually 0 or 1 - what in logic is sometimes called *truth function*. Boolean functions usually consists of a number of Boolean variables joined by the Boolean connectives AND and OR, as we can see in this section.

5 Taken from [73], page 125.

Indeed, formulas have no sharing mechanisms: this means that $B$ has to be written twice. That is why circuits are generally (and exponentially) more compact than the corresponding formulas. This is even more evident if we increase the complexity of the Boolean function in question.

Such compactness, as we have already mentioned, is the reason why computer hardware is based on circuits in the first place - rather than on formulas or other structures that bear no advantage in sharing.

As a result, switching from formulas to cirquents becomes syntactically and semantically necessary when developing resource logics.
Indeed, fine-level resource-semantical approaches intrinsically require the ability to account for the possibility of resource sharing; ability that, alas, neither linear logic nor other formula/sequent-based approach do not - and cannot - possess.

Cirquents are more general than Boolean circuits[6]: every Boolean circuit is a cirquent but not vice versa. Out of these two cirquents,

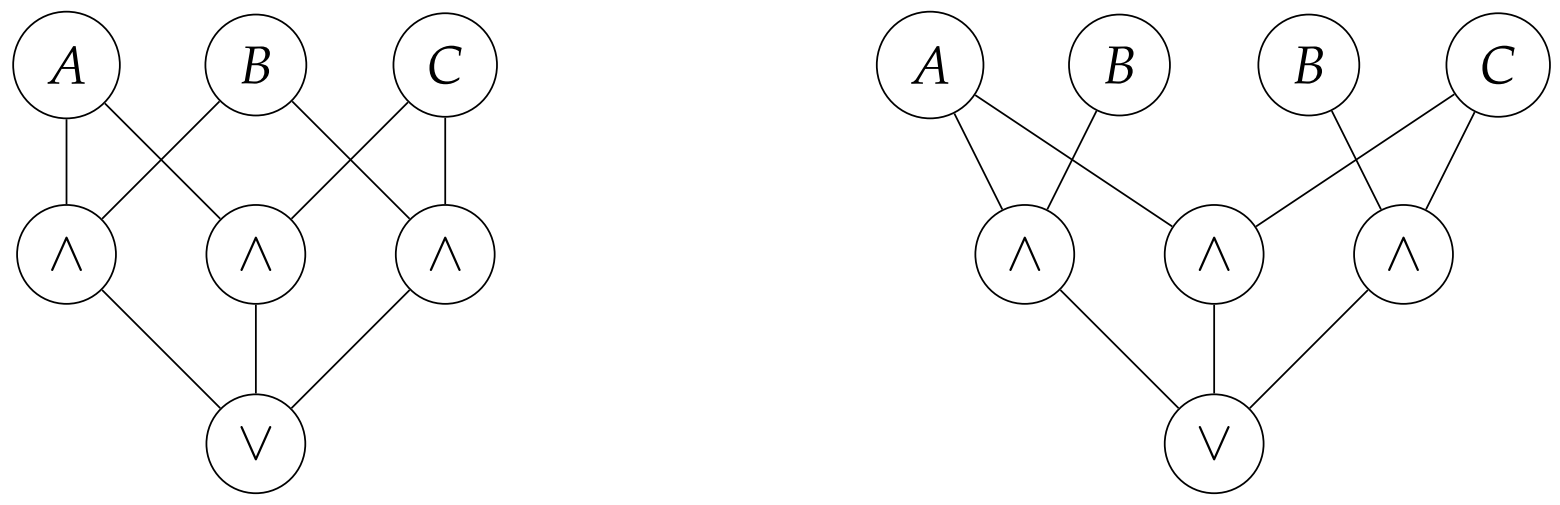

only the left one is a Boolean circuit in the proper sense.

The cirquent on the right is not how Boolean circuits are normally drawn, as it has two copies of $B$. Indeed, from the classical point of view, writing (the "same" input) $B$ twice makes no sense.
Thus, we conclude that the circuit on the right is a direct reading of the formula, while the circuit on the left is a most economical representation of the same Boolean function.

On the other hand, from the point of view of CC, the above cirquents are

6 Roughly put, Boolean circuits are graphs that compute a Boolean function from input to output using nodes and edges.

simply different: not only syntactically, but also semantically.

Linear logic, viewing the nodes as resources rather than just Boolean values, would agree in stating that the two "$B$" nodes of the right circuit, even though having the same type, are two different individual resources. However, what linear logic generally fails to account for, as already mentioned, is the possibility of resource sharing, which is what differentiates it from the semantical framework of CC.

Allowing shared resources in Cirquent Calculus means refining the otherwise crude approach of linear logic. This, however, does not mean that resources should be accurately book-kept: a shared resource is not a duplicated one.

As Japaridze puts it ([42], 496):

> Imagine Victor has $10,000 on his bank account. One day he decides to give his wife access to the account. From now on the $10,000 is shared. Two persons can use it, either at once, or portion by portion. Yet, this does not turn the $10,000 into $20,000, as the aggregate possible usage remains limited to $10,000.

The ability to compress formulas and proofs is not the only — in fact, not even the primary — appeal of Cirquent Calculus. Generality, flexibility and expressiveness are other, more fundamental, advantages to point out.

Cirquent Calculus is more general than the calculus of structures (Guglielmi et al. [28], 2007): the former is more general than hypersequent calculus (first introduced by Pottinger [88], 1983, and then independently developed by Avron [2], 1987), while the latter, in turn, is more general than sequent calculus (Gentzen).
Indeed, each framework of this hierarchy allows a successful axiomatisation of certain logics that other, traditional ones fail to tame.

To summarise CC's strengths[7], the main reasons behind its formulation have been the following:

7 We suggest having a look at Japaridze's example shown in [49], page 1000 in order to have a further grasp of CC's advantages.

- **To axiomatise CoL**: as Japaridze explains, it was proven in [12] that axiomatising even the most basic fragment of CoL through traditional means was impossible in principle. As a non-traditional approach, Cirquent Calculus offers an adequate proof system for the otherwise untameable CoL;
- **To have a more fine-grained logic of resources than linear logic**: here "resources" are meant in a more general and abstract sense than just computational resources studied in CoL. Indeed, resource sharing is a ubiquitous phenomenon; however, linear logic still lacks a suitable mechanism to account for it;
- **To provide a more efficient proof system for classical logic (and beyond) than traditional ones**: as shown in [49], Cirquent Calculus is indeed advantageous in that it achieves an exponential speedup of analytic proofs over cut-free sequent calculus or other analytic proof systems.

---

*Classical and linear logics according to CC*

To understand the relationship between Cirquent Calculus and classical logic and linear logics, let us consider the formula:

$$(A \wedge B) \vee (B \wedge C)$$

and ask which of the following two cirquents is a more adequate or reasonable representation of its meaning:

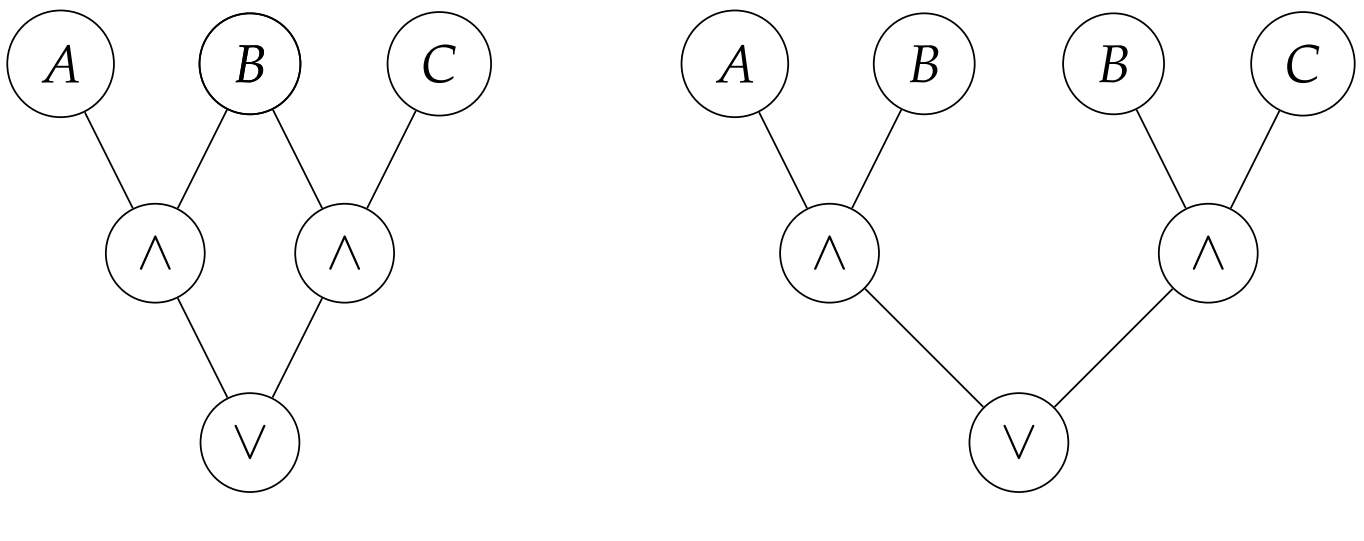

Obviously, the answer depends on who you ask.

According to classical logic, it is the cirquent on the left: even though in the formula we see $B$ twice, those two occurrences are semantically the same in every relevant sense, so there is no point in not combining them.

On the other hand, a linear logician would say that the meaning of the formula is captured by the cirquent on the right: the formula has two copies of $B$, with each copy standing for a separate resource, even though both resources have the same type $B$.

As mentioned before, from the point of view of CC, classical and linear logics are two faulty extremes. Their resource-semantical stances on logical expressions can be schematically summarised as follows:

- Classical logic: **everything** is *shared* that can be shared. Namely, each occurrence of the same subcomponent stands for the same copy of the same resource;
- Linear logic: **nothing** is *shared* at all. Namely, each occurrence of the same subcomponent stands for a different copy of the same resource.

But how about **mixed** (real life!) cases, where some same-type resources may be shared between different subresources while some may not?

Take the following cirquent, for instance. The two conjunctions both share $A$ and $C$, however each one has its own $B$.
Only Cirquent Calculus makes it possible to account for such cases.

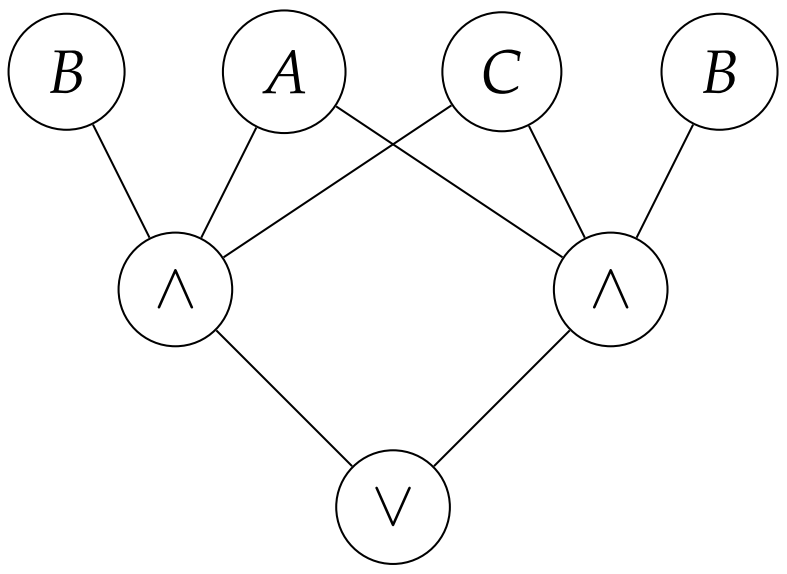

This is why Japaridze feels that Cirquent Calculus has the potential to improve linear logic's proof system - and that of any other resource-conscious logic.

As he writes in [49], CC is able to overcome the "the insufficiency of the expressive power of linear logic or other formula-based approaches to developing resource logics".

In addition, as mentioned in the previous subsection, this sharing mechanism makes it possible to achieve exponential-magnitude compressions of formulas and proofs - both in CoL and in the old, graciously well-mannered lady Classical Logic.

Let us now carefully examine how CC and sequent calculus[8] build different proofs for a one same formula.

---

*Cirquent vs. sequent calculi*

A number of Cirquent Calculus systems of various degrees of expressiveness have been developed thus far.
The earliest, simplest and most basic one, first introduced in [42], is system CL5 - which we will later overview in its full swing.

CL5, as Japaridze cares to inform us before our exploration, operates with special, *shallow* sorts of cirquents - even though CC has evolved into a deep-inference system on the whole. Every such cirquent is of height 2, having a conjunctive root with disjunctive children and $(\neg, \wedge, \vee)$-formulas as grandchildren.

With $F, G, H$ standing for some arbitrary (not necessarily atomic) formulas, here is an example of a shallow cirquent:

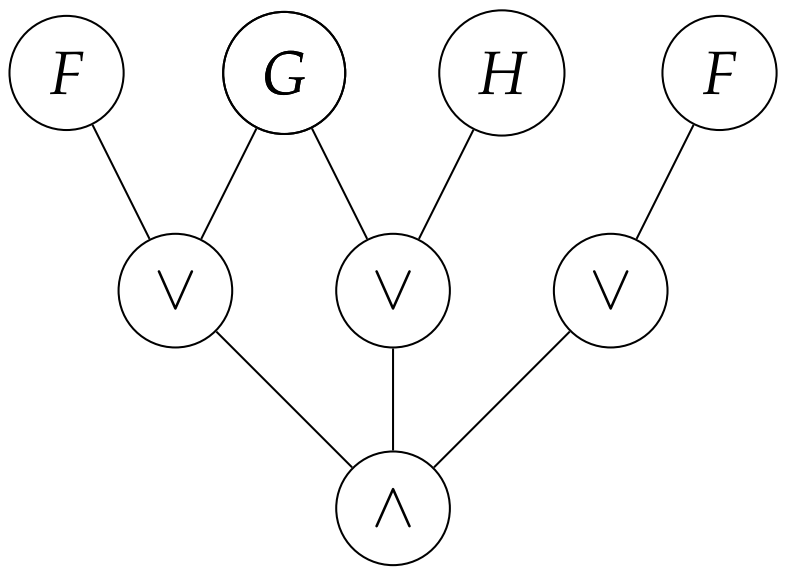

8 In the following subsection, with "sequent calculi" we will specifically refer to the one-sided ones, named *à la Tait*. Here, a sequence is a finite set $\Gamma$ (sometimes multiset, other times list) of formulas thread together through disjunction. Indeed, in Gentzen's original calculus a sequent is an object of the type: $\Gamma \Rightarrow \Delta$; typically, a one-sided sequent is defined as $\neg\Gamma, \Delta$. What makes it preferable to the original one is that Tait's calculus actually needs less rules.

For economy of language, as we will later see when introducing all of CoL's axiomatisations, we will write a CL5 cirquent without showing the (always) conjunctive root and using bullets (•), called **groups**, instead of disjunctive gates.
Here is the same example in new guise:

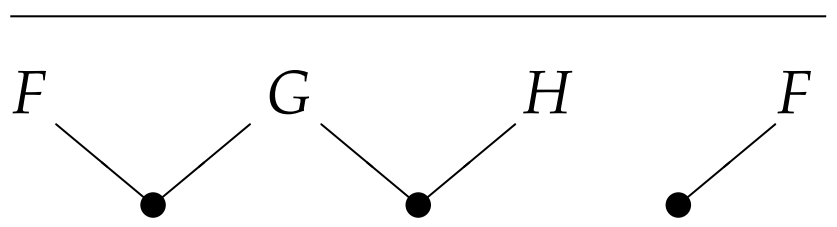

A group is said to **contain** the **oformulas** (meaning the particular occurrences of formulas) it is connected to through an **arc**. The horizontal line on the top is to show that this is just one cirquent rather than two - as it may be otherwise perceived.

A formula $F$ in isolation, also known as **singleton**, is understood as the cirquent with a single group that contains just $F$:

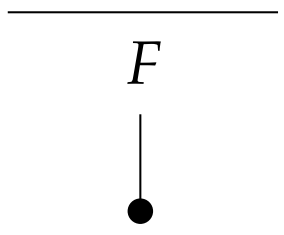

Proof-theoretically, a cirquent can be thought of as the collection of potentially-subformulas-sharing leaf sequents that appear in the corresponding sequent calculus proof tree.

Below is a side-by-side example of the CC (specifically, CL5) representation of the target formula proven in sequent calculus.

On the right, we have a fragment of the sequent calculus proof tree where the formula $G$ is implicitly assumed to be shared between the two leaf sequents $\langle F, G \rangle$ and $\langle G, H \rangle$. The formula $F$, while also occurring in two leaf sequents, is not assumed to be shared between them.

The corresponding cirquent on the left, whose three groups represent the three leaves of the proof tree, brings to light these implicit assumptions on the presence and absence of sharing.

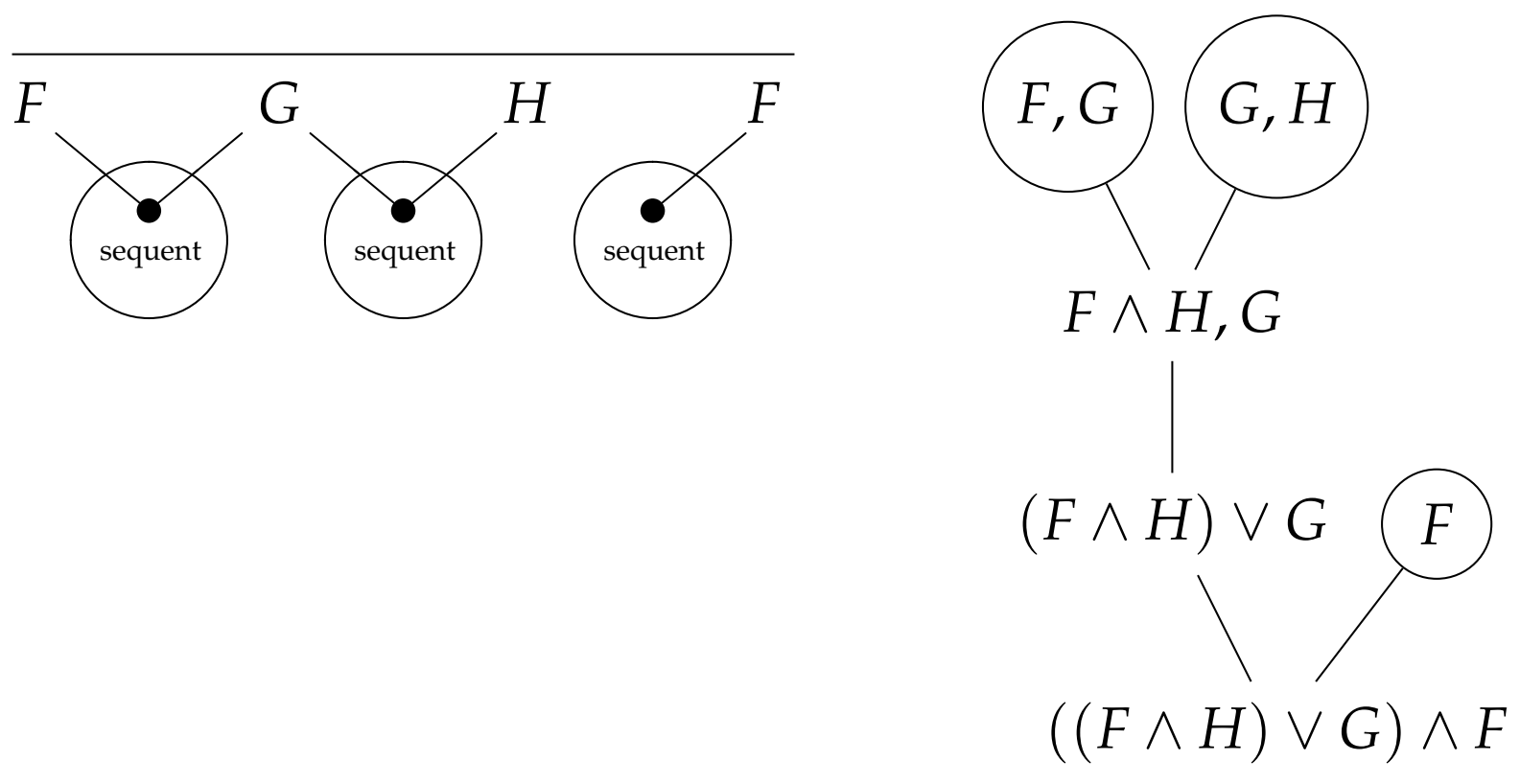

The inference rules of CL5 will be explained through the following sections. However, for preliminary and intuitive insights[9] on the expressive power of CC, let us compare a fragment of the Cirquent Calculus proof of $((F \wedge H) \vee G) \vee F$ with the corresponding portion of the sequent calculus proof of the same formula:

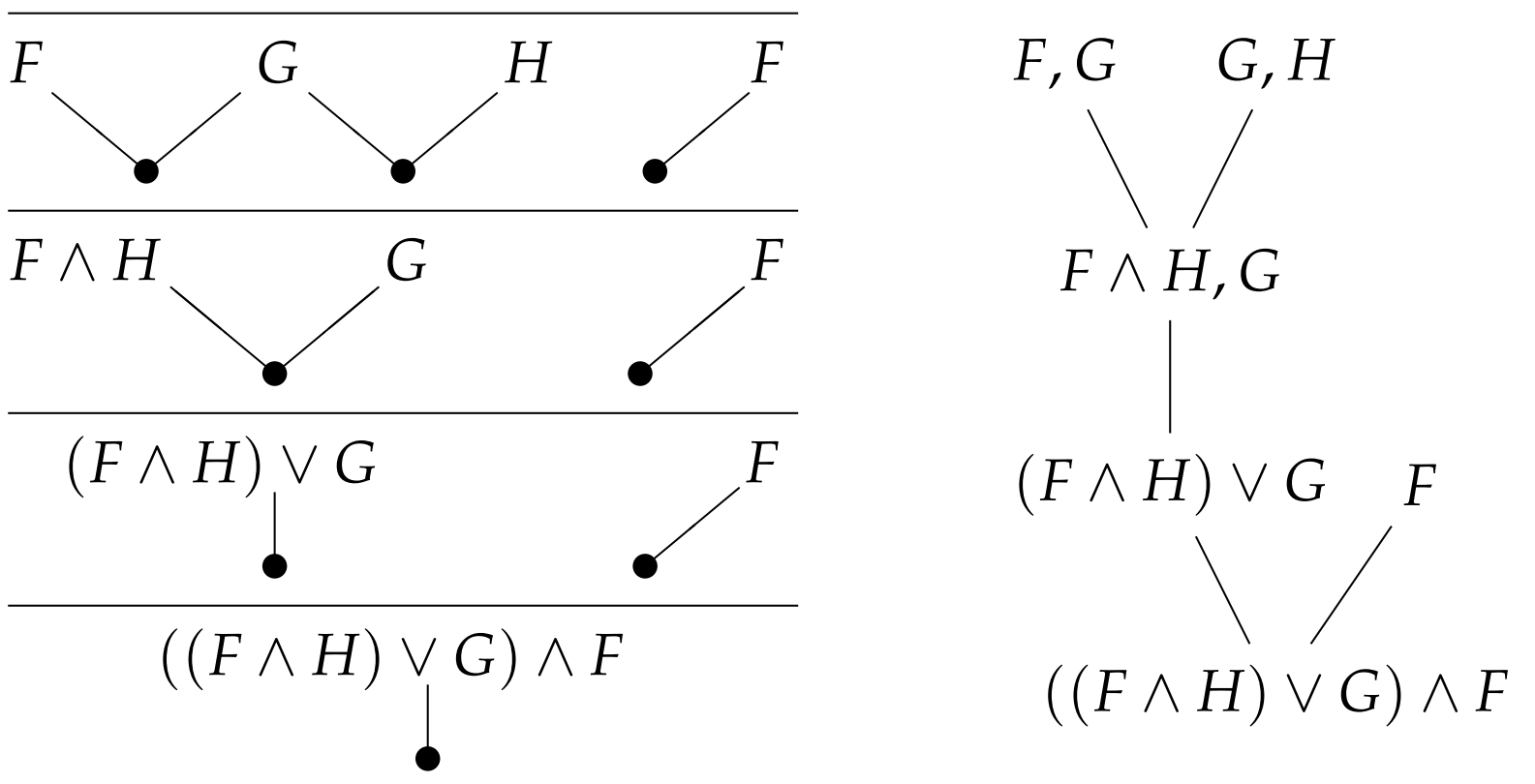

According to a bottom-up (conclusion to premises) approach, the rightmost proof begins with the sequent of the target formula.

On the left we see the corresponding target cirquent with a single group. The number of leaves in the sequent calculus proof tree is thus 1, and so is the number of groups at the corresponding step of the CC proof. This situation will persist throughout the rest of steps.

In the sequent calculus proof, due to the main connective $\wedge$ in $((F \wedge H) \vee$

9 Per usual with our *wok-pan* method.

$G) \wedge F)$, the target sequent is split into two, namely $\langle (F \wedge H) \vee G \rangle$ and $\langle F \rangle$, when proceeding upwards.
Correspondingly, the single group of our CC proof splits into two groups, one containing $(F \wedge H) \vee G$ and the other $F$.

Going further, the third step of the right-side proof replaces $\vee$ with a comma in $(F \wedge H) \vee G$.
In the other proof, the single formula $(F \wedge H) \vee G$ of the left group correspondingly splits into $F \wedge H$ and $G$.

Nothing remarkable has happened so far. Indeed, it is when transitioning to the top level of either proof that things get interesting.

Due to the conjunctive formula $F \wedge H$, the sequent $\langle F \wedge H, G \rangle$, together with the corresponding cirquent group, splits in two: one splinter taking the formula $F$ with it and the other taking $H$.

Which splinter can inherit $G$?
Classical logic says that both corresponding branches of the sequent calculus proof tree can simultaneously take $G$, because there is no difference between multiple $G$s and a single $G$.
On the other hand, the resource-conscious linear logic insists that only one of the branches can take $G$ with it.

Cirquent Calculus, just as CoL does in its neighbourhood, reconciliates the two extremes by letting $G$ be inherited by both splinter groups, all the while explicitly showing that $G$ is shared between them - hence we still have a single occurrence rather than two copies of $G$.

Let us now get more acquainted with the notion of deep-inference.

---

*Deep-inference: reaching*

In order to reconnoitre every single nook and cranny that comes our way, we shall briefly elucidate what we mean by **deep-inference**[10].

Deep-inference is a relatively young approach which has given rise to new,

10 Much of this part references http://alessio.guglielmi.name/res/cos/diom.html, which offers a brilliant summary of the notion we are trying to unfold.

alternative formalisms of proof theory (such as **calculus of structures**[11] and **open deduction**, which subsumes it).

Indeed, we call **shallow-inference** the traditional type calculi such as Hilbert-Frege systems, natural deduction (NK), sequent calculus (LK) and so on.

In these formalism, inference rules can only be applied at the root of formulas (or sequents or other structures).
As a result, deductions are rigidly constrained[12] and a great deal of readily available information contained in formulas is not used - as a matter of fact, it is unreachable, out of our depth.

Deep-inference, on the other hand, makes it so that inference rules can be applied *anywhere*, even in the deepest levels. This way, all the information is available to deduction - rich!

As shown by Guglielmi:

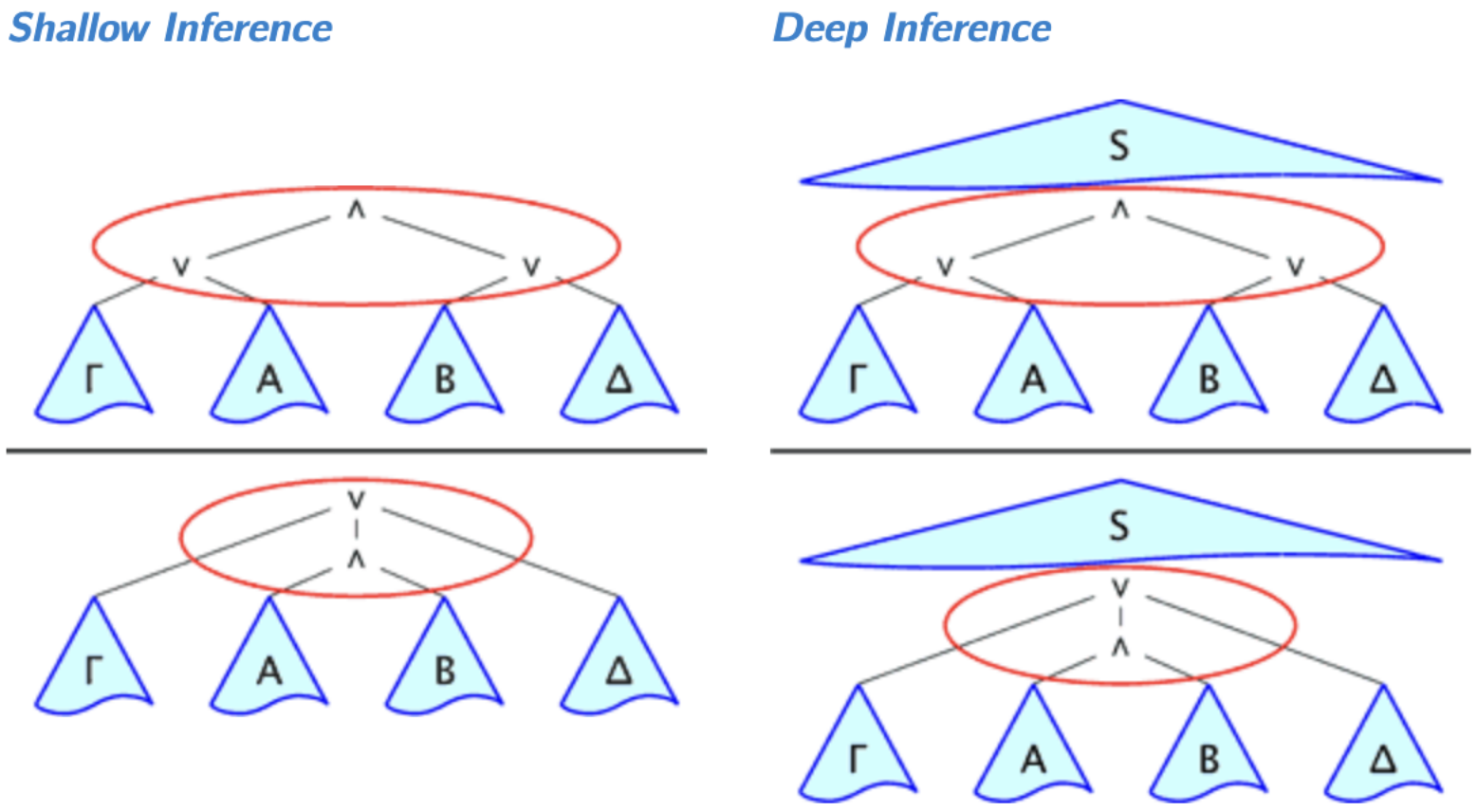

11 Developed by the same Alessio Guglielmi et al. in [27], 2001.
12 For instance, if we dispose of two proofs $\Phi = A \Rightarrow B$ and $\Psi = C \Rightarrow D$, then their compositions $\Phi \wedge \Psi = (A \wedge C) \Rightarrow (B \wedge D)$ and $\Phi \vee \Psi = (A \vee C) \Rightarrow (B \vee D)$ are valid proofs. Problem is that while $\Phi \wedge \Psi$ can be represented with Gentzen systems, $\Phi \vee \Psi$ cannot, or it only can with severe limitations. Gentzen formalisms and deep-inference differ in that the latter allows composition of proofs by the same connective of formulas, while the former struggles to.

On the left we have the sequent calculus inference:

$$\frac{\Gamma, A \qquad B, \Delta}{\Gamma, A \wedge B, \Delta}$$

The red circled part of the diagram is the only area that can be inspected by the inference rule, meaning the root of the entire structure.

However, on the right side, we show an example taken from calculus of structures. Here we have:

$$\frac{S\{(\Gamma \vee A) \wedge (B \vee \Delta)\}}{S\{\Gamma \vee (A \wedge B) \vee \Delta\}}$$

Indeed, the inference rule can be applied anywhere, since $S$ is an additional schematic variable.

In other words, being deep-inference means being able to deduce inside a formula at any depth, as opposed to the shallow-inference of sequent calculus, where rules only see the roots of formula trees.

Indeed, this can be exploited in many ways: for instance, to obtain new normalisations other than cut elimination; to model quantum causal evolution; to shorten analytic proofs by exponential factors compared to Gentzen; and so much more.

In our case, deep-inference allows us to, as Guglielmi writes, "express logics that cannot be expressed in Gentzen". CC is, indeed, a deep-inference calculus for such reason.

We may ask ourselves why was not deep-inference exploited before. The answer probably lies in the fact that deep-inference breaks the core of traditional normalisation techniques.

For example, in a typical cut elimination step, one needs to deal with the following situation:

$$\frac{\dfrac{\Gamma, A \qquad \Gamma, B}{\Gamma, A \wedge B} \qquad \dfrac{\overline{A}, \overline{B}, \Delta}{\overline{A} \vee \overline{B}, \Delta}}{\Gamma, \Delta}$$

We know how to normalise this piece of proof because we know that the formulas $A \wedge B$ and $\bar{A} \vee \bar{B}$ are taken apart at their root by the shallow-inference rules.

However, this is not true in the case of deep-inference. This crucial piece of information is missing, so new techniques for normalisation need to be developed.

Concerning the calculus of structures and open deduction, we have fairly general, new techniques which not only prove cut elimination, but also give us access to many other fine-grained normalisations that are simply impossible in shallow-inference.

As a result, CC has allowed us to axiomatise many fragments of CoL which have a deep-inference outline. In [49], just after having introduced CL5, Japaridze explores the possibility of deeper versions of CC, thus developing CL8, which we will later see.

Indeed, CL8 permits cirquents of arbitrary depths and forms, which naturally invites inference rules that modify cirquents at any level rather than only around the root as is the case in sequent calculus.

In the same paper, Japaridze points out that:

> [...] it is not only cirquent calculus where the subformula property is no longer meaningful. The same holds for deep-inference systems in general, such as the calculus of structures.

Thus, we are faced with a new philosophical issue.
Indeed, the term "analytic" seems to assume a more relaxed sense in deep-inference, simply meaning the absence of cut, substitution, extension and other rules.

The common undesirable feature of these rejected rules is that, when moving from a conclusion to a premise, they introduce some new components, as opposed to the rules deemed analytic[13] in [27], which merely regroup some already existing components without creating new ones.

What "components" or "regrouping" should exactly mean here, however,

13 As Japaridze points out, all rules of CL8 would also qualify as analytic by similar standards.

certainly does require some additional and probably non-trivial explanations.

To summarise, there appears to be no well-agreed-upon concept of analiticity in the literature after the advent of deep-inference.

It is now time to master the powerful tool we have analysed thus far and learn how it works.

### 2.1.2 *Formal instruction booklet*

*CC through ARS*

Even though CC was initially brought up for CoL, Japaridze has developed another semantics for it, namely ARS - **Abstract Resource Semantics**.

We briefly showcase its main key points in a somewhat formal approach - the reader is advised to have a look at [42], chapter 8, for a more in-depth presentation, for which, alas, we do not have enough space here.

Indeed, ARS was first introduced for an earlier shallow class of cirquents - CL5, as seen in [42]. Its generalisation is presented in [49], where ARS is defined as a conservative extension of the semantics of classical logic from circuits to all cirquents.

As Japaridze explains, ARS is a "real"[14] semantics of resources, formalising the resource philosophy traditionally (and, as argued in [42], somewhat wrongly) associated with linear logic and its variations.

Briefly, the starting point of this semantics is the concept of a **truth assignment** for a given cirquent $C$. This is a function that assigns one of two values - *true* or *false* - to each **port**[15] of $C$. Any function $f$ is a

14 "Real" as in dealing with resources similarly to how they are dealt with in real life.

15 A port is $P$ or $-P$, where $P$ is an atom called the **type** of the port. A node labelled with a literal $L$ is said to be an ***L*-port**. Furthermore, a port which is just an atom is said to be an **output**, while a port which is a "$-$" prefixed atom is said to be an **input**. The I/O status of a port is said to be the **gender** of that port. The two genders *input*, *output* are said to be **opposite**.

legitimate truth assignment, including the cases when $f$ assigns different truth values to ports that have identical labels.

Intuitively this is perfectly meaningful in the world of resources. For instance, a slot of a vending machine that only accepts 25*c* coins (thus, a 25*c*-port) may receive a true coin, while another 25*c*-port may receive a false coin or no coin at all.

Among the basic notions used in ARS is that of **atomic resource**. This is an undefined concept, as Japaridze explains: we can only point at some examples of what might be intuitively considered as an atomic resource. For instance, these can be a specified amount of money; the electric power of a specified voltage and amperage; a certain task performed by a computer, such as providing Internet browsing capabilities; and so on.

We agree that atomic resources are nothing but propositional letters, i.e. atoms of the language underlying Cirquent Calculus[16].

Stemming from atomic resources, Japaridze introduces **compound resources**[17].
For instance, \$2 can be treated as an atomic resource, but it may as well be understood as a combination of \$1 and \$1 — specifically, the combination \$1 $\wedge$ \$1, with $\alpha \wedge \beta$ generally being the resource having which intuitively means having both $\alpha$ and $\beta$.

One thing to point out is that, when talking about resources, we always have two parties in mind: the resource provider and the resource user. Correspondingly, every entity that we call a resource comes in two flavours, depending on who is providing the resource. An example of this is provided in [42].

We call **interface** a finite sequence of ports. A particular occurrence of a port (input, output) in an interface will be referred to as an **oport** (**oinput**, **ooutput**).

16 Clearly, this is some abuse of concepts. Indeed, strictly speaking, the atoms of our language are variables ranging over atomic resources, rather than atomic resources as such. Similar terminological liberties are extended to the concepts formally defined below.

17 "Of course, whether a resource is considered atomic or thought of as a combination of some more basic resources depends on the degree of abstraction or encapsulation we choose in a given treatment", [49], page 1010.

As we did with oformulas in the context of cirquents, we usually refer to an oport by the name of the corresponding port (as in "the oport $P$"), even though different oports may be identical as ports.

By a **situation** for a formula $F$ we mean an assignment **s** of truth values 0/1 to its oatoms: specifically, $s$ is a function of the type $\{1, \ldots, n\} \to \{0, 1\}$.

Each truth assignment for a cirquent extends from its ports to all gates and the cirquent itself in the following, expected way:

- A disjunctive gate (node) is true iff it has at least one true child;
- A conjunctive gate (node) is true iff so are all of its children;
- The cirquent is true iff so is its root.

An **allocation** for a given cirquent $C$ is an unordered pair $\{a, b\}$ of ports of $C$ with opposite labels (labels $P$ and $\neg P$ for some — the same — atom $P$).

Finally, an **arrangement** for $C$ is any set of pairwise disjoint allocations for $C$. We call the condition requiring all allocations to be disjoint the **monogamicity condition**.

Indeed, **abstract resources** are here represented through cirquents. They are roughly the same to Abstract Resource Semantics as Boolean functions to the semantics of classical logic.

**Definition 2.1.2.1**
An **abstract resource** — henceforth simply "resource" — is a pair $\alpha = (\mathrm{Int}^\alpha, \mathrm{Tfn}^\alpha)$, where:

1. $\mathrm{Int}^\alpha$, called the interface of $\alpha$, is an interface;
2. $\mathrm{Tfn}^\alpha$, called the truth function of $\alpha$, is a function that sends every situation $s$ for $\mathrm{Int}^\alpha$ to 0 or 1, such that the following monotonicity condition is satisfied:
   - Whenever $s \leq_{Int^\alpha} s'$, we have $\mathrm{Tfn}^\alpha(s) \leq \mathrm{Tfn}^\alpha(s')$.

When $\mathrm{Tfn}^\alpha(s) = 1$ (respectively $= 0$), we say that $\alpha$ is true (respectively false) in situation $s$.

Furthermore, "output of $\alpha$", "oport of $\alpha$" and "situation for $\alpha$" mean "output of $\mathrm{Int}^{\alpha}$", "oport of $\mathrm{Int}^{\alpha}$", "situation for $\mathrm{Int}^{\alpha}$", and so on.

Let us briefly talk validity.

Let $C$ be a cirquent, $f$ a truth assignment for $C$, and $\alpha$ an arrangement for $C$.

We say that $f$ is **consistent** with $\alpha$ iff, for every allocation $\{a, b\} \in \alpha, f(a) = f(b)$. That is, if ports $a$ and $b$ are allocated to each other (meaning that $\{a, b\} \in \alpha$), a truth assignment consistent with $\alpha$ should assign opposite truth values to a and b.

We say that $\alpha$ is **validating** (for $C$) iff $C$ is true under every truth assignment consistent with $\alpha$.

**Definition 2.1.2.2**
We say that a cirquent is a **valid** (in abstract resource semantics) iff there is a validating arrangement for it.

Validity in our sense and validity (*tautologicity*) in the classical sense mean the same for circuits, hence also for formulas of classical logic understood as circuits.
As a matter of fact, as already mentioned, ARS is a conservative extension of classical semantics from circuits to all cirquents.

**Lemma 2.1.2.3**
A cirquent is valid iff it is an instance of a valid circuit.

Therefore, as shown with respect to CL8 in [49], a cirquent is provable iff it is valid in Abstract Resource Semantics.

In conclusion, Abstract Resource Semantics is not far from CoL's semantics - they both claim to be semantics of resources. ARS and CoL's semantics have been shown ([42]) to validate the same principles.

Indeed, ARS is a more general purpose framework, with its basic notions such as (atomic) resources, their truth values, or allocations being open to various specific interpretations.

Computability Logic offers one such interpretation and its semantics can be seen as a materialisation of Abstract Resource Semantics. A resource in CoL has a very specific meaning: it is a derived concept defined in terms of some other, more basic entities (specifically, games[18]).
The same can be said about other semantical concepts.

While being less general than ARS, however, CoL's semantics is still general enough to validate anything that ARS deems as such.

---

*CC through CoL's semantics*

As already mentioned, cirquents can be seen as generalised sequents by imposing circuit-style structures on their oformulas.

In a preliminary attempt to recognise some familiar meaning in cirquents, it might be helpful to think of them as Boolean circuits of depth 2. Here, in a top-down account, every oformula serves as an input, while all first-level gates — representing ogroups — are $\vee$- gates and the only second-level gate, connected to each first-level gate, is an $\wedge$-gate.

Additionally, cirquents are more general than Boolean circuits in that, when dealing with certain non-classical logics, the former can have non-Boolean gates as well.

Furthermore, we define the notion of **hyperformulas**[19]. Visually, these are the same as formulas, with the only difference that some subformulas in an hyperformula may be overlined (double overlines are not allowed).

Hyperformulas, just like formulas, are understood as cirquents. To translate a hyperformula $F$ into a corresponding cirquent, one should first ignore all overlines in $F$ and translate it into a tree-like cirquent according to the earlier prescriptions; then, merge all subcirquents that correspond to (originate from) the same, identical overlined subformulas of $F$.

18 Indeed, successfully providing the resource (truth value 1 in ARS) corresponds to winning the game in CoL.
19 Specifically, we introduce the syntactic embodiment of the notion that will later come in handy while climbing up the golden-grassed peaks.

Here is an example. The hyperformula:

$$(Q \vee \bar{R}) \wedge (\bar{R} \wedge Q)$$

is translated into the cirquent:

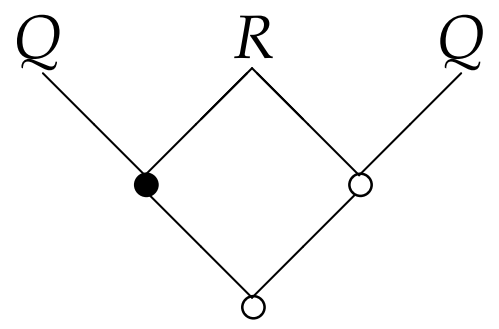

Indeed, the two occurrences of $R$ are considered to be "the same" as they are both overlined. On the other hand, the two occurrences of $Q$ did not merge as they were not overlined.

Let us introduce a few, final definitions of CoL's semantics in order to have a complete picture of CC and be ready for our ascent.

We fix some infinite collection of finite alphanumeric expressions called **atoms**, and use $p, q, p_1, p_2, p(3,6), q_7(1,1,8), \ldots$ as metavariables for them. If $p$ is an atom, then the expressions $p$ and $\neg p$ are said to be **literals**.

Let us agree that a **graph** is a directed acyclic multigraph with countably many nodes and edges, where the outgoing edges of each node are arranged in a fixed left-to-right order. Each node is labelled with either a literal or one of CoL's operators.

Since the sets of nodes and edges are countable, we assume that they are always subsets of $\{1, 2, 3, ...\}$. For a node $n$ of the graph, the string representing $n$ in the standard decimal notation is said to be the **ID number**, or just ID, of the node. Similarly for the ID of the edges.

When there is an edge from a node $a$ to a node $b$, we say that $b$ is a **child** of $a$ and $a$ is a **parent** of $b$.

The relations of **descendant** and **ancestor** are the transitive closures of the relations of child and parent, respectively. The meanings of some other standard relations such as **grandchild**, **grandparent** and so on should also be clear.

The **outdegree** of a node of a graph is the quantity of outgoing edges of that node, which can be finite or infinite. Since there are only countably many edges, any two infinite outdegrees are equal.

Similarly, the **indegree** of a node is the quantity of the incoming edges of that node.

We say that a graph is **well-founded** iff there is no infinite sequence $a_1, a_2, a_3, \ldots$ of nodes where each $a_i$ is a parent of $a_{i+1}$. Of course, any graph with finitely many nodes is well-founded.

As Japaridze explains, we say that a graph is **effective** iff the following basic predicates and partial functions characterising it are recursive: "$x$ is a node", "$x$ is an edge", "the label of node $x$", "the outdegree of node $x$", "the $y$th outgoing edge of node $x$", "the origin of edge $x$", "the destination of edge $x$".

**Definition 2.1.2.4**
A **cirquent** is an effective, well-founded graph satisfying the following two conditions:

1. Ports have no children;
2. There is a node, called the root, which is an ancestor of all other nodes in the graph.

We say that a cirquent is **finite** iff it has only finitely many edges (and hence nodes); otherwise it is **infinite**.
A cirquent is tree-like iff the indegree of each of its non-root node is 1.

Indeed, as already mentioned, the nodes labelled $\wedge$ or $\vee$ we call $\wedge$-gates and $\vee$- gates.
Graphically, we represent ports through the corresponding literals, $\vee$-gates through $\vee$- inscribed circles, and $\wedge$-gates through $\wedge$-inscribed circles.

We agree that the direction of an edge is always upward, which allows us to draw lines rather than arrows for edges. It is understood that the official order of the outgoing edges of a gate coincides with the (left to right) order in which the edges are drawn.

Also, typically we do not show nodes IDs unless necessary. In most cases,

what particular IDs are assigned to the nodes of a cirquent is irrelevant - such an assignment can be chosen arbitrarily. Similarly, edges IDs are usually irrelevant, thus will never be shown.

By a **(perfect) interpretation** we mean a function $*$ that sends each atom $p$ to one of the values (elementary games) $p* \in \{\top, \bot\}$. It immediately extends to a mapping from all literals to $\{\top, \bot\}$ by stipulating that $(\neg p)* = \neg(p*)$; that is, * sends $\neg p$ to $\top$ iff it sends $p$ to $\bot$.

Each interpretation * induces the predicate of truth as defined below. This definition, as well as similar other ones, silently and essentially relies on the fact that we only consider well-founded graphs.

**Definition 2.1.2.5**
Let $C$ be a cirquent and * an interpretation. With "port" and "gate" below meaning those of $C$, and "true" to be read as "true with respect to", we say that:

- An $L$-port is true iff $L* = \top$ (any literal $L$);
- A $\vee$-gate is true iff so is at least one of its children (thus, a childless $\vee$-gate is always true;
- A $\wedge$-gate is true iff so is are all of its children (thus, a childless $\wedge$-gate is always true;

Finally, we say that the cirquent $C$ is true iff so is its root.

**Definition 2.1.2.6**
Let $C$ be a cirquent and * an interpretation. We define $C*$ to be the elementary game the (only) legal run $\langle\rangle$ of which is won by $\top$ iff $C$ is true with respect to *.

We say that two cirquents $C_1$ and $C_2$ are **extensionally identical** iff, for every interpretation *, $C_1^* = C_2^*$.
Finally, we say that a cirquent $C$ is valid iff, for any interpretation *, $C* = \top$.

---

*Selectional gates: first generalisation*

These definitions concerned the simplest type of cirquents. We now explore some different sort of gates while extending and generalising cirquents and their semantics.

In doing so, we will not have enough space to deeply cover every description and notion that Japaridze carefully lays out in [55]. However, we will present the core, basic ones which will help us better understand CC (and CoL's fragments) rules.

We extend the concept of cirquents defined in the previous section by allowing, along with $\vee$ and $\wedge$, the following six additional possible labels for gates: $\veebar$, $\barwedge$, $\bigtriangledown$, $\bigtriangleup$, $\sqcup$, $\sqcap$.
Gates labelled with any of these new symbols we call **selectional gates**, while the old $\vee$ and $\wedge$ we call **parallel gates**.

Selectional gates are subdivided into three groups:

- $\{\veebar, \barwedge\}$, referred to as **toggling gates**;
- $\{\bigtriangledown, \bigtriangleup\}$, referred to as **sequential gates**;
- $\{\sqcup, \sqcap\}$, referred to as **choice gates**.

The eight kinds of gates are also divided into the following two groups:

1. $\{\vee, \veebar, \bigtriangledown, \sqcup\}$, termed **disjunctive gates** (or simply **disjunctions**);
2. $\{\wedge, \barwedge, \bigtriangleup, \sqcap\}$, termed **conjunctive gates** (or simply **conjunctions**).

Thus, $\wedge$ is to be referred to as "parallel conjunction", $\sqcup$ as "choice disjunction" and so on.

**Definition 2.1.2.7**
Let $C$ be a cirquent, * an interpretation, and $\Phi$ a position. $\Phi$ is a legal position of the game $C*$ iff, with a "gate" below meaning a gate of $C$, the following conditions are satisfied:

- Each labmove of $\Phi$ has one of the following forms:
  1. $\top g.i$, where $g$ is a $\veebar-$, $\bigtriangledown-$ or $\sqcup-$gate and $i$ is a positive integer not exceeding the outdegree of $g$;
  2. $\bot g.i$, where $g$ is a $\barwedge-$, $\bigtriangleup-$ or $\sqcap-$gate and $i$ is a positive integer not exceeding the outdegree of $g$;
- Whenever $g$ is a choice gate, $\Phi$ contains at most one occurrence of a labmove of the form $\wp g.i$;

- Whenever $g$ is a sequential gate and $\Phi$ is of the form $\langle \ldots, \wp g.i, \ldots, \wp g.i, \ldots \rangle$, we have $i < j$.

The intuitive meaning of a move of the form $g.i$ is selecting the $i$th outgoing edge — together with the child pointed at by such an edge — of the selectional gate $g$.

In a disjunctive selectional gate, a selection is always made by player $\top$; in a conjunctive selectional gate, a selection is always made by player $\bot$.

The difference between the three types of selectional gates lies in how many selections and in what order these can be made in the same gate. In a choice gate, a selection can be made only once.

In a sequential one, selections can be reconsidered any number of times only in the left-to-right order - once an edge is chosen, no lower ID can be (re)selected afterwards.

In a toggling gate, selections can be reconsidered any number of times and in any order. This includes the possibility to select the same edge over and over again.

**Definition 2.1.2.8**
In the context of a given cirquent $C$ and a legal run $\Gamma$ of $C$, we will say that a selectional gate $g$ is **unresolved** iff either no moves of the form $g.j$ have been made in $\Gamma$, or infinitely many such moves have been made[20]. Otherwise $g$ is **resolved** and, where $g.i$ is the last move of the form $g.j$ made in $\Gamma$, the child pointed at by the $i$th outgoing edge of $g$ is said to be the **resolvent** of $g$.

After a few generalisations of previously seen definitions, we come across the extension of validity:

**Definition 2.1.2.9**
Let $C$ be a cirquent in the sense of the present or any of the subsequent sections. We say that:

---

20 The latter, of course, may not be the case if $g$ is a choice gate, or a sequential gate with a finite outdegree.

1. $C$ is **weakly valid** iff, for any interpretation *, there is an HPM $\mathcal{M}$ such that $\mathcal{M}$ wins the game $C*$;

2. $C$ is **strongly valid** iff there is an HPM $\mathcal{M}$ such that, for any interpretation *, $\mathcal{M}$ wins the game $C*$.

When $\mathcal{M}$ and $C$ satisfy the second clause, we say that $\mathcal{M}$ is a **uniform solution** of $C$.

---

*Clustering: second generalisation*

We may take a step further and generalise the previous cirquents by adding an extra parameter to them, called **clustering selectional gates**.

This is a partition of the set of all selectional gates into subsets, called **clusters**, satisfying the condition that all gates within any given cluster have the same label (all are $\sqcup-$gates, or $\forall-$gates and so on) and the same outdegree.

Due to this condition, we can talk about the outdegree of a cluster meaning the common outdegree of its elements, or the type of a cluster meaning the common type (label) of its elements.

An additional condition that we require to be satisfied is that the question of whether any two given selectional gates are in the same cluster has to be decidable.

Just like nodes do, each cluster also has its own ID.

It may help the reader's intuition to think of each cluster as a single gate-style physical device rather than a collection of individual gates.

Namely, a cluster consisting of $n$ gates of outdegree $m$, as a single device, would have $n$ outputs and $m$ $n$-tuples of inputs.

The following figure depicts this new kind of "gate" with $n = 3$, $m = 2$ and $\sqcup$-type of the cluster:

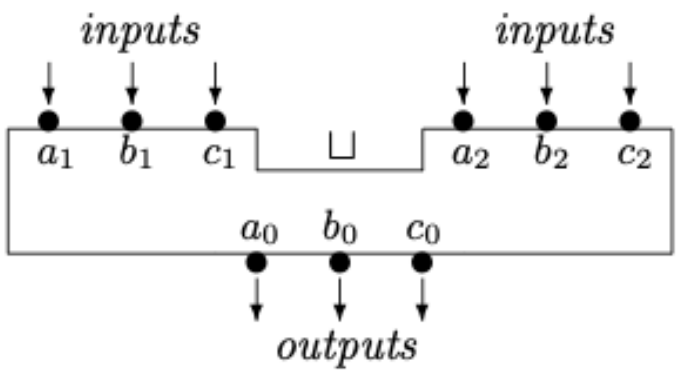

Figure 19: Clusters as generalised gates.

We can think of it as a switch that can be set to one of the positions 1 or 2 (otherwise no signals will pass through it).

Setting it to 1 simultaneously connects the three input lines $a_1, b_1$ and $c_1$ to the output lines $a_0, b_0$ and $c_0$, respectively (the three lines are parallel, isolated from each other, so that no signal can jump from one line to another).

Similarly, setting the switch to 2 connects $a_2, b_2$ and $_c2$ to $a_0, b_0$ and $c_0$, respectively.

Thus, we have an "either $(a_1, b_1, c_1)$ or $(a_2, b_2, c_2)$" kind of a switch; combinations such as $(a_1, b_2, c_1)$ are not available.

Another example of cluster representation is provided in [55], page 20. Given the following cirquent with clustered selectional gates:

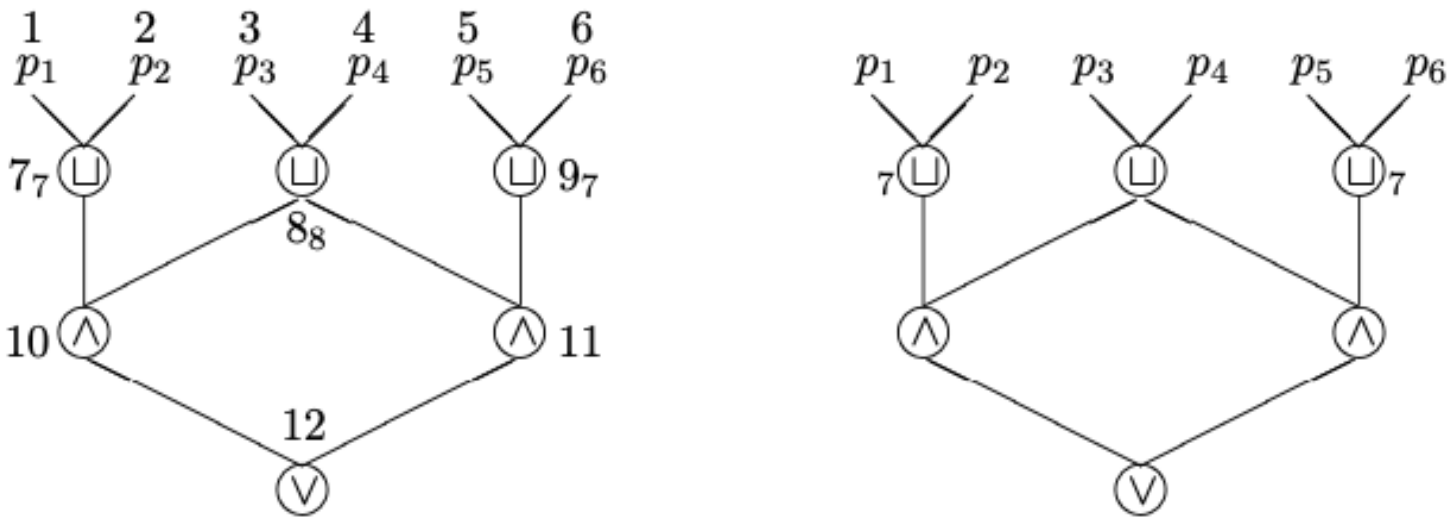

Figure 20: On the left all IDs are shown, while on the right only clusters IDs are indicated. Indeed, we have two clusters: one, cluster 8, containing gate 8 and the other, cluster 7, containing gates 7 and 9.

Here is an alternative representation:

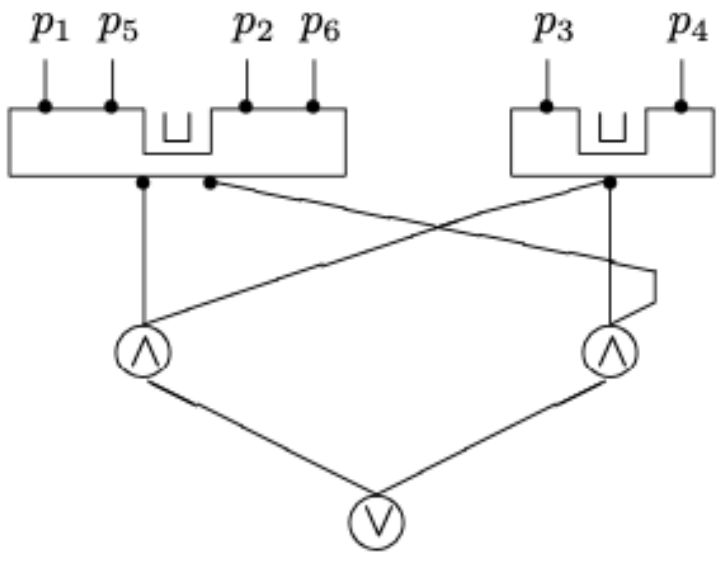

Figure 21: Indeed, clusters are nothing but generalised gates.

Cirquents of the previous section should be viewed as special cases of these new ones: namely, as cases where each selectional gate forms its own, single-element cluster.

As a result, the semantical concepts that Japaridze now introduces conservatively generalise those of the previous section.

For instance, let us take a look at the new, expanded definition of **legal position**:

**Definition 2.1.2.10**
Let $C$ be a cirquent, * an interpretation and $\Phi$ a position. $\Phi$ is a legal position of the game $C^*$ iff, with a "cluster" below meaning a cluster of $C$, the following conditions are satisfied:

- Each labmove of $\Phi$ has one of the following forms:
  1. $\top c.i$, where $c$ is a $\curlyvee-$, $\bigtriangledown-$ or $\sqcup-$cluster and $i$ is a positive integer not exceeding the outdegree of $c$;
  2. $\bot c.i$, where $c$ is a $\curlywedge-$, $\bigtriangleup-$ or $\sqcap-$cluster and $i$ is a positive integer not exceeding the outdegree of $c$;
- Whenever $c$ is a choice cluster, $\Phi$ contains at most one occurrence of a labmove of the form $\wp c.i$;
- Whenever $c$ is a sequential gate and $\Phi$ is of the form $\langle \ldots, \wp c.i, \ldots, \wp c.i, \ldots \rangle$, we have $i < j$.

Similarly, Japaridze extends all definitions we have touched upon - such as interpretation and unresolved/resolved clusters.

He then proceeds to elaborate an IF logic[21] version of CC, which may come as a further generalisation attempt. We will not go too much into detail, as we will highlight only a few key intuitions.

As Japaridze explains,

> Other than the claimed fact that the cirquents of the previous section achieve the full expressive power of IF logic, there are no good reasons to stop at those cirquents.

Indeed, we have already clustered selectional and $\vee-$gates: so why not do the same with the remaining $\wedge-$type of gates?

Naturally, the semantics of clustered $\wedge-$gates has to be symmetric to the one of clustered $\vee-$gates.

By a **metaselection** for a cirquent $C$ we will mean a partial function $f$ from $\vee-$clusters of $C$ to the set of positive integers, such that, for any $\vee-$cluster $c$, whenever defined, $f(c)$ does not exceed the outdegree of $c$.

This way, Japaridze manages to extend the notion of truth to that of **meta-truth**, which is a conservative generalisation of its earlier counterparts.

Namely, a universally quantified metaselection $h$ should be associated with $\wedge-$clusters, as we associate an existentially quantified metaselection $f$ with $\vee-$clusters.

These generalised cirquents have two main features.
As for the first one, the set of all gates is partitioned into clusters, with

21 As already mentioned, independence-friendly logic was introduced by Jaakko Hintikka and Gabriel Sandu in [34], 1989, as an extension of classical first order logic. It is based upon a game theoretical approach in which we can efficiently account for informational independence in games with incomplete information (the ones in which, as Japaridze mentioned with blind quantifiers, some moves must be made without depending on earlier ones). We express this through an independence marker "//" which applies to connectives. Given two operators $\alpha$ and $\beta$, "$\alpha//\beta$" means that the game for the occurrence of $\beta$ within the scope of $\alpha$ must be played without knowledge of the choices made for $\alpha$. For instance, $(\forall x//\exists y)\exists y\phi(x,y)$ asserts that for any choice of value for $x$ by the *falsifier* (Environment), the *verifier* (Machine) can find a value for $y$ that does not depend on the value of $x$, such that the game $\phi(x,y)$ comes out true (it is won by Machine).

each cluster satisfying the conditions of same type and same outdegree - just as before.

As for the second, there is an additional parameter we call **ranking**. This is a partition of the set of all parallel clusters into a finite number of subsets, called **ranks**[22]. These are arranged in linear order, with each rank satisfying the condition that all included clusters must have the same type - however, not necessarily the same outdegree.

A rank containing $\wedge-$clusters is said to be **conjunctive**, while a rank containing $\vee-$clusters is **disjunctive**.

Since ranks are linearly ordered, we can refer to them as "first rank", "second rank" and so on - or even just rank 1 and rank 2. Instead of "cluster $c$ is in the $i$th rank", we say "$c$ is of (or has) rank $i$".

At the end of this ever-pushing generalising excursion we have been on[23], we may provide a complete and final definition of cirquent.

Indeed, we are going to extend part of what has been defined thus far. As Japaridze mentions in [42],

> The version of cirquent calculus presented in this paper captures the most basic yet only a modest fraction of the otherwise very expressive language of computability logic. [...] Extending the cirquent-calculus approach so as to accommodate incrementally expressive fragments of CL is a task for the future. The results of the present paper could be seen just as first steps on that long road.

As a matter of fact, we still need a definition which actually encompasses *every* CoL operator that we currently know of.

Hence, we are going to provide a brief sketch of it by partially extending the inductive definition provided in [68], page 4.

Blind quantifiers, recurrences, implications and refutations will be left to

22 In other words, as Xu defines them in [101], "Ranks are superior consoles of a certain subset of clusters, with all such consoles arranged in a linear order indicating in what order selections by the consoles should be made".

23 Almost a preparatory uphill circuit, one may observe.

to define - since some further, possibly long examination needs to be put in place beforehand.

**Definition 2.1.2.11**
A **cirquent** is defined inductively as follows:

- $\top$ and $\bot$ are cirquents;
- Each literal is a cirquent;
- If $A$ and $B$ are cirquents, then $(A) \wedge (B)$ is a (parallel conjunctive) cirquent;
- If $A$ and $B$ are cirquents, then $(A) \vee (B)$ is a (parallel disjunctive) cirquent;
- If $A$ and $B$ are cirquents, then $(A) \curlywedge (B)$ is a (toggling conjunctive) cirquent;
- If $A$ and $B$ are cirquents, then $(A) \curlyvee (B)$ is a (toggling disjunctive) cirquent;
- If $A$ and $B$ are cirquents, then $(A) \triangle (B)$ is a (sequential conjunctive) cirquent;
- If $A$ and $B$ are cirquents, then $(A) \bigtriangledown (B)$ is a (sequential disjunctive) cirquent;
- If $A$ and $B$ are cirquents, then $(A) \sqcap (B)$ is a (choice conjunctive) cirquent;
- If $A$ and $B$ are cirquents, then $(A) \sqcup (B)$ is a (choice disjunctive) cirquent;
- If $A$ and $B$ are cirquents and $c$ is a parallel conjunctive cluster, then $(A) \wedge^c (B)$ is a cirquent;
- If $A$ and $B$ are cirquents and $c$ is a parallel disjunctive cluster, then $(A) \vee^c (B)$ is a cirquent;
- If $A$ and $B$ are cirquents and $c$ is a toggling conjunctive cluster, then $(A) \curlywedge^c (B)$ is a cirquent;
- If $A$ and $B$ are cirquents and $c$ is a toggling disjunctive cluster, then $(A) \curlyvee^c (B)$ is a cirquent;

- If $A$ and $B$ are cirquents and $c$ is a selectional conjunctive cluster, then $(A) \triangle^c (B)$ is a cirquent;
- If $A$ and $B$ are cirquents and $c$ is a selectional disjunctive cluster, then $(A) \triangledown^c (B)$ is a cirquent;
- If $A$ and $B$ are cirquents and $c$ is a choice conjunctive cluster, then $(A) \sqcap^c (B)$ is a cirquent;
- If $A$ and $B$ are cirquents and $c$ is a choice disjunctive cluster, then $(A) \sqcup^c (B)$ is a cirquent.
- If $A$ is a cirquent, $x$ is a variable and $c$ is a parallel conjunctive cluster, then $\bigwedge^c x(A)$ is a cirquent;
- If $A$ is a cirquent, $x$ is a variable and $c$ is a parallel disjunctive cluster, then $\bigvee^c x(A)$ is a cirquent;
- If $A$ is a cirquent, $x$ is a variable and $c$ is a toggling conjunctive cluster, then $⩓^c x(A)$ is a cirquent;
- If $A$ is a cirquent, $x$ is a variable and $c$ is a toggling disjunctive cluster, then $⩔^c x(A)$ is a cirquent;
- If $A$ is a cirquent, $x$ is a variable and $c$ is a sequential conjunctive cluster, then $\bigtriangleup^c x(A)$ is a cirquent;
- If $A$ is a cirquent, $x$ is a variable and $c$ is a sequential disjunctive cluster, then $\bigtriangledown^c x(A)$ is a cirquent;
- If $A$ is a cirquent, $x$ is a variable and $c$ is a choice conjunctive cluster, then $\bigsqcap^c x(A)$ is a cirquent;
- If $A$ is a cirquent, $x$ is a variable and $c$ is a choice disjunctive cluster, then $\bigsqcup^c x(A)$ is a cirquent.

Let us now learn the rules of this deep-inference calculus in order to glance at some examples and then start hiking towards CoL's peaks[24].

24 For the record, the *stratus* is still hovering motionless upon the distant crests.

### 2.1.3 *Playing by the rules*

Generally, Cirquent Calculus systems differ in what logical atoms and operators their underlying formal languages include - and, consequently, in what rules of inference are allowed.

The rules that we are going to introduce in this section are the basic (and basal) ones, which give rise to all the other ones we will tackle further on.

We explain these rules in terms of inserting arcs, swapping oformulas and similar other visual modifications. These descriptions are rather clear; we agree with Japaridze in affirming that translating them into rigorous formulations is hardly necessary.

We need to concur on some additional terminology first.

We will call **adjacent oformulas** of a given cirquent two oformulas $F$ and $G$, where $G$ appears next to (to the right of) $F$ in the pool[25] of the cirquent. Thus, we say that $F$ **immediately precedes** $G$ and that $G$ **immediately follows** $F$.
This is similarly done for **adjacent ogroups** as well.

By **merging two adjacent ogroups** $\Gamma$ and $\Delta$ in a given cirquent $C$, we mean replacing in $C$ the two ogroups $\Gamma$ and $\Delta$ by the one ogroup $\Gamma \cup \Delta$ - the rest of the cirquent is left unchanged.

The resulting cirquent will thus only differ from $C$ in that it will have one • where $C$ had the two adjacent •s; the arcs of this new • will point exactly to the oformulas to which the arcs of one or both of the old •s were pointing.

For instance, the right cirquent of the following figure is the result of merging ogroups number 2 and 3 in the left cirquent:

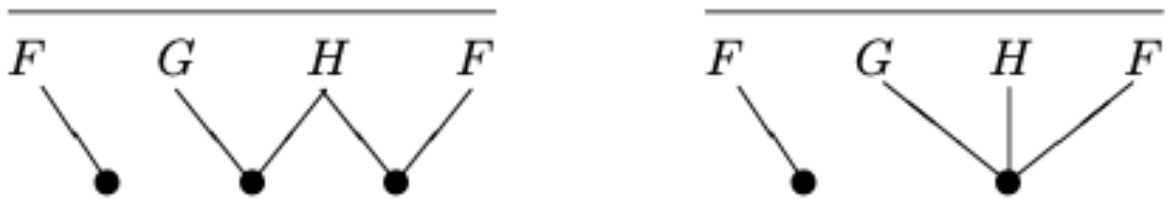

25 Where $k$ is a natural number, by a $k$-ary **pool** we mean a sequence $\langle F_1, \ldots, F_k \rangle$ of oformulas.

Likewise, **merging two adjacent oformulas** $F$ and $G$ into $H$ means replacing those two oformulas by the one oformula $H$, and redirecting to it all arcs that were pointing to $F$ or $G$.

For instance, the right cirquent of the following pair is the result of merging, in the left cirquent, the first $F$ and $G$ into $H$:

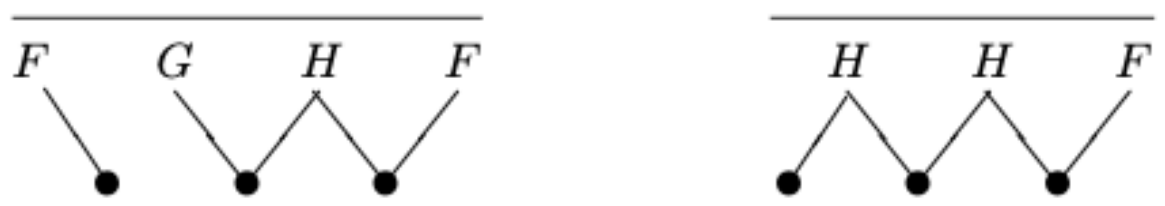

Let us now visualise the actual core rules of every CC system - the ones whence every other stems from. Most of them borrow their names from the structural rules of sequent calculus; the last two rules are non-structural and concern the parallel conjunction and disjunction introductions.

We are going to stick to the basic versions provided in [42], which will change over time thanks to the new definitions of overgroups and undergroups introduced in [72] for CL15[26].

---

*Axioms (A)*

Axioms are rules with no premises. There are two axioms, called the **empty cirquent axiom** and the **identity axiom**.

The former introduces the **empty cirquent** $(\langle\rangle, \langle\rangle)$ in which both the pool and the structure are empty.

The latter is, just like the rest of the rules, a *scheme* of rules - since $F$ can be an arbitrary formula. The identity axiom, on the other hand, introduces the cirquent $(\langle\{1,2\}\rangle, \langle\neg F, F\rangle)$.

26 As CL15 will be our main focus in the third chapter, we will get into the depths of such notions once we successfully map and name CoL's mountains.

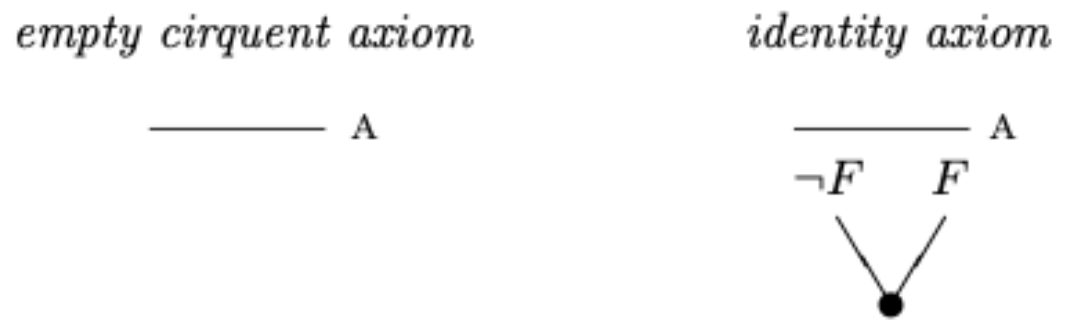

---

*Mix (M)*

This rule takes two premises. The conclusion is obtained by simply putting one premise next to the other, thus creating one cirquent out of the two, as illustrated below:

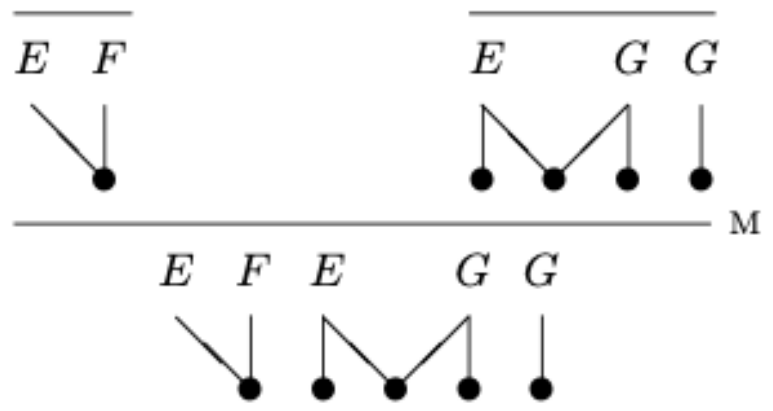

---

*Exchange (E)*

This, and all of the remaining rules, takes a single premise.
The exchange rule comes in two flavours: **oformula exchange** and **ogroup exchange**.

The conclusion of oformula (respectively ogroup) exchange is the result of swapping in the premise two adjacent oformulas (respectively ogroups) and correspondingly redirecting all arcs.

The following oformula exchange example swaps $F$ with $G$; and the ogroup exchange example swaps ogroup number 2 with ogroup number 3:

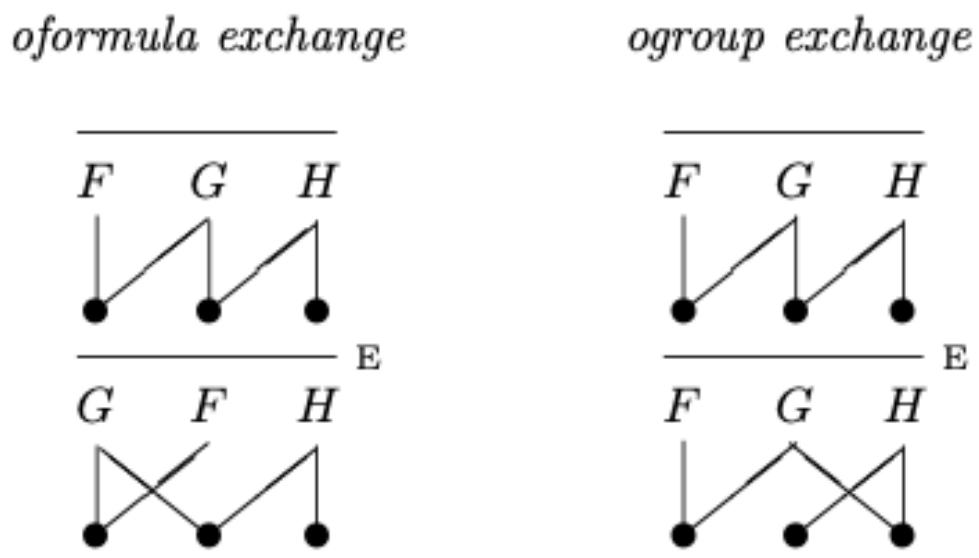

The presence of oformula exchange essentially allows us to treat the pool of a cirquent as a multiset rather than a sequence of formulas.

Similarly, the presence of ogroup exchange makes it possible to see the structure of a cirquent as a multiset rather than a sequence of groups.

---

*Weakening (W)*

This rule comes in two flavours: **ogroup weakening** and **pool weakening**.

In the first case we obtain the conclusion from inserting a new arc between an existing ogroup and an existing oformula of the premise.

In the second case, the conclusion is the result of inserting a new oformula anywhere in the pool of the premise.

As you can see:

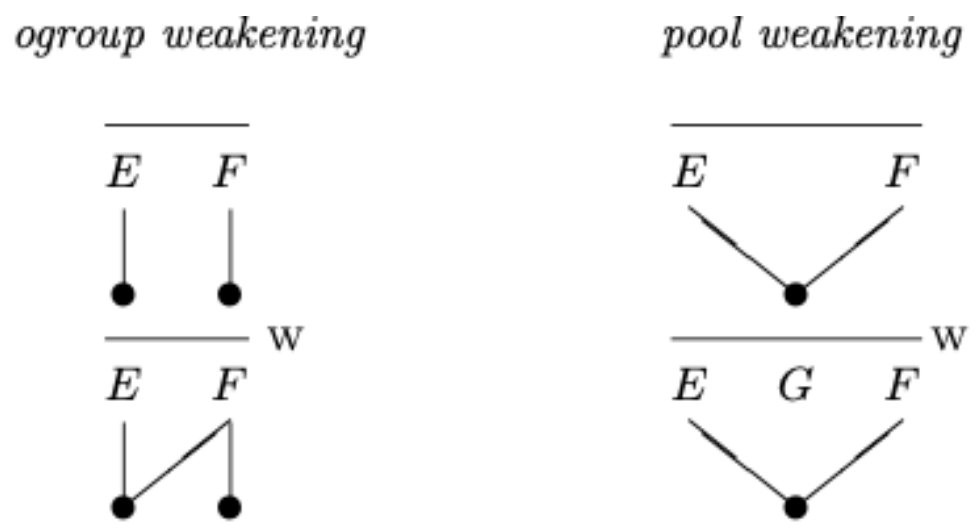

---

*Duplication (D)*

This rule comes in two versions as well: **downward duplication** and **upward duplication**.

The conclusion (respectively premise) of downward (respectively upward) duplication is the result of replacing in the premise (respectively conclusion) some ogroup Γ by two adjacent ogroups that, as groups, are identical with Γ.

Visually:

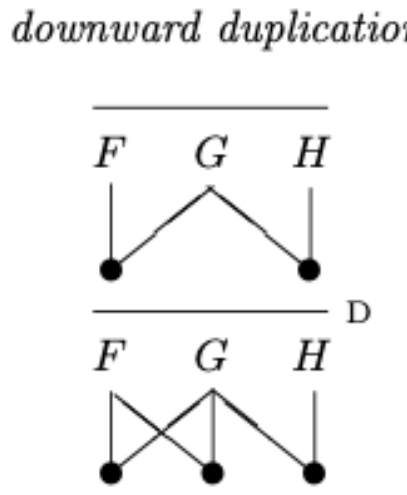

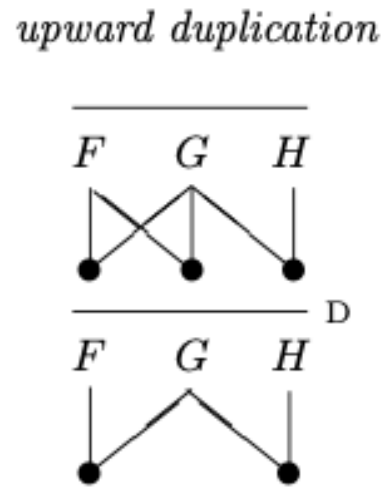

The presence of duplication, together with ogroup exchange, further allows us to think of the structure of a cirquent as a *set* rather than a *sequence* or *multiset* of groups.

---

*Contraction (C)*

The premise of this rule is a cirquent with two identical, adjacent oformulas $F$, $F$.

The conclusion is obtained by merging those two oformulas into $F$.

The following two examples illustrate such application:

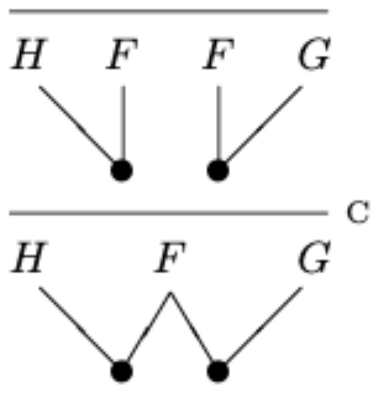

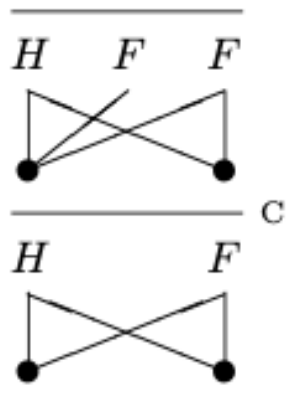

---

*$\vee$−introduction ($\vee$)*

The conclusion of this rule is obtained by merging in the premise some two adjacent oformulas $F$ and $G$ into $F \vee G$.

We say that this application of the rule **introduces** $F \vee G$.

Below are three illustrations:

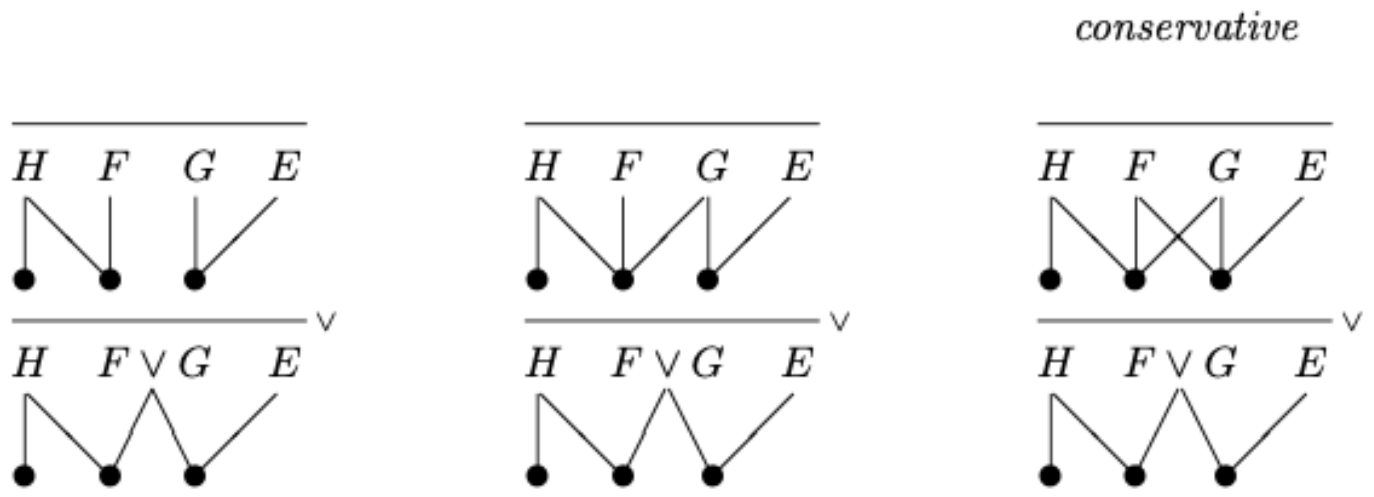

In what we call **conservative** $\vee$−introduction (the rightmost example), which is a special case of $\vee$−introduction, whenever an ogroup of the conclusion contains the introduced $F \vee G$, the corresponding ogroup of the premise contains both $F$ and $G$.

In the general case (the other two examples), this is not necessary. What is always necessary, however, is that if an ogroup of the conclusion contains the introduced $F \vee G$, then the corresponding ogroup of the premise should contain at least one of the oformulas $F$, $G$.

As we have been doing thus far, we use the expression "the corresponding ogroup" to mean that the present rule does not change the number or order of ogroups: it only modifies the contents of some of those ogroups.

Thus, to ogroup number $i$ of the conclusion corresponds ogroup number $i$ of the premise, and vice versa.

The same applies to the rules of oformula exchange, weakening and contraction.
In an application of ogroup exchange that swaps ogroups number $i$ and number $i + 1$, to ogroup number $i$ of the premise corresponds ogroup number $i+1$ of the conclusion, and vice versa.

Conversely, to ogroup number $i+1$ of the premise corresponds ogroup number $i$ of the conclusion, and vice versa; to any other ogroup number $j$ of the premise corresponds ogroup number $j$ of the conclusion and vice versa.

Finally, in an application of (M), to ogroup number $i$ of the first premise corresponds ogroup number $i$ of the conclusion, and vice versa; and, where $n$ is the number of the ogroups of the first premise, to ogroup number $i$ of the second premise corresponds ogroup number $n+i$ of the conclusion, and vice versa.

---

*$\wedge$−introduction ($\wedge$)*

The premise of this rule is a cirquent with adjacent oformulas $F$ and $G$, such that the following two conditions are satisfied:

- No ogroup contains both $F$ and $G$;
- Every ogroup containing $F$ is immediately followed by an ogroup containing $G$, and every ogroup containing $G$ is immediately preceded by an ogroup containing $F$.

The conclusion is obtained from the premise by merging each ogroup containing $F$ with the immediately following ogroup (containing $G$); then, in the resulting cirquent, by merging $F$ and $G$ into $F \wedge G$. We say that the the rule **introduces** $F \wedge G$.

Below are three examples for the simple case when there is only one ogroup in the conclusion that contains the introduced $F \wedge G$:

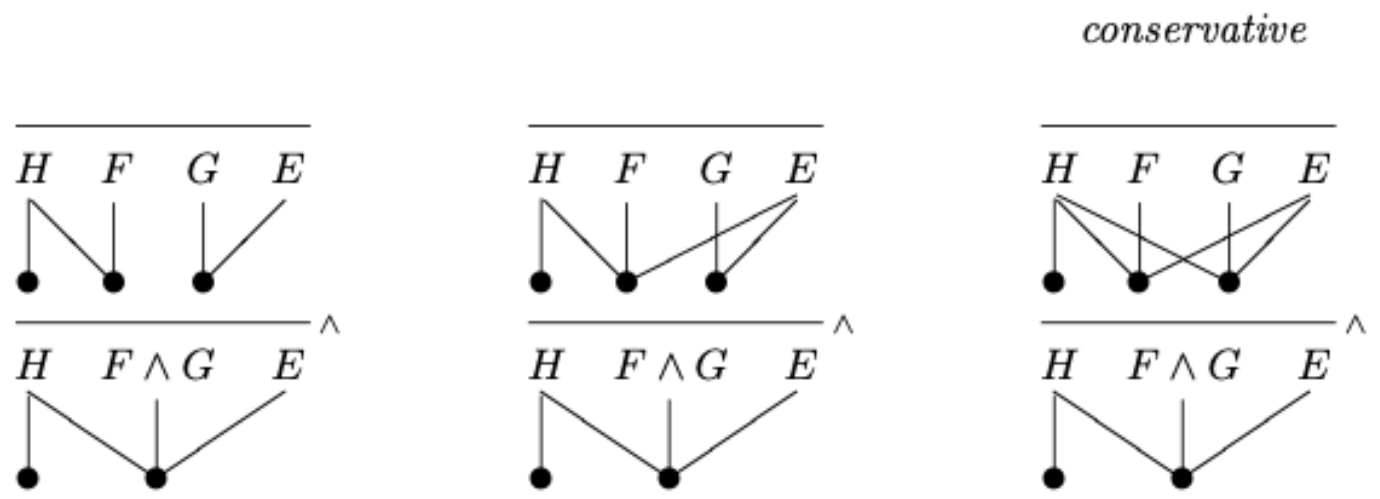

Perhaps this rule is easier to comprehend through the bottom-up approach.

Indeed, to obtain a premise from the conclusion (where $F \wedge G$ is the introduced conjunction), we split every ogroup $\Gamma$ containing $F \wedge G$ into two adjacent ogroups $\Gamma^F$ and $\Gamma^G$, where $\Gamma^F$ contains $F$ (but not $G$), and $\Gamma^G$ contains $G$ (but not $F$); only every $\neq (F \wedge G)$ oformula of $\Gamma$ should be included in either $\Gamma^F$ , or $\Gamma^G$, or both.

In the **conservative** $\wedge-$introduction, all of the non-$F \wedge G$ oformulas of $\Gamma$ should be included in both $\Gamma^F$ and $\Gamma^G$.

The following is an example of an application in a little bit more complex case, where the conclusion has two ogroups containing the introduced conjunction.

This is not a conservative $\wedge-$introduction. To make it conservative, we should add two more arcs to the premise: one connecting ogroup number 3 with $J$, and one connecting ogroup number 4 with $E$:

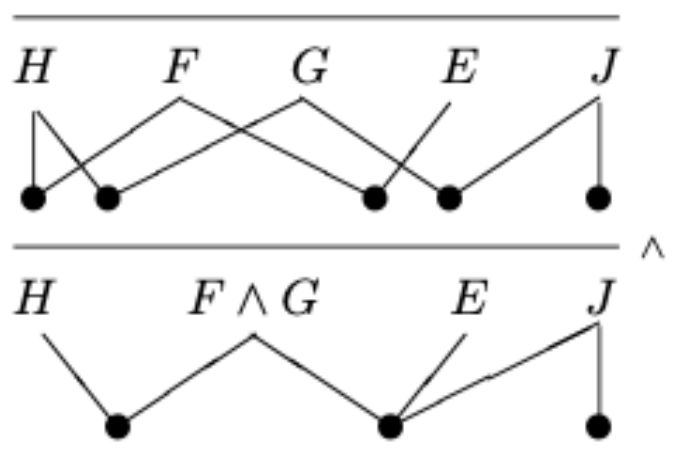

---

*Cirquent Calculus systems*

As Japaridze explains when introducing CC for the first time, a **Cirquent Calculus system** is any subset of the set of these eight rules.

The full set of such rules is named **CCC** ("Classical Cirquent Calculus"), while the one that has all rules minus (C) is called CL5.

As previously mentioned, many other rules have been introduced, making it possible to axiomatise ever more expressing fragments of CoL.

As of 2024, we dispose of 17 different axiomatisations. We are going to later introduce them, together with their own rules.

For now, let us have a look at some properties of CC systems in general.

---

*Derivability, provability and admissibility*

Let $S$ be a Cirquent Calculus system, and $C$, $A_1, \dots, A_n$ (possibly $n = 0$) any cirquents.

A **derivation** of $C$ from $A_1, \dots, A_n$ in $S$ is a tree of cirquents with $C$ at its root, where each node is a cirquent that either follows from its children by one of the rules of $S$, or is among $A_1, \dots, A_n$ and has no children.

A derivation is usually required to come with a **justification**, meaning an indication of *which* rule makes any given cirquent $C_{i+1}$ follow from $C_i$ and *where* it is applied.

A derivation of $C$ in $S$ from the empty set of cirquents is said to be a **proof** of $C$ in $S$. Of course, if $S$ does not contain axioms, then there will be no proofs in it.

When a proof of a cirquent $A$ exists, we say that $A$ is **provable** in a certain CC system $S$ and write $S \vdash A$.

An (atomic-level) **instance** of a cirquent is the result of renaming (all, some or no) atoms in it.

Different occurrences of the same atom are required to be renamed into the same atom; however, it is also possible that different atoms are renamed into the same atom.

Most importantly, we can show that if a cirquent is an instance of a more general one, and this one has already been proven, then it is provable.

**Lemma 2.1.3.1**
If a cirquent is provable, then so are all of its instances[27].

27 Proven in [49], page 1003.

Also, from [42], we know that:

**Lemma 2.1.3.2**
If a given Cirquent Calculus system proves a cirquent $C$, then it also proves every instance of $C$.

Furthermore, by a **transition** we mean any binary relation $\mathcal{T}$ on cirquents. When $A\mathcal{T}B$, we say that $B$ follows from $A$ by $\mathcal{T}$, and call $A$ and $B$ the premise and the conclusion of the given application of the transition, respectively.

Transitions are the same as rules of inference, only in a more relaxed sense.

Of course, every rule $R$ of inference induces — and can often be identified with — a transition $\mathcal{T}$, such that $B$ follows from $A$ by $\mathcal{T}$ iff $B$ follows from $A$ by $R$. As Japaridze explains, we may not always be very terminologically strict in differentiating between transitions and rules.

A transition is said to be **strongly admissible** in a given system if, whenever $B$ follows from $A$ by that transition, there is also a derivation of $B$ from $A$.

On the other hand, a transition is **weakly admissible** iff, whenever $B$ follows from $A$ by that transition and $A$ is provable in the system, $B$ is also provable.

However, since transitions are properly explained through CL8 rules, which are slightly different from the other ones, we are now going to move on through our instruction booklet in order to tackle this mountain range. The reader is strongly advised to look at [49], page 1005 for examples of transitions once we have reached peak CL8.

---

*Truth*

Here we show some important facts concerning truth in CC systems.

By a classical model we mean a function $M$ that assigns a truth value — *true* (1) or *false* (0) — to each atom, and extends to compound formulas in the standard classical way.

The traditional concepts of truth and tautologicity naturally extend from formulas to groups and cirquents.

Let $M$ be a model, and $C$ a cirquent. We say that a group $\Gamma$ of $C$ is true in $M$ iff at least one of its oformulas is so. $C$ is true in $M$ if every group of $C$ is so.

"False", as always, means "not true".
Finally, $C$ or a group $\Gamma$ of it is a **tautology** iff it is true in every model.

Identifying each formula $F$ with the singleton cirquent $(\langle\{1\}\rangle, \langle F\rangle)$, our concepts of truth and tautologicity of cirquents preserve the standard meaning of these concepts for formulas.

Indeed, a cirquent is tautological if and only if all of its groups are so. Moreover, a cirquent containing the empty group is always false, while a cirquent with no groups, such as the empty cirquent $(\langle\rangle, \langle\rangle)$, is always true.

The following theorems and lemmas have been proven in [42], page 509.

**Lemma 2.1.3.3**
All of the rules of CCC preserve truth in the top-down direction — that is, whenever the premise(s) of an application of any given rule is (are) true in a given model, so is the conclusion. Taking no premises, (the conclusions of) axioms are thus tautologies.

**Lemma 2.1.3.4**
The rules of mix, exchange, duplication, contraction, conservative $\vee$- introduction and conservative $\wedge$-introduction preserve truth in the bottom-up direction as well — that is, whenever the conclusion of an application of such a rule is true in a given model, so is (are) the premise(s).

**Theorem 2.1.3.5**
A cirquent is provable in CCC iff it is a tautology.

Furthermore, a cirquent is said to be **binary** iff no atom has more than two occurrences in it.

A binary cirquent is said to be **normal** iff, whenever it has two occurrences of an atom, one occurrence is negative and the other is positive.

A **binary tautology** (respectively **normal** binary tautology) is a binary (respectively normal binary) cirquent that is a tautology in the sense of the previous section.

This terminology also extends to formulas understood as cirquents.

**Lemma 2.1.3.6**
A cirquent is an instance of some binary tautology iff it is an atomic-level instance of some normal binary tautology.

**Lemma 2.1.3.7**
The rules of Mix, Exchange, Duplication, $\wedge-$introduction and $\vee-$introduction preserve binarity and normal binarity in both top-down and bottom-up directions.

**Lemma 2.1.3.8**
Weakening preserves binarity and normal binarity in the bottom-up direction.

As a result, a cirquent is provable in CL5 iff it is an instance of a binary tautology - indeed, it has no contraction rule.

---

*Proof of Blass' Principle*

To have a taste of how rule application works, we may as well have a look at the CL5-derivation of Blass' Principle, meaning:

$$((\neg P \vee \neg Q) \wedge (\neg R \vee \neg S)) \vee ((P \vee R) \wedge (Q \vee S))$$

What is interesting about this formula, as we have previously mentioned, is that, while valid in CoL and, correspondingly, provable in CL5, it is not provable in linear logic or even the stronger version of it known as affine logic.

The rules-deleting performed by linear and affine logics, according to Japaridze, has entailed some sort of impoverishment: indeed, by expelling formulas they might as well have expelled "deeply hidden resource-semantically irreproachable derivative principles".

As he writes in [42], page 492:

> [. . . ] by merely deleting the offending rule of contraction without otherwise trying to first appropriately re(de)fine ordinary sequent calculus, has thrown out the baby with the bath water. Among the innocent victims expelled together with contraction is Blass's principle.

Here we prove the such principle:

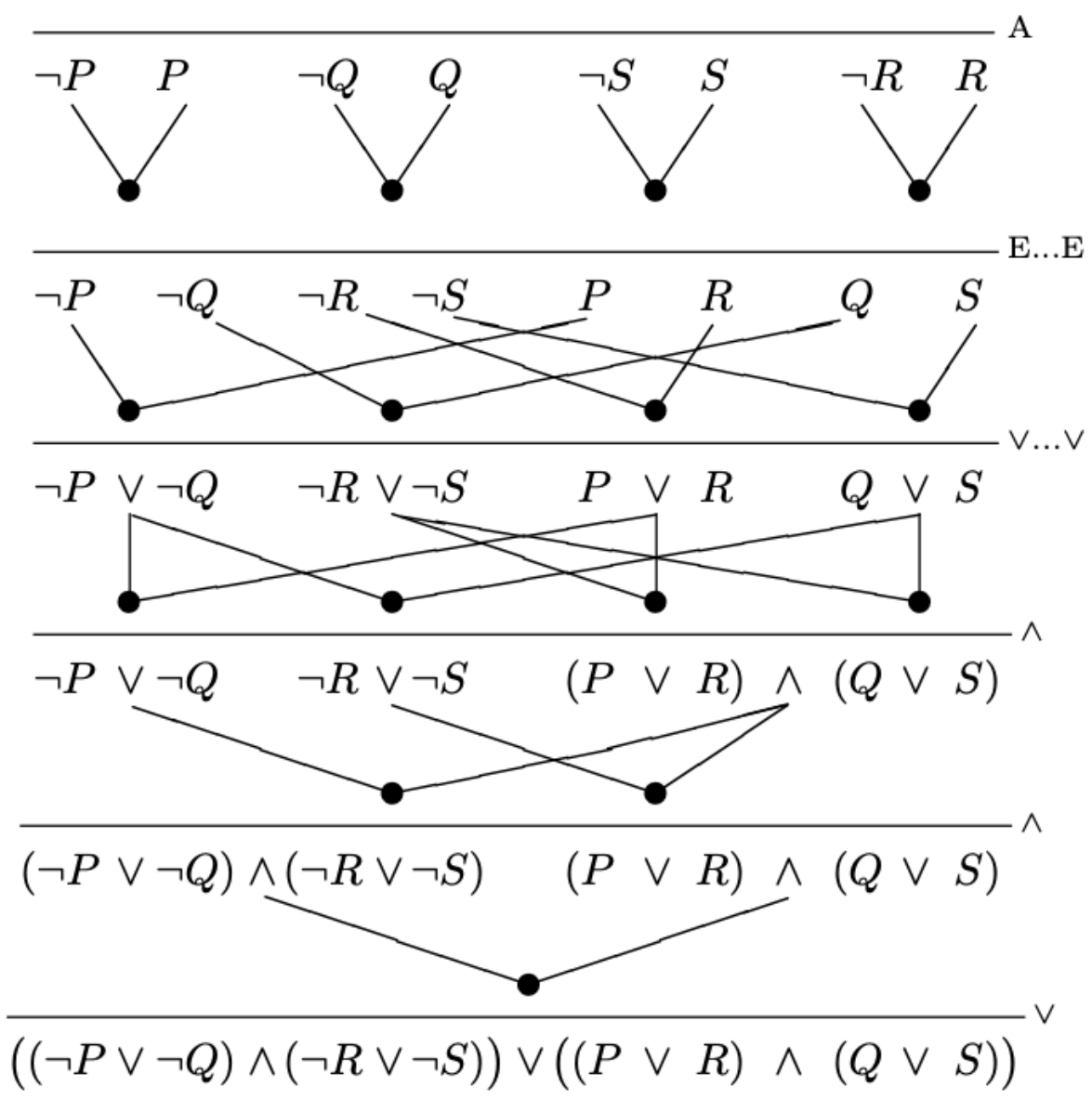

---

Our preliminary CC learning session has now come to an end.

Indeed, it is time to finally inspect the 17 peaks of Computability Logic using our Cirquent Calculus equipment.

## 2.2 POINTING THE FINGER

Geared up and ready to ascent, our intent will be to first name - as in *point the finger to* - every elevation and then explore it while taking notes on its structure and properties.

We do not know whether CC is able to account for all of this land. Indeed, as of yet, an axiomatisation of the set of valid formulas of the full language of CoL has not been found so far. Indeed, we still do not know if such axiomatisation exists in principle.

However, axiomatisations do exist for numerous fragments; we obtain these by disallowing certain operators and imposing some restrictions on formulas.

We are now fully prepared to start climbing these carpeted rocky sides. Once back from our topographic excursion, we will present a brief overview of our findings and piece them all together.

### 2.2.1 *Mountain view*

After a few-weeks long expedition, in which we have managed to keep track of everything we have come across, we are back in the planes of this golden green grassland.

Indeed, as we have been told by the few locals of CL17, the low *stratus* has been there for quite some time already - years, apparently. It will probably go away sooner or later; at present, we did not manage to go any further.

Most probably, there are many more mountains and fragments to explore; however, since *to every thing there is a season, and a time to every purpose under the heaven* (Eccl. 3, 1), this has yet to be proven only with clear skies. Surely enough, a clean visibility will soon open the path to new peaks and axiomatisations of CoL.

In the meantime, here is an overviewing table of what we have gathered thus far.

We show these fragments alongside their signature, meaning the language spoken therein.

In addition, we further provide three fundamental traits for each axiomatisation: the type of atoms (game letters) allowed; its order; and, finally, the type of deductive systems it is defined in.

| | Signature | Order | Atoms | Deductive system |
|---|---|---|---|---|
| CL1 | $\{\neg, \wedge, \vee, \sqcap, \sqcup\}$ | Propositional | Elementary | Brute force |
| CL2 | $\{\neg, \wedge, \vee, \sqcap, \sqcup\}$ | Propositional | General | Brute force |
| CL3 | $\{\neg, \wedge, \vee, \sqcap, \sqcup, \bigsqcap, \bigsqcup, \forall, \exists\}$ | First | Elementary | Brute force |
| CL4 | $\{\neg, \wedge, \vee, \sqcap, \sqcup, \bigsqcap, \bigsqcup, \forall, \exists\}$ | First | General | Brute force |
| CL5 | $\{\neg, \wedge, \vee\}$ | Propositional | Elementary | Cirquent Calculus |
| CL6 | $\{\neg, \wedge, \vee\}$ | Propositional | General | Cirquent Calculus |
| CL7 | $\{\rightarrow\}$ | Propositional | General | Gentzen-style |
| CL8 | $\{\neg, \wedge, \vee\}$ | Propositional | General | Cirquent Calculus |
| CL9 | $\{\neg, \wedge, \vee, \sqcap, \sqcup, \triangle, \triangledown\}$ | Propositional | General | Brute force |
| CL10 | $\{\neg, \wedge, \vee, \sqcap, \sqcup, \triangle, \triangledown\}$ | Propositional | Elementary | Brute force |
| CL11 | $\{\wedge, \vee, \sqcap, \sqcup, \triangle, \triangledown, \bigsqcap, \bigsqcup, \forall, \exists\}$ | First | General | Brute force |
| CL12 | $\{\wedge, \vee, \sqcap, \sqcup, \triangle, \triangledown, \bigsqcap, \bigsqcup, \forall, \exists,$ external $\circ\!-\}$ | First | Elementary | Brute force |
| CL13 | $\{\neg, \wedge, \vee, \sqcap, \sqcup, \triangle, \triangledown, ⩚, ⩛\}$ | Propositional | General | Brute force |
| CL14 | $\{\neg, \wedge, \vee, \sqcap, \sqcup, \triangle, \triangledown, ⩚, ⩛\}$ | Propositional | Elementary | Brute force |
| CL15 | $\{\neg, \wedge, \vee, ⫰, ⫯\}$ | Propositional | General | Cirquent Calculus |
| CL16 | $\{\neg, \wedge, \vee, \sqcap, \sqcup\}$ | Propositional | Elementary | Cirquent Calculus |
| CL17 | $\{\neg, \wedge, \vee, \sqcap, \sqcup, \bigsqcap, \bigsqcup\}$ | First | Elementary | Cirquent Calculus |

Table 1: A first glance at CoL's fragments up to date.

All of these, starting from CL1, are conservative extensions of classical logic. Indeed, CC has been of the utmost importance from CL5 onwards - the "most unaxiomatisable" fragment.

As the reader may recall from our expedition, we have logged that a few of these fragments have a "Brute force" deductive system.

This means that the system formally uses Gentzen-style objects, such as formulas and sequents; however, it does so in a *brute force* manner - a more direct, almost game-like approach, as explained by Japaridze in [67], page 50.

Indeed, brute force systems operate with traditional object while using new rules, which are directly derived from the underlined game semantics. This will get clearer once we review the brute force fragments together.

On the other hand, CL7 is the only Gentzen-style fragment[28] of CoL,

28 Together with $Int^+$, which is CoL's fragment of propositional intuitionistic logic (without $\top$ and $\bot$), meaning $\{\sqcap, \sqcup, \circ\!-\}$. This fragment is not officially included in our table since it is not an individual peak of the 17, but rather an additional elevation which sits right on the slopes of CL7 itself. More on this later.

meaning a sequent calculus system in the traditional sense.

A few notational conventions.

Just like Japaridze does in a few papers, we refrain from writing $\rightarrow$ in every signature[29], since it can be defined as $A \rightarrow B = \neg A \vee B$.

As for $\neg$, we are to apply it only to atoms. The sole exception is using De Morgan abbreviations (e.g. $\neg(A \wedge B) = \neg A \vee \neg B$).

Indeed, Japaridze uses different symbols in different papers[30] to mean the same thing. For formal coherence's sake, we are going to use only $\Rightarrow$ when defining rules of inference, with $\mathcal{P} \Rightarrow F$ meaning "from premise(s) $\mathcal{P}$ conclude $F$".

This is going to be a brief overview of said fragments - indeed, we do not have enough space to cover every theorem, proof, lemma and definition provided in the papers so far. As a matter of fact, it is a lot of fragments and a copious amount of information[31].

However, we feel that it may be useful to the reader to have a summarising view of the topic, in order to later fill in the blanks by having a look at the references provided[32] if needed.

We will also provide a few syntactic examples of some of these fragments; in case they are a bit too long and require further analysis, the reader will be directed to have a look at the specific references.

In some generalising fragments we will avoid to present yet other examples, since what usually changes is just the type of atoms allowed.

We will now systematically recount our steps in an all-encompassing reviewing endeavour.

29 Except for CL7, as it does not have any other operator.
30 For instance, $\rightsquigarrow$, $\mapsto$ and $\vdash$.
31 A volcanic production! All the more thrilling.
32 Meaning the original guidelines Japaridze gave us beforehand.

---

*CL1*

As introduced in [40], CL1 is a propositional fragment with the following signature: $\{\neg, \wedge, \vee, \sqcap, \sqcup\}$.

The language of this logic is obtained by augmenting the one of classical propositional logic with two additional operators, $\sqcap$ and $\sqcup$.
Indeed, the $\sqcap, \sqcup-$free fragment of CL1 is exactly classical propositional logic.

Since it is a propositional system, only nullary games are allowed.

Recalling the definitions we have provided when approaching CoL's formal semantics, let us view CL1's rules.

Logic CL1 is given by the following two rules:

1. $\vec{H} \Rightarrow F$, where $F$ is stable and $\vec{H}$ is the smallest set of formulas such that, whenever $F$ has a positive (respectively negative) surface occurrence of a subformula $G_1 \sqcap \ldots \sqcap G_n$ (respectively $G_1 \sqcup \ldots \sqcup G_n$), $\vec{H}$ contains, for each $i \in \{1, \ldots, n\}$, the result of replacing this occurrence in $F$ by $G_i$;

2. $H \Rightarrow F$, where $H$ is the result of replacing in $F$ a negative (respectively positive) surface occurrence of a subformula $G_1 \sqcap \ldots \sqcap G_n$ (respectively $G_1 \sqcup \ldots \sqcup G_n$) by $G_i$ for some $i \in \{1, \ldots, n\}$.

Axioms are not explicitly stated, but note that the set $\vec{H}$ of premises of Rule 1 can be empty, in which case the conclusion $F$ acts as an axiom. Indeed, if $F$ is an elementary formula, then the only way to prove $F$ in CL1 is to derive it by Rule 1 from the empty set of premises.

In particular, this rule is applicable when $F$ is stable, which for an elementary $F$ means that $F$ is a classical tautology.

Consequently, we say that every classically valid formula is an elementary formula derivable in CL1 by Rule 1 from the empty set of premises[33].

33 Indeed, as already mentioned, CL1 $\setminus \{\sqcap, \sqcup\}$ = CL (as in Classical Logic).

It is now time for a syntactic bite. Let us prove that:

$$\mathrm{CL1} \vdash ((p \to q) \sqcap (p \to r)) \to (p \to (q \sqcap r)).$$

| | | |
|---|---|---|
| 1. | $(p \to q) \to (p \to q)$ | from $\varnothing$ by Rule 1; |
| 2. | $((p \to q) \sqcap (p \to r)) \to (p \to q)$ | from 1. by Rule 2; |
| 3. | $(p \to r) \to (p \to r)$ | from $\varnothing$ by Rule 1; |
| 4. | $((p \to q) \sqcap (p \to r)) \to (p \to r)$ | from 3. by Rule 2; |
| 5. | $((p \to q) \sqcap (p \to r)) \to (p \to (q \sqcap r))$ | from 2. and 4. by Rule 1. |

CL1 is provably sound, complete and decidable.

For its soundness and completeness proofs, see [39], page 320. Japaridze proves the former through the construction of EPMs, while, on the other hand, he proves CL1's completeness using its complementary logic CL1′ (with complementary rules) - where CL1 $\nvdash F$ iff CL1′ $\vdash F$.

---

*CL2*

As explained in [41], CL2 has the same signature of CL1. The difference lies in the type of atoms allowed.

Indeed, we say that CL2 is a generalisation of CL1 in that it uses general atoms, while CL1 is elementary-base.

Hence, we may conclude that there are two fragments of CL2[34]; specifically, the general-base fragment of CL2 is the set of all general-base theorems of CL2, while the elementary-base fragment of CL2 is the set of all elementary-base theorems of CL2, which is nothing but the set of theorems of CL1.

The rules of inference of CL2 are the ones of CL1[35] plus one additional rule 3.

34 Thanks to the soundness and completeness proofs provided in the same paper.
35 These are now applied to any CL2-formulas rather than just elementary-base ones.

Here are such rules:

1. $\vec{H} \Rightarrow F$, where $F$ is stable and $\vec{H}$ is the smallest set of formulas such that, whenever $F$ has a positive (respectively negative) surface occurrence of a subformula $G_1 \sqcap \ldots \sqcap G_n$ (respectively $G_1 \sqcup \ldots \sqcup G_n$), $\vec{H}$ contains, for each $i \in \{1, \ldots, n\}$, the result of replacing this occurrence in $F$ by $G_i$;
2. $H \Rightarrow F$, where $H$ is the result of replacing in $F$ a negative (respectively positive) surface occurrence of a subformula $G_1 \sqcap \ldots \sqcap G_n$ (respectively $G_1 \sqcup \ldots \sqcup G_n$) by $G_i$ for some $i \in \{1, \ldots, n\}$.
3. $F' \Rightarrow F$, where F' is the result of replacing in $F$ two - one positive and one negative - surface occurrences of some general atom by a nonlogical elementary atom that does not occur in $F$.

As an example, the following is a CL2-proof of $P \wedge P \rightarrow P$:

| | | |
|---|---|---|
| 1. | $p \wedge P \rightarrow p$ | from $\varnothing$ by Rule 1; |
| 2. | $P \wedge P \rightarrow P$ | from 1. by Rule 3. |

Being a general-base conservative extension of CL1, CL2 formulas can have elementary, general and also hybrid atoms - thus introducing hyperformulas[36].

It is also worth mentioning that CL2, just like CL1, is also provably decidable[37].

All theorems and lemmas on validity, soundness and completeness of CL1 are also valid for CL2 (with due different proofs), carefully adjusted to fit the general-baseness of the system.

---

*CL3*

Indeed, as shown in [44], CL3 axiomatises the most basic first order fragment of CoL - meaning a finite-depth, elementary-base, brute force

36 Which we have previously graphically introduced in full CC fashion.

37 This is done through a brute force algorithm. Basically, this type of algorithm checks every possible solution until the correct one is found (if, indeed, there even is a correct one) by sequentially applying the rules of the fragment until all the possible paths have already been taken.

fragment.

CL3 is a conservative extension of classical predicate calculus in the language which, alongside the (appropriately generalised) logical operators of classical logic, contains propositional connectives and quantifiers representing the so called Choice operations.
Hence, the signature is: $\{\neg, \wedge, \vee, \sqcap, \sqcup, \bigsqcap, \bigsqcup, \forall, \exists\}$.

The atoms of this language are interpreted as elementary problems, i.e. predicates in the standard sense.

Curiously enough, Japaridze not only proves that CL3 is sound and complete, but he also shows that its classical-quantifier-free fragment is decidable - all the while remaining first order.

These are the three rules of inference[38]:

1. $\vec{H} \Rightarrow F$, where $F$ is stable and $\vec{H}$ is a set of formulas satisfying the following conditions:
   - Whenever $F$ has a positive (respectively negative) surface occurrence of a subformula $G_1 \sqcap \ldots \sqcap G_n$ (respectively $G_1 \sqcup \ldots \sqcup G_n$), $\vec{H}$ contains, for each $i \in \{1, \ldots, n\}$, the result of replacing this occurrence in $F$ by $G_i$;
   - Whenever $F$ has a positive (respectively negative) surface occurrence of a subformula $\bigsqcap x G(x)$ (respectively $\bigsqcup x G(x)$), $\vec{H}$ contains the result of replacing that occurrence in $F$ by $G(y)$ for some variable $y$ not occurring in $F$.
2. $H \Rightarrow F$, where $H$ is the result of replacing in $F$ a negative (respectively positive) surface occurrences of a subformula $G_1 \sqcap \ldots \sqcap G_n$ (respectively $G_1 \sqcup \ldots \sqcup G_n$) by $G_i$ for some $i \in \{1, \ldots, n\}$;
3. $H \Rightarrow F$, where $H$ is the result of replacing in $F$ a negative (respectively positive) surface occurrence of a subformula $\bigsqcap x G(x)$ (respectively $\bigsqcup x G(x)$) by $G(t)$ for some term $t$ such that $t$ is not bound in $H$.

Just as in CL1, axioms are not explicitly stated.
However, the set $\vec{H}$ of premises of Rule 1. can be empty, in which case

38 For the record, rules 2. and 3. were first introduced with $F'$ and not $H$ - a little notational hiccup that we have fixed by sticking to $H$, just as in all the other fragments.

the conclusion $F$ acts as an axiom.

Here is an example of a CL3-proof for $\sqcap x \sqcup y(p(x) \vee \neg p(y))$:

| | | |
|---|---|---|
| 1. | $p(z) \vee \neg p(z)$ | from $\varnothing$ by Rule 1; |
| 2. | $\sqcup y(p(z) \vee \neg p(y))$ | from 1. by Rule 3; |
| 3. | $\sqcap x \sqcup y(p(x) \vee \neg p(y))$ | from $\varnothing$ by Rule 1. |

---

*CL4*

Introduced in [45], CL4 is a generalisation of CL3 that shares the same signature.

What makes CL4 more expressive is the presence of two sorts of atoms in its language: elementary and general ones.
For such reason, just like CL2, it includes hyperformulas.

Indeed, we say that CL4 conservatively extends CL3, with the latter being the general-atom-free fragment of the former.

Removing the blind, classical quantifiers from CL4's language is shown to yield a decidable logic, despite it still being first order (just like CL3). Hence, the $\forall, \exists-$free fragment of CL4 is decidable.

The rules of inference of CL4 are the three rules 1, 2 and 3 of CL3[39] plus one additional rule 4.

Thus[40], we have:

1. $\vec{H} \Rightarrow E$, where $E$ is stable and $\vec{H}$ is a set of formulas satisfying the following conditions:
   - Whenever $E$ has a positive (respectively negative) surface occurrence of a subformula $G_1 \sqcap \ldots \sqcap G_n$ (respectively $G_1 \sqcup \ldots \sqcup G_n$),

39 Now applied to any CL4-formulas rather than just CL3-formulas.

40 For continuity's sake, we preferred to expose these rules as provided in [45]. The more intuitive and simplified version of these rules is provided in [67], page 62: here Rule 1. is called "Wait"; Rule 2. is "$\sqcup-$Choose"; Rule 3. is "$\bigsqcup-$Choose"; and Rule 4. is "Match".

$\vec{H}$ contains, for each $i \in \{1, \ldots, n\}$, the result of replacing this occurrence in $E$ by $G_i$;

- Whenever $E$ has a positive (respectively negative) surface occurrence of a subformula $\sqcap x G(x)$ (respectively $\sqcup x G(x)$), $\vec{H}$ contains the result of replacing that occurrence in $E$ by $G(y)$ for some variable $y$ not occurring in $E$.

2. $H \Rightarrow E$, where $H$ is the result of replacing in $E$ a negative (respectively positive) surface occurrences of a subformula $G_1 \sqcap \ldots \sqcap G_n$ (respectively $G_1 \sqcup \ldots \sqcup G_n$) by $G_i$ for some $i \in \{1, \ldots, n\}$;

3. $H \Rightarrow E$, where $H$ is the result of replacing in $E$ a negative (respectively positive) surface occurrence of a subformula $\sqcap x G(x)$ (respectively $\sqcup x G(x)$) by $G(t)$ for some term $t$ such that $t$ is not bound in $H$;

4. $H \Rightarrow E$, where $H$ is the result of replacing in $E$ two - one positive and one negative - surface occurrences of some $n$-ary general letter by an $n$-ary non-logical elementary letter that does not occur in $E$.

The following is a CL4-proof of $\sqcap x \sqcup y(P(x) \to P(y))$:

| | | |
|---|---|---|
| 1. | $p(z) \to p(z)$ | from $\varnothing$ by Rule 1; |
| 2. | $P(z) \to P(z)$ | from 1. by Rule 4; |
| 3. | $\sqcup y(P(z) \to P(y))$ | from 2. by Rule 3; |
| 4. | $\sqcap x \sqcup y(P(x) \to P(y))$ | from $\varnothing$ by Rule 1. |

Again, for soundness and completeness results have a look at [45].

---

*CL5*

As shown in [42], if we remove the contraction rule from the full collection of CCC rules, we obtain a sound and complete system for the basic fragment CL5, previously thought to be the "most unaxiomatisable" - one of the reasons behind CC's development.

Indeed, with CL5 we have a sound and complete system for $\{\neg, \wedge, \vee\}$, which is the very core of CoL. We may also show that is remains sound and complete also with respect to ASR.

As Japaridze points out, deleting the rule of contraction in ordinary sequent calculus, results in the strictly *weaker* affine logic.
Indeed, being complete, CL5 is stronger than incomplete affine logic.

The set of theorems of CL5 admits an alternative non-deductive characterisation, according to which this is the set of all binary tautologies[41] and their substitutional instances.

What is interesting (and all the more "convincing") is that the set of CL5-theorems has already emerged in several, past unrelated contexts.

The earliest relevant piece of literature Japaridze mentions is [74], 1963, where Stanisław Jaśkowski considers binary tautologies as a solution to characterise the provable formulas of a certain deductive system.
Furthermore, as Japaridze observes, also Blass has come across the same class of formulas twice.

Since the rules have been already introduced in the previous section, we might continue on with our overview.
As for a syntactic example, the earlier shown proof of Blass' Principle is, indeed, a CL5-proof.

---

*CL6*

Presented in [42], CL6 is a more expressive CC system which generalises CL5: indeed, CL6-formulas contain both general and elementary atoms.

The set of rules of CL6 is obtained from that of CL5 by adding $\top$[42] as another axiom, plus the rule of contraction limited to elementary formulas.
As a result, we have three axioms (empty cirquent, identity and $\top$) and seven rules (M, E, W, D, C, $\vee$, $\wedge$).

Moreover, in [98], 2012, Wenyan Xu managed to prove soundness and

41 With binary tautologies we mean tautologies of classical propositional logic in which no propositional letter occurs more than twice.
42 Understood as a singleton cirquent.

completeness for this conservative extension of both CL5 and classical logic.

---

*CL7*

As introduced in [52], CL7 is, indeed, a complete and sound axiomatisation of $\mathrm{Int}^{\supset}$[43], meaning the implicative fragment of Heyting's intuitionistic calculus.

The only difference lies in the abandonment of the contraction rule inside a Gentzen-style calculus[44]. This means that CL7 is the implicative fragment of affine logic, with the minimalistic signature: $\{\rightarrow\}$.

We say that CL7-sequent is a pair $\Gamma \Rightarrow F$, where $\Gamma$ is a (possibly empty) multiset of CL7-formulas and $F$ is a CL7-formula.

As usual, when $\Gamma$ and $\Delta$ are multisets of formulas and $F$ is a formula, we shall write "$\Gamma, \Delta$" for $\Gamma \cup \Delta$ and "$\Gamma, F$" for $\Gamma \cup \{F\}$.

The axioms of CL7 are all CL7-sequents of the form:

$$\Gamma, F \Rightarrow F.$$

Furthermore, we only have the following two rules of inference (namely, **Left** $\rightarrow$ and **Right** $\rightarrow$ introductions):

$$\frac{\Gamma, F \Rightarrow G \qquad \Delta \Rightarrow E}{\Gamma, \Delta, E \rightarrow F \Rightarrow G} \qquad\qquad \frac{\Gamma, E \Rightarrow F}{\Gamma \Rightarrow E \rightarrow F}$$

Furthermore, if we replace $\supset$ of $\mathrm{Int}^{\supset}$ with $\rightarrow$, we obtain CoL's fragment of propositional intuitionistic logic, namely $\mathrm{Int}^{+}$.

This fragment sits right on the slopes of CL7, since it can be obtained by simply adding the rule of contraction plus $\sqcap, \sqcup$ and other four rules of inference (with $i \in \{1, 2\}$):

$$\frac{\Gamma, E_1 \Rightarrow F \qquad \Gamma, E_2 \Rightarrow F}{\Gamma, E_1 \sqcup E_2 \Rightarrow F} \qquad\qquad \frac{\Gamma \Rightarrow E_i}{\Gamma \Rightarrow E_1 \sqcup E_2}$$

43 With $\supset \in \{ \succ\!\!-, \circ\!\!-, \circ\!\!-^{\aleph_0} \}$.
44 Indeed, we can show that adding the contraction rule to CL7 yields an unsound system.

$$\frac{\Gamma, E_i \Rightarrow F}{\Gamma, E_1 \sqcap E_2 \Rightarrow F} \qquad \frac{\Gamma \Rightarrow E_1 \quad \Gamma \Rightarrow E_2}{\Gamma \Rightarrow E_1 \sqcap E_2}$$

Moreover, the full propositional intuitionistic logic Int can be obtained from $\text{Int}^+$ by allowing $\perp$ in the language and adding the axiom $\Gamma, \perp \Rightarrow F$. Such a system is sound but incomplete with respect to our semantics.

---

*CL8*

CL8 is introduced in [49] as the first deep-inference CC system for CoL. As Japaridze points out, it happens to coincide with the logic induced by ARS; furthermore, it is a conservative extension of classical logic, so that CL8 is also a (provably) sound and complete alternative system. Its signature is $\{\neg, \wedge, \vee\}$.

CL8 allows cirquents of arbitrary depths and forms, which means that inference rules can modify cirquents at any level rather than only around the root - as is the case of shallow sequent calculus.

CL8 defines **parameters** as nodes that are affected by the application of inference rules from premise to conclusion. In particular:

- $a_1, \ldots, a_m$, said to be **central parameters**, are pairwise distinct nodes, each one being a node of either the premise or the conclusion or both;
- Each $\Pi_i$, said to be a **peripheral parameter**, is a set of nodes not containing any central parameters, such that every $b \in \Pi_i$ is a parent or a child of some central parameter in either the premise or the conclusion or both.

Thus, the role of peripheral parameters is to list all of the central parameters' parents and children which are not central parameters themselves. Moreover, peripheral parameters are also useful to divide those parents or children into groups for reference purposes.

The single-node cirquent $\circ$ (conjunctive gate), i.e. $\top$, is the (only) axiom of CL8.

Concerning the inference rules[45], these come in two flavours: **restructuring** ones and **main** ones.

The former ones can be applied both in top-down and bottom-up directions. Specifically, these are[46]:

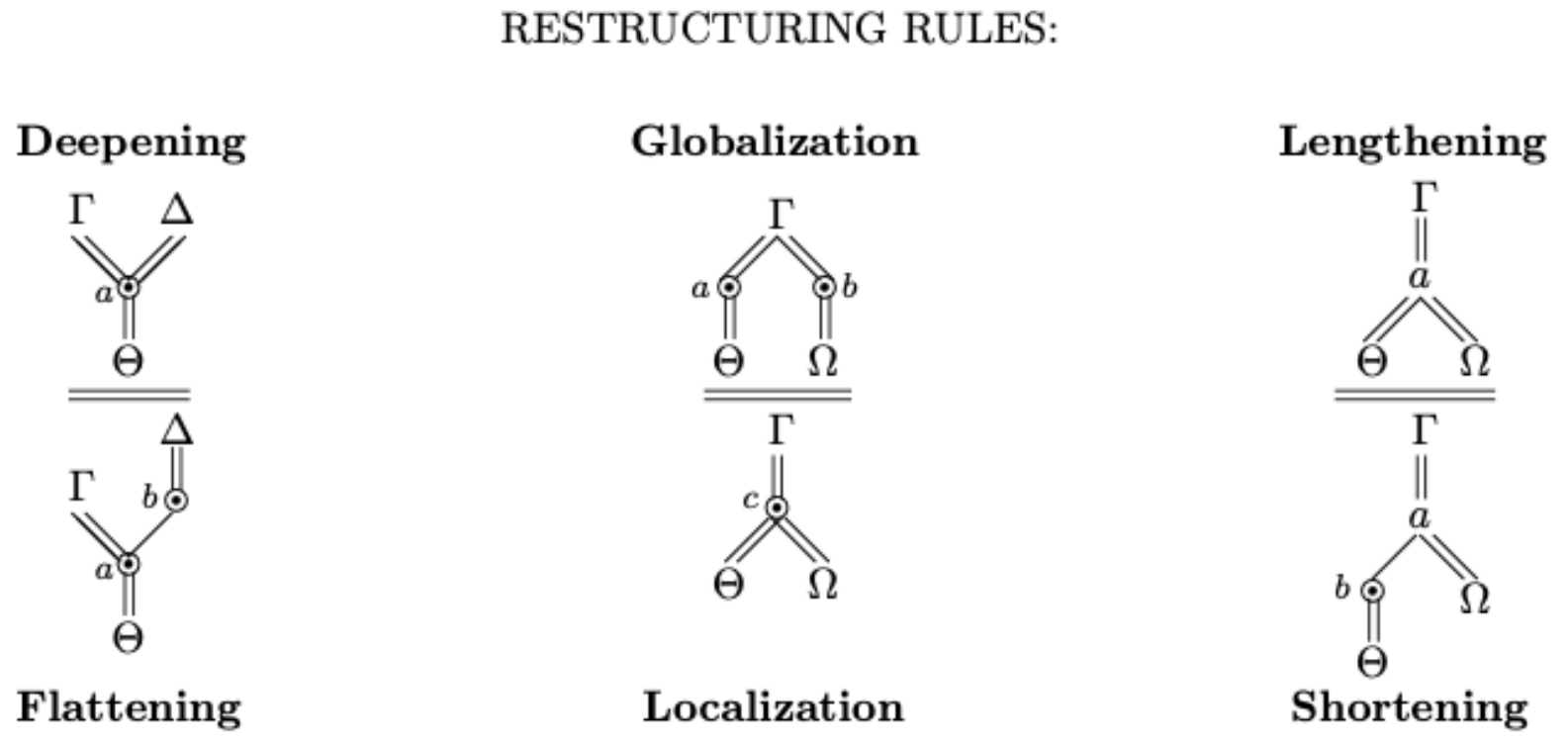

1. **Deepening**: we take two central parameters $a, b$ and three peripheral ones $\Gamma, \Delta, \Theta$. If a cirquent has a gate $b$ with exactly one parent $a$ such that $b$ and $a$ are of the same type (both conjunctive, or both disjunctive), then the $a$ premise can be obtained by deleting $b$ and connecting its children $\Delta$ directly to $a$;

2. **Flattening**: symmetrical to Deepening, obtained by interchanging premise with conclusion;

3. **Localisation**: if a cirquent has two conjunctive or two disjunctive gates $a, b$ with exactly the same children $\Gamma$ (but not necessarily the same parents), then a premise can be obtained by merging $a$ and $b$ and calling the resulting node $c$. Here "merging" means that $c$ has the same type and same children as $a$ and $b$ have, and the set of the parents of $c$ is the union of those of $a$ and $b$;

4. **Globalisation**: symmetrical to Localisation, obtained by interchanging premise with conclusion;

5. **Lengthening**: if a cirquent has a gate $b$ with exactly one child $a$, then a premise can be obtained by deleting $b$ and connecting $a$ directly to the parents $\Theta$ of $b$;

45 Which are graphically represented in a different way than usual in [49].
46 See [49], page 993 for examples.

6. **Shortening**: symmetrical to Lengthening, obtained by interchanging premise with conclusion.

Furthermore, since $\odot$ is a variable over $\{\bullet, \circ\}$, each restructuring rule comes in two versions, one for $\bullet$ (meaning disjunctive gate) and the other for $\circ$ (which, as already mentioned, means conjunctive gate). Thus, we have 12 restructuring rules in total.

On the other hand, we have:

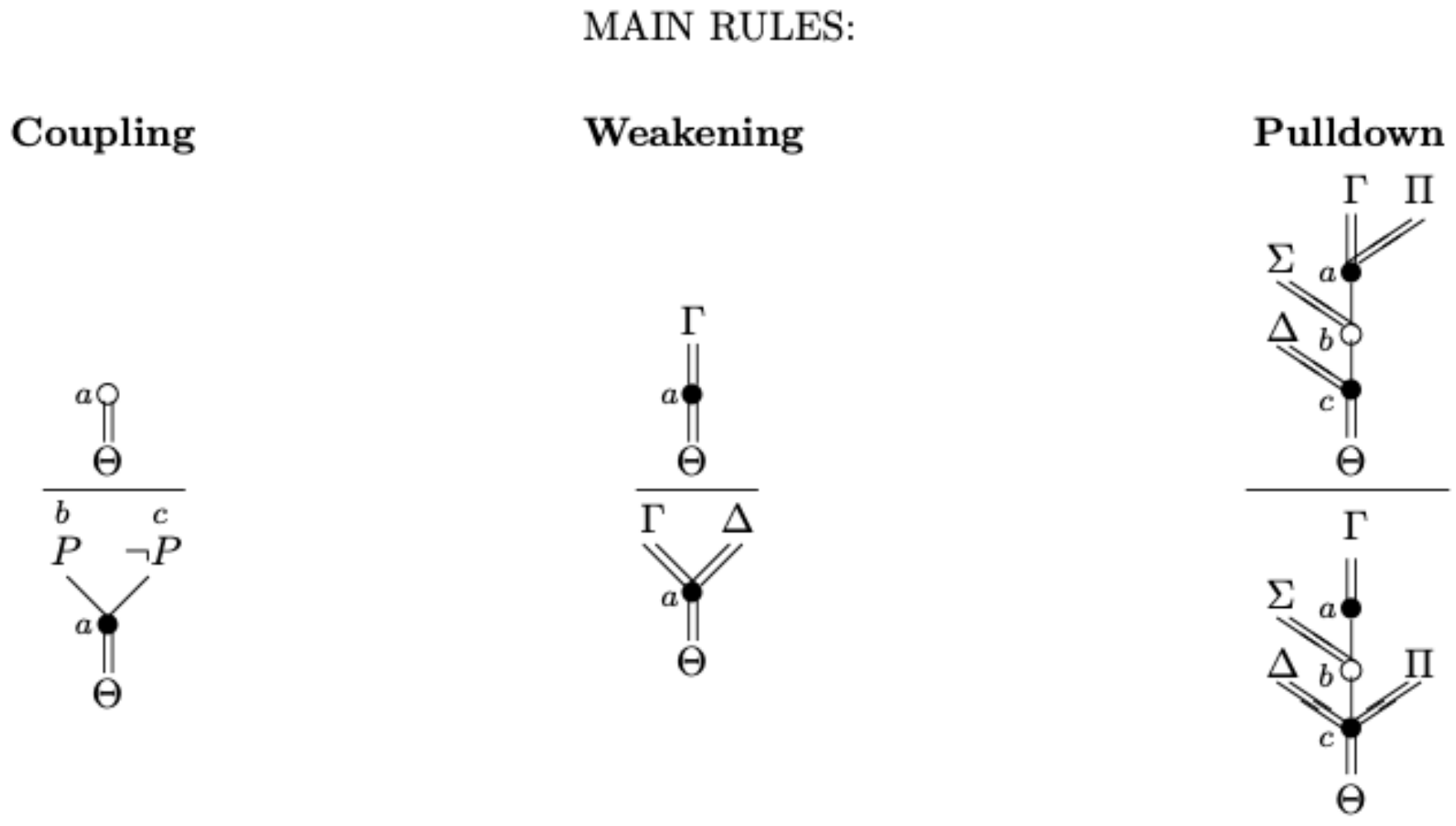

Which mean:

1. **Coupling**: if a cirquent has a childless conjunctive gate $a$, then a conclusion can be obtained by making $a$ a disjunctive gate and adding two children $b$ and $c$ which are ports with opposite labels (with $b$ and $c$ not occurring in the premise);

2. **Weakening**: a premise can be obtained from the conclusion by deleting arcs from a disjunctive gate $a$ to some children $\Delta$ of it.

3. **Pulldown**: when the conclusion (as well as the premise) has a disjunctive gate $a$ with a single conjunctive parent $b$, which, in turn, has a single disjunctive parent $c$, a premise can be obtained by passing some (any) children $\Pi$ from $c$ to $a$.

In order to acquire a syntactic taste of these rules, the reader is strongly advised to look at the example provided in [49], from page 999 onwards - a long but useful one.

In the same paper, page 1008, we find proof for the soundness and completeness of this fragment with respect to ARS.

---

*CL9*

In [50], Japaridze introduces the provably sound and complete CL9, whose signature is $\{\neg, \wedge, \vee, \sqcap, \sqcup, \triangle, \triangledown\}$.

Indeed, CL9 is a brute-force, general-base fragment. The set of all its elementary-base theorems will coincide with CL10, as we will see later on: thus, CL9 is a generalisation of the latter.

This fragment is given by the following four rules of inference.
Indeed, there are no axioms; however, the Wait rule can act as such when the set of its premises is empty.

Thus, we have:

1. **Wait**: $\vec{H} \Rightarrow F$, where $F$ is stable and $\vec{H}$ is the smallest set of formulas satisfying the following two conditions:
   - Whenever $F$ has a surface occurrence of a subformula $G_1 \sqcap \ldots \sqcap G_n$, $\vec{H}$ contains, for each $i \in \{1, \ldots, n\}$, the result of replacing that occurrence in $F$ by $G_i$;
   - Whenever $F$ has a surface occurrence of a subformula $G_0 \triangle G_1 \triangle \ldots \triangle G_n$ , $\vec{H}$ contains the result of replacing that occurrence in $F$ by $G_0 \triangle G_1 \triangle \ldots \triangle G_n$.
2. **Choose**: $H \Rightarrow F$, where $H$ is the result of replacing in $F$ a surface occurrence of a subformula $G_1 \sqcup \ldots \sqcup G_n$ by $G_i$ for some $i \in \{1, \ldots, n\}$;
3. **Switch**: $H \Rightarrow F$, where $H$ is the result of replacing in $F$ a surface occurrence of a subformula $G_0 \triangledown G_1 \triangledown \ldots \triangledown G_n$ by $G_1 \triangledown \ldots \triangledown G_n$;
4. **Match**: $H \Rightarrow F$, where $H$ is the result of replacing in $F$ two — one positive and one negative — surface occurrences of some general atom by a non-logical elementary atom that does not occur in $F$.

Here is an example of a CL9-proof of $P \sqcup Q \rightarrow P \triangledown Q$:

| | | |
|---|---|---|
| 1. | $\neg q \vee q$ | from $\varnothing$ by Wait; |
| 2. | $\neg Q \vee Q$ | from 1. by Match; |
| 3. | $\neg Q \vee (P \triangledown Q)$ | from 2. by Switch; |
| 4. | $\neg p \vee (p \triangledown Q)$ | from $\varnothing$ by Wait; |
| 5. | $\neg P \vee (P \triangledown Q)$ | from 4. by Match; |
| 6. | $(\neg P \sqcap \neg Q) \vee (P \triangledown Q)$ | from 3. and 5. by Wait. |

Furthermore, other than being sound and complete, we know that CL9 is decidable through a brute force decision algorithm proof.

For more on the subject, check out [50], which we will keep as reference for the next two fragments.

---

*CL10*

As already mentioned, CL10 is nothing but the brute-force, elementary-base fragment of CL9 obtained by restricting the language to elementary-base formulas and deleting the Match rule[47]. It was first introduced to prove the soundness and completeness of CL9.

As a result, CL10 inherits soundness from CL9, while its completeness proof[48] is provided in [50], page 1464.

---

*CL11*

CL11 is the first order extension of CL9; we may observe that it is also a conservative extension of CL4, obtained by augmenting its language with $\triangle$ and $\triangledown$.

Indeed, its signature is $\{\neg, \wedge, \vee, \sqcap, \sqcup, \triangle, \triangledown, \bigsqcap, \bigsqcup, \forall, \exists\}$.

47 Indeed, we may say that CL1 is a $\triangle$, $\triangledown$ −free counterpart of CL10.
48 This uses a modification of CL10, namely CL10°, which is the restriction of CL9 to exclusively formulas that do not contain any general or hybrid atoms.

This brute-force, general-base fragment is given by the following five rules of inference:

1. **Wait**: $\vec{H} \Rightarrow F$, where $F$ is stable and $\vec{H}$ is a set of formulas satisfying the following three conditions:
   - Whenever $F$ has a surface occurrence of a subformula $G_1 \sqcap \ldots \sqcap G_n$, $\vec{H}$ contains, for each $i \in \{1, \ldots, n\}$, the result of replacing that occurrence in $F$ by $G_i$;
   - Whenever $F$ has a surface occurrence of a subformula $G_0 \bigtriangleup G_1 \bigtriangleup \ldots \bigtriangleup G_n$ , $\vec{H}$ contains the result of replacing that occurrence in $F$ by $G_1 \bigtriangleup \ldots \bigtriangleup G_n$;
   - Whenever $F$ has a surface occurrence of a subformula $\bigsqcap x G(x)$, $\vec{H}$ contains the result of replacing that occurrence in $F$ by $G(y)$, where $y$ is a variable not occurring in $F$.
2. $\sqcup$**-Choose**: $H \Rightarrow F$, where $H$ is the result of replacing in $F$ a surface occurrence of a subformula $G_1 \sqcup \ldots \sqcup G_n$ by $G_i$ for some $i \in \{1, \ldots, n\}$;
3. $\bigsqcup$**-Choose**: $H \Rightarrow F$, where $H$ is the result of replacing in $F$ a surface occurrence of a subformula $\bigsqcup x G(x)$ by $G(t)$, where $t$ is a term with no bound occurrence in F;
4. **Switch**: $H \Rightarrow F$, where $H$ is the result of replacing in $F$ a surface occurrence of a subformula $G_0 \bigtriangledown G_1 \bigtriangledown \ldots \bigtriangledown G_n$ by $G_1 \bigtriangledown \ldots \bigtriangledown G_n$;
5. **Match**: $H \Rightarrow F$, where $H$ is the result of replacing in $F$ two — one positive and one negative — surface occurrences of some $n$-ary general letter $P$ by an $n-$ary non-logical elementary letter $p$ that does not occur in $F$.

Here is an example of how these rules actually work.

Let us prove in CL11 that:

$$\bigsqcap x \bigsqcup y (q(x) \longleftrightarrow p(y)) \wedge \bigsqcap x(\neg p(y) \bigtriangledown p(x)) \rightarrow \bigsqcap x(\neg q(x) \bigtriangledown q(x)).$$

Step 1.

$\neg(q(z) \longleftrightarrow p(v)) \vee \neg p(v) \vee q(z)$

from $\emptyset$ by Wait;

Step 2.

$\neg(q(z) \longleftrightarrow p(v)) \vee \neg p(v) \vee (\neg q(z) \triangledown q(z))$

from 1. by Switch;

Step 3.

$\neg(q(z) \longleftrightarrow p(v)) \vee (p(v) \vartriangle \neg p(v)) \vee (\neg q(z) \triangledown q(z))$

from 2. by Wait;

Step 4.

$\neg(q(z) \longleftrightarrow p(v)) \vee \bigsqcup x(p(x) \vartriangle \neg p(x)) \vee (\neg q(z) \triangledown q(z))$

from 3. by $\bigsqcup$-Choose;

Step 5.

$\neg(\bigsqcap y(q(z) \longleftrightarrow p(y))) \vee \bigsqcup x(p(x) \vartriangle \neg p(x)) \vee (\neg q(z) \triangledown q(z))$

from 4. by Wait;

Step 6.

$\neg(\bigsqcap y \bigsqcup x(q(x) \longleftrightarrow p(y))) \vee \bigsqcup x(p(x) \vartriangle \neg p(x)) \vee (\neg q(z) \triangledown q(z))$

from 5. by $\bigsqcup$−Choose;

Step 7.

$\neg(\bigsqcap y \bigsqcup x(q(x) \longleftrightarrow p(y))) \vee \bigsqcup x(p(x) \vartriangle \neg p(x)) \vee \bigsqcap x(\neg q(x) \triangledown q(x))$

from 6. by Wait.

As Japaridze points out, there is every reason to expect that the already known soundness and completeness theorems for CL9 and CL4 extend to their common generalisation CL11 - however, this has yet to be shown.

On the other hand, we already know that the $\forall, \exists$-free fragment of CL11 is provably decidable.

---

*CL12*

CL12 is one of the most important fragments for applicational purposes. Indeed, as explained in [53], it is a reasonable, computationally meaningful and constructive alternative to classical logic and, as such, a fertile basis for applied theories (such as PA, Peano Arithmetic, which we will later show in the fourth chapter).

As shown in [67], page 67, CL12 is a brute-force, elementary-base fragment of CoL, whose signature is $\{\neg, \wedge, \vee, \sqcap, \sqcup, \bigsqcap, \bigsqcup, \forall, \exists\}$.

The language of CL12 is incomparable to that of CL4. First of all because it is elementary-base; secondly, because it allows unireiterated $\circ\!-$, as CL4 does not.
The very presence of such $\circ\!-$ is what makes CL12 a very powerful tool for constructing CoL-based applied theories[49].

CL12 has six rules of inference[50]. We will introduce them according to the formulation given in [67] - which is slightly different from the one originally provided in [63].

Axioms are not explicitly stated; however, the set of premises of the Wait rule can be empty, in which case its conclusion acts as an axiom.

Thus, we have:

49 For this reason, CL12 is also the best studied fragment of CoL, especially from the complexity-theoretic point of view.

50 Where $F[E]$ means a formula $F$ together with a single fixed surface occurrence of a subformula $E$. Using this notation sets a context, in which $F[H]$ means the result of replacing in $F[E]$ that occurrence of $E$ by $H$. Moreover, $F[H_0 \sqcup H_1]$ means a formula $F$ together with a pool of osubformulas.

1. $\sqcup$**-Choose**: $G \circ\!\!- F[H_i] \Rightarrow G \circ\!\!- F[H_0 \sqcup H_1]$, where $i \in \{1, \ldots, n\}$;
2. $\sqcap$**-Choose**: $G, E[H_i], K \circ\!\!- F \Rightarrow G, E[H_0 \sqcap H_1], K \circ\!\!- F$ where $i \in \{1, n\}$;
3. $\bigsqcup$**-Choose**: $G \circ\!\!- F[H(t)] \Rightarrow G \circ\!\!- F[\bigsqcup x H(x)]$, where $t$ is either a constant or a variable with no bound occurrences in the premise;
4. $\bigsqcap$**-Choose**: $G, E[H(t)], K \circ\!\!- F \Rightarrow G, E[\bigsqcap x H(x)], K \circ\!\!- F$ where $t$ is either a constant or a variable with no bound occurrences in the premise;
5. **Replicate**: $G, E, K, E \circ\!\!- F \Rightarrow G, E, K \circ\!\!- F$;
6. **Wait**: $G_1 \circ\!\!- F_1, \ldots, G_n \circ\!\!- F_n \Rightarrow K \circ\!\!- E$, with $(n \geq 0)$, where $K \circ\!\!- E$ is stable and the following four conditions are satisfied:
   - Whenever the conclusion has the form $K \circ\!\!- E[H_0 \sqcap H_1]$, both $K \circ\!\!- E[H_0]$ and $K \circ\!\!- E[H_1]$ are among the premises;
   - Whenever the conclusion has the form $L, J[H_0 \sqcup H_1], M \circ\!\!- E$, both $L, J[H_0], M \circ\!\!- E$ and $L, J[H_1], M \circ\!\!- E$ are among the premises;
   - Whenever the conclusion has the form $K \circ\!\!- E[\bigsqcap x H(x)]$, for some variable $y$ not occurring in the conclusion, $K \circ\!\!- E[H(y)]$ is among the premises;
   - Whenever the conclusion has the form $L, J[\bigsqcup x H(x)], M \circ\!\!- E$, for some variable $y$ not occurring in the conclusion, $L, J[H(y)], M \circ\!\!- E$ is among the premises.

For instance, here is a CL12-proof of $\bigsqcap x \bigsqcup y (p(x) \to p(y))$:

| | | |
|---|---|---|
| 1. | $p(s) \to p(s)$ | from $\varnothing$ by Wait; |
| 2. | $\bigsqcup y(p(s) \to p(y))$ | from 1. by $\bigsqcup$−Choose; |
| 3. | $\bigsqcap x \bigsqcup y(p(x) \to p(y))$ | from 2. by Wait. |

Soundness and completeness results are published in [63].

---

*CL13*

As explained in [67], CL13's signature is $\{\neg, \wedge, \vee, \sqcap, \sqcup, \triangle, \triangledown, \curlywedge, \curlyvee\}$: indeed, it contains all kinds of conjunction and disjunction seen so far.

CL13 is a brute-force, general-based fragment, as observed in [54]. Its soundness and completeness proofs are provided in the same paper, from page 986 onwards.
Furthermore, it is also provably decidable[51].

There are six rules of inference. We introduce them according to the brute-force formulation given in [54] and [67].

Axioms are not explicitly stated; however, the set of premises of the ($\curlywedge$) rule can be empty, in which case its conclusion acts like an axiom.

Thus, we have:

1. ($\curlywedge$): $\vec{H} \Rightarrow F$, where $F$ is a stable quasielementary formula, and $\vec{H}$ is the smallest set of formulas satisfying the following condition:
   - Whenever $F$ has a surface osubformula $E0 \curlyvee E1$, $\vec{H}$ contains, for both $i \in \{0, 1\}$, the result of replacing in $F$ that osubformula by $E_i$;
2. ($\curlyvee$): $H \Rightarrow F$, where $F$ is a quasielementary formula, and $H$ is the result of replacing in $F$ a surface osubformula $E \curlyvee G$ by $E$ or $G$.
3. ($\sqcap\triangle$): $|F|, \vec{H} \Rightarrow F$, where $F$ is a non-quasielementary formula (note that otherwise $F = |F|$), and $\vec{H}$ is the smallest set of formulas satisfying the following two conditions:
   - Whenever $F$ has a semisurface osubformula $G_0 \sqcap G_1$, $\vec{H}$ contains, for both $i \in \{0, 1\}$, the result of replacing in $F$ that osubformula by $G_i$;
   - Whenever $F$ has a semisurface osubformula $E \triangle G$, $\vec{H}$ contains the result of replacing in $F$ that osubformula by $G$.
4. ($\sqcup$): $H \Rightarrow F$, where $H$ is the result of replacing in $F$ a semisurface osubformula $E \sqcup G$ by $E$ or $G$;

51 It can be shown through a brute force decision algorithm.

5. ($\triangledown$): $H \Rightarrow F$, where $H$ is the result of replacing in $F$ a semisurface osubformula $E \triangledown G$ by $E$ or $G$;

6. (M): $H \Rightarrow F$, where $H$ is the result of replacing in $F$ two - one positive, one negative - semisurface occurrences of come general atom $P$ by a nonlogical elementary atom $p$ which does not occur in $F$.

For instance, here is a CL13-proof of $P ⩚ Q \to P \triangle Q$:

| | | |
|---|---|---|
| 1. | $\neg p \vee p$ | from $\varnothing$ by ($⩚$); |
| 2. | $(\neg p ⩛ \bot) \vee p$ | from 1. by ($⩛$); |
| 3. | $\neg q \vee q$ | from $\varnothing$ by ($⩚$); |
| 4. | $(\neg p ⩛ \neg q) \vee q$ | from 3. by ($⩛$); |
| 5. | $(\neg p ⩛ \neg Q) \vee Q$ | from 4. by (M); |
| 6. | $(\neg p ⩛ \neg Q) \vee (p \triangle Q)$ | from 2. and 5. by ($\sqcap \triangle$); |
| 7. | $(\neg p ⩛ \neg Q) \vee (P \triangle Q)$ | from 6. by (M). |

---

*CL14*

Indeed, as explained in [54], CL14 is the brute-force fragment of CL13 obtained by restricting its language to elementary-base formulas and deleting the (M), "Mix" rule.

It was first introduced as a tool for proving the soundness and completeness of CL13.

Consequently, CL14 inherits soundness, while its completeness is shown through its dual, $\overline{CL14}$. This one has the same language of CL14, with its five rules obtained by replacing $\top$ with $\bot$.

---

*CL15*

First introduced in [60], CL15 is a provably sound and complete CC system whose signature is $\{\neg, \wedge, \vee, ⫰, ⫯\}$. Other than $\neg$ being only applied to atoms[52] and writing $\neg E \vee F$ as $E \to F$, we may also define $E \circ\!\!- F$ as

52 As always, we can tip-toe this condition around by using De Morgan abbreviations.

$⫰E \rightarrow F.$

We will later get into the depths of this fragment, in order to provide a proof for its decidability.
However, for the time being, we may as well introduce the notions of **overgroups** and **undergroups**, which are vital for a thorough understanding of its rules of inference.

**Definition 2.2.1.1** A **CL15-cirquent** is a triple $C = (\vec{F}, \vec{U}, \vec{O})$ where:

1. $\vec{F}$ is a nonempty finite sequence of CL15-formulas, whose elements are said to be the oformulas of C. Here the prefix "o" is for "occurrence", and is used to mean a formula together with a particular occurrence of it in $\vec{F}$. So, for instance, if $\vec{F} = \langle E, G, E \rangle$, then the cirquent has three oformulas even if only two formulas;

2. Both $\vec{U}$ and $\vec{O}$ are nonempty finite sequences of nonempty sets of oformulas of $C$. The elements of $\vec{U}$ are said to be the **undergroups** of $C$, and the elements of $\vec{O}$ are said to be the **overgroups** of $C$. As in the case of oformulas, it is possible that two undergroups or two overgroups are identical as sets (have identical contents), yet they count as different undergroups or overgroups because they occur at different places in the sequence $\vec{U}$ or $\vec{O}$. Simply "group" will be used as a common name for undergroups and overgroups;

3. Additionally, every oformula is required to be in at least one undergroup and at least one overgroup.

The presence of overgroups and undergroups slightly changes the graphical representation of the inference rules, as we will later see in detail. However, CL15 has ten rules, namely:

1. Axiom (A);

2. Exchange (E);

3. Weakening (W);

4. Contraction (C);

5. Duplication (D);

6. Merging (M);

7. Disjunction Introduction (∨);
8. Conjunction Introduction (∧);
9. Recurrence Introduction (⫰);
10. Corecurrence Introduction (⫯);

Although long lists of names are not ideal to anyone's perception, this brief presentation was meant to acquire the flavour of CL15, whose rules and characteristics we are going to properly unravel later on[53].

To this day, CL15 is provably sound and complete (see [60] and [61]), while its decidability has yet to be proven (so far, we just know that it is recursively enumerable for the way it has been defined). More on this in the next chapter.

---

*CL16*

First introduced in [68], CL16 is a propositional brute-force fragment of CoL, whose logical vocabulary consists of negation and parallel and choice connectives. CL16 only allows elementary atoms, meaning moveless games, in its clustered cirquents.

In the same paper Japaridze provides proofs for its soundness and completeness.

CL16 has ten rules of inference. The only axiom of this fragment is $\top$.

The first seven rules come in two versions, between which Japaridze later differentiates by suffixing the name of the rule with "(a)" for the first version and "(b)" for the second one.

The last rule needs two premises, while all the other ones need only one. The rules are written schematically, with $A, B, C, D$ (possibly with indices)

53 For the purpose of keeping everything *afloat* and present, as we have agreed in the introduction of the present work, it is best to get into the depths of CL15 in the next chapter. Firstly, because repetition helps retaining new information; secondly, because we do not wish the reader to go back and forth these pages everytime a rule that is applied in chapter 3 needs to be recalled.

acting as variables for subcirquents; $a, b, c$ as variables for clusters; and $X, Y$ as variables for "structures".

The names of these rules have been chosen according to the conclusion-to-premises approach.

Thus, we have:

1. **Commutativity**: $X[B \vee A] \Rightarrow X[A \vee B]$ and $X[B \wedge A] \Rightarrow X[A \wedge B]$;
2. **Associativity**: $X[A \vee (B \vee C)] \Rightarrow X[(A \vee B) \vee C)]$ and $X[A \wedge (B \wedge C)] \Rightarrow X[(A \wedge B) \wedge C)]$;
3. **Identity**: $X[A] \Rightarrow X[A \vee \bot]$ and $X[A] \Rightarrow X[A \wedge \top]$;
4. **Domination**: $X[\top] \Rightarrow X[A \vee \top]$ and $X[\bot] \Rightarrow X[A \wedge \bot]$;
5. **Choosing**: $X[A_1, \ldots, A_n] \Rightarrow X[A_1 \sqcup^c B_1, \ldots, A_n \sqcup^c B_n]$ and $X[B_1, \ldots, B_n] \Rightarrow X[A_1 \sqcup^c B_1, \ldots, A_n \sqcup^c B_n]$, where $A_1 \sqcup^c B_1, \ldots, A_n \sqcup^c B_n$ are all $\sqcup^c-$rooted subcirquents of the conclusion;
6. **Cleansing**: $X[Y[A] \sqcap^c C] \Rightarrow X[Y[A \sqcap^c B] \sqcap^c C]$ and $X[C \sqcap^c Y[B]] \Rightarrow X[C \sqcap^c Y[A \sqcap^c B]]$;
7. **Trivialisation**: $X[\top] \Rightarrow X[\neg p \vee p]$, where $p$ is an elementary letter;
8. **Quadrilemma**: $X\big[\big((A \wedge (C \sqcap^b D)) \sqcap^a (B \wedge (C \sqcap^b D))\big) \sqcap^c \big(((A \sqcap^a B) \wedge C) \sqcap^b ((A \sqcap^a B) \wedge D)\big)\big] \Rightarrow X\big[(A \sqcap^a B) \wedge (C \sqcap^b D)\big]$, where $c$ does not occur in the conclusion;
9. **Splitting**: $A, B \Rightarrow A \sqcap^c B$, where either $A$ nor $B$ has an occurrence of $c$.

For a syntactic example, the reader is suggested to have a look at the long, useful one provided in [68], page 372.

---

*CL17*

In [71] Japaridze presents the "final" fragment of CoL (so far). Indeed, CL17 is the first order extension of CL16 with choice quantifiers: $\{\neg, \wedge, \vee, \sqcap, \sqcup, \bigsqcap, \bigsqcup\}$.
It is provably sound, complete and decidable - proofs in [71].

CL17 is the first CC system with quantifiers. Due to its expressiveness, unlike its predecessor CL16, the present system CL17 can be used as a logical basis for constructive applied theories.

This fragment, just as its predecessor, makes use of clusters.

Indeed, all rules of CL16 are also rules of CL17, even if renamed. The only axiom is, again, $\top$. Moreover, the "cleansing" sort of rules were first used by Xu in [100].

In total we have 19, quite dense rules: 18 of these are the same as the ones of CL16, renamed and readjusted for the new operators. A useful and rich example is provided in [71], page 773.

---

This ends our brief overview of what we know about CoL's fragments so far. We have successfully climbed over these elevated grounds while keeping track of all the information we have managed to gather.

However, we have come down those carpeted rocky sides with a certain *je ne sais quoi* that is making us hungry for more.
Indeed, we have had a hunch about coming close to a specific property of CL15 that has yet to be proven; still, since our expedition was to be a general, perusal one, we decided to continue onto CL16 for the time being.

We are, now, to go back to this specific fragment and discover what we think could be a hidden cave waiting for us right in its innermost crevices.

Let us climb up CL15 yet again and assess its decidability - as to further acquire knowledge of this silent, all-concealing grassland.

# 3 THERE AND BACK AGAIN: A DECIDABILITY TALE

## 3.1 AN UNCERTAIN OPENING

**Open Problem 6.2** *Is* **CL15** *decidable?*

This is how Japaridze unfurls a new path for our curiosity to follow; it does so in [60] and [72][1].

Indeed, sometimes the best hikes are the ones in which we discover something new.

Surely the reader knows the feeling of glancing at a small, quiet footpath that did not seem to be there before. The thrill of the adventure, or however we want to call it, enhances the landscape, almost making everything high definition: crisp and sharp.

This is why we want to go back. We have a hunch of discovering something new, which, let us not forget, is the main purpose of our expedition. If we ought to map everything that we see, making sure nothing is left behind - and more than that, we should always account for any unexpected turns.

That, we think, is the duty of every explorer; and as such, let us go back.

As Dersu Uzala himself would say, *walk together, work together*: while we are climbing back to CL15, let us review what we know so far and go further into it.

### 3.1.1 *Into the depths of CL15*

As previously mentioned, we decided to leave a thorough analysis of CL15 to this chapter, in order to help the reader easily intertwine a strong

1 Also in http://www.csc.villanova.edu/~japaridz/CL/OpenProblems.pdf, which is an updated list of all the open problems of Computability Logic one may wish to work on.

braid of thought.

We are going to extensively review each notion, in order to get to the top (and bottom) of this grand peak.

---

*A specific dialect*

Let us start from the very basic notion of **CL15-formulas**, which we will simply call *formulas* from now on.

CL15-formulas are formulas of the language of CoL which only contain nonlogical 0-ary general gameframe letters[2], plus $\neg, \wedge, \vee, ⫰, ⫯$.

Per usual, $\neg$ is only allowed to be applied to atoms. However, shall we write $\neg E$ for a non-atomic $E$, we understand this as a standard De Morgan abbreviation:

- $\neg\neg F = F$;
- $\neg(F \wedge G) = \neg F \vee \neg G$;
- $\neg(F \vee G) = \neg F \wedge \neg G$;
- $\neg ⫰ F = ⫯ \neg F$;
- $\neg ⫯ F = ⫰ \neg F$.

Similarly, according to each operator's definition:

- $F \rightarrow G$ should be understood as $\neg F \vee G$;
- $F \circ\!\!- G$ should be understood as $⫯ \neg F \vee G$;
- $\circ\!\neg F$ should be understood as $⫯ \neg F$[3].

Indeed, CL15 is built in CC, whose formalism goes beyond formulas. Consequently, we will repeat the definition of a **CL15-cirquent** (henceforth simply *cirquent*), in order to have it nearer if needed.

**Definition 2.2.1.1**
A **CL15-cirquent** is a triple $C = (\vec{F}, \vec{U}, \vec{O})$ where:

---

2 There are no $\top$ and $\bot$.
3 Since there is no $\bot$.

1. $\vec{F}$ is a nonempty finite sequence of CL15-formulas, whose elements are said to be the **oformulas** of C. Here the prefix "o" is for "occurrence", and is used to mean a formula together with a particular occurrence of it in $\vec{F}$. So, for instance, if $\vec{F} = \langle E, G, E \rangle$, then the cirquent has three oformulas even if only two formulas;

2. Both $\vec{U}$ and $\vec{O}$ are nonempty finite sequences of nonempty sets of oformulas of $C$. The elements of $\vec{U}$ are said to be the **undergroups** of $C$, and the elements of $\vec{O}$ are said to be the **overgroups** of $C$. As in the case of oformulas, it is possible that two undergroups or two overgroups are identical as sets (they have identical contents), yet they count as different undergroups or overgroups because they occur at different places in the sequence $\vec{U}$ or $\vec{O}$. Simply "group" will be used as a common name for both;

3. Additionally, every oformula is required to be in at least one undergroup and at least one overgroup.

While oformulas are not the same as formulas, we may often identify an oformula with the corresponding formula and, for instance, say "the oformula $E$" if it is clear from the context which of the possibly many occurrences of $E$ is meant.

Since cirquents are here defined through undergroups and overgroups, we will be drawing diamond-like structures.
Indeed, we represent cirquents using three-level diagrams such as the one shown below:

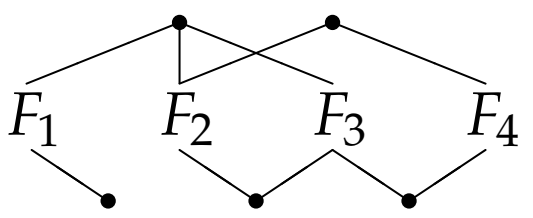

This diagram represents the cirquent with four oformulas ($F_1, F_2, F_3, F_4$), three undergroups ($\{F_1\}, \{F_2, F_3\}, \{F_3, F_4\}$) and two overgroups ($\{F_1, F_2, F_3\}$, $\{F_2, F_4\}$).

Each group is represented by a $\bullet$, where the **arcs** (lines connecting the $\bullet$'s with oformulas) are pointing to the oformulas that the given group contains.

*The Rules of the Road*

As we have already seen, CL15 has ten rules of inference.
The first one takes no premises, meaning it serves as an axiom.
All other rules take a single premise.

As already mentioned, we will explain them in terms of graphical modifications[4].

After all, with the introduction of overgroups, it is interesting to see how the cirquent-representation has changed compared to CCC's graphs.

**Axiom** (A)
The conclusion of this premiseless rule looks like an array of $n$ (with $n \geq 1$) "diamonds" where the oformulas are $\neg F$ and $F$, for some formula $F$.

Here is the case for $n = 3$:

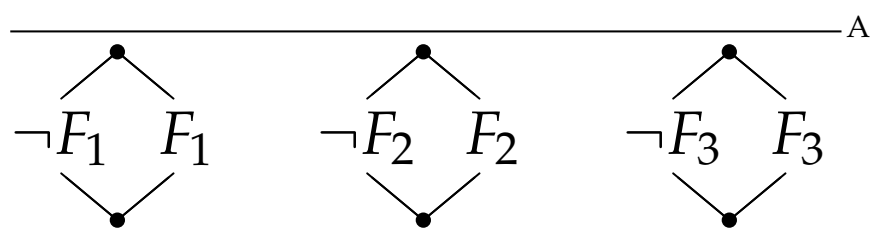

**Exchange** (E)

This rule comes in three versions: **undergroup** exchange, **oformula** exchange and **overgroup** exchange.
These allow us to swap any two adjacent objects (undergroups, oformulas or overgroups) of a cirquent, while preserving all oformulas, groups and arcs.

In the next page are three examples, one per each sort of Exchange. Obviously, the upper cirquent is the premise and the lower cirquent is the conclusion of the application of the rule:

4 Which are obviously more helpful to our *practical* knowledge than any rigorous formulation could be.

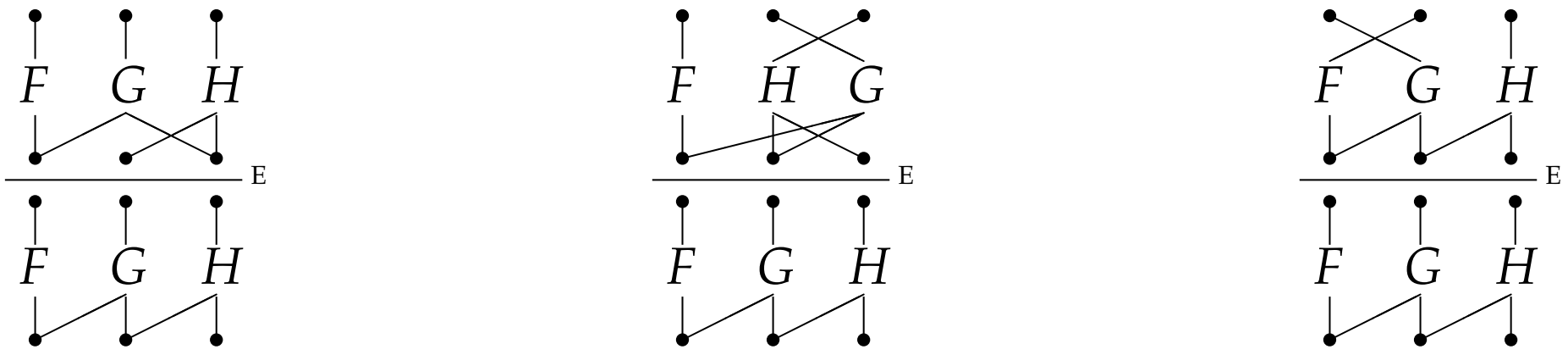

Exchange essentially allows us to treat all three components $(\vec{F}, \vec{U}, \vec{O})$ of a cirquent as multisets rather than sequences.

---

**Weakening** (W)

The premise of this rule is obtained from the conclusion by deleting an arc between some undergroup $U$ with $\geq 2$ elements and some oformula $F$.

Furthermore, if $U$ was the only undergroup containing $F$, then $F$ should also be deleted (to satisfy condition 3. of Definition 2.2.1.1), together with all arcs between $F$ and overgroups.

If such a deletion makes some overgroups empty, then they should also be deleted (to satisfy condition 2. of Definition 2.2.1.1).

Below are three examples:

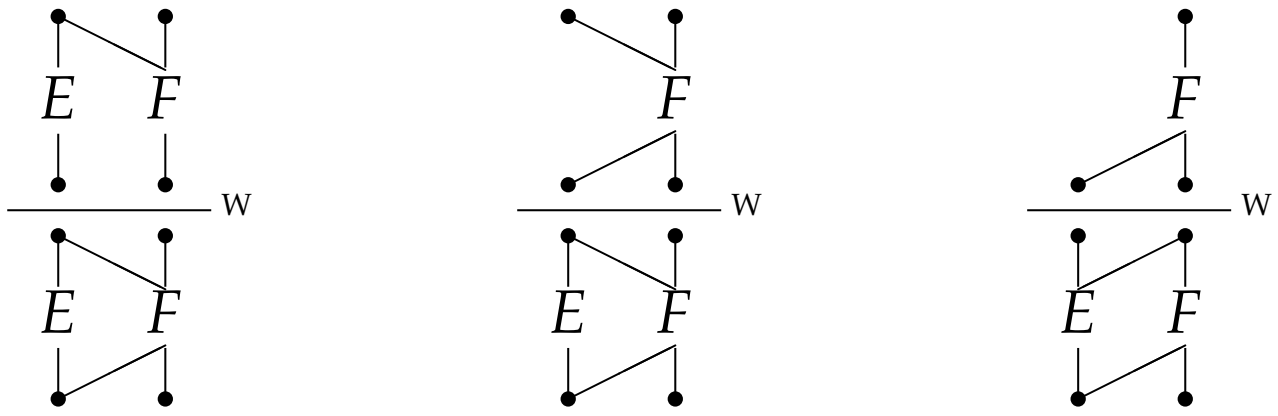

---

**Contraction** (C)

The premise of this rule is obtained by replacing an oformula ⫯$F$ by two adjacent oformulas ⫯$F$ and ⫯$F$ in the conclusion; then, by including them in exactly the same undergroups and overgroups in which the original oformula was before.

For instance:

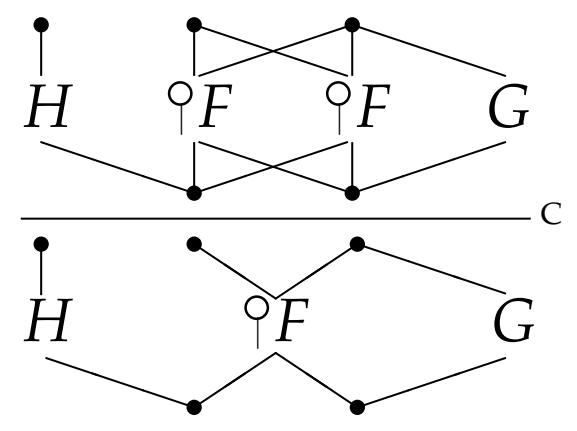

---

**Duplication** (D)

This rule, too, comes in a dual capacity: **undergroup** duplication and **overgroup** duplication.

The conclusion of the former is the result of replacing, in the premise, some undergroup $U$ with two adjacent undergroups whose contents are identical to that of $U$.

The same goes for the latter, with respect to overgroups.

For example:

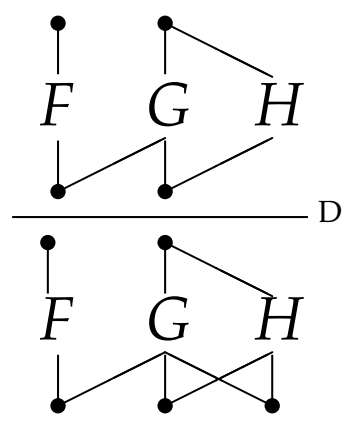

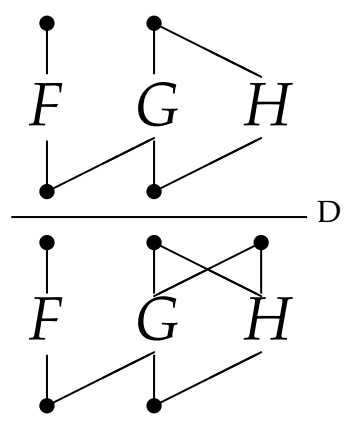

---

**Merging** (M)

In a top-down view, this rule merges any two adjacent overgroups.

For instance:

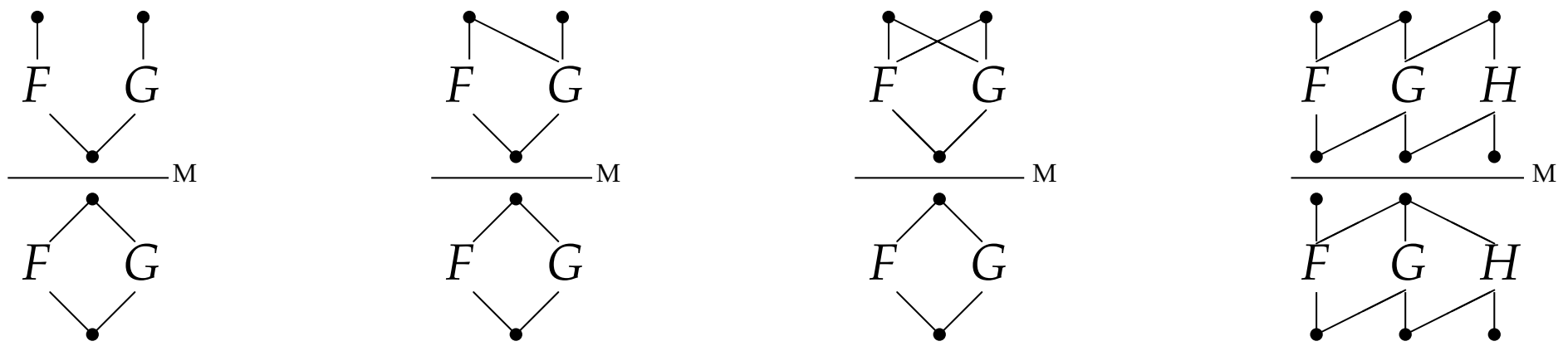

---

**Disjunction Introduction** ($\vee$)

The premise of this rule is obtained by replacing an oformula $F \vee G$ with two adjacent oformulas $F, G$ in the conclusion; then, by including both of them in exactly the same undergroups and overgroups in which the original oformula was.

As illustrated below:

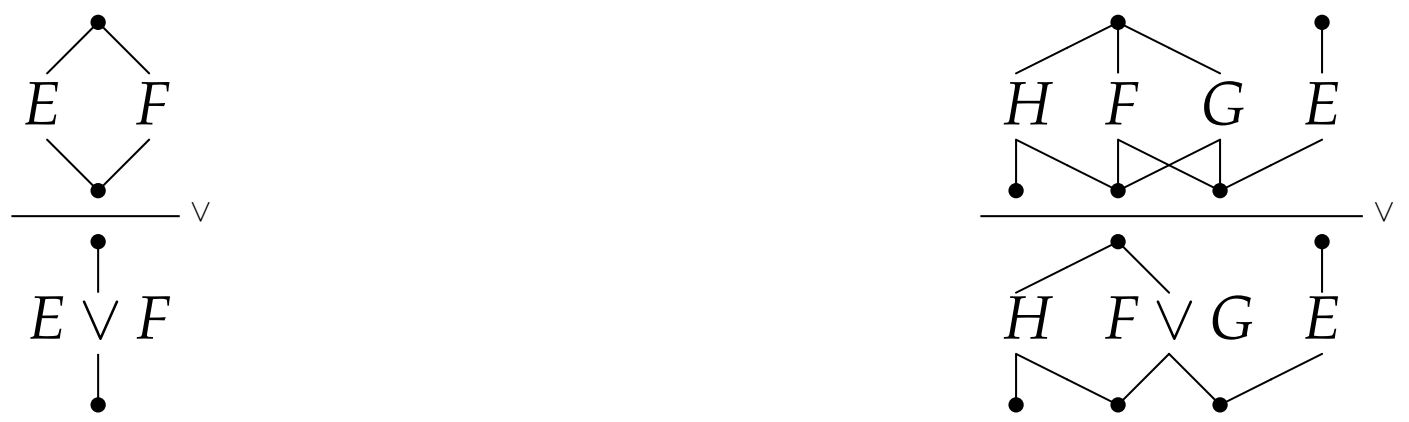

---

**Conjunction Introduction** ($\wedge$)

The premise of this rule is obtained from the conclusion by picking an arbitrary oformula $F \wedge G$ and applying the following two steps:

1. We first replace $F \wedge G$ with two adjacent oformulas $F, G$ and include both of them in exactly the same undergroups and overgroups in which the original oformula was;

2. Consequently, we replace each undergroup $U$ originally containing $F \wedge G$ (and now containing $F, G$ instead) with the two adjacent undergroups $U \setminus \{G\}$ and $U \setminus \{F\}$.

Below are three examples:

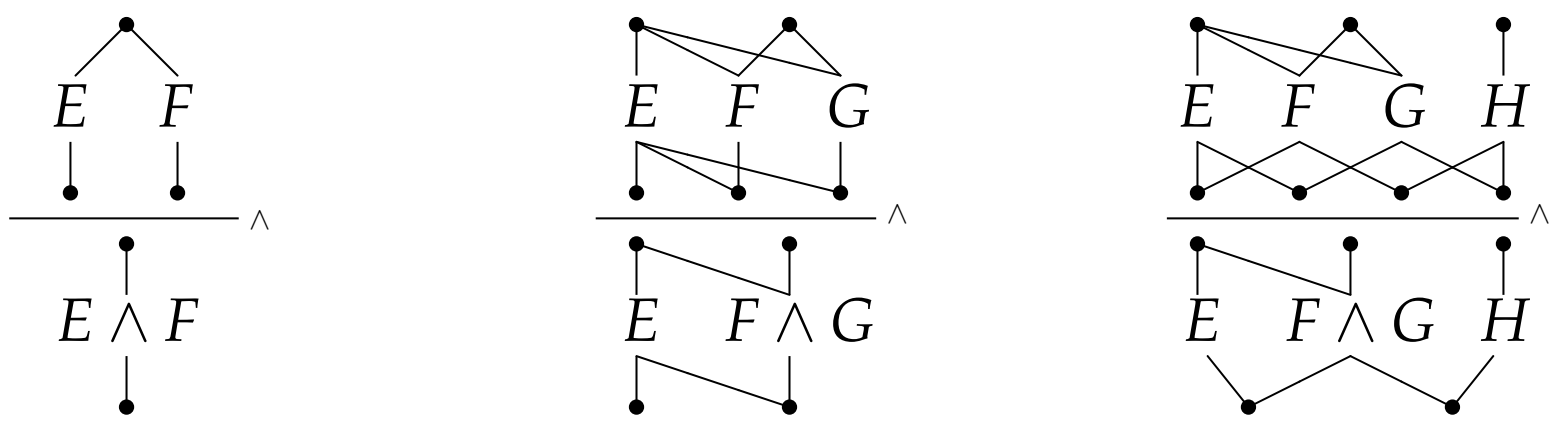

**Recurrence Introduction** (⫰)

The premise of this rule is obtained from the conclusion by replacing an oformula ⫰$F$ with $F$ (while preserving all arcs); then inserting, anywhere in the cirquent, a new overgroup that contains $F$ as its only oformula.

For instance:

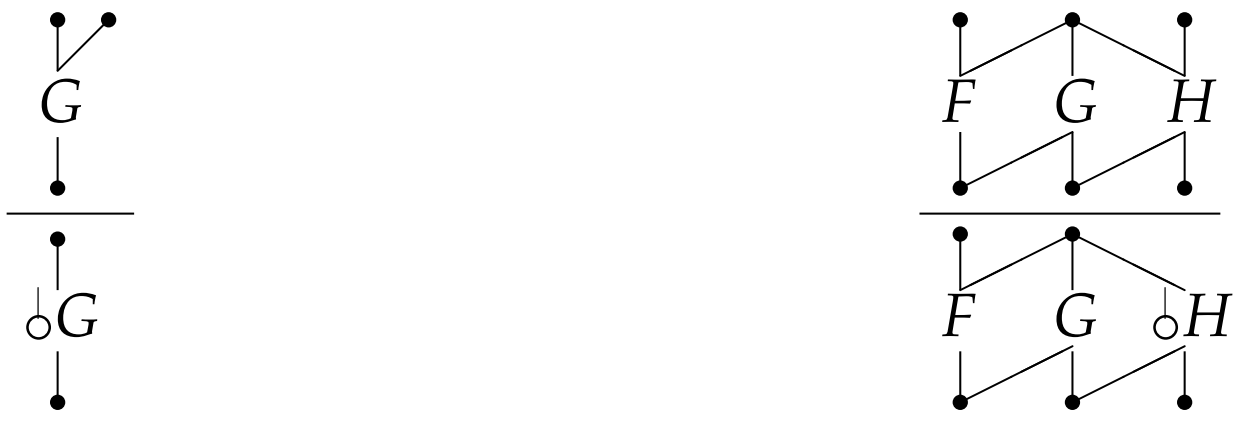

**Corecurrence Introduction** (⫯)

The premise of this rule is obtained by replacing an oformula ⫯$F$ with $F$ in the conclusion; then including $F$ in any (potentially 0) number of already existing overgroups in addition to those in which the original oformula ⫯$F$ already was.

As seen below:

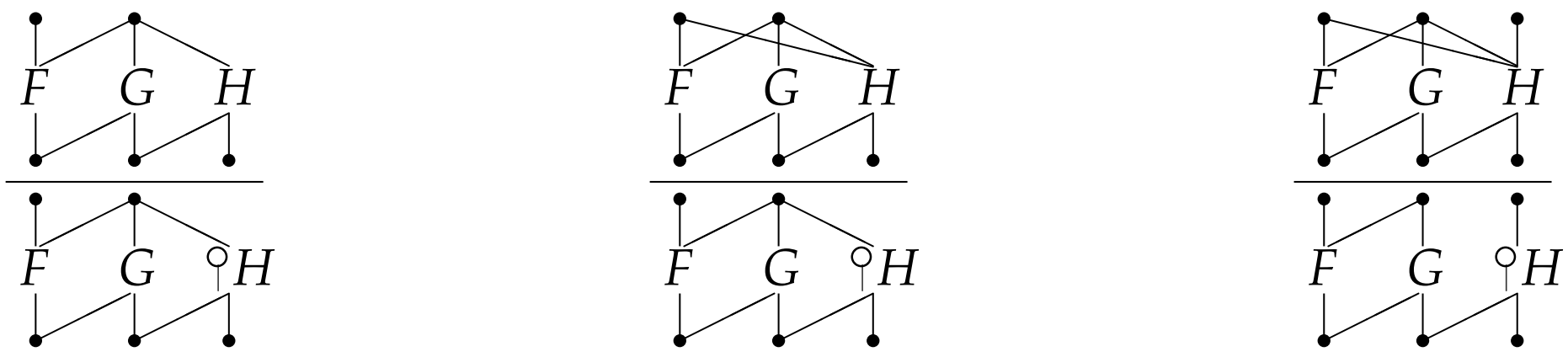

Let us now try and prove some formulas in CL15, in order to get the hang of these rules and their applications.

This should help vivify our comprehension of this system and, thus, facilitate our understanding of the proof we are about to introduce in just a few pages.

---

*Testing the lie of the land*

First, a basic (ontological) definition.

We say that a **CL15-proof** of a cirquent $C$ is a sequence of cirquents ending in $C$, where:

- the first cirquent is the conclusion of (an instance of) Axiom;
- every subsequent cirquent follows from the immediately preceding cirquent by one of the rules of CL15.

A proof of a formula $F$ is understood as a proof of the cirquent $(\langle F \rangle, \{F\}, \{F\})$.

We are now ready to touch and feel these chiselled walls, starting from the example provided below.

Indeed, we ought to find a CL15-proof of:

$$⫯⫰F \to ⫰⫯F,$$

which means

$$⫰⫯\neg F \vee ⫰⫯F.$$

Proof sketched in the following page.

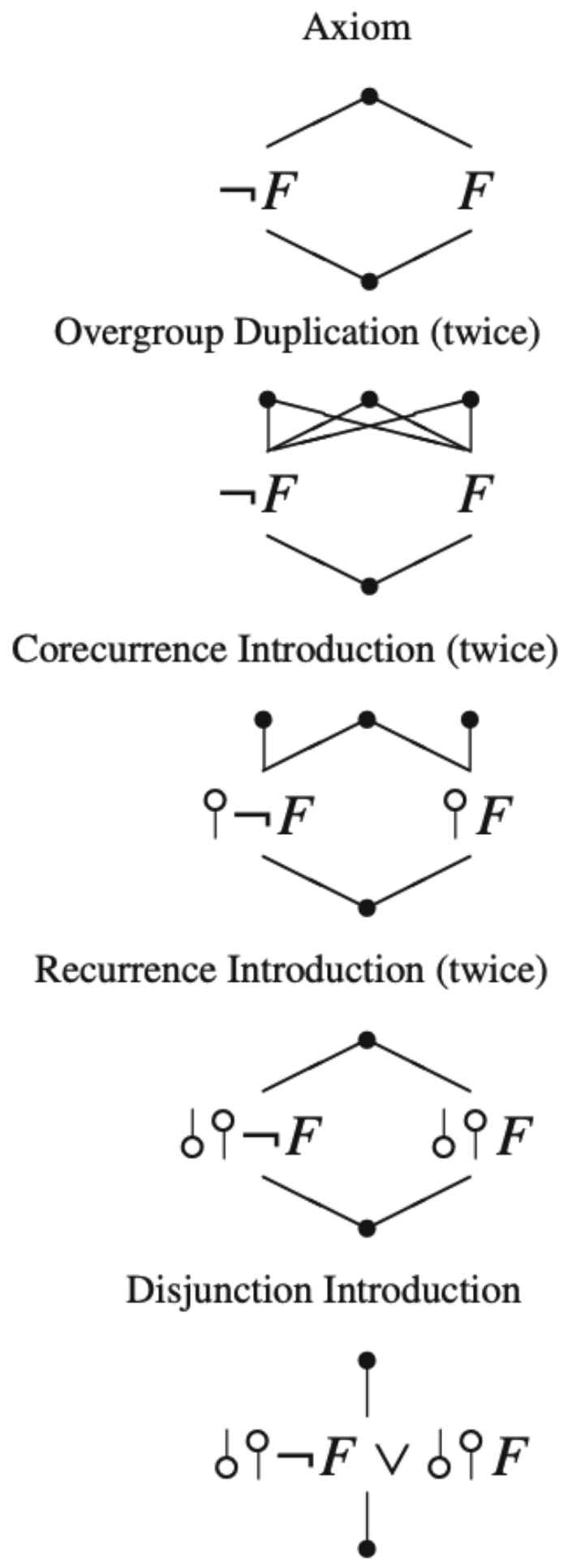

Alternatively, to save space, the cirquents can be arranged horizontally and separated with $\Longrightarrow$, overwritten with the symbolic names of the rules used. When such a name is duplicated, as in DD, it means that the rule was applied twice rather than once.

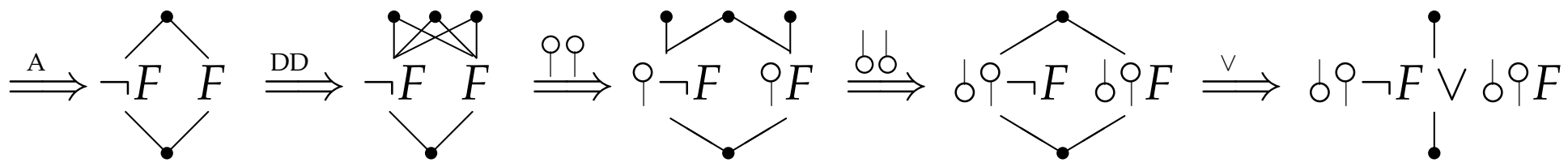

Another example, which makes use of Contraction, is:

$$\text{⫰}F \rightarrow \text{⫰}F \wedge \text{⫰}F,$$

which means:

$$⫯\neg F \vee (⫰F \wedge ⫰F).$$

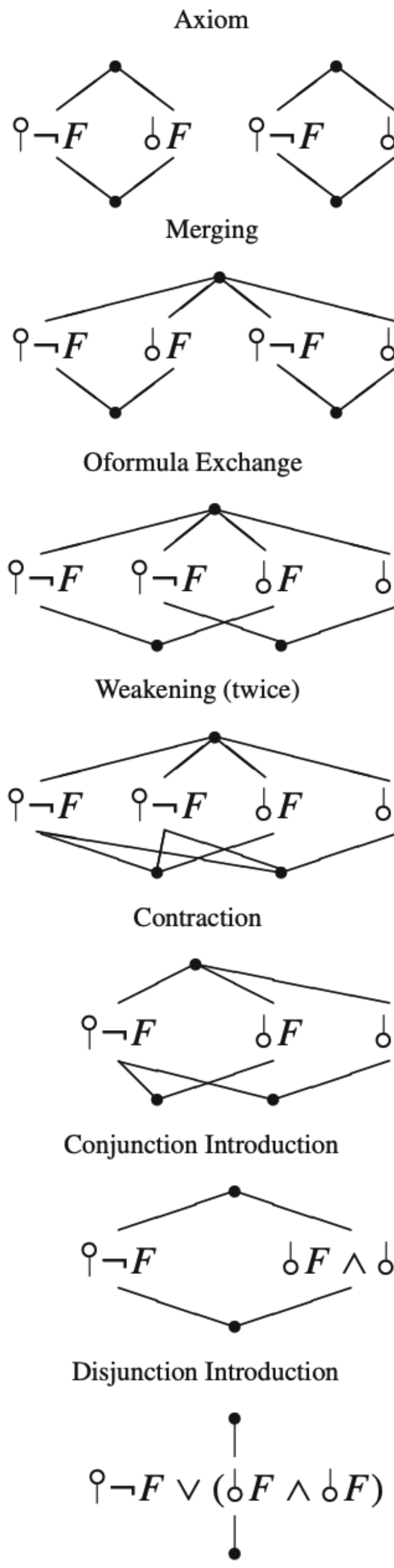

One last, final example we provide here is the CL15-proof[5] of:

$$⫰F \to ⫰⫰F,$$

which means:

$$⫯\neg F \vee ⫰⫰F.$$

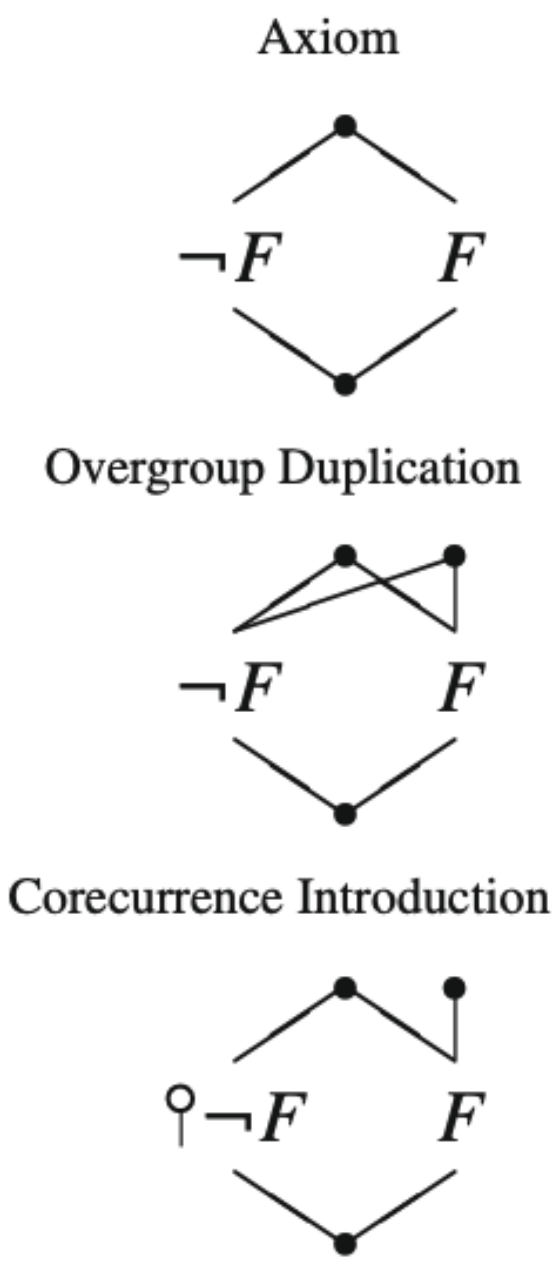

5 Alas, due to spacing reasons, the proof is divided between this page and the next one. Giving that we are following a bottom-up approach, per usual, the reader has to start from the following page and then come back here to finish with Axiom.

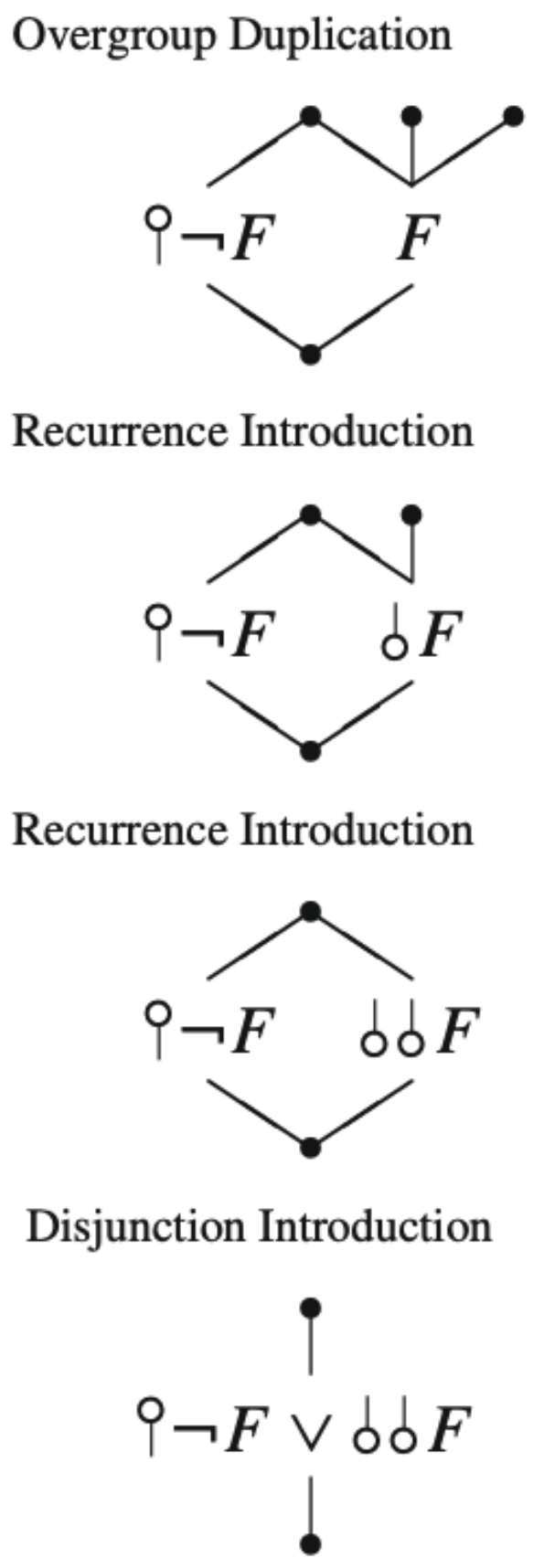

The reader is welcomed to have a look at [60], page 21 onwards, to have other few examples to taste, if needed.

---

*Refining semantics*

We are now going to briefly touch upon Japaridze's proof of CL15's soundness and completeness.

In order to properly understand the theorems to come, we need to introduce (and recall) some useful notions.
On such occasion, we will also provide a quick overview of how CC's semantics is adjusted to fit CL15's physique, with overgroups and undergroups.

As a result, not all the notions here presented will be *actually* used in the two proofs that we are trying to see; however, it never hurts to be given the full-picture before having to zoom-in on a certain detail.

First, we say that a **constant** (respectively **unary**) **interpretation** is one which interprets all atoms as constant (respectively unary) games.

In addition, we jog the reader's memory by summoning back[6] Definition 1.2.3.1 regarding **uniform** and **multiform validity**.

We re-present it here lest the reader should go back a "few" pages:

**Definition 1.2.3.1**
We say that a sentence $F$ is:

- **logically** (or **uniformly**) **valid** iff there is an HPM $\mathcal{M}$ such that, for every interpretation * admissible for $F$, $\mathcal{M}$ solves $F^*$. Such an $\mathcal{M}$ is said to be a **logical** (or **uniform**) **solution** of $F$;
- **extralogically** (or **multiformly**) **valid** iff for every interpretation * admissible for $F$, there is an HPM $\mathcal{M}$ such that $\mathcal{M}$ solves $F^*$.

Correspondingly, we say that a system $\Gamma$ is **uniformly complete** iff each uniformly valid sentence $F$ is deducible in the system itself. Namely, we might say, in our own notation[7]:

$$\Gamma \vDash_u F \longrightarrow \Gamma \vdash_u F.$$

In similar fashion, we say that a system $\Gamma$ is **multiformly complete** iff each multiformly valid sentence $F$ is deducible in the system itself. Again, in our own notation[8]:

$$\Gamma \vDash_m F \longrightarrow \Gamma \vdash_m F.$$

Next, in order to prove the soundness and completeness of this CC system, Japaridze introduces some semantical adjustments specifically made for CL15-cirquents. Many of these recall what we already know so far about cirquents in general.

Indeed, overgroups can be seen as generalised $\overset{\circ}{\mid}$s, with the main difference being that an overgroup can be shared by several oformulas

6 Dramatic tones for dramatic settings (getting higher and higher!).
7 With the $u$ subscript standing for "uniformly".
8 With the subscript $m$ standing for "multiformly"

(arguments).

Similarly, undergroups are disjunctions prefixed with generalised ⫰s, with the main difference being that undergroups may have shared arguments with each other.

Finally, we may see the whole cirquent as the conjunction of its undergroups.

To briefly define our semantics formally, we need the following notational convention.

Let $\Omega$ be a run. Then:

- where $\alpha$ is a move, we will be using the notation

  $$\Omega^{\alpha}$$

  to mean the result of deleting from $\Omega$ all moves and corresponding labels, except those that look like $\alpha\beta$ for some move $\beta$; then further deleting the prefix "$\alpha$" from such moves. For instance, $\langle \top 0.\beta, \bot .1\gamma, \bot 0.\delta\rangle^{0.} = \langle \top\beta, \bot\delta\rangle$;

- where $x$ is an infinite bitstring, we will be using the notation

  $$\Omega^{\preceq x}$$

  to mean the result of deleting from $\Omega$ all moves and corresponding labels, except those that look like $u.\beta$ for some move $\beta$ and some finite initial segment $u$ of $x$, and then further deleting the prefix "$u.$" from such moves. For instance, $\langle \top 00.\alpha, \bot 001.\beta, \bot 0.\delta\rangle^{\preceq 000\ldots} = \langle \top\alpha, \bot\delta\rangle$.

Consequently, we know that given a run $\Omega$, a decimal numeral $a$ for a positive integer and a nonempty sequence $\vec{x} = x_1, \ldots, x_n$ of $n$ infinite bitstrings, we have that:

$$\Omega^{\preceq a;\vec{x}}$$

is the result of deleting from $\Omega$ all moves (together with their labels) except those that look like $a; u_1, \ldots, u_n.\beta$ for some move $\beta$ and some finite initial segments $u_1, \ldots, u_n$ of $x_1, \ldots, x_n$; then, further deleting the prefix "$a; u_1, \ldots, u_n.$" from such moves.

For instance, if $x = 000\ldots$ and $y = 111\ldots$ then:

$$\langle \top 3; 00, 1.\alpha, \bot 3; 001, 11.\beta, \bot 5; 00, 1.\delta, \top 3; 0, 111.\gamma\rangle^{\preceq 3; x, y} = \langle \top\alpha, \top\gamma\rangle.$$

This notation will be useful to refer to specific threads of a certain game, as we will see below.

Throughout this paper, the letter $\epsilon$ is used to denote the *empty bitstring*. The latter is a prefix of every bitstring.

Next, Japaridze gives a new definition of constant games in CL15:

**Definition 3.1.1.1**
Consider a constant interpretation $^*$ (in the old, ordinary sense) and a cirquent:

$$C = (\langle F_1, \ldots, F_k\rangle, \langle U_1, \ldots, U_m\rangle, \langle O_1, \ldots, O_n\rangle)$$

with $k$ oformulas, $m$ undergroups and $n$ overgroups. Then $C^*$ is the *constant game* defined as follows, with $\Gamma$ ranging over all runs and $\Omega$ ranging over the legal runs of $C^*$:

1. $\Gamma \in \mathbf{Lr}^{C^*}$ iff the following two conditions are satisfied:
   a) Every move of $\Gamma$ looks like $a; \vec{u}.\alpha$, where $\alpha$ is some move, $a \in \{1, \ldots, k\}$, and $\vec{u} = u_1, \ldots, u_n$ is a sequence of $n$ finite bitstrings such that the following condition is satisfied:

      whenever an overgroup $O_j (1 \leq j \leq n)$ does not contain the oformula $F_a$, $u_j = \epsilon$;

   b) For every $a \in \{1, \ldots, k\}$ and every sequence $\vec{x}$ of $n$ infinite bitstrings, $\Gamma^{\preceq a; \vec{x}} \in \mathbf{Lr}^{F_a^*}$;
2. $\mathbf{Wn}^{C^*}\langle\Omega\rangle = \top$ iff, for every $i \in \{1, \ldots, m\}$ and every sequence $\vec{x}$ of $n$ infinite bitstrings, there is an $a \in \{1, \ldots, k\}$ such that the undergroup $U_i$ contains the oformula $F_a$ and $\mathbf{Wn}^{F_a^*}\langle\Omega^{\preceq a; \vec{x}}\rangle = \top$.

Thus, intuitively, when $C$ and $^*$ are as above, a (legal) run $\Omega$ of $C^*$ consists of parallel plays of a continuum of threads of each of the games $F_a^* (1 \leq a \leq k)$.

Namely, every thread of such an $F_a^*$ is $\Omega^{\preceq a; \vec{x}}$ for some array $\vec{x} = x_1, \ldots, x_n$ of $n$ infinite bitstrings.

In the context of a fixed $\Omega$, we may refer to $\Omega^{\preceq a;\vec{x}}$ as the **thread of $\vec{x}$ of $F_a^*$**.

For an undergroup $U_i$, we say that $\top$ is the **winner in** $U_i$ iff, for every array $\vec{x}$ of $n$ infinite bitstrings, there is an oformula $F_a$ in $U_i$ such that the thread $\vec{x}$ of $F_a^*$ is won by $\top$.
Furthermore, $\top$ wins the overall game $C^*$ iff it wins in all undergroups of $C$.

Basically, condition 1. of the above definition means that, for any array $\vec{x} = x_1, \dots, x_n$ of infinite bitstrings, only some of the elements of $\vec{x}$ are really *relevant* to any given oformula $F_a$ of the cirquent.

In particular, an element $x_j$ of $\vec{x}$ is relevant if the overgroup $O_j$ contains $F_a$.

This relevance/irrelevance distinction entails that given an array $\vec{y}$ which only differs from another one, $\vec{x}$, in the "irrelevant" elements, then, as it is easy to see from condition 1. plus the fact that $\epsilon$ is a prefix of every bitstring, we have that $\Omega^{\preceq a;\vec{x}} = \Omega^{\preceq a;\vec{y}}$.

We then redefine uniform (and multiform) validity in CL15 as such:

**Definition 3.1.1.2**
We say that a cirquent $C$ is **uniformly valid** iff there is a machine $\mathcal{M}$, called a **uniform solution** of $C$, such that, for every *constant* interpretation $^*$, $\mathcal{M}$ wins $C^*$.

Similarly, $C$ is **multiformly valid** iff there is a machine $\mathcal{M}$, called a **multiform solution** of $C$, that for every *constant* interpretation $^*$, $\mathcal{M}$ wins $C^*$.

We are now ready to have a look at the soundness and completeness proofs of CL15, in order to better frame the question of decidability.

### 3.1.2 *A penny for your properties*

Let us introduce the most important theorem of CL15:

**Theorem 3.1.2.1**
For any formula $F$, the following conditions are equivalent:

1. $\text{CL15} \vdash F$;
2. $F$ is uniformly valid;
3. $F$ is multiformly valid.

Furthermore:

a. The implication 1. $\Rightarrow$ 2. holds in the strong sense that there is an effective procedure which takes any CL15-proof of any formula $F$ and constructs a uniform solution of $F$;

b. The implication 2. $\Rightarrow$ 1. holds in the strong sense that, if $\text{CL15} \nvdash F$, then, for every HPM $\mathcal{H}$, there is a constant interpretation $^*$ such that $\mathcal{H}$ fails to compute $F^*$;

c. The implication 3. $\Rightarrow$ 1. holds in the strong sense that, if $\text{CL15} \nvdash F$, then there is a unary interpretation $^\dagger$ such that $F^\dagger$ is not computable.

Since proving this theorem requires a serious number of pages, which, at the moment, we may not dispose of, here is a brief outline of the ideas behind such proof.

---

*Building Soundness*

First of all, Japaridze proves 1. $\Rightarrow$ 2. (soundness), in the form of clause a., by showing that it is a direct corollary of the following lemma[9]:

**Lemma 3.1.2.2**
There is an effective procedure which takes any CL15-proof of any formula $F$ and constructs a machine $\mathcal{M}$ such that, for any constant interpretation $^*$, $\mathcal{M}$ wins $F^*$.

9 Proven by Japaridze in [60] through an induction on the lenghts of CL15-proofs.

As a result, we may focus exclusively on constant games, where the winning strategies can fully ignore the valuation tape - as its content is irrelevant.

Next, Japaridze shows that:

**Lemma 3.1.2.3**
There is an effective function $f$ from machines to machines such that, for every machine $\mathcal{M}$, formula $F$ and interpretation $^*$, if $\mathcal{M}$ wins ⫰$F^*$, then $f(\mathcal{M})$ wins $F^*$.

He proves this by establishing that affine logic is sound with respect to uniform validity; given that affine logic, indeed, proves ⫰$P \rightarrow P$, this formula is uniformly valid. This means that there is a machine $\mathcal{N}_0$ that wins ⫰$F^* \rightarrow F^*$ for any formula $F$ and interpretation $^*$.

In addition, since the computability of static games is closed under Modus Ponens in the strong sense[10], it is clear that the function $f(\mathcal{M})$ defined by $f(\mathcal{M}) = h(\mathcal{N}_0, \mathcal{M})$ satisfies the promise of the lemma.

Subsequently, Japaridze shows that:

**Lemma 3.1.2.4**
There is an effective function $g$ from machines to machines such that, for every machine $\mathcal{M}$, formula $F$ and constant interpretation $^*$ if $\mathcal{M}$ wins $(F^{\clubsuit})^*$, then $g(\mathcal{M})$ wins $F$.

Here, $F^{\clubsuit}$ means the cirquent $(\langle F\rangle, \langle\{F\}\rangle, \langle\{F\}\rangle)$ for a certain formula $F$. Visually this means the following cirquent:

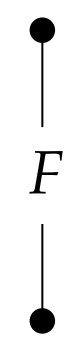

10 Meaning that any pair $(\mathcal{N}, \mathcal{M})$ of machines can be effectively converted into a machine $h(\mathcal{N}, \mathcal{M})$ such that, for any static games $A$ and $B$, if $\mathcal{N}$ wins $A$, then $h(\mathcal{N}, \mathcal{M})$ wins $B$.

Indeed, to prove this we just need to show[11] that, employing this last lemma, the games $(⫰F)^*$ and $(F^{\clubsuit})^*$ are essentially the same, with only a minor technical difference in the forms of their legal moves.

While every legal move of $(F^{\clubsuit})^*$ looks like $1;\omega.\alpha$, for some finite bitstring $\omega$ and move $\alpha$, the corresponding move of $(⫰F)^*$ simply looks like $\omega.\alpha$ instead; and vice versa.

This means that if $\mathcal{M}$ wins $(F^{\clubsuit})^*$, then an "essentially-the-same" strategy $g(\mathcal{M})$ wins $(⫰F)^*$.

We say that a rule of CL15, other than Axiom, is **uniformly-constructively sound** iff there is an effective procedure that takes any instance $(A, B)$ (meaning a particular premise-conclusion pair) of the rule, any machine $\mathcal{M}_A$ and returns a machine $\mathcal{M}_B$ such that, for any constant interpretation $^*$, whenever $\mathcal{M}_A$ wins $A^*$, $\mathcal{M}_B$ wins $B^*$.

Then, of course, as long as $\mathcal{M}_A$ is a uniform solution of $A$, $\mathcal{M}_B$ is a uniform solution of $B$.

As for Axiom, by its uniform-constructive soundness we simply mean existence of an effective procedure that takes any instance $B$ of the conclusion of Axiom and returns a uniform solution $\mathcal{M}_B$ of $B$.

Japaridze then proves that:

**Theorem 3.1.2.5**
All rules (including Axiom) of CL15 are uniform-constructively sound.

Furthermore:

**Theorem 3.1.2.6**
Every cirquent provable in CL15 is uniformly valid. Furthermore, there is an effective procedure that takes an arbitrary CL15-proof of an arbitrary cirquent $C$ and constructs a uniform solution of $C$.

This comes directly from Theorem 3.1.2.5 by induction on the lenghts of CL15-proofs.

11 See [60], page 31, for the complete proof.

As we can see, Lemma 3.1.2.2 (our goal) is an immediate corollary of Theorem 3.1.2.6 and Lemma 3.1.2.4.
Our only remaining duty is to prove Theorem 3.1.2.5, which is done by showing the uniform-constructive soundness of all rules of CL15 one by one.

Japaridze does so by building[12] an $\mathcal{M}_B$ machine from a $\mathcal{M}_A$ machine and stating that the former wins $B$ - meaning the conclusion of the rule is derived in CL15, iff the latter wins $A$ - meaning the premise of an arbitrary instance of the same rule is deducible in CL15.

In each non-axiom case, as Japaridze explains, it will be implicitly assumed that $\mathcal{M}_A$ wins $A$.
Our construction of the corresponding $\mathcal{M}_B$ will never depend on this assumption, though; only the subsequent conclusion that $\mathcal{M}_B$ wins $B$ will depend on it.

Finally, $\mathcal{M}_B$ is always implicitly assumed to be an EPM - and so will also be $\mathcal{M}_A$, unless specified otherwise.

The reader is strongly advised to have a look at [60], page 205 onwards, for the proof of each rule.
Here we only present the Axiom case, which is the most intuitive.

Assume that $B$ is an axiom. Namely:

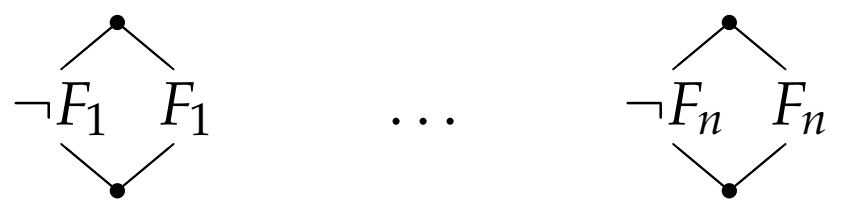

The EPM $\mathcal{M}_B$ that wins $B$ works as follows.
It keeps granting permission.

Every time the adversary makes a move $\alpha;\vec{\omega}.\alpha$, where $1 \leq \alpha \leq 2n$ and $\vec{\omega}$ is an array of $n$ finite bitstrings[13], $\mathcal{M}_B$ responds by the move $b;\vec{\omega}.\alpha$ where $b$ is $a+1$ if $a$ is odd, and $a-1$ if $a$ is even.

Basically, $\mathcal{M}_B$ acts as a copycat between the two oformulas/games of each thread in each diamond.

12 As in *constructive* soundness.
13 Note that every legal move of $B$ should indeed look like this.

Namely, when $a, b$ are as above and $\Gamma$ is any run generated by $\mathcal{M}_B$, with an array $\vec{x}$ of $n$ infinite bitstrings, we have that $\Gamma^{\preceq a;\vec{x}} = \neg\Gamma^{\preceq b;\vec{x}}$.

It is therefore obvious that $\Gamma$ is a $\top$-won run of $B$, meaning that $\mathcal{M}_B$ wins $B$. ■

Let us now review the completeness proof Japaridze provides in the subsequent paper, [61].

---

*Completeness: a fair duel*

So far, concerning Theorem 3.1.2.1, we have managed to show that 1. $\Rightarrow$ 2. - meaning that CL15 is, indeed, sound.

We may also add that, since uniform validity is stronger than multiform validity, implication 2. $\Rightarrow$ 3. is trivial.

At this point, we need to show that 3. $\Rightarrow$ 1. (multiform completeness), in the form of c. and b. (uniform completeness).

First, we fix an arbitrary formula $\mathbb{F}_0$ and assume that:

$$\text{CL15} \nvdash \mathbb{F}_0.$$

We need to prove that $\mathbb{F}_0$ is not uniformly valid; this result, then, must be extended from uniform to multiform (in)validity.

As Japaridze explains, in order to do this, we are going to show that there is an effective *counterstrategy*[14] $\mathcal{E}$ such that, when Environment plays according to $\mathcal{E}$, no HPM wins $\mathbb{F}_0^*$ for an appropriately selected constant interpretation $^*$.

Of course, we have never defined the concept of an Environment's effective strategy. As explained in [61], page 214 onwards, this is just like a Machine's strategy which can be understood simply as an EPM[15].

We may say that from now we will host a fairer version of games, in

14 Because we need to find an effective procedure that guarantees any HPM $\mathcal{M}$, meaning $\top$, to lose $\mathbb{F}_0$! That is why $\bot$ has to act effectively - and it can, since it can act *any* way.
15 As the chapter title of [61] reads, it is now "Machines against machines".

which both contestants start out and progress through the same means[16].

For such purpose, Japaridze introduces a number of new notions which would require a whole entire separate work to get into with the appropriate care.
For instance, from a bird's eye view, he presents the definitions of: enumeration games; prompts; unit trees; resolutions; visibility; domination; hypercirquents; and many more.

Let us only address the first one in order to have a look at the main lemmas.

**Enumeration games** are games in (every instance of) which any natural number $a$ is a legal move by either player $\wp$ at any time; plus, there are no other legal moves.

Either player, then, basically enumerates a set of numbers, meaning moves during the play.

The winner in a (legal) play of a given instance of an enumeration game only depends on the two sets enumerated this way.
That is, it only matters *what* moves have been eventually made, regardless of *when* (in what order) and *how many times* (only once or repetitively) those moves were made.

As a result, an **enumeration interpretation** $^*$ is one that sends every atom $P$ to an enumeration game $P^*$.
From now on, we will limit our consideration to enumeration interpretations.[17]

Moreover, $e[\mathbb{F}_0^*]$ will be the game into which it turns after a constant enumeration interpretation $^*$ is applied to it.

16 Indeed, Environment now acts according to an effective procedure, since it has to avoid any scenario in which Machine wins. There is no philosophical repercussion of sorts on the nature of $\perp$ itself, since, as we have mentioned numerous times, it can act and play in whichever manner it "prefers".

17 Consequently, the set of legal runs of $e[\mathbb{F}_0^*]$ does not depend on the valuation $e$ or the interpretation $^*$, since all instances of all enumeration games have the same set of legal runs. Hence, we may already address legal moves and runs of $e[\mathbb{F}_0^*]$ without having yet defined the interpretation $^*$ and having specified the valuation $e$.

Thus, to prove uniform completeness[18] Japaridze first manages to show, thanks to many lemmas and notions[19] introduced beforehand, that:

**Lemma 3.1.2.7**
There is a constant enumeration interpretation $^*$ such that the HPM $\mathcal{H}$ does not win $\mathbb{F}_0^*$. Namely, $\mathcal{H}$ loses this game against $\mathcal{E}$ on valuation $e$.

Now, recalling that $\mathbb{F}_0^*$ was an arbitrary fixed CL15-unprovable formula, this lemma means that the implication 2. $\Rightarrow$ 1.[20] of our Theorem 3.1.2.1, in the strong form of clause b., holds[21].

Now for the extension to multiform completeness[22].

A transition from uniform completeness to the multiform one is already rather standard in the literature on CoL.

In other CC systems, multiform completeness has been proven through a diagonalisation-style idea.

Basically, let:

$$\mathcal{H}_0, \mathcal{H}_1, \mathcal{H}_2, \ldots$$

be a fixed list of all HPMs arranged in lexicographic order of their descriptions. We call this list $H$.
We then pick a variable $x$ and let:

$$e_0, e_1, e_2, \ldots$$

be the valuations such that, for each $n \geq 0, e_n$ assigns $n$ to $x$ and assigns 0 to all other variables. We call this list $V$.

18 Meaning the fact that 2. $\Rightarrow$ 1. in Theorem 3.1.2.1
19 Specifically, hypermodels and hyperliterals.
20 Which is non other than $\neg$1. $\Rightarrow$ $\neg$2.
21 As Japaridze points out, whenever CL15 does not prove a formula $F$, there is simply no strategy, whether effective or not, that wins $F^*$ for every constant and enumeration interpretation $^*$. Indeed, $F$ has no uniform solution even if we do not require uniform solutions to be effective! The entire argument that leads to Lemma 3.1.2.7 would work just as well if $\mathcal{H}$ was an HPM with an arbitrary oracle. However, CoL takes no interest in non-effective strategies, as explained in [59], page 629: a problem $A$ is computable by an HPM with an oracle for a function $f(x)$ iff the problem $\sqcap x \sqcup y(y = f(x)) \circ\!\!- A$ is computable in the ordinary sense - meaning without any oracles.
22 Meaning the 3. $\Rightarrow$ 1. step from our Theorem 3.1.2.1

Remember that, even though $\mathcal{H}$ and $e$ have been fixed so far, they were an arbitrary HPM and an arbitrary valuation.

This means that $\mathcal{H}$ and $e$ just as well could have been treated (as we are going to do from now on) as a variable ranging over HPMs and a variable ranging over valuations, respectively.

Everything we have showed about the fixed $\mathcal{H}$ and $e$ is automatically valid also for $\mathcal{H}$ and $e$ as variables.
Namely, considering only (*HPM*, *valuations*) pairs of the form $(\mathcal{H}_n, e_n)$, Lemma 3.1.2.7 can now be re-written as follows:

**Lemma 3.1.2.8**
For every natural number $n$ there is a constant enumeration interpretation $^{*n}$ - which we choose and fix for the purposes of this section - such that the HPM $\mathcal{H}_n$ from the list $H$ does not win the game $\mathbb{F}_0^{*n}$ against our EPM $\mathcal{E}$ on the valuation $e_n$ from the list $V$.

We now define an enumeration interpretation $^{\dagger}$ such that, for any atom $P$, $P^{\dagger}$ is a unary game that depends on no variables other than $x$.

Namely, we let $P^{\dagger}$ be the game such that, for any valuation $e$, $e[P^{\dagger}] = P^{*n}$ (where $n$ is the value assigned to the variable $x$ by $e$, and $^{*n}$ is the interpretation fixed in Lemma 3.1.2.8 for $n$).

Clearly, $^{\dagger}$ is a unary interpretation, as promised in clause c. of Theorem 3.1.2.1; to complete our proof of such clause, we need to show that no HPM wins $\mathbb{F}_0^{\dagger}$.

Indeed, consider an arbitrary HPM $\mathcal{H}$. Since all HPMs are on the list $H$, we have $\mathcal{H} = \mathcal{H}_n$ for some fixed $n$.

Next, consider the valuation $e_n$ from the list $V$. By our choice of $^{\dagger}$, for every atom $P$, we have $e_n[\mathbb{F}_0^{\dagger}] = \mathbb{F}_0^{*n}$.

By Lemma 3.1.2.8, $\mathcal{H}_n$, i.e. $\mathcal{H}$, does not win $\mathbb{F}_0^{*n}$ against $\mathcal{E}$ on $e_n$.

Thus, $\mathcal{H}$ does not win $e_n[\mathbb{F}_0^{\dagger}]$ against $\mathcal{E}$ on $e_n$.

However, not winning $e_n[\mathbb{F}_0^{\dagger}]$ against $\mathcal{E}$ on $e_n$ obviously means the same as not winning $\mathbb{F}_0^\dagger$ against $\mathcal{E}$ on $e_n$.

Thus, $\mathcal{H}$ does not win $\mathbb{F}_0^\dagger$ as desired.

This concludes both soundness and completeness proofs through Theorem 3.1.2.1.
Here is a quick recap.

Our goal was to show that, in our own notation:

$$1.\ \text{CL15} \vdash F \iff 2.\ \text{CL15} \vDash_u F \iff 3.\ \text{CL15} \vDash_m F.$$

Japaridze first showed that 1. ⇒ 2. - soundness - in the (strong) form of clause a. of the same theorem.

Then, he proceeded to observe that 2. ⇒ 3. is rather trivial, since uniform validity is stronger than multiform one.

Afterwards, he managed to prove that 2. ⇒ 1. - uniform completeness - in the (strong) form of clause b. of the same theorem.

He then extended his findings to multiform completeness, meaning he showed 3. ⇒ 1. in the (strong) form of clause c. of the same theorem.

One can conclude by simply applying transitivity and obtaining that 1. ⇒ 3. follows as a logical consequence. ■

Some final remarks.

We know that our system CL15 becomes incomplete if $⫰, ⫯$ are replaced with either $⅄, ⋎$ or $⫰^{\aleph_0}, ⫯^{\aleph_0}$[23], as explained in [60], page 27.

Furthermore, Xu and Liu [99] have shown that CL15 remains sound with $⅄, ⋎$ instead of $⫰, ⫯$.
Completeness, however, is lost in this case since, for instance, as shown in [57], the formula $F \wedge ⅄(F \rightarrow F \wedge F) \rightarrow ⅄F$ is logically valid while $F \wedge ⫰(F \rightarrow F \wedge F) \rightarrow ⫰F$ is not.

Of course, the set of CL15-theorems is recursively enumerable. At this

23 Where $⋎$ and $⫯^{\aleph_0}$ are dual to $⅄$ and $⫰^{\aleph_0}$ in the same sense as $⫯$ is dual to $⫰$.

point, however, we do not have an answer to its decidability.

Let us try and analyse the problem in order to evaluate a possible solution (meaning, a possible CL15-proof of its problem of provability).

## 3.2 STRATEGY OUTLINE

### 3.2.1 *Preliminary considerations*

Almost overwhelming, this view. No picture taken, spoiled landscape. The steppe from this height is just something else.

At last, we are back to the place we have been thinking of ever since climbing down this mountain.

Before heading straight into this supposed cave, we should probably review the question of decidability and why we think there might be a possible proof somewhere around here, time-sewn between the fine wavy creases of this ever-concealing range.

We say that a certain set is **decidable** when there is an effective procedure (algorithm, Turing Machine, total function) which can "decide" whether a given element belongs to that set or not.

As we already know, decidability is at the core of every *Entscheidungsproblem* - namely, in its apparently deciding disposition through the outputs "yes" and "no" to a given query.

In this specific case, our goal is to show that the problem of provability of CL15 is decidable: we have to find an effective procedure that decides whether a certain formula belongs to CL15 (as in the set of CL15-theorems) or not - in other words, whether the formula is CL15-provable or not.
Thus:

$$\text{CL15} \vdash F \Longleftrightarrow F \in \text{CL15}.$$

We may imagine a proof-search algorithm that branches out like a tree in search of every possible CL15-proof of a certain formula $F$, which is put at the very root of this tree.

Intuitively, the idea would be to go along each branch and check if they all end, i.e. if after a finite number of steps the procedure cannot apply any more rules of inference and, thus, the algorithm ends (*converges*).

In such eventuality, if the last rule applied was Axiom, then the formula in question is provable in CL15 and belongs to its set of theorems; otherwise, the algorithm terminates and decides that formula $F$ cannot be derived in CL15 - i.e. it does not belong to its set of theorems.

We are going to gradually proceed through steps, as Japaridze himself would suggest.

First, we consider CL15 without the Contraction rule.

Indeed, all the rules of CL15 guarantee that the size/complexity of the premise is lower than that of the conclusion (meaning, in a bottom-up approach, the rules do not add new information)[24]; in other words, these rules have a finite number of premises for any conclusion on which they can be applied.

However, Contraction is the only one that acts differently.

Indeed, as the reader may have already realised, this rule "contracts" two formulas into one in a *top-down* view: this means that, when going the other way round, the rule *adds* premises to the one conclusion to which it is applied - even though it does not add completely "new" information, from an intensional point of view.

This becomes tricky to handle in a decidability proof: indeed, Contraction may be applied lots of times, potentially infinitely many, leading its branch to never terminate.

As a result, we will first (attempt to) prove that CL15 *without* Contraction,

24 This corresponds to the Subformula Property first proposed in LK, meaning Gentzen's sequent calculus for classical logic. Still, the thorough reader would be quick to ask: did we not assess that CC was a deep-inference calculus and, as such, could not guarantee this property? True. The more thorough reader, however, would be quick to answer that, just like CL5, CL15 uses *shallow* cirquents, with a conjunctive root and disjunctive nodes - we mentioned this in *Refining semantics*. Thus, as we can clearly see from the rules themselves, CL15 can account for the subcirquent property, even though Contraction does not act so strictly.

which we now call $CL15^C$, is decidable.
Only then, we will try to find an algorithm that effectively decides whether a given formula belongs to the set of CL15-theorems.

Let us now provide some useful definitions for how we are going to tackle this cave exploration. We ought to take out our mappings and finger-follow the directions we will agree upon in the following minutes.

---

*Tracing a new path*

Here is the plan.

First of all, we should define the notion of *proof-search tree*, which is the main engine of our decidability proof.

Intuitively, a proof tree is a tree whose nodes are logical formulas (in our case, cirquents); it is a nice way of visualising the steps that are taken in searching for a proof of some query - a sort of declarative view of the trace of a computation.

Given a certain proof system (in our case, $CL15^C$), a proof tree then represents all possible proofs (and absences of thereof) of the root formula, applying the system's rules at each step[25].
A closed tree means a proof for the given formula.

Here we outline a basic definition for a proof-search tree for CL15[26]. We may use the same jargon adopted for tree-like proofs of CoL in general: *child* node, *parent* node and *grandparent* node.

**Definition 3.2.1.1**
Given a cirquent $F$ in the CL15-language of Computability Logic, we define a **proof-search tree for** $F$ as a tree where each node is a finite list of cirquents

$$F_1, F_2, \ldots, F_n$$

as such:

25 Thus, we call it proof-*search* tree as we need to find a proof for the root formula amongst all possible branches of computation.
26 Which is also valid for $CL15^C$, since it only regards the *structure* of our procedure and not it contents in particular.

- The root of the tree is given by the cirquent $F$;
- A child of a node, where $\Delta_1$ and $\Delta_2$ are (possibly empty) lists of cirquents:

  $$\Phi = \Delta_1, F, \Delta_2$$

  can be formed by a list:

  $$\Delta_1, F', \Delta_2$$

  where

  $$\frac{F'}{F}$$

  is an instance of a certain rule of CL15 (with the exception of Axiom, which leaves the child node completely empty);
- A child of a node, where $\Delta_1$ and $\Delta_2$ are (possibly empty) lists of cirquents:

  $$\Phi = \Delta_1, F, \Delta_2$$

  can be formed by a list:

  $$\Delta_1, F', F'', \Delta_2$$

  where

  $$\frac{F' \qquad F''}{F}$$

  is an instance of a certain rule of CL15.

We now need to define another fundamental notion.

Indeed, we are going to limit our proof-search only to what we call *reasonable* branches: these are CL15 (respectively CL15$^{C}$) derivations of $F$ in which there are no two essentially identical cirquents[27].

As a matter of fact, due to circularity, unreasonable branches do not lead anywhere new - they become redundant in the tree. On top of that, unreasonable branches could also endlessly loop[28]. Thus, these can be discarded without any loss of information or computational content.

---

27 Beware: as the reader should recall, the two formulas $A \vee (B \wedge B)$ and $A \vee B$, for instance, are **not** essentially identical cirquents! Indeed, we are still working in CoL, not in classical logic. This means that we can still apply all the other rules, such as ($\wedge$) and (W). By two essentially identical cirquents we mean, for example, $A \vee (B \wedge B)$ and $(B \wedge B) \vee A$.

**Definition 3.2.1.2**
We say that a branch is **reasonable** iff there are no two cirquents essentially identical - meaning they bear the same identical overgroups, oformulas and undergroups.

Hence, our proof-search machine will be instructed to disregard unreasonable branches and continue working on the other ones.

Finally, we define the notion of *recurrence* (or *exponential*) *complexity*.

We call the operators ⫰ and ⫯ *branching recurrence* operators, or *exponentials*[29].

Thus, we say that a branching recurrence (sub)cirquent is of the form ⫰$F$ or ⫯$F$. In the former case we have a ⫰-cirquent, while in the latter we have a ⫯ one.

**Definition 3.2.1.3**
Given a cirquent $F$, we say that the **recurrence** (or **exponential**) **complexity** of $F$, meaning $\mathcal{C}(F)$, is the number of occurrences of branching recurrence operators ⫰, ⫯ in that cirquent.

We call CL15$^{C}$-rules the list[30] of all rules of inference of CL15$^{C}$. These are fed into the algorithm, which, during the computation, checks for their possible application in the following order[31]:

1. $\vee$ (Disjunction introduction);
2. $\wedge$ (Conjunction introduction);
3. ⫰ (Branching recurrence introduction);

28 Just imagine a branch in which the Exchange rule is applied infinitely many times! When improperly used, *abused*, E leads to an endless and repetitive branch.

29 As Japaridze defines them in [67], page 5: "[...] ⫰ and ⫯, which are CoL's counterparts of the exponentials !, ? of linear logic".

30 We chose to fetch the algorithm a list, rather than a set, to facilitate rule application - indeed, no choosing of sorts here.

31 To guarantee efficiency and least amount of time/space possible. We may expect a substantial majority of unreasonable looping branches, then, developing towards the rightmost side of our structure - since Exchange is the last rule applied each time.

4. ⫯ (Branching corecurrence introduction);
5. A (Axiom);
6. D (Duplication);
7. W (Weakening);
8. M (Merging);
9. E (Exchange).

We are now ready to build our effective proof-search procedure for the decidability of CL15$^{C}$. Let us grab our things, put these maps away and proceed for the path we have traced together.

### 3.2.2 *Stepping into the cave*

It is getting darker by the minute inside this cave. We can faintly see where to put our feet and walk steadily: the floor is a bit humid, almost slippery at times.

Be careful not to fall on the wet rock: we need to be able to map the most out of here.

On that note, let us start right away.

---

*A snap-hook start*

Given a certain cirquent $F$, we want to know whether CL15$^{C} \vdash F$ or CL15$^{C} \nvdash F$.

Here is an effective way to proceed[32]:

1. Is there any $\rightarrow$, ⟜ or ⊸¬ in $F$?[33]
   - yes $\Rightarrow$ rewrite $F$ using corresponding definition, repeat 1.
   - no $\Rightarrow$ go to 2.

---

32 In step 1. with "corresponding definition" we mean $A \rightarrow B = \neg A \vee B$; $A$ ⟜ $B$ = ⫰$A \rightarrow B$; and ⊸¬$F$ = ⫯$\neg F$.

33 By these symbols we mean the " " version of them, their *names*: the algorithm is asked to syntactically individuate them (and not their meanings!), if present.

2. Is $\mathcal{C}(F) = 0$?
   $\begin{cases} \text{yes} \Rightarrow \text{go to 3.} \\ \text{no} \;\Rightarrow \text{go to 4.} \end{cases}$

3. Is there a cirquent $F$, in the current node $\Phi$, on which a $\text{CL15}^{\text{C}}$-rule can be applied?
   $\begin{cases} \text{yes} \Rightarrow \text{apply rule and go to 6.} \\ \text{no} \;\Rightarrow \text{go to 5.} \end{cases}$

4. Is there a cirquent $F$, in the current node $\Phi$, on which a $\text{CL15}^{\text{C}}$-rule can be applied?
   $\begin{cases} \text{yes} \Rightarrow \text{apply rule and go to 2.} \\ \text{no} \;\Rightarrow \text{go to 5.} \end{cases}$

5. Was the last rule applied Axiom (A)?
   $\begin{cases} \text{yes} \Rightarrow \text{stop: CL15}^c \vdash F \text{ is computable} \\ \text{no} \;\Rightarrow \text{stop: CL15}^c \vdash F \text{ is not computable} \end{cases}$

6. Is the cirquent obtained essentially identical to a previous one?
   $\begin{cases} \text{yes} \Rightarrow \text{go back to parent node, discard branch and repeat 3.} \\ \text{no} \;\Rightarrow \text{repeat 3. on current node} \end{cases}$

We know that this process takes a finite amount of time and space to be computed. In fact, given a certain formula $F$, there are only two possible cases:

- If $F$ is an axiom of $\text{CL15}^{\text{C}}$, then $F$ is provable (finite number of steps);
- Otherwise, it still takes a finite number of steps to reach a point of no possible rule-applications: the process always terminates with a "yes" or "no" output to the question "Does $F$ belong to the set of $\text{CL15}^{\text{C}}$-theorems?".

Furthermore, as the reader may have already realised, there are no unreasonable branches taken into consideration: whenever the algorithm stumbles upon one, it goes back to the parent node, discards the branch and repeats the process by following new branches with a different rule-application order.

The reader may try and carry out a few tests to be somewhat convinced that this should work.

---

*Skid avoiding*

Now, the big question is: does this still hold with Contraction in the picture?

Clearly there is no problem for all the formulas that do not contain an occurrence of ⫯ (or ¬⫰), since Contraction can only be applied in such case.

We need to prove that every reasonable branch always terminates, i.e. the algorithm always converges in the branches that do not contain two essentially identical cirquents.

It may be argued that once Contraction is consequently iterated many times at each node of the branch, sooner or later will probably yield the same results, thus making the whole branch unreasonable and, therefore, discardable.

However, proving this seems to be quite out of our reach for the time being. Alternatively, we may proceed this way.

Can we prove that the application of Contraction may be bounded[34] without lessening the deductive power of CL15?

Here are some preliminary considerations.

Upon some scrutiny, the rule of Contraction seems to be particularly useful when, in our cirquent, we have an uneven number of positive and negative occurrences of a certain oformula[35].

34 As a matter of fact, the Contraction rule here is *already* bounded to ⫯ - a bit like Girard has done in [24].

35 Just as in the second example provided in *Testing the lie of the land* - if the reader has some difficulty in recalling it, we really do recommend going back for another look.

Indeed, in order to prove a certain formula/cirquent, we need to be able to conclude our branch by applying rule A.
This one, as the reader recalls, needs one positive and one negative occurrence of the same oformula - thus, a pair of disjuncted formulas which both share one undergroup and one overgroup.

Contraction, then, seems to be effective when there is an odd number of the same oformulas to begin with (for instance, two occurrences of $\female F$ and one for $\female \neg F$, as in the above-mentioned example).

When we apply C, this one brings a new $\female F$ (or $\female \neg F$, depending on what is needed) to the following step, allowing us to proceed in the A-direction[36].

Consequently, applying the C rule to a cirquent in which we already have an even number of both positive and negative occurrences of a certain oformula *leads nowhere useful* - we obtain an odd number of occurrences, which complicates things[37].

As a result, we feel that applying multiple times the same C rule leads nowhere useful - or even worse, nowhere new.

Indeed, if the cirquent needs another positive/negative occurrence of a $\female$-oformula, we only have to provide it through one application of the rule.

If it already has the same number of positive and negative occurrences, there is no need to apply it.

If we apply it two or more times, we get to the same starting point by just applying Weakening as many times as the undergroups of the oformula we want to delete.

---

36 This does not mean that Contraction generates two identical cirquents! The reader must always pay attention to what is contained in each overgroup and undergroup of our cirquent. We cannot, then, discard any Contraction-branch, since the rule does not necessarily entail two essentially identical cirquents - at least not after having just applied it once or twice.

37 Indeed, we may say that Contraction acts like a *phármakon*: a remedy, a poison, a scapegoat, so powerful both in the good and in the bad. We need to handle it finely and carefully.

For such reasons, we are going to limit our Contraction rule application.

At first, one may think that not applying C in two (or more) consecutive nodes could do the trick.

However, we can show that there is always a valid sequence of applications that generates an endless (still apparently reasonable, nonetheless) branch: just think of alternating C and ⫯[38].

Following this long back-and-forth to which the reader may have something to object, we feel that the best shot we have at maintaining the C rule without sacrificing decidability is through a restriction, a bound on C itself.

Namely, the application of Contraction should be limited to **one use per branch**.

As this measure may seem a bit drastic at first, it sure avoids endless branches causing trees to never close.
We are going to show that for any branch in which we need to use C, we get the desired results by just applying it once: indeed, one Contraction is always enough.

Any iterated application of C would be an unuseful, unbeneficial move: it would blow us further away from the finish (inference) line, both from a complexity and derivation-length point of view.

In the case of recreating A's conditions, two applications of C would reproduce the even or odd number of initial occurrences, which would call for another application - not useful at all; or, alternatively, it would lead back to the same cirquent[39] using other rules, since many applications of C can be replaced with other strategies in which Contraction is only used once.

Another purpose for the application of C would seem to be simplifying the amount of overgroups created by branching recurrence operators[40],

---

38 And for three nodes, for example, we may think of C - ⫯ - W.

39 Obtained in the meantime by the application of a C-less strategy.

40 As seen in the first example of *Testing the lie of the land*.

with the further application of Duplication, if possible[41].

However, in such a situation, only one use of Contraction is sufficient, since the ♀-rule already includes an unlimited (potentially zero) amount of new arcs that connect the oformula to already-present overgroups[42].

This means that C can be used only once to obtain the same results of terminating branches that include more than one application of Contraction!

For such purpose, we only need one (or none) application of C: thus, our limitation does not seem to dreadfully reduce the expressive power of CL15.

We may compress all of this in one single line. We use $CL15^{\overline{C}}$ to mean CL15 with bounded Contraction.

**Theorem 3.2.2.1**
A cirquent $F$ is derivable in CL15 iff $F$ is derivable in $CL15^{\overline{C}}$.

Indeed, we need to prove that $CL15 \vdash F \Longleftrightarrow CL15^{\overline{C}} \vdash F$.

The "$\Leftarrow$" direction is trivial: since the two systems have the same rules, surely CL15 contains also proofs in which Contraction is used only once.

The tricky, slippery part, then, is trying to prove the "$\Rightarrow$" direction, going from CL15 to $CL15^{\overline{C}}$.

Let us try and show this by induction on the exponential complexity of $F$.

41 These are all detailed cases considerations to better understand how Contraction actually works and in which cases we may need to use it. Note that if we cannot use D and we are far from potentially closing our proof, C is unuseful, as we have already pointed out.

42 Plus, we believe that the "mixed" case in which we may use C both to simplify overgroups and create A's conditions may be tackled in the same one application of C, if we can use D.

*Cautiously crossing*

We need to prove that:

**Lemma 3.2.2.2**
Any cirquent derivable in CL15 is derivable in CL15$^{\overline{C}}$.

**Tentative proof**

Let us proceed by induction on $\mathcal{C}(F)$.

We agree that with "positive occurrence of ⫯" (respectively ⫰) in a cirquent, we mean ⫯$A$ (respectively ⫰$A$), where $A$ can be either a positive or negative oformula.
The same goes for "negative occurrence", which means $\neg$⫯$A$ (respectively $\neg$⫰$A$), with $A$ either a positive or negative oformula.

Given a cirquent $F$ which is derivable in CL15, we have that:

1. If $\mathcal{C}(F) = 0$, then $F$ is derivable in CL15$^{\overline{C}}$, since no branching-recurrence rules are needed;
2. If $\mathcal{C}(F) > 0$, then $F$ contains at least one occurrence of an exponential operator. We analyse all possible cases of branching-operators occurrences:
   a) If $\mathcal{C}(F) = 1$, there is only one positive (negative) occurrence of either ⫰ or ⫯. The possible situations may be the following:
      i. $F$ contains only one positive occurrence of ⫰ (meaning a negative one of ⫯): then $F$ is derivable in CL15$^{\overline{C}}$, since the rule of Contraction is not applicable;
      ii. $F$ contains only one negative occurrence of ⫰ (meaning a positive one of ⫯): a proof for $F$ can be found without applying Contraction. Indeed, if we iterate it without a bound it would never bring $\mathcal{C}(F)$ to 0 - which is the necessary condition for us to close the tree. On the other hand, if we use it only once and then apply other rules, it would eventually loop and recreate the same cirquent obtained from the initial one by only applying ⫯$-$introduction. At

the end, we have that $\mathcal{C}(F) = 0$; as a result, since we assumed that $F$ is derivable in CL15, and $\mathcal{C}(F) = 0$, for step 1. and for the subformula property (SP), we know that $F$ is derivable in CL15$^{\overline{C}}$. The same goes for when we apply C more than once;

b) If $\mathcal{C}(F) > 1$, there may be any number of positive (negative) occurrences of either ⫰ or ⫯. We may be in the following situations:

   i. $F$ contains only positive occurrences of ⫰ (meaning negative ones of ⫯): then $F$ is derivable in CL15$^{\overline{C}}$, since the rule of Contraction is not applicable;

   ii. $F$ contains only negative occurrences of ⫰ (meaning positive ones of ⫯[43]): this means that, in addition to the other rules, we can also apply C and ⫯−introduction. We consider the following cases:

      A. If our initial $\mathcal{C}(F)$ is an even number (specifically, $\mathcal{C}(F)/2 = n + m$, for some $n, m \in \mathbb{N}$ and $m = 0$), there may be three different situations:

         - Our positive ⫯s are attached to $n$ positive and $n$ negative oformulas (e.g. we have one ⫯$\neg F$ and one ⫯$F$ connected to the same undergroup[44]): this is an optimal situation, since, even though there could be other oformulas in the cirquent, the premises for Axiom seem to emerge from under the derivation[45]. As a result, applying Contraction leads nowhere new or useful (indeed, there is no other ⫯ or ⫰[46]); if we apply it even only once, we will see that, just as before, if we then apply other rules, it would eventually loop and recreate the same cirquent obtained from the initial one by only applying ⫯−introductions, which bring $\mathcal{C}(F)$ to zero. Since we assumed that $F$ is deriv-

43 Since, for De Morgan's laws, we know that $\neg$⫰$A$ = ⫯$\neg A$.

44 This also goes for the case of different undergroups: indeed, Contraction would not bring a more Axiom-like structure, since it would create two identical copies of the same positive or negative oformula, while only adding further complexity.

45 Since, in order to close the branch, we need to apply Axiom to a pair of positive and negative instances of the same oformula.

46 In such case, which we are now going to analyse, Contraction can be crucial to show that $F$ is indeed derivable in CL15$^{\overline{C}}$.

able in CL15, and $\mathcal{C}(F) = 0$, for step 1. and SP, we know that $F$ is derivable in $CL15^{\overline{C}}$. The same goes for when we apply C more than once;

- Our positive ⚲s are attached to a $\neq n$ number of positive and a $\neq n$ number of negative oformulas[47] (e.g. we have two $⚲\neg F$ and one $⚲F$ connected to the same undergroup[48]: here, Contraction becomes a powerful tool which adds the one oformula needed to recreate the Axiom conditions (for instance, in the above-mentioned situation, Contraction allows us to add another $⚲F$ to our cirquent and obtain two positive and two negative occurrences of the same oformula). Then, by applying many ⚲−introductions, we reach $\mathcal{C}(F) = 0$ and an optimal situation for closing the branch through Axiom[49]. As usual, since $F$ is derivable in CL15, for step 1. and SP, we know that $F$ is derivable in $CL15^{\overline{C}}$;
- If there are no two same oformulas to which our ⚲ are attached (or there is one oformula different from a pair of positive and negative ones), then Contraction only adds complexity without creating anything useful (no Axiom-like conditions). In such case, we can only proceed with the ⚲−introductions, bringing the overall complexity down to 0 and, if the formula $F$ is derivable in CL15, then it is derivable in $CL15^{\overline{C}}$;

B. Conversely, if our initial $\mathcal{C}(F)$ is an odd number (specifically, $\mathcal{C}(F)/2 = n + m$, for some $n, m \in \mathbb{N}$ and $m > 0$), we have that:

- Our ⚲ are attached to a different number of positive and negative oformulas (e.g. we have two $⚲\neg F$ and one $⚲F$). In this case we just have to apply Contraction for the usual reasons and make $\mathcal{C}(F)$ even. Then we will find ourselves in the first case of an even complexity, with the same number $n$ of positive and

47 Clearly in the case of $\mathcal{C}(F) = 2$ this cannot happen.
48 Just as before, this also goes for the case of different undergroups.
49 This also works when we have other different oformulas in our cirquent.

negative oformulas - thus, bringing $\mathcal{C}(F)$ to 0 and proving that $F$ is derivable in CL15$^{\overline{C}}$;

- If there are no two same oformulas to which our ⫯ are attached (or there is one oformula different from a pair of positive and negative ones), then Contraction only adds complexity without creating anything useful (no Axiom-like conditions). In such case, we can only proceed with the ⫯−introductions, bringing the overall complexity down to 0 and, if the formula $F$ is derivable in CL15, then it is derivable in CL15$^{\overline{C}}$;

iii. $F$ contains both negative and positive occurrences of ⫰ (meaning positive and negative occurrences of ⫯, respectively): this means that, other than the usual rules, we may either apply ⫯−introduction (in which case, the new $\mathcal{C}(F)$ is the previous one $-1$) or C (in which case, the new $\mathcal{C}(F)$ is the previous one $+1$).
The Contraction rule is only necessary when it can produce Axiom-like conditions: this is done by not taking into account the positive occurrences of ⫰ (since, as already mentioned, we cannot apply the C rule on them and, thus, are not relevant here) and reducing this case to the ii. one, by only taking into account the ⫯ occurrences.
Indeed, in order to bring $\mathcal{C}(F)$ back to 0, regardless of it being an even or odd number, we can first use the ⫰−introduction(s) and then the ⫯−one(s) which is able to connect the given ⫯−oformula to any overgroup the ⫰−oformulas have produced through the corresponding rule[50]. By then applying Duplication, if possible[51], we have a good canvas onto which paint our Axiom rule - and, thus, close our branch. Since $F$ is derivable in CL15, for step 1. and SP, we then know that $F$ is derivable in CL15$^{\overline{C}}$.

All other cases can be reduced to these ones.

■

50 Indeed, the ⫯−introduction rule is really powerful: we only need one positive occurrence to eventually eliminate, through Duplication, if possible, all the new overgroups that any number of positive occurrences of ⫰s have created. This is why Contraction is not needed.

51 Otherwise Contraction is still not a good option, since the best one would be to not connect the ⫯−oformula to any overgroup already present.

We have thus presented a possible proof of Lemma 3.2.2.2, according to which CL15 is equal to CL15$^{\overline{C}}$.

In other words, we now know that given a model $\mathcal{M}$ of CL15 and a model $\mathcal{M}'$ for CL15$^{\overline{C}}$, $\mathcal{M} \equiv \mathcal{M}'$ - indeed, the sets of the theorems of each system have the same extension and, therefore, are isomorphic[52].

This means that no deductive power is lost[53], since the situations in which we have to use C only require one application of such rule.

Indeed, applying the Contraction rule more than once can lead to endless branches or even possibly looping ones.

We may now provide a proof for the decidability of CL15$^{\overline{C}}$ (hence, of CL15).

---

*On with the flashlight!*

Given a certain cirquent $F$, we want to know whether it is provable in CL15. Here is an effective way to proceed.

We initialise a counter $K$ of applications of C to 0. $K'$ means to add 1 to the ongoing $K$ (successor); while $K^-$ means to subtract 1 from the ongoing $K$ (predecessor).

1. Is there any $\rightarrow$, $\circ\!\!-\!\!-$ or $\circ\!\!\neg$ in $F$?[54]
$$\begin{cases} \text{yes} \Rightarrow \text{rewrite } F \text{ using corresponding definition, repeat 1.} \\ \text{no } \Rightarrow \text{go to 2.} \end{cases}$$
2. Is $\mathcal{C}(F) = 0$?
$$\begin{cases} \text{yes} \Rightarrow \text{go to 3.} \\ \text{no } \Rightarrow \text{go to 4.} \end{cases}$$

52 In Ehrenfeucht–Fraïssé terms, re-formulated by Hodges, Eloise is the last one to make a move in the back and forth game (see [16], [20] and [35] for more on the subject - a very interesting one in game semantics!).

53 As in: the two systems, meaning their models, verify the same set of formulas.

54 Again, we are referring to the names of these symbols and not their meanings.

3. Is there a cirquent $F$, in the current node $\Phi$, on which a CL15-rule can be applied?
$$\begin{cases} \text{yes} \Rightarrow \text{apply rule and go to 9.} \\ \text{no} \ \Rightarrow \text{go to 5.} \end{cases}$$
4. Is there a cirquent $F$, in the current node $\Phi$, on which a CL15-rule can be applied?
$$\begin{cases} \text{yes} \Rightarrow \text{apply rule and go to 6.} \\ \text{no} \ \Rightarrow \text{go to 5.} \end{cases}$$
5. Was the last rule applied Axiom (A)?
$$\begin{cases} \text{yes} \Rightarrow \text{stop: CL15} \vdash F \text{ is computable} \\ \text{no} \ \Rightarrow \text{stop: CL15} \vdash F \text{ is not computable} \end{cases}$$
6. Is the cirquent obtained essentially identical to a previous one?
$$\begin{cases} \text{yes} \Rightarrow \text{go back to parent node, discard branch and repeat 4.} \\ \text{no} \ \Rightarrow \text{go to 7.} \end{cases}$$
7. Was the last rule applied Contraction (C)?
$$\begin{cases} \text{yes} \Rightarrow K', \text{go to 8.} \\ \text{no} \ \Rightarrow \text{go to 2.} \end{cases}$$
8. Is $K > 1$?
$$\begin{cases} \text{yes} \Rightarrow \text{go back to parent node, discard branch, } K^-, \text{ repeat 4.} \\ \text{no} \ \Rightarrow \text{go to 4.} \end{cases}$$
9. Is the cirquent obtained essentially identical to a previous one?
$$\begin{cases} \text{yes} \Rightarrow \text{go back to parent node, cut off branch, delete applied rule} \\ \qquad\quad \text{from the set of CL15-rules and repeat 3.} \\ \text{no} \ \Rightarrow \text{go to 3.} \end{cases}$$

Indeed, from 7. we go directly back to 4. either way, since, as we know, Contraction only applies to cirquents of $\mathcal{C}(F) > 0$ and does nothing to bring that number down - indeed, it makes it go 1 up!

Just for the same reasons we have exposed before, this process is a finite-time computation.

---

*Roughly sketched*

Trying to draw a quick summary of the whole chapter, we know that CL15 is sound and complete.

Furthermore, we have put forward a tentative proof for the decidability of $CL15^{C}$ (meaning CL15 without the C rule) and, correspondingly, of the set of CL15-theorems that do not contain any occurrence of ⫯ and ¬⫰.

As for CL15 on the whole, we attempted to prove that it is decidable, since the equivalent CL15 with the bounded C rule is decidable itself.

We hope that the reader will have some questions, or at least some criticism to offer. Just as the Word should hurt in some way in order to be fertile, we need to *stumble* against the rock of this cave to be able to precisely map it - some scratches always help retaining memory[55].

In the meantime, having explored this secluded hollow for its most part, we might as well leave and climb back down.

Did we obtain what we came here for? Perhaps. Indeed, we managed to get a rough sketch of our findings inside the cave; we hope that what has been mapped can be of help to those who will come and attempt to go even further[56].

Still, we may be pleased with ourselves for not having slipped on some slithery, humid flatstone - or at least, it would seem.

Let us now head back, again, for one last time.

55 Stumbling upon some unexpected rocks on our path makes the journey all the more memorable. We consider the act of understanding as a fertile grasping of concepts, which, ideally, should have a rough surface to grab on - almost scratchy, sometimes hurting. We mean *grasp* as in *grab* and *hold* with the same hand we used for naming things. Indeed, we probably understand most when we live or touch something - since interaction, again, makes comprehension work wonders. Thus, in order to ease one's apprehension, it is fruitful to introduce notions in their bumpy, rocky versions (after all, we are in the mountains), as we have tried to do in this section with the long back-and-forth reasoning. Roughness guarantees memory - and correct memory guarantees the most rock solid of comprehensions.

56 Even though we will probably be back soon ourselves to check it out even better!

# 4

# SILENTLY GROWING: APPLICATIONS

Far down our journey, we have come to know, touch and call[1] this steppe pretty well. Our mappings have been traced, the confines have been tried, the highs and lows all inspected with great care.

Of course, there is still more to come; however, we may rest for a while[2] and remind ourselves why this steppe is really this *fertile*.

It is a question of composition, to be exact; a chemistry matter, with different elements coming into play. Computability Logic is the soil together with its yellow-green carpet: the *fruit of the earth* without the *work of human hands* part.

As such, we may recall *why*[3] we came all the way down here to inspect and chart away. Indeed, this land was tipped-off as a most fertile one, not for the primal purpose of *using* it, but to witness how it can sponta-

1 As in *calling*, not naming - it is a deeper sense of acquaintance, one that has already been known prior to any first encounter.

2 A methodological suggestion. We are out in this vast naked land, a green-contoured void that fills one's being. It is only appropriate for us to rest in silence, as it is one of the most fruitful things we can welcome - obviously, we do not mean any antonym of interaction, here exalted from a logical point of view. This silence is different, a physical one. It is of a thick and weightless kind, even though spacially extended; it means abandonment, an act of self-overcoming in which what is felt as necessary suddenly becomes superfluous; what is considered frightful now becomes powerless - *almost* in the same tones as "The Lord will fight for you, and you have only to be silent" (Ex. 14, 14), which, by all means, does not reference the typical quietness that first comes to mind. Let us then fill ourselves up with this silence, which is not an absence of expression, nor an upset petty move that speaks volumes, but rather a constructive (without structure) filling process - a pouring one, if you will. This unusual sort of quietness does not necessarily entail any stillness of mind, let us be reasonable; however, it always helps to dive deeper into one's thoughts, which can bring about ever-new realisations. We may already list this as one of the reasons why this remote geographical area is, indeed, fertile. *And when the sun beats down, and I'm lying on the bench, I can always hear...*

3 Almost a full-circle moment, since we started out with the "puerile question of *why*" of CoL. However, now we ask *why* of *us* in CoL - and, thus, *what* can we do from here: the three topical moments of any geographical exploration.

neously help us in different ways.

So far, we have identified two main areas of potential (and actual) applications of CoL - surely, there are many more to be found.

First of all, we may observe the smiling case of PA.
Indeed, Computability Logic's soil composition allows Peano Arithmetic to sprout up through the grassland. To this day, we have detected 11 different varieties of these plants, which we will get into in the next section (when taking out our phytology notes).

Secondly, CoL's materials offer us new tools to harvest the ever-growing field of Artificial Intelligence in a clean, new fashion, obtaining more crops than ever before.

Let us now get into the topic of potential applications while resting in this empty-space, wind-excepting silence[4].

## 4.1 clarithmetic, a botanic study

As explained in [81], the Georgian steppe is mainly dotted by *Bothriochloeta ischaemum*, meaning the commonly named Yellow bluestem. We have seen lots of these everywhere, since it can be found all around the world.

However, the curious thing about the Georgian steppe Yellow bluestem is that it grows specifically on foothills and lower mountain belts, approximately 450-900 metres above sea level - just where we have been all this time.

These plant communities develop on slopes with various exposure and inclination values, mainly on (grey) cinnamonic soils - meaning very

4 (Meta-footnote disclaimer) Clearly, the reader may not be in the vicinity of a vast, empty natural space. In such case, a similar sort of peculiar silence can be nurtured through very specific sound frequencies. For instance, we strongly recommend *In the Steppes of Central Asia* by Alexander Borodin; otherwise, perhaps even better, one can put Mahler's *Symphony No.5 in C Sharp Minor IV Adagietto - Sehr langsam* on, while muting everything else off. In any case, we leave the choice to the reader while glancing away at the yellow-grassed prairie once again.

deep, well drained soils formed in pyroclastic flows of volcanic ash and pumice that we can find beneath terraces, footslopes, and backslopes of mountains[5].

Indeed, we stumbled upon some spontaneous specimens of the *Bothriochloeta ischaemum* just at the feet of our well-inspected mountain ridge. However, upon a closer look, these specimens did not resemble any of the plant communities we have come to classify so far[6].

Thus, on our way down, we took some notes on these curious looking *Bothriochloeta* varieties.

Here is our brief report[7].

---

*The clarithmetic community*

As Japaridze explains in [56], *clarithmetic* is a generic name for formal number theories, similar to Peano Arithmetic, based on Computability Logic rather than classical or intuitionistic logics.

In these theories, just as in CoL, formulas represent interactive computational problems: their "truth" is understood as the existence of an algorithmic solution.

Imposing various complexity constraints on such solutions yields various versions of clarithmetic[8].

5 In Georgia the total area of grey cinnamonic soils is 5,8% (402000 ha). These soils are distributed in the south-east of Marneuli, Gardabani, Sagarejo and other districts. In particular, grey cinnamonic soils are characterised by the presence of carbonates near the surface. See [80] for further information.

6 Such as *Bothriochloetum gramino-mixtoherbosum, Botriochloetum festucoso gramino-mixtoherbosum, Bothriochloetum festucosum* and so on.

7 Just as we did for our mountain climbing notes, we are going to provide a general overview for the reader to reason upon and, maybe, further deepen by looking into the specific references we will point out along the way.

8 Many complexity classes of computation, meaning sets of problems that take a similar range of space and time to solve, can be characterised by different versions of arithmetic. The systems of bounded arithmetic, as introduced in [29] and [79], are, in a sense, the closest thing to clarithmetic. Indeed, just like our theories, these systems control computational complexity through explicit bounds that are attached to quantifiers (usually in induction or similar postulates). The best known alternative line of research, primarily developed by recursion theorists, controls computational complexity via type

Of all CoL's fragments we have managed to inspect, these theories seem to grow mainly on the mountain slope of CL12.

Indeed, as explained in the opening of [53], CL12 presents a *reasonable, computationally meaningful, constructive alternative to classical logic as a basis for applied theories* - unsurprisingly, it is the most studied fragment of all, as we have already mentioned.

On such particularly fertile grounds we identify the first variety: **CLA1**.

Unlike its classical-logic-based counterpart PA, CLA1 is not merely about what arithmetical facts are true, but, rather, about what arithmetical problems can be actually computed or effectively solved.

Specifically, every CLA1-formula expresses a number-theoretic computational problem, rather than just a true/false fact. Moreover, every theorem expresses a problem that has an algorithmic solution, while every proof encodes such solution.

Together with some nonlogical axioms, we may also be equipped with some nonlogical rules of inference: these preserve the property of computability, as, for instance, the *constructive induction rule* of CLA1 does - which we will show in just a few lines.

In addition, the soundness of the underlying axiomatisation of CoL - which is in its strong form of uniform constructive soundness, as the reader recalls - guarantees that every theorem $T$ of the theory also has an algorithmic solution which can be effectively constructed from a proof of $T$.

As we will repeatedly realise every now and then, all of this makes CoL a problem-solving tool even for PA.

---

information instead - see [3], [92] and [82] in particular. On the logical side, we should also mention *bounded linear logic* [24] and *light linear logic* [25] by Girard.

Let us have a brief look at this new plant community[9], in order to introduce the reader to the fertility of CoL[10].

---

*CLA1*

As introduced in [53], CLA1 is a first model example of sound clarithmetic theories.

Its language is comprised of the language of CL12[11] minus all nonlogical predicate symbols (leaving only "="), all constants (leaving only "0") and all function symbols (leaving three: successor " $'$ ", sum "$+$" and product "$\times$").

Branching implication $\circ\!-$ is hereby replaced with its stronger version **dfb-reduction**, as in *double-finite branching* $\bullet\!\!\nrightarrow$: just like $\circ\!-$, it allows to use the resources (i.e. the antecedent) any *finite* number of times, which makes it stronger and easier to handle than original brimplication[12].

We may also define two similar, weaker versions: **sfb-reduction** (*single-finite branching*) $\bullet\!\!\rightharpoondown$ and **fb-reduction** (*finite-branching*) $\bullet\!-$ [13].

The first one states that $\top$ is automatically proclaimed the winner when it prevails in the succedent, even after infinitely many replications of the antecedent.

The second one is further weaker, in that Machine's finitely or infinitely replications are completely ignored when determining the winner. Indeed, only the copies of the resources (of the antecedent) that emerged as a

9 Which, as the reader may already imagine, it is not a finished one: we are confident in the many varieties that seem to potentially grow all around this particular lush portion.

10 For this section, we will generally make reference to [67], in addition to the individual papers that will be mentioned for each theory.

11 Which, to jog the reader's memory, presents the following signature: $\{\wedge, \vee, \sqcap, \sqcup, \triangle, \triangledown, \bigsqcap, \bigsqcup, \forall, \exists,$ external $\circ\!-\}$. As a result, PA's language is augmented by adding all of these operators.

12 Japaridze mentions this new operator only once, in [53], with respect to CLA1. It is not clear whether this also goes for the other theories of clarithmetic; however, since all the subsequent theories are in some way "based" upon this one, it may reasonable to conclude that this type of reduction is still included in the language of clarithmetics.

13 However, we will be concerned with dfb-reduction only - also because sfb and fb reductions have not been developed much further.

result of finitely many replications are taken into consideration.

Expressions of the form $\vec{A} \bullet\!\!\not\!- B$ we call sequents. If $\top$ has an algorithmic winning strategy for the game represented by such an expression, we say that $B$ is *dfb-reducible* to $\vec{A}$.

Let us have a look at a plain example of how CLA1 is able to solve a problem by dfb-reducing it to another one.

The following sequent represents the problem of dfb-reducing the one of computing a cube value to the one of computing both multiplication and the definition of cube:

$$\forall w(w^3 = (w \times w) \times w), \sqcap z \sqcap u \sqcup v(v = z \times u) \bullet\!\!\not\!- \sqcap x \sqcup y(y = x^3).$$

Below is a winning strategy, which succeeds for any possible meaning of $x^3$ and $z \times u$: this means that we have an algorithmic solution which, in fact, is also a logically valid sequent.

First Machine waits until Environment picks a value $n$ for $x$.
This brings the game down to:

$$\forall w(w^3 = (w \times w) \times w), \sqcap z \sqcap u \sqcup v(v = z \times u) \bullet\!\!\not\!- \sqcup y(y = n^3).$$

Machine now replicates the second resource of the antecedent, resulting in:

$$\forall w(w^3 = (w \times w) \times w), \sqcap z \sqcap u \sqcup v(v = z \times u), \sqcap z \sqcap u \sqcup v(v = z \times u) \bullet\!\!\not\!- \sqcup y(y = x^3).$$

After this, Machine specifies both $z$ and $u$ as $n$ in the second resource of the antecedent, further bringing the game down to:

$$\forall w(w^3 = (w \times w) \times w), \sqcup v(v = n \times n), \sqcap z \sqcap u \sqcup v(v = z \times u) \bullet\!\!\not\!- \sqcup y(y = x^3).$$

Environment will have to respond by choosing a (correct) value $m$ for $v$ in the same component.
Thus the game evolves to:

$$\forall w(w^3 = (w \times w) \times w), (m = n \times n), \sqcap z \sqcap u \sqcup v(v = z \times u) \bullet\!\!\not\!- \sqcup y(y = x^3).$$

Machine then specifies $z$ and $u$ in the third resource of the antecedent, which means:

$$\forall w(w^3 = (w \times w) \times w), (m = n \times n), \bigsqcup v(v = m \times n) \mathrel{\bullet\!\!\not\!\!-} \bigsqcup y(y = x^3).$$

Again, Environment will have to selecting a value $k$ for $v$:

$$\forall w(w^3 = (w \times w) \times w), (m = n \times n), (k = m \times n) \mathrel{\bullet\!\!\not\!\!-} \bigsqcup y(y = x^3).$$

Finally, Machine specifies $y$ as $k$ and, having brought the play down to the following true elementary game, celebrates victory:

$$\forall w(w^3 = (w \times w) \times w), (m = n \times n), (k = m \times n) \mathrel{\bullet\!\!\not\!\!-} k = x^3.$$

Indeed, as already mentioned, our strategy does not depend on the meanings of $x^3$ and $z \times u$, since the succedent of the above sequent is a logical consequence (in the classical sense[14]) of the formulas of the antecedent.

The axioms of CLA1 are the ones of PA plus one scheme of sentences[15]:

1. $\forall x(0 \neq x')$
2. $\forall x \forall y(x' = y' \to x = y)$
3. $\forall x(x + 0 = x)$
4. $\forall x \forall y(x + y' = (x + y)')$
5. $\forall x(x \times 0 = 0)$
6. $\forall x \forall y(x \times y' = (x \times y) + x)$
7. $\bigsqcap x \bigsqcup y(y = x')$
8. $\forall(F(0) \wedge \forall x(F(x) \to F(x')) \to \forall x F(x))$, where $F(x)$ is an elementary formula.

CLA1 is equipped with a natural deduction system[16]: the above axioms serve as rules of inference that take no premises.

14 Meaning a classical *semantical* sense. Indeed, given a formula $A$ of a certain language $\mathcal{L}$ and a set $\Sigma$ of formulas, we say that $\Sigma \vDash A$, i.e. $A$ is a *logical consequence* of $\Sigma$, iff every model $\mathcal{M}$ of $\Sigma$ is also a model of $A$ (in other words, every formula of $\Sigma$ plus the one of $A$ are true in $\mathcal{M}$).

15 From now on, when defining rules of inference, Japaridze uses $\bigsqcap F$ as in the $\bigsqcap$−closure of the formula $F$, i.e. $\bigsqcap x_1, \ldots, \bigsqcap x_n F$, where $x_1, \ldots, x_n$ are the free variables of $F$ listed in their lexicographic order. Similarly we may also define $\forall F$, $\exists F$ and $\bigsqcup F$.

16 Visually represented through a "slightly improved" Fitch-style - see page 589 of [53] for further examples other than the one below.

In addition to these, CLA1 has two other rules: **Logical Consequence** (LC) and **Constructive Induction** (CI).

The former derives from the notion of logical consequence in CL12.

We say that a CL12-formula $F$ is a *logical consequence* of CL12-formulas $E_1, \ldots, E_n$, where $(n \geq 0)$, iff CL12 proves the sequent $E_1, \ldots, E_n \circ\!\!-\, F$[17].

Thus:

**Logical Consequence** (LC)

$$E_1, \ldots, E_n \Rightarrow \sqcap F,$$

where $\sqcap F$ is a CL12-logical consequence of $E_1, \ldots, E_n$.[18]

As for CI, we show it in the natural deduction garments Japaridze clothed it in:

**Constructive Induction** (CI)

$$
\begin{array}{c|l}
 & F(0) \\
 & \vdots \\
 & \begin{array}{|l}
 F(x) \\ \hline
 \vdots \\
 F(x')
 \end{array} \\
 & \vdots \\
\bullet & \sqcap x F(x)
\end{array}
$$

where $x$ is a fresh variable and $\bullet$ indicates the conclusion of the rule.

We are not going to formally define CLA1 any further, as we are giving just a sample test for the reader to bite into clarithmetics for a moment.

17 Remember that CL12 uses $\circ\!\!-$ in its sequents, as declared in its signature.
18 Indeed, $\Rightarrow$ maintains the same usual meaning of "from premise(s) $\mathcal{P}$ conclude $F$".

In case the reader is interested in a more in-depth presentation of CLA1, we suggest having a look at [53].

---

*CLA2*

As briefly introduced in the same paper, CLA2 extends CLA1 to the full language of CL12, keeping constants together with an infinite supply of fresh predicate and function symbols for each arity ($0, =, ', +, \times$, however, still maintain a special status amongst these).

The axioms and rules of CLA2 are virtually the same of CLA1, with the only exception that Axiom 8. and the two inference rules are no longer limited to CLA1-formulas.

This allows greater flexibility and convenience. For instance, CLA2 is able to compute primitive recursive functions.

---

*CLA3*

Still in the same paper, CLA3 is briefly introduced as a restriction of CLA1, obtained by forbidding $\forall$ and $\exists$[19].

The axioms of CLA3 are always the same but with choice quantifiers in front. There is no induction rule. As seen below:

1. $\sqcap x(0 \neq x')$
2. $\sqcap x \sqcap y(x' = y' \rightarrow x = y)$
3. $\sqcap x(x + 0 = x)$
4. $\sqcap x \sqcap y(x + y' = (x + y)')$
5. $\sqcap x(x \times 0 = 0)$
6. $\sqcap x \sqcap y(x \times y' = (x \times y) + x)$
7. $\sqcap x \sqcup y(y = x')$

19 As a result, CLA3 becomes a "perfectly constructive" system of arithmetic.

In addition, LC and CI are still valid rules, only now restricted to $\forall, \exists-$free formulas.

As Japaridze explains, there is an interesting way to express the notion of "truth" (= computability) in PA through the language of CLA3 (unlike CLA1).

Indeed, one could show that the arithmetical complexity of the predicate of "truth" for CLA3-formulas is $\Sigma_3$[20].
Replacing here classical quantifiers with their choice counterparts yields a formula which, in a sense, expresses this predicate of CLA3 in the same language.

This unusual phenomenon, as Japaridze observes in [53], may yield unusual consequences in the metatheory of CLA3 and, generally, in further metainvestigations.

---

*CLA4*

First presented in [56], CLA4 is a sound and complete theory of clarithmetic for polynomial time computability - meaning it proves only formulas that are PTIME[21] computable, unlike CLA1.

CLA4 is, indeed, a first model for complexity-oriented versions of clarithmetic. Indeed, it is also complete in a certain reasonable sense Japaridze calls *extensional completeness*[22].

The language of CLA4 is the same as that of CLA1.

20 Where, recalling what the reader may already know, a $\Sigma_n^0$ formula is equivalent to a formula that begins with some existential quantifiers and alternates $n-1$ times between series of existential and universal quantifiers.
21 Roughly, PTIME is the complexity class that contains all decision problems solvable by a deterministic Turing Machine (meaning one that always has a unique next move for any given state and input symbol) using a polynomial amount of computation *time*, or polynomial time.
22 Which means that every number-theoretic PTIME computational problem is represented by some theorem of CLA4. Taking into account that there are many ways to represent the same problem, extensional completeness is weaker than what can be called *intensional completeness*, according to which any formula representing an (efficiently) computable problem is provable. For instance, Gödel takes into consideration intensional rather than extensional incompleteness.

For any variable $x$ Japaridze defines its **polynomial sizebound**[23], a sort of limit to the size of computation of $x$. This is then used to define **polynomially bounded** CLA4-formulas.

The axioms of CLA4 are 8, plus a scheme of sentences that is able to generate possibly infinite axioms.

1. $\forall x(0 \neq x)$
2. $\forall x \forall y(x' = y' \rightarrow x = y)$
3. $\forall x(x + 0 = x)$
4. $\forall x \forall y(x + y' = (x + y)')$
5. $\forall x(x \times 0 = 0)$
6. $\forall x \forall y(x \times y' = (x \times y) + x)$
7. $\sqcap x \sqcup y(y = x')$
8. $\sqcap x \sqcup y(y = x\mathbf{0})$[24]
9. $\forall(F(0) \wedge \forall x(F(x) \rightarrow F(x')) \rightarrow \forall x F(x))$ for each elementary formula $F(x)$.

As for the rules of inference, CLA4 has a logical one (LC) and a nonlogical one, meaning CLA4-Induction. This only uses polynomially-bounded formulas:

$$\frac{\sqcap(F(0)) \qquad \sqcap(F(x) \rightarrow F(x\mathbf{0})) \qquad \sqcap(F(x) \rightarrow F(x\mathbf{1}))}{\sqcap(F(x))}$$

For the proofs of soundness and (extensional) completeness with respect to PTIME computability, see [56].

---

*CLA5*

As briefly mentioned at the end of [56], and further properly introduced in [64], CLA5 is a sound and complete[25] theory wih respect to polynomial

23 See [56], page 1334 for the formal definition.

24 Where $x\mathbf{0}$ means the **binary 0-successor** of $x$ - basically an identity function. Indeed, a number $a\mathbf{0}$ (e.g. $2a$) is a binary 0-successor of $a$, while $a\mathbf{1}$ (e.g. $2a + 1$) is the binary 1-successor of $a$.

25 As in the sense of extensionally complete, just like CLA4.

space computability.

Indeed, CLA5 captures the set of PSPACE[26] solvable interactive number-theoretic problems (while CLA4 focused on polynomial time).

The axiomatisation of CLA5 is obtained from that of CLA4 by deleting the scheme of sentences and replacing the rule of induction with CLA5-induction, as seen below:

$$\frac{\sqcap(F(0)) \qquad \sqcap(F(x) \to F(x'))}{\sqcap(F(x))}$$

Furthermore, the nonlogical axioms are those of PA plus an additional one:

$$\sqcap x \sqcup y(y = x')$$

---

*CLA6*

In the same paper [64], Japaridze introduces CLA6 as a provably sound and extensionally complete theory with respect to elementary recursive time computability[27].

It shares a lot with CLA5: same language, same axioms, same logical rule LC. In addition, it replaces the rule of induction with another one, namely CLA6-Induction, which is syntactically the same as CLA5.

The main difference between CLA5 and CLA6, then, lies in the requirements of such rule: while the induction rule of the former needs the formula $F(x)$ to be polynomially bounded, the induction rule of the latter has the weaker requirement that $F(x)$ should be *exponentially bounded*.

This means that the size of any variable (and, thus, of every formula) has an exponential limit set. This limit is expressible through the composition of different variables with our special operations: $0, ', +, \times$.

26 Similarly to PTIME, PSPACE is the set of all decision problems that can be solved by a deterministic Turing machine using a polynomial amount of *space*.

27 Which coincides with elementary recursive space.

As a result, CLA6 is a theory of elementary recursive (time/space) computability; indeed, it can account for recursive definitions of function letters from CL12's language.

---

*CLA7*

As for CLA7, Japaridze introduces it in the same paper. This sound and extensionally complete theory is built just as CLA5 and CLA6 are, equipped with its own induction rule.

The main difference between this and the other two theories lies, again, in the requirements of such rule. While the induction rules of CLA6 and CLA7 need the formula $F(x)$ to be polynomially or exponentially bounded, the induction rule of CLA7 imposes no restrictions on $F(x)$ at all.

As a result, CLA7 is a theory of primitive recursive (time/space) computability, since it can account for primitive recursive definitions of any function letter of CL12's language.

---

*CLA8*

In [62], Japaridze presents three other new clarithmetic theories.
First of these is CLA8, which is proven to be sound and extensionally complete with respect to PA-provably recursive time computability.

This means that an arithmetical problem $A$ has a $\tau$-time solution for some PA-provably recursive function $\tau$ iff $A$ is represented by some theorem of CLA8.

The language of CLA8 is just the same as the previous theories.
Its axiomatisation is obtained from that of CLA7 by adding the single new rule of inference, which we call **Finite Search** (FS).

For any elementary formula $F(x)$, we have:

$$\frac{\sqcap(F(x) \sqcup \neg F(x)) \qquad \sqcap\exists x F(x)}{\sqcap \bigsqcup x F(x)}$$

where $F(x)$ is an elementary formula.

Just as CLA5, the nonlogical axioms of CLA8 are those of PA plus an additional one:

$$\sqcap x \sqcup y(y = x').$$

There are no logical axioms.

Furthermore, the only logical rule of inference is LC, while the nonlogical ones are FS, as mentioned, and CLA7-Induction.

---

*CLA9*

This theory is shown[28] to be sound and intensionally complete with respect to constructively PA-provable computability.

This means that a sentence $X$ is a theorem of CLA9 iff, for some particular HPM $\mathcal{M}$, PA proves that $\mathcal{M}$ computes the problem represented by $X$.

CLA9 is obtained from CLA7[29] by replacing the FS rule with the **Infinite Search** (IS) one:

$$\frac{\sqcap(F(x) \sqcup \neg F(x))}{\sqcap(\exists x F(x) \to \sqcup x F(x))}$$

where $F(x)$ is any elementary formula.

Indeed, this rule merely "modifies" Finite Search by changing the status of $\exists x F(x)$ from being a premise of the rule to being an antecedent of the conclusion.

---

*CLA10*

Still in [62], Japaridze introduces CLA10 as a sound and intensionally complete theory with respect to not-necessarily-constructively PA-provable computability.

28 In the same paper as CLA8, namely [62].
29 Of which shares the induction rule.

This means that a sentence $X$ is a theorem of CLA10 iff PA proves that $X$ is computable, even though PA does not "know" of any particular machine $\mathcal{M}$ that computes $X$.

This theory is obtained from CLA7 by adding the above-mentioned IS rule plus the following one, which we call **Constructivisation**:

$$\frac{\exists F(x)}{\sqcup F(x)}$$

where $F(x)$ is any elementary formula containing no free variables other than $x$.

Indeed, if an $x$ satisfies $F(x)$, this can be "computed" (generated) even if we do not know what particular machine "computes" it. By all means, this is a non-constructive justification.

---

*CLA11*

As introduced in [65] and [66], this is the "newest" clarithmetic development we have formulated thus far - also known as the first example of *Tunable clarithmetic*.

CLA11 differs from the other theories since it is developed in a quite different fashion: that of parameters. Unlike its predecessors, then, it is a *scheme* of clarithmetical theories rather than a particular theory.

Indeed, we call this scheme $\mathrm{CLA11}^{P_1,P_2,P_3}_{P_4}$.

By tuning the three parameters $P_1, P_2, P_3$ in an essentially mechanical manner, one automatically obtains sound and complete theories with respect to a wide range of target *tricomplexity* classes (i.e. combinations of time (set by $P_3$), space (set by $P_2$) and so called amplitude (set by $P_1$) complexities[30]).

This theory is sound[31] in the sense that every theorem $T$ of the system

30 Indeed, $P_1$ determines the *amplitude complexity* of the class of problems captured by the theory, meaning the complexity measure concerned with the sizes of Machine's moves relative to the sizes of Environment's ones.

31 See [66] for the soundness proof.

represents an interactive number-theoretic computational problem with a solution from the given tricomplexity class; furthermore, such solution can be automatically extracted from a proof of $T$.

On the other hand, we say it is (extensionally) complete[32] in the sense that every interactive number-theoretic problem with a solution from the given tricomplexity class is represented by some theorem of the system.

If we turn the $P_4$ parameter, at the cost of sacrificing recursive axiomatisability (but not simplicity or elegance), the above extensional completeness can be strengthened to intensional completeness, according to which every formula representing a problem with a solution from the given tricomplexity class is a theorem of the system[33].

In other words, $P_1, P_2, P_3$ are sets of terms (or *pseudoterms*[34]) used as bounds for certain quantifiers in certain postulates, while $P_4$ is a set of formulas that act as supplementary axioms. All of these parameters have easy-to-satisfy regularity conditions.

Without getting too much into its (intricate) formal characterisation, let us have a quick look at the axioms and rules of inference of this scheme of theories. We will try to informally explain these core concepts as simple as possible.

First, we must set the parameters to a certain generic value.

$P_1, P_2, P_3$ will become a *boundclass triple*, meaning a triple
$R = (R_{amplitude}, R_{space}, R_{time})$ of boundclasses. These are sets of bounds[35] closed under variable renaming.
As for $P_4$, this will be tuned to a certain set $A$ of sentences.

Thus, every boundclass triple $R$ and set $A$ of sentences induces the **theory** $\mathbf{CLA11}_A^R$ that we define as follows.

---

32 See [65] for the completeness proof.

33 Basically $P_4$ contains or entails all true sentences of PA.

34 As explained for all clarithmetic theories, an $n-$ary $(n \geq 0)$, pseudoterm - **pterm** for short, is an elementary formula $p(y, x_1, \ldots, x_n)$ with all free variables as shown and one of such variables - $y$ in the present case - designated as the *value variable* of the pterm, such that PA proves $\forall x_1, \ldots, \forall x_n \exists! y\ p(y, x_1, \ldots, x_n)$.

35 Here, **bound** means a pterm $p(x_1, \ldots, x_n)$, with all free variables as shown, which satisfies the monotonicity condition.

The extra-PA axioms of $CLA11^R_A$, with $x$ and $y$ below being arbitrary two distinct variables, are:

1. $\sqcap x \sqcup y(y = x')$
2. $\sqcap x \sqcup y(y = |x|)$[36]
3. $\sqcap x \sqcap y(Bit(y,x) \sqcup \neg Bit(y,x))$[37]
4. All sentences of $A$.

In the naturally occurring cases, the set $A$ will typically be either empty or consisting of all true sentences of PA.

$CLA11^R_A$ has two nonlogical rules of inference other than classic LC: $R$-**Induction** and $R$-**Comprehension**. Before introducing them, we need to define the notion of *bounded formula*.

Let $F$ be a formula and $B$ a boundclass. We say that $F$ is $B-$**bounded** iff every $\sqcap-$subformula (respectively $\sqcup-$subformula) of $F$ has the form:

$$\sqcap|z| \leq b|s|H \text{ (respectively } \sqcup|z| \leq b|s|H)$$

where $z$ and all items of the tuple $s$ are pairwise distinct variables not bound by $\forall$ or $\exists$ in $F$, and $b|s|$ is a **bound from** $B$.
Japaridze does not explicitly account for the role of $H$ here; we may see it as a formula which is either premise or conclusion of a certain deduction.

Now, back to our rules of inference, this is how we represent them[38]:

**$R$-Induction**

$$\frac{\sqcap F(0) \qquad \sqcap(F(x) \rightarrow F(x'))}{\sqcap(x \leq b|s| \rightarrow F(x))}$$

36 Where $|x|$ refers to the length of the binary numeral for $x$.
37 Where $Bit(y,x) =_{def} (x)_y = 1$. This means that given the pterm/function $(x)_y$, which stands for the $y$th least significant bit of the binary representation of $x$ (= the $y$th bit from the right, where the bit count starts from 0 rather than 1), it is equal to 1 (thus $1 \leq y \leq |x|$ - otherwise, it would have been 0).
38 With the requirement of having to deal exclusively with sentences, meaning formulas with no free occurrences of variables.

where $x$ and all items of the tuple[39] $s$ are pairwise distinct variables, $F(x)$ is an $R_{space}$-bounded formula and $b(s)$ is a bound from $R_{time}$.

**$R$-Comprehension**

$$\frac{\sqcap(p(y) \sqcup \neg p(y))}{\sqcap(\bigsqcup |x| \leq b|s| \forall y < b|s|(Bit(y,x) \longleftrightarrow p(y)))}$$

where $x, y$ and all items of the tuple $s$ are pairwise distinct variables, $p(y)$ is an elementary formula not containing $x$ and $b(s)$ is a bound from $R_{amplitude}$.

---

This ends our brief presentation of the peculiar *Bothriochloeta* varieties we have found in our fertile steppe.

We may now move on to overview the second potential application of CoL outside its borders: Artificial Intelligence systems.

## 4.2 COL, THE AI SCYTHE

As we have carefully seen so far, CoL is a problem solving tool: it can identify what the solution to a given problem is and also provide the *actual* implementation of such solution[40].

This constructive nature provides precious materials for assembling new and valuable tools worth in many areas of applications - one of which, indeed, is Artificial Intelligence.

Let us examine the advantages of using CoL in such field.

---

*Knowledgebase systems*

A **knowledgebase system** is a computer system for knowledge management. It provides the means for collection, organisation and retrieval of knowledge.

39 As in, an ordered list of elements.
40 As the reader recalls, the *what* and the *how*. Indeed, CoL cuts away two problems in one scythe-like blow - which means more crops for us.

There is no clear distinction between knowledgebase systems and *database systems*. We may say that a knowledgebase system is a database system with some higher degree of "intelligence" than primitive database systems.

In other words, we call knowledgebase a database that usually has some AI components. *Expert systems* are the most common examples of knowledgebases heavily relying on AI.

As advanced sorts of databases, knowledgebases should allow complex logical expressions for storage and retrieval (for given queries) of knowledge; ideally, it should also be able to apply logic for automated deduction and problem-solving.

As a matter of fact, these knowledgebases are applied logic systems; consequently, we may ask ourselves on what logic should they be based upon.

Typically, classical logic is not sufficient. Some augmentations of it have been sought, for instance through the addition of epistemic constructs (such as the modal "know that" operator).
However, CoL does not need any augmentations of the sort, since it already is a logic of knowledge. Here we sketch a quick answer as to *why*.

Below are some advantages of CoL as a knowledgebase-logic over the traditional approaches.

- **Constructiveness**: a good logic for knowledgebase systems should be able to differentiate between truth and actual ability to tell/find what is true. For instance, $\forall x \exists y (y = Father(x))$ is very different from $\sqcap x \sqcup y (y = Father(x))$[41]: yet, classical logic fails to account for this thick difference;
- **Interactivity**: indeed, most knowledgebase and information systems are interactive (recall the medical diagnosis example, of which we give a free sample in the next few pages). CoL, which is designed to be a logic of interactive tasks, is a well-suited formal framework for them and an appealing alternative to more orthodox approaches;

41 By now, the reader should be aware of why.

- **Resource-consciousness**: clearly, a well-fitting logic for knowledge-base systems should be resource-conscious. Traditional approaches to these systems, however, are unable to see the important and relevant difference between $A$ and $A \wedge A$.

To provide a complete overview of CoL as a query logic, consider[42] the formula $D$:

$$\sqcap x(\sqcup s Smpt(x,s) \wedge \sqcap t(Pst(x,t) \sqcup \neg Pst(x,t)) \rightarrow \sqcup y Has(x,y)),$$

where:

- $Smpt(x,s)$ is the predicate "patient $x$ has the set $s$ of symptoms";
- $Pst(x,s)$ is the predicate "patient $x$ tests positive for test $t$";
- $Has(x,s)$ is the predicate "patient $x$ has disease $y$".

As Japaridze explains, $D$, while still overly simplified, is a more realistic problem for a real medical diagnostic system to handle than the problem $\sqcap x \sqcup y Has(x,y)$, solving which would require the ability to diagnose an arbitrary patient without any additional information on the person itself.

Here is a possible scenario of "playing" this "game":

1. At the beginning, the system (Machine) is waiting for the user (Environment) to specify a patient to be diagnosed. After the user enters the patient's name $X$, the game is brought down to:
$$\sqcup s Smpt(X,s) \wedge \sqcap t(Pst(X,t) \sqcup \neg Pst(X,t)) \rightarrow \sqcup y Has(X,y);$$
2. The system waits until the user enters also $X$'s symptoms, which we call $S$:
$$Smpt(X,S) \wedge \sqcap t(Pst(X,t) \sqcup \neg Pst(X,t)) \rightarrow \sqcup y Has(X,y);$$

42 Example provided in lecture notes number 15 (a full circle moment, if we will, since we mentioned this particular case right at the beginning of this work). Again, we may savour the potential real-life relevance of CoL right through this very simplified instance.

3. Based on all the information received, the system selects a test $T$ to perform on $X$:

$$Smpt(X,S) \wedge (Pst(X,T) \sqcup \neg Pst(X,T)) \rightarrow \bigsqcup y Has(X,y);$$

4. The user (specifically, the doctor) performs test $T$ on the patient and enters that it is positive:

$$Smpt(X,S) \wedge Pst(X,T) \rightarrow \bigsqcup y Has(X,y);$$

5. The system reports back that patient $X$ has a disease $Y$:

$$Smpt(X,S) \wedge Pst(X,T) \rightarrow Has(X,Y).$$

Machine wins either if patient $X$ indeed has disease $Y$, or if Environment has lied/erred while reporting the required information.

The single occurrence of $\bigsqcap t(Pst(X,t) \sqcup \neg Pst(X,t))$ in the antecedent of $D$ means that the system can request testing only once; if $n$ tests were potentially required, we should insert the $\wedge-$conjunction of $n$ identical conjuncts in the antecedent.

Still, if the system needed a potentially unbounded number of tests, then we should write:

$$\bigsqcap x(\bigsqcup s Smpt(x,s) \wedge ⅄ \bigsqcap t(Pst(x,t) \sqcup \neg Pst(x,t)) \rightarrow \bigsqcup y Has(x,y))$$

thus further weakening $D$[43].

Replacing the main quantifier $\bigsqcap x$ with $\forall x$ would strengthen $D$: the system would then be able to diagnose a patient purely on the basis of its symptoms and test results, without knowing who the patient really is. However, we need medical histories to avoid a potentially harmful misdiagnosis - hence, we cannot actually replace the quantifier.

On the other hand, substituting $\bigsqcap x$ with $\bigwedge x$ would be another way to strengthen $D$: the system would then be able to diagnose all patients rather than any particular one[44].

At this point, one could observe that $D$ is not a uniformly valid formula:

---

43 As it requires stronger user-provided information - in other words, $D$ becomes more dependent from external resources.

44 Clearly, effects of at least the same strength would be achieved by just prefixing $D$ with $⅄$ or $⫰$.

should not a logic-based system be able to solve only uniformly valid problems?

Yes. Indeed, the logical problem that our imaginary system has solved is not really $D$. Rather, it is $KB \to D$, where $KB$ is all the additional extralogical knowledge plus resources that the system possesses.

It is this very $KB$ we call *knowledgebase* of the knowledgebase system.

Formally, such a $KB$ is a finite (multi)set of sentences. We identify it with the $\wedge-$conjunction of its elements, so that $KB$ can also be thought of a single, probably very long formula.

Knowledgebase is not to be confused with knowledgebase systems. The latter is just a logic-base problem-solving software of universal utility that initially comes to the user without any extralogical knowledge whatsoever.

Indeed, built-in extralogical knowledge would make it no longer universally applicable: one thing can be true in the world of a potential user and false in the world of another one.

Thus, it is the user who selects and maintains a certain $KB$ for the system, filling it with all the informational resources that are considered relevant, correct and maintainable by the user itself.

CoL's formalism can be considered as a highly declarative programming language; as a result, the process of creating a given $KB$ would simply mean to program inside it.

Consequently, a deductive system for CoL should be thought of as a compiler, or interpreter, for such programming language.

A strong asset of this language would be the full absence of the **program-verification problem**[45]; more importantly, it would remove the problem-solving burden from the human programmer - which, in turn, would only have to *state* problems, rather than *solve* them.

45 Meaning the problem of ensuring that specific programs or sub-programs meet their design requirements - in other words, checking for their correct behaviour in real-time applications.

A *KB* can contain anything; its elements we call **resources**.

To each resource $R \in KB$ we associate, at least conceptually, its *provider*, meaning an agent that solves query $R$ *for* the system - as in playing the game $R$ *against* the system itself.

Physically, the provider could be either a computer program allocated to the system; a network server having the system as a client; another knowledgebase system; and so on.

We should not think of providers as part of the system itself: the latter only sees *what* resources are available to it, without knowing or caring about *how* the corresponding providers come up with their solutions[46].

Indeed, the system's only responsibility is to correctly solve queries for the user - as long as none of the providers fail to accomplish their tasks.

Of course, when $R$ is elementary, the provider has nothing to do: its successfully playing $R$ against the system simply means that $R$ is true.

We may further observe that the system plays each game $R \in KB$ in the role of $\bot$; hence, the game that the system is playing against $R$'s provider, according to its point of view, is $\neg R$ and not $R$.

Assume $KB = R_1 \wedge \ldots \wedge R_n$ and let us now try to visualise a system solving a certain problem $F$ for the user.

The designer would probably select an interface where the user only sees the moves made by the system in $F$, and hence gets the illusion that the system is just playing $F$. However, as we have already mentioned, the game that the system is actually playing is $KB \rightarrow F$, i.e. $\neg R_1 \vee \ldots \vee \neg R_n \vee F$.

Indeed, the system is not only interacting with the user in $F$, but, in parallel, also with its resource providers (against whom, as we already know, plays $\neg R_1, \ldots, \neg R_n$).

As long as those providers do not fail, the system loses each of the games

46 Indeed, the system does not really take into consideration the supposed righteousness of its providers' practice.

$\neg R_1, \ldots, \neg R_n$.
This means that the system wins its play over the "big game" $\neg R_1 \vee \ldots \vee \neg R_n \vee F$ iff it wins it in the $F$ component, i.e. successfully solves $F$.

Thus, the system's ability to solve a problem/query $F$ is reducible to its ability of generating a solution for $KB \to F$.
What would give the system such an ability is built-in knowledge of CoL - in particular, a uniform-constructively sound axiomatisation of it.

According to the uniform-constructive soundness property, it would be sufficient for the system to find a proof of $KB \to F$; this would then allow the system to (effectively) construct a machine $\mathcal{M}$ and run it on $KB \to F$ with guaranteed success.

As a matter of fact, simple soundness (as in extralogical, multiform soundness) is not enough for knowledgebase systems to function[47].
We may identify two reasons behind this:

1. Extralogical validity of a formula $E$ only implies that, for every admissible interpretation $^*$, a solution for the problem $E^*$ exists. However, it may be the case that different interpretations require different solutions: thus, choosing the right solution for $E^*$ requires knowledge of the actual interpretation (i.e. the meaning) of its atoms. Our assumption is that the system has no extralogical knowledge: this means that it has no knowledge of the interpretation $^*$. Indeed, a solution generated by the system for $E^*$ should be successful under any possible admissible interpretation $^*$. In other words, it should be a logical solution for $E$;

2. With simple soundness, it may be the case that the system *knows* that a solution for $E^*$ exists: however, it is not able to actually *find* such solution. On the contrary, uniform constructive soundness guarantees that a (uniform) solution for every provable formula not only exists, but can also be effectively extracted from a proof.

As for completeness of the built-in logic, it is a desirable but not necessary condition - unlike uniform constructive soundness.

So far, a complete axiomatisation of this type has only been found for certain fragments of CoL, such as the fragment limited to the language

47 Even if every provable formula was known to be logically valid.

of CL4. We can still certainly succeed in finding ever stronger axiomatisations that are uniform constructively sound even if not necessarily complete[48].

However, when it comes to practical applications in the proper sense, the logic employed is likely far from complete.

For instance, the popular classical-logic-based systems/programming languages are incomplete: not because a complete axiomatisation for classical logic is not known, but because, (un)fortunately, efficiency often comes at the expense of completeness.

Nevertheless, even CL4, despite the absence of recurrence operators in it, is already very powerful, as Japaridze observes.

We are going to taste this query-solving logic in the next subsection.

---

*Systems for planning and action*

We can further extend the notion of knowledgebase systems to that of systems for **planning and action**.

Roughly, the formal semantics in such applications remains the same; the only thing that changes is the underlying philosophical assumption, according to which the truth values of predicates and propositions are fixed or predetermined.

In CoL-based planning systems, these values are considered something that interacting agents can actually manage; indeed, predicates or propositions here stand for **tasks** rather than **facts**.

As first mentioned in [38], 2001[49], Japaridze believes that his new "logic of tasks"[50] could be a possible alternative to the existing logics of planning and action used in AI.

Indeed, this embryonic version of CoL manages to avoid two major prob-

48 As Japaridze explains, one of such axiomatisations is affine logic, while another one is intuitionistic logic.
49 When Computability Logic was still in a germinal state.
50 Which we have already mentioned in the first pages of the present work.

lems most planning logics face: the **Frame problem**[51] and the **Knowledge preconditions** one.

Regarding the former, we tend to assume that our actions usually change only a small part of the world and leave the rest unaffected.

Nevertheless, when talking about the effects of a certain action, we need not only to take into account what is directly changed, but also what our action does *not* affect: otherwise, we may miss something and rush to the wrong conclusion.
This, however, may result in dramatic increase in the representational and inferential complexity of planning problems.

Concerning the second problem, on the other hand, most AI planners work on the assumption that they have complete knowledge of the world, which, in actual planning situations, is rarely the case.

Moreover, certain pieces of information can be acquired only from how the environment reacts to a certain action we have planned; as a result, our next action strategy may depend on those same reactions which are now part of our knowledge.

The original formalism of the most popular planning logic — **situation calculus** — fails to capture these nuances.

Some researchers[52] have approached the problem by extending the language of situation calculus with special means of knowledge representation.

This, however, significantly overburdens the calculus and messes the formalism up a bit; furthermore, the corresponding logics typically lose the property of semidecidability.

McCarthy and Hayes [84] were the first ones to recognise this knowledge preconditions problem. Ultimately, this one boils down to the following question[53]: how can an agent reason that (s)he knows how to perform an action? Indeed, a good planning-logic should explain under what

51 For more information on the matter, the reader is advised to have a look at [91].
52 See [30], [36] and [86] for instance.
53 As explained in the opening lines of [87].

circumstances such agent is aware of being able to perform said action.

When our logic is used for planning, a planning situation is represented by a certain task $\gamma$, which usually has the form of $\alpha \to \beta$.
Here, $\beta$ is a description of the goal for the planning agent (initially named *slave* in [38]), while $\alpha$ is a description of the resources the agent possesses.

The agent's knowledge is about (the state of) task $\gamma$, so there is no need in having a special language or some semantical constructs representing knowledge. Indeed, we do not really need to distinguish between physical resources and informational ones.

Actions of the planning agent are represented by replacing, for any $i \in \{1,2\}$, $\alpha_1 \sqcap \alpha_2$ (or $\sqcap x\alpha(x)$) with $\alpha_i$ (or $\alpha(a)$) in the negative parts of $\gamma$.

These replacements only affect the subformulas (resources) in which they are made, which automatically neutralises the frame problem.

Furthermore, reactions of the environment are represented by replacing, for any $i \in \{1,2\}$, $\alpha_1 \sqcap \alpha_2$ (or $\sqcap x\alpha(x)$) with $\alpha_i$ (or $\alpha(a)$) in the positive parts of $\gamma$.
Since the agent has full knowledge of the state of the task, these reactions are visible to him: as a result, the knowledge update problem is naturally taken care of.

Indeed, per usual, finding a uniform solution to $\alpha \to \beta$ means finding a proof of it in the underlying uniform-constructive axiomatisation of the logic.

Then, as we already know, an actual solution for the problem (meaning a winning strategy) can be automatically obtained from the proof.

Let us consider a clear example of planning problem which we solve using CL4[54].

---

54 It goes without saying that more serious applications require a more expressiveness power than that of CL4 - probably even more than that of CoL. Adding sequential operators to such language would greatly improve the applicability of CoL in systems for planning and action.

Just the other day, as the reader may recall, our small airplane[55], used for the initial land inspection we completed in the first chapter, needed its oil to be changed, in order to keep the piston engine well-greased and cooled down in its innermost parts[56].

Hence, let us go back and examine this task from a *planning and action* point of view inside Computability Logic's framework[57].

Indeed, there are several engine oils we could use - the important thing is that the chosen one satisfies the specifics provided by the maker of the aircraft[58].

Thus, the domain of our universe of discourse is the set of all engine oils; $a$ and $b$ are two constants denoting two elements of the domain.

Here is our task:

**Goal** $(G)$: Fill the airplane radiator with a safe sort of engine oil.

*Safe*, here, means an oil that, other than checking all the boxes required by the manual, it does not contain any acids *before*[59] even being put into the engine.

Thus, we may formalise $G$ as such:

$$\exists x(Safe(x) \wedge Fill(x))$$

Let us assume that these are our informational and physical resources for achieving the goal:

---

55 Namely, a cream coloured Vulcanair P-68C with muted red accents. Also known as Partenavia P.68 from its Italian manufacturer, it is one of the standard light aircrafts available for observational uses; it was first introduced in 1971.

56 Usually this process should be carried out every 50 hours of engine work. The aircraft operating manual contains information on the capacity of the oil cup, specifying the minimal quantities of oil needed to satisfy lubrication and cooling. For instance, in engines that carry up to 8kg of oil, the minimum quantity to guarantee a good lubrication is 2kg, while the minimal quantity for a safe cooling is usually around 6kg.

57 The reader will observe how the frame problem and the knowledge preconditions one remain out of the scene in CoL-based planning systems.

58 For instance, the chosen oil must have specific density, viscosity and fluidity values.

59 Indeed, during the operation of most engines, acids are released as by-products and can enter the oil. However, these should be neutralised by the alkalis contained in the additives of oils.

**Resource/knowledgebase**

What we *know*:

1. Engine oil is safe iff it does not contain acid:
$$\forall x(Safe(x) \longleftrightarrow \neg Acid(x));$$
2. At least one of the engine oils $a, b$ is safe:
$$Safe(a) \vee Safe(b).$$

What we *have* or *can do*:

3. A piece of litmus paper:
$$\sqcap x(Acid(x) \sqcup \neg Acid(x));$$
4. Fill the airplane radiator with any one engine oil:
$$\sqcap x(Fill(x)).$$

The planning problem can be successfully solved without any additional informational or physical resources, since the formula:

$$1. \wedge\ 2. \wedge\ 3. \wedge\ 4. \rightarrow G$$

is logically valid.

Here is a strategy, meaning a solution to such problem.

1. Use the litmus paper to find out whether the engine oil $a$ contains acid (move $\top 0.2.a$):
$$\forall x(Safe(x) \longleftrightarrow \neg Acid(x)) \wedge (Safe(a) \vee Safe(b)) \wedge \sqcap x(Acid(x) \sqcup \neg Acid(x)) \wedge \sqcap x(Fill(x)) \rightarrow \exists x(Safe(x) \wedge Safe(x) \wedge Fill(x));$$
2. Observe the result and, depending on it (move $\perp 0.2.0$ or $\perp 0.2.1$), go to step 3. or 5. :
$$\forall x(Safe(x) \longleftrightarrow \neg Acid(x)) \wedge (Safe(a) \vee Safe(b)) \wedge (Acid(a) \sqcup \neg Acid(a)) \wedge \sqcap x(Fill(x)) \rightarrow \exists x(Safe(x) \wedge Safe(x) \wedge Fill(x));$$
3. Fill the radiator with $b$ (move $\top 0.3.b$):
$$\forall x(Safe(x) \longleftrightarrow \neg Acid(x)) \wedge (Safe(a) \vee Safe(b)) \wedge Acid(a) \wedge \sqcap x(Fill(x)) \rightarrow \exists x(Safe(x) \wedge Safe(x) \wedge Fill(x));$$

4. Wash your hands:

$$\forall x(Safe(x) \longleftrightarrow \neg Acid(x)) \wedge (Safe(a) \vee Safe(b)) \wedge Acid(a) \wedge Fill(b) \rightarrow \exists x(Safe(x) \wedge Safe(x) \wedge Fill(x));$$

5. Fill the radiator with $a$ (move $\top 0.3.a$):

$$\forall x(Safe(x) \longleftrightarrow \neg Acid(x)) \wedge (Safe(a) \vee Safe(b)) \wedge \neg Acid(a) \wedge \sqcap x(Fill(x)) \rightarrow \exists x(Safe(x) \wedge Safe(x) \wedge Fill(x));$$

6. Wash your hands:

$$\forall x(Safe(x) \longleftrightarrow \neg Acid(x)) \wedge (Safe(a) \vee Safe(b)) \wedge \neg Acid(a) \wedge Fill(a) \rightarrow \exists x(Safe(x) \wedge Safe(x) \wedge Fill(x)).$$

As you can see, the notorious problems we have discussed at the start of this section fail to appear.

Alternatively, avoiding *ad hoc* attempts to solve the problem, one can find a radiator-filling strategy through the CL4-proof of:

$$\forall x(Safe(x) \longleftrightarrow \neg Acid(x)) \wedge (Safe(a) \vee Safe(b)) \wedge \sqcap x(Acid(x) \sqcup \neg Acid(x)) \wedge \sqcap x(Fill(x)) \rightarrow \exists x(Safe(x) \wedge Safe(x) \wedge Fill(x))$$

as provided below.

This is guaranteed by the uniform-constructive soundness of CL4.

**Step 1.**

$\forall x(Safe(x) \longleftrightarrow \neg Acid(x)) \wedge (Safe(a) \vee Safe(b)) \wedge \neg Acid(a) \wedge Fill(a)$
$\rightarrow \exists x(Safe(x) \wedge Fill(x))$

from $\varnothing$ by Rule 1;

**Step 2.**

$\forall x(Safe(x) \longleftrightarrow \neg Acid(x)) \wedge (Safe(a) \vee Safe(b)) \wedge \neg Acid(a)$
$\wedge \sqcap x(Fill(x)) \rightarrow \exists x(Safe(x) \wedge Fill(x))$

from 1. by Rule 3;

**Step 3.**

$\forall x(Safe(x) \longleftrightarrow \neg Acid(x)) \wedge (Safe(a) \vee Safe(b)) \wedge Acid(a) \wedge Fill(b)$
$\rightarrow \exists x(Safe(x) \wedge Fill(x))$

from ∅ by Rule 1;

**Step 4.**

$\forall x(Safe(x) \longleftrightarrow \neg Acid(x)) \wedge (Safe(a) \vee Safe(b)) \wedge Acid(a) \wedge \sqcap x(Fill(b)) \rightarrow \exists x(Safe(x) \wedge Fill(x))$

from 3. by Rule 3;

**Step 5.**

$\forall x(Safe(x) \longleftrightarrow \neg Acid(x)) \wedge (Safe(a) \vee Safe(b)) \wedge (Acid(a) \sqcup \neg Acid(x)) \wedge \sqcap x(Fill(x)) \rightarrow \exists x(Safe(x) \wedge Fill(x))$

from 2. and 4. by Rule 1;

**Step 6.**

$\forall x(Safe(x) \longleftrightarrow \neg Acid(x)) \wedge (Safe(a) \vee Safe(b)) \wedge \sqcap x(Acid(a) \sqcup \neg Acid(x)) \wedge \sqcap x(Fill(x)) \rightarrow \exists x(Safe(x) \wedge Fill(x))$

from 5. by Rule 3.

Indeed, this is how we are going to fly back right where we started our exploratory journey from - which, alas, is now coming to an end.

# 5 CONCLUSION

Thanks to the reader's help in changing the oil of our P-68C, we are ready for our flight back.

Surely, we are not going home empty-handed; however, many more concealed angles of this fertile steppe still need to be investigated[1].

While taking off for one last time, let us review what have we managed to accomplish so far.

---

These tumbling valleys, over which the reader can roll its eyes from the right-side window, were once unknown to us.

Indeed, we started our journey eager to map[2], track, touch and chart away this remote land.

The initial un-knowledge, which paired well with our cautious approach, soon became fertile *otherness* for these hills and mountains to clash onto.

We remember preparing for our aerial-perusal survey and asking ourselves *why* Computability Logic was first developed - even though this soon became a *what* and *how* question, if the reader recalls.

From there, we warmed up according to the suggested FDP, while glancing at the neighbouring grounds and focusing on the crucial notion of interaction.

Sweeping over these wide carpets made us realise the extent of what was

1 Indeed, as Arsen'ev would say in our Kurosawa, *man's measure is dwarfed by the vastness of nature*.
2 We are fully aware that "a map is not a territory", as Alfred Korzybski simply put it in [78]. In fact, we are not trying to act *als ob*: our sketches are simply sketches. The main goal was to *physically* acquaint ourselves with, not to *scan*, this unexplored vale.

waiting for us[3]. From there, we started sketching the things we saw from the high skies, preparing a preliminary map for us to use on the ground.

Flying through the steppe, we gradually made our way to Semantics. Stretching out a parabolic trajectory from informal to formal notions, we hosted in our laps the rudiments of local dialects[4], soon gliding in Syntax-core, where the "zoo operator" was ready to be spotted by us.

Landing on this fertile soil, rightly equipped for the journey ahead, we started the second chapter of our expedition *uphill*. The reader surely remembers the great hike on the mountain range, yes, that one over there, with the stratus hovering on the distant peaks.

We touched and felt those 17 axiomatisations of CoL thanks to the great tool Japaridze provided us with: Cirquent Calculus. We learned how to properly use it to befittingly inspect all of the fragments, which we *named* with the power of our own legs and fingers.

The third chapter smelt dark and dingy. We prepared a further hike on that range, specifically on CL15, in order to assess the presence of a cavernous property we initially brushed off: that of decidability.

After thoroughly retracing Japaridze's steps concerning the soundness and completeness proofs of this fragment, we reasoned out a *potential* proof, a rough sketch, a rich attempt to show that yes, indeed, the cave exists: CL15 is decidable.

We came back down tired and weary. Resting in the *silent* embrace of such vast grassland, we recalled *why* we were there in the first place: *why* we were sent to get to know this remote patch.

The fourth chapter, then, saw us taking out our notes about the springing clarithmetics community, temporarily capped with CLA11. We also provided a brief account of the potential benefits Computability Logic is

3 If we were to clash together Max Weber [97] and Norbert Elias [17], we would (also) obtain the notion of *foresight*, as in the *distance* between *us* and the *things* around us. Indeed, mentioning Poe's short story *The fishermen in the maelstrom*, according to Elias, we have to *detach* ourselves in order to acquire "reality-adequate" knowledge. In our case, we may see this first detachment from the ground as a preliminary focusing of sorts, which enabled us to better understand what was expecting us.

4 Which helped us erect our *Tower of Behirut*, as the reader may recall.

able to provide to AI systems - both knowledgebase and planning/action ones.

And here we are[5].

There is only one conclusion that we could possibly crochet at the end of this long journey: the steppe is, indeed, ნაყოფიერი - as in *fertile*, productive.

CoL is expressive and constructive. It is a problem-solving tool soaked in the classical and intuitionistic waters of logic; it manages to bring peace in the neighbourhood, while enhancing the traditional resource (linear) semantics to account for subresources-sharing. In addition, it can further expand both traditional arithmetic approaches and Artificial Intelligence - areas in which there is *a lot* of research still to be carried out![6]

Indeed, we feel that this is not the end of a journey, but a first chapter to a longer one. There are many open problems we need to face; numerous properties we need to unveil; and original new applications we have to dig up.

In Japaridze's own words[7]:

> [. . . ] despite having been evolving for 15 years already, CoL, due to its ambitiousness, still remains at an early stage of development, with more open questions than answered ones. A researcher who decides to join the project will find enough interesting material to be occupied with for many years to come. Students are especially encouraged to **try**.

5 We would also like to underline something that we feel important for the reader's full *grasp* of this work. As it may already be quite evident, we believe that *space* plays a crucial role in concept-grasping. Indeed, notions are naturally easier to touch when they are *spatialised*: the three(or whatever)-dimensional asset works wonders when trying to augment our comprehension even more. Any spacial-theory certainly works more in our heads than in reality; here we are stating that it actually *enhances* understanding. Everything is space; intertwined, braided, physical, rough, organic extension. We hope you have enjoyed being part of this space for the time we had - and that maybe we will come back here together, someday.

6 As true explorers like Dersu, we shall not let these slide away un-perused.

7 Taken from [69], page 118. Bold effect added here.